%% file: template.tex
\documentclass[journal]{vgtc}                % final (journal style)

\ifpdf%                                % if we use pdflatex
  \pdfoutput=1\relax                   % create PDFs from pdfLaTeX
  \usepackage{graphicx}                % allow us to embed graphics files
  \DeclareGraphicsExtensions{.pdf,.png,.jpg,.jpeg} % for pdflatex we expect .pdf, .png, or .jpg files
\else%                                 % else we use pure latex
  \usepackage{graphicx}                % allow us to embed graphics files
  \DeclareGraphicsExtensions{.eps}     % for pure latex we expect eps files
\fi%https://www.overleaf.com/project/63288225c9328c4d29d12cc2

\graphicspath{{figures/}{pictures/}{images/}{./}} % where to search for the images

\usepackage[dvipsnames,table, svgnames]{xcolor}
\usepackage{tabu}                      % only used for the table example
\usepackage{booktabs}                  % only used for the table example
\usepackage[normalem]{ulem}
\usepackage{textgreek} % needed for bartlett1954note bib reference
\usepackage{eurosym}
\usepackage{ccicons}
\usepackage{xspace}
\usepackage{cuted}
\usepackage{flushend}
\usepackage{multirow}

\useunder{\uline}{\ul}{}
\newcommand{\eg}{e.\,g.}
\newcommand{\ie}{i.\,e.}

\definecolor{antiquefuchsia}{rgb}{0.57, 0.36, 0.51}
\newcommand{\hty}[1]{\textcolor{RoyalBlue}{#1}} %changes according to reviewers' questions.
\newcommand{\harozissue}[1]{\textcolor{red}{#1}}
\newcommand{\todo}[1]

\renewcommand{\hty}[1]{#1}
\renewcommand{\harozissue}[1]{#1}

\newlength{\picturewidth}
\newlength{\pictureheight}

\onlineid{1229}

\title{\textls[-3]{Design Characterization for Black-and-White Textures in Visualization}}
\AtEndPreamble{\hypersetup{pdftitle={Design Characterization for Black-and-White Textures in Visualization}}}

\author{%
  \authororcid{Tingying He}{0000-0002-9670-5587},
  \authororcid{Yuanyang Zhong}{0000-0002-0548-7423},
  \authororcid{Petra Isenberg}{0000-0002-2948-6417},
  \authororcid{Tobias Isenberg}{0000-0001-7953-8644}
}

\authorfooter{
\item Tingying He (\raisebox{-.5pt}{\includegraphics[height=6pt]{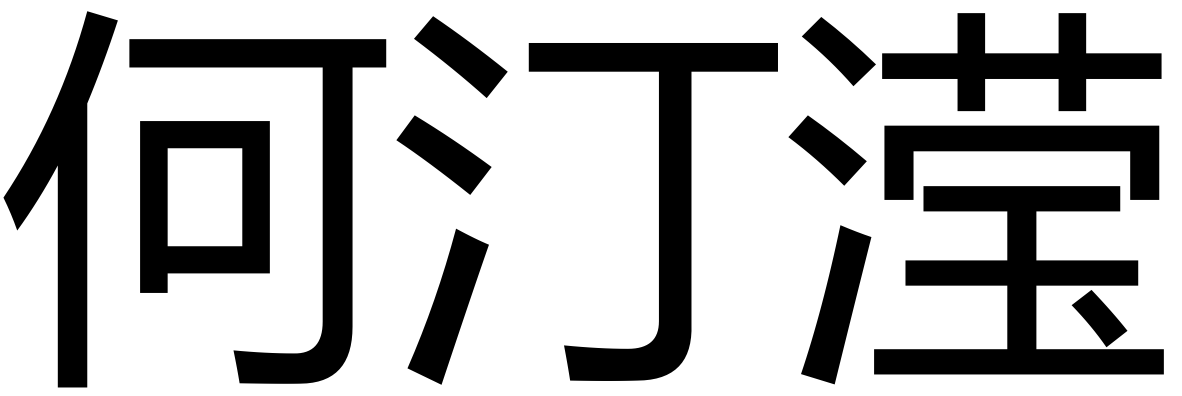}}), Petra Isenberg, and Tobias Isenberg are with Université Paris-Saclay, CNRS, Inria, LISN, France. E-mail: \{tingying.he\,$|$\,petra.isenberg\,$|$\,tobias.isenberg\}@inria.fr.
\item Yuanyang Zhong (\raisebox{-.5pt}{\includegraphics[height=6pt]{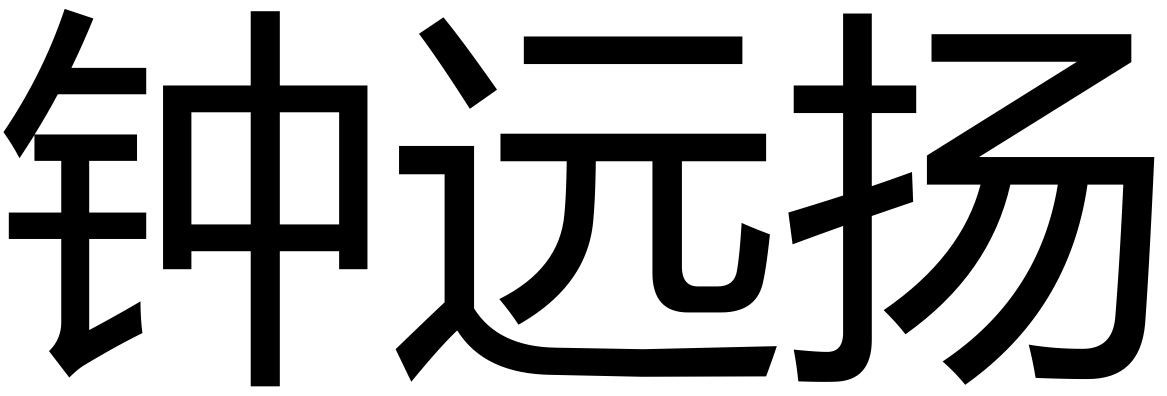}}) is with Tencent Technology (Shenzhen) Company Limited, China. E-mail: zoniaczhong@tencent.com.
}

\shortauthortitle{He \MakeLowercase{\textit{et al.}}: Black-and-white textures for visualization}
\abstract{%
We investigate the use of 2D black-and-white textures for the visualization of categorical data and contribute a summary of texture attributes, and the results of three experiments that elicited design strategies as well as aesthetic and effectiveness measures. 
Black-and-white textures are useful, for instance, as a visual channel for categorical data on low-color displays, in 2D/3D print, to achieve the aesthetic of historic visualizations, or to retain the color hue channel for other visual mappings. We specifically study how to use what we call geometric and iconic textures. Geometric textures use patterns of repeated abstract geometric shapes, while iconic textures use repeated icons that may stand for data categories. We parameterized both types of textures and developed a tool for designers to create textures on simple charts by adjusting texture parameters. \hty{30 visualization experts used our tool and designed 66 textured bar charts, pie charts, and maps.} We then had 150 participants rate these designs for aesthetics. Finally, with the top-rated geometric and iconic textures, our perceptual assessment experiment with 150 participants revealed that 
textured charts perform about equally well as non-textured charts, and that there are some differences depending on the type of chart.
} % end of abstract

\keywords{Aesthetics, textures, icons, black and white, visualization, visual representations, categorical data, design, perception.}

\CCScatlist{ % not used in journal version
 \CCScat{K.6.1}{Management of Computing and Information Systems}%
{Project and People Management}{Life Cycle};
 \CCScat{K.7.m}{The Computing Profession}{Miscellaneous}{Ethics}
}

\teaser{
\setlength{\pictureheight}{32mm}
\centering
		\includegraphics[height=\pictureheight]{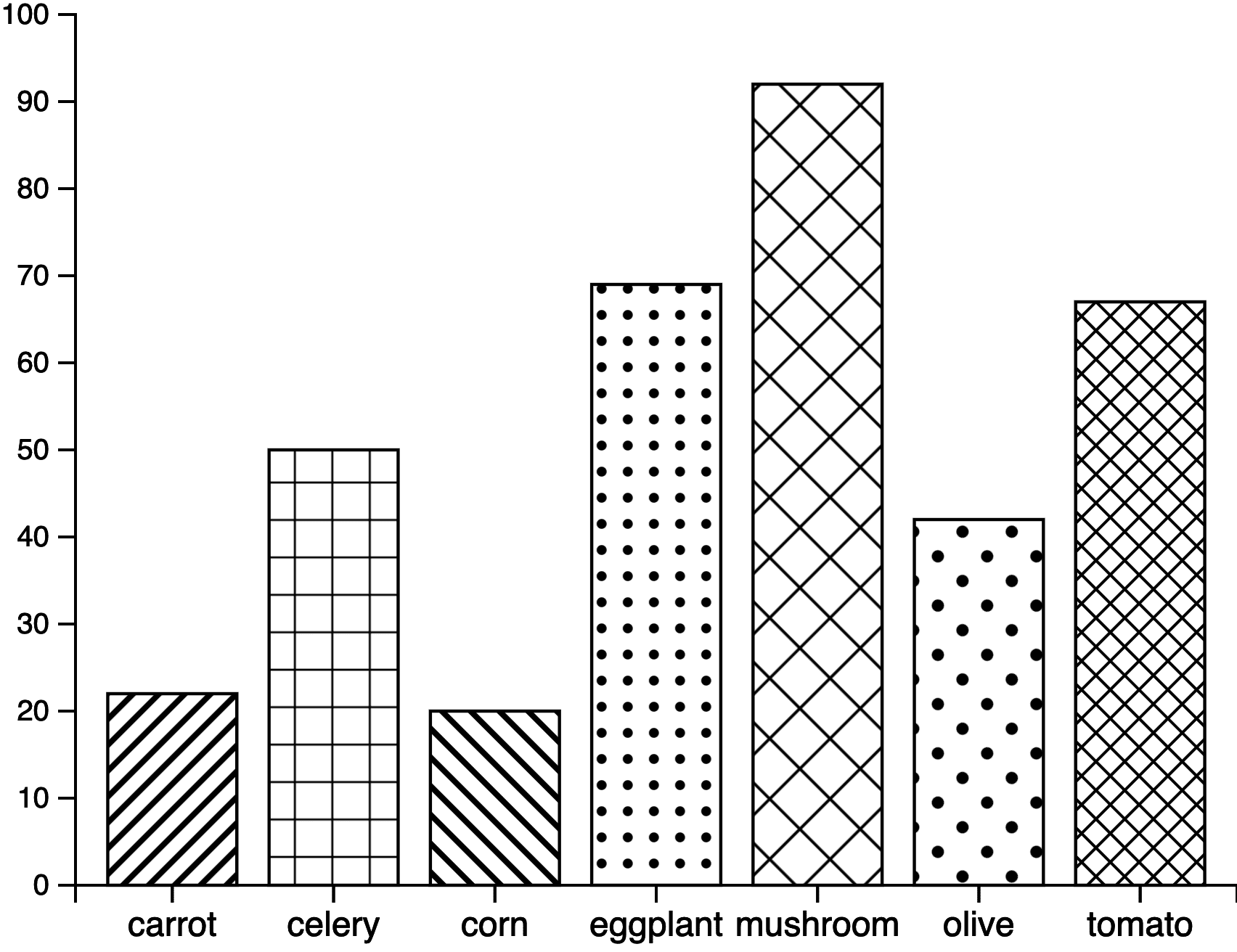}\hspace{3mm}%
            \includegraphics[height=\pictureheight]{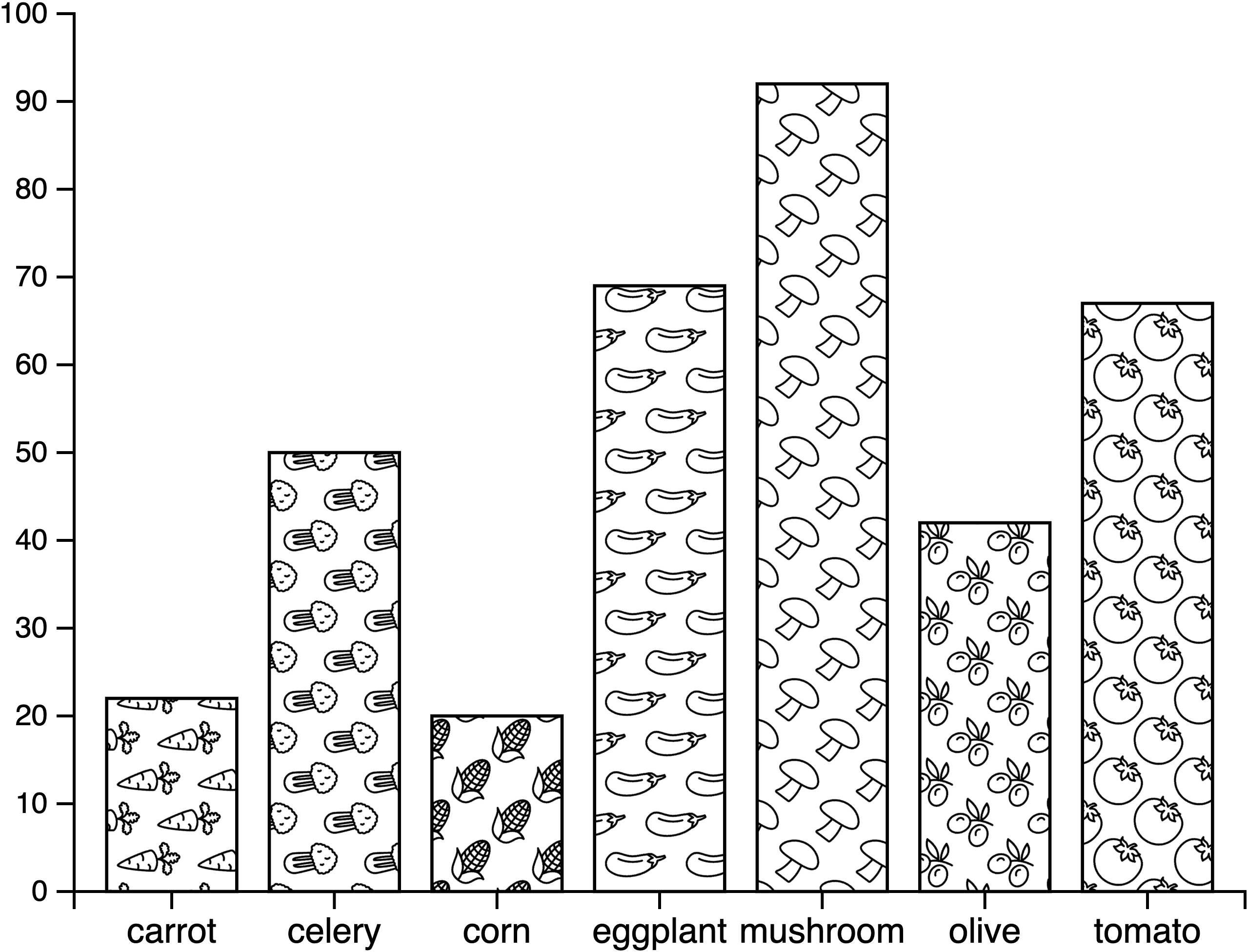}\hspace{3mm}%
		\includegraphics[height=\pictureheight]{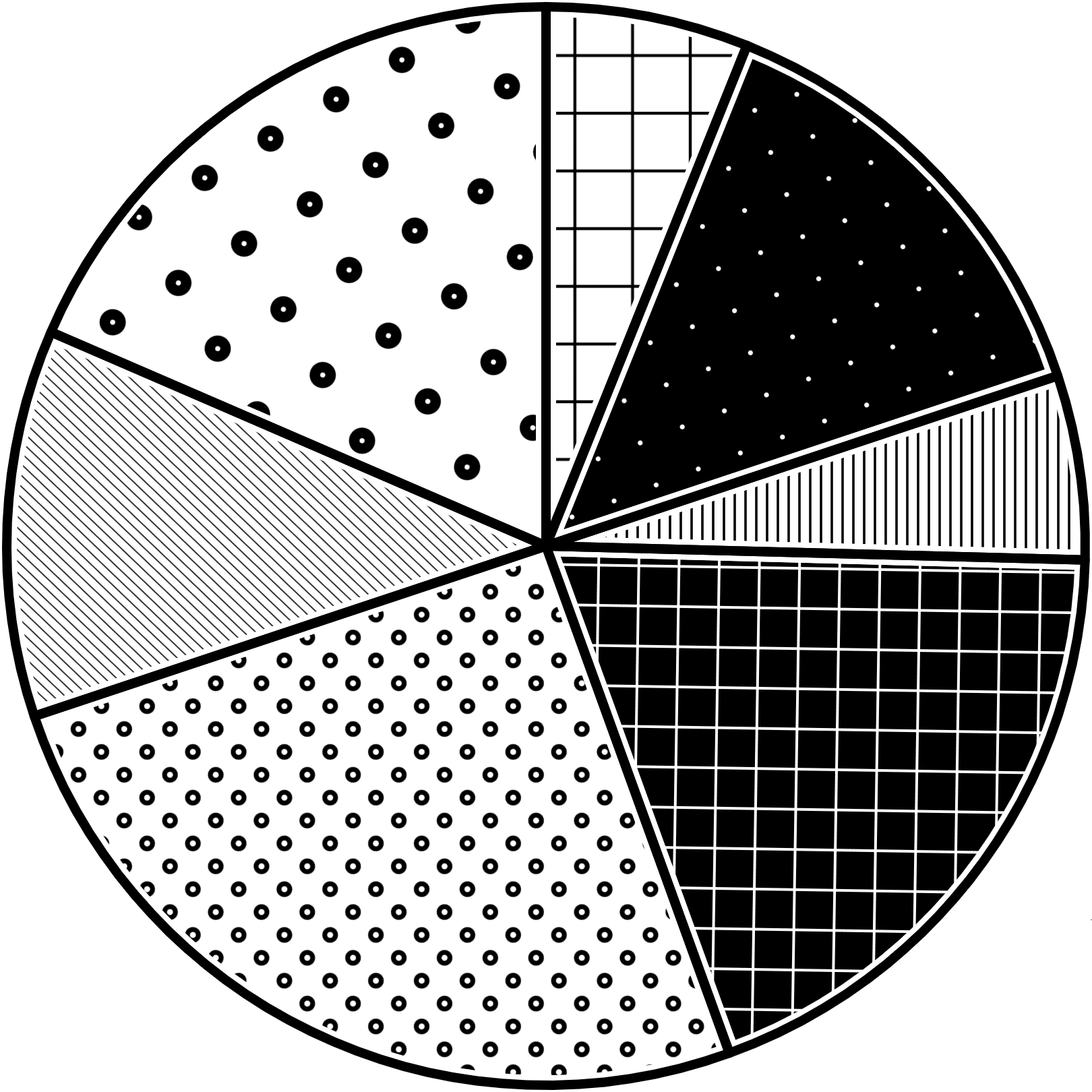}\hspace{3mm}%
		\includegraphics[height=\pictureheight]{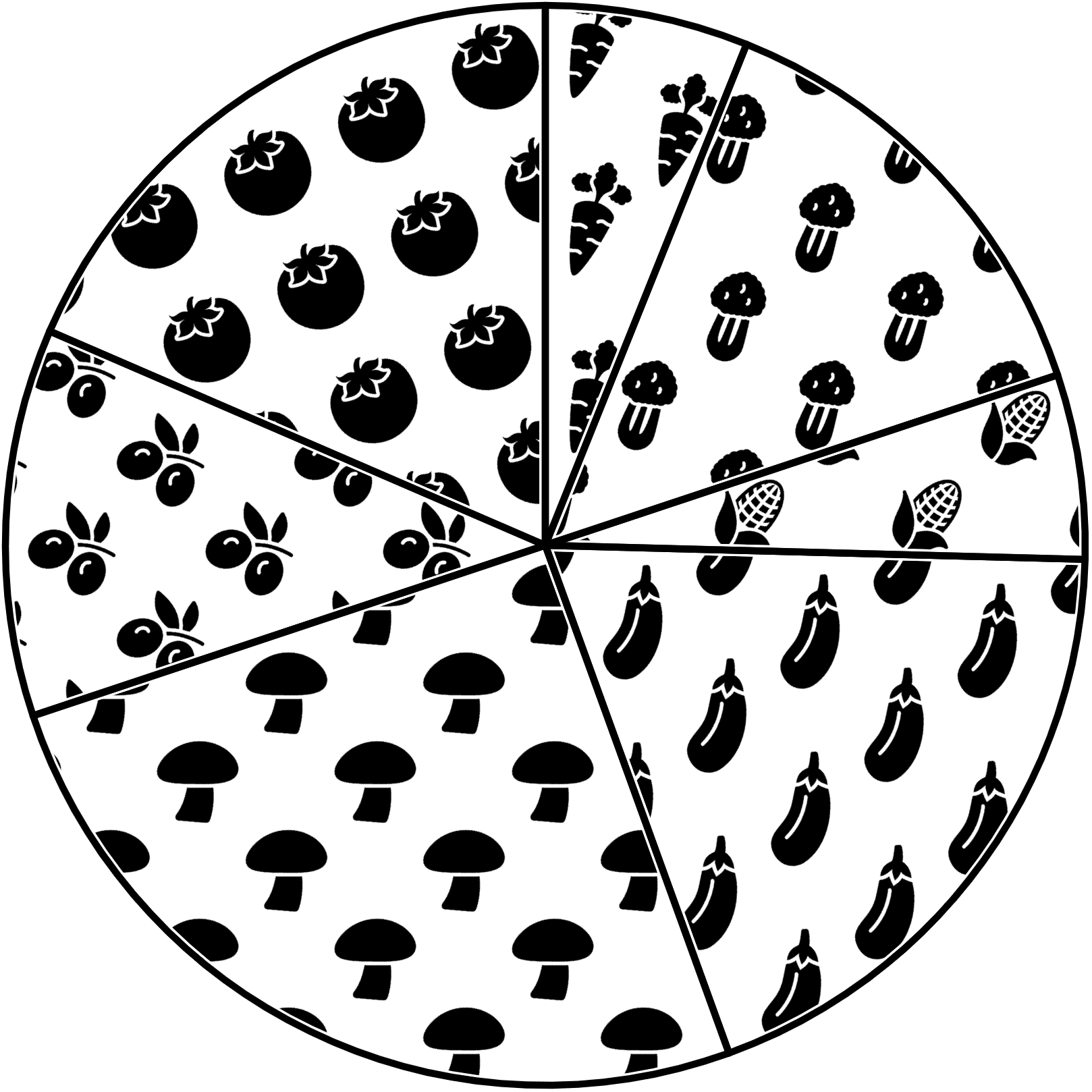}%\hspace{4mm}%
    \caption{The bar chart, pie chart designs with geometric and iconic textures with the highest ratings in Experiment 2.}
    \label{fig:teaser}
  \label{fig:subfigures}
}

\newcommand{\inlinevis}[3]{\raisebox{#1}[0pt][0pt]{\includegraphics[height=#2]{#3}}}

\newlength{\plotheight}
\newlength{\figraisecaptionoffset}
\newlength{\figraisecaptionoffsetwithextra}
\newlength{\figaftercaptionoffset}
\newlength{\figaftercaptionextraoffset}

\begin{document}


% \firstsection{Introduction}

\maketitle


\input{sections/01_introduction}

\input{sections/02_related_work}
\input{sections/design_characterization}

\input{sections/experiment1_design}

\input{sections/experiment2_rate}

\input{sections/experiment3_perception}

\input{sections/discussion}

%\clearpage

\acknowledgments{We thank all visualization experts who participated in Experiment 1, as their invaluable insights and expertise served as a crucial cornerstone of our work. We also thank all participants of the other two experiments. \harozissue{We thank Steve Haroz for his suggestions on the data analysis of Experiment 3.}}

\section*{Supplemental Material Pointers}

The pre-registrations for our three experiments can be found at \href{https://osf.io/r4z2p/}{\texttt{osf.io/r4z2p}}, \href{https://osf.io/nyru7/}{\texttt{osf.io/nyru7}}, and \href{https://osf.io/8cy62/}{\texttt{osf.io/8cy62}}, respectively. We also share our study results, analysis scripts, and additional material (appendix, video) at \href{https://osf.io/n5zut/}{\texttt{osf.io/n5zut}}. \hty{All source code can be found at \href{https://github.com/tingying-he/design-characterization-for-black-and-white-textures-in-visualization}{\texttt{github\discretionary{}{.}{.}com\discretionary{/}{}{/}tingying\discretionary{}{-}{-}he\discretionary{/}{}{/}design\discretionary{}{-}{-}characterization\discretionary{}{-}{-}for\discretionary{}{-}{-}black\discretionary{}{-}{-}and\discretionary{}{-}{-}white\discretionary{}{-}{-}textures\discretionary{}{-}{-}in\discretionary{}{-}{-}visualization}}.}

\section*{Images/graphs/plots/tables/data license/copyright}
With the exception of those images from external authors whose licenses/copyrights we have specified in the respective figure captions, we as authors state that all of our own figures, graphs, plots, and data tables in this article (\ie, those not marked) are and remain under our own personal copyright, with the permission to be used here. We also make them available under the \href{https://creativecommons.org/licenses/by/4.0/}{Creative Commons At\-tri\-bu\-tion 4.0 International (\ccLogo\,\ccAttribution\ \mbox{CC BY 4.0})} license and share them at \href{https://osf.io/n5zut/}{\texttt{osf.io/n5zut}}.

\bibliographystyle{abbrv-doi-hyperref-narrow}
\bibliography{abbreviations,template}

%\end{document}

\input{sections/appendix}

\end{document}

%% file: sections/01_introduction.tex
\section{Introduction}
\label{sec:introduction}

Texture is a powerful visual channel with broad application potential for nominal data. Texture is selective, associative, and it has a theoretically infinite number of instantiations \cite{Bertin:1998:SG,bertin:1983:semiology}. In our work we focus on a specific type, black-and-white textures, which have several potential benefits. Black-and-white visuals can improve visualization expressivity when a device's color display capabilities are limited, for example for e-ink displays. Textures can also be used instead of color to avoid unwanted data-to-color associations or to avoid problems related to color-blindness. Encoding categorical data in black-and-white also allows us to extend visualization techniques to target groups with more severe forms of visual impairments: black-and-white visualizations can be turned into embossed representations that can be touched and felt. In addition, visualizations with few colors can be used in physical display environments such as knitting, embroidery, or for 3D printing.

\begin{figure}
    \centering
		\includegraphics[height=0.35\columnwidth]{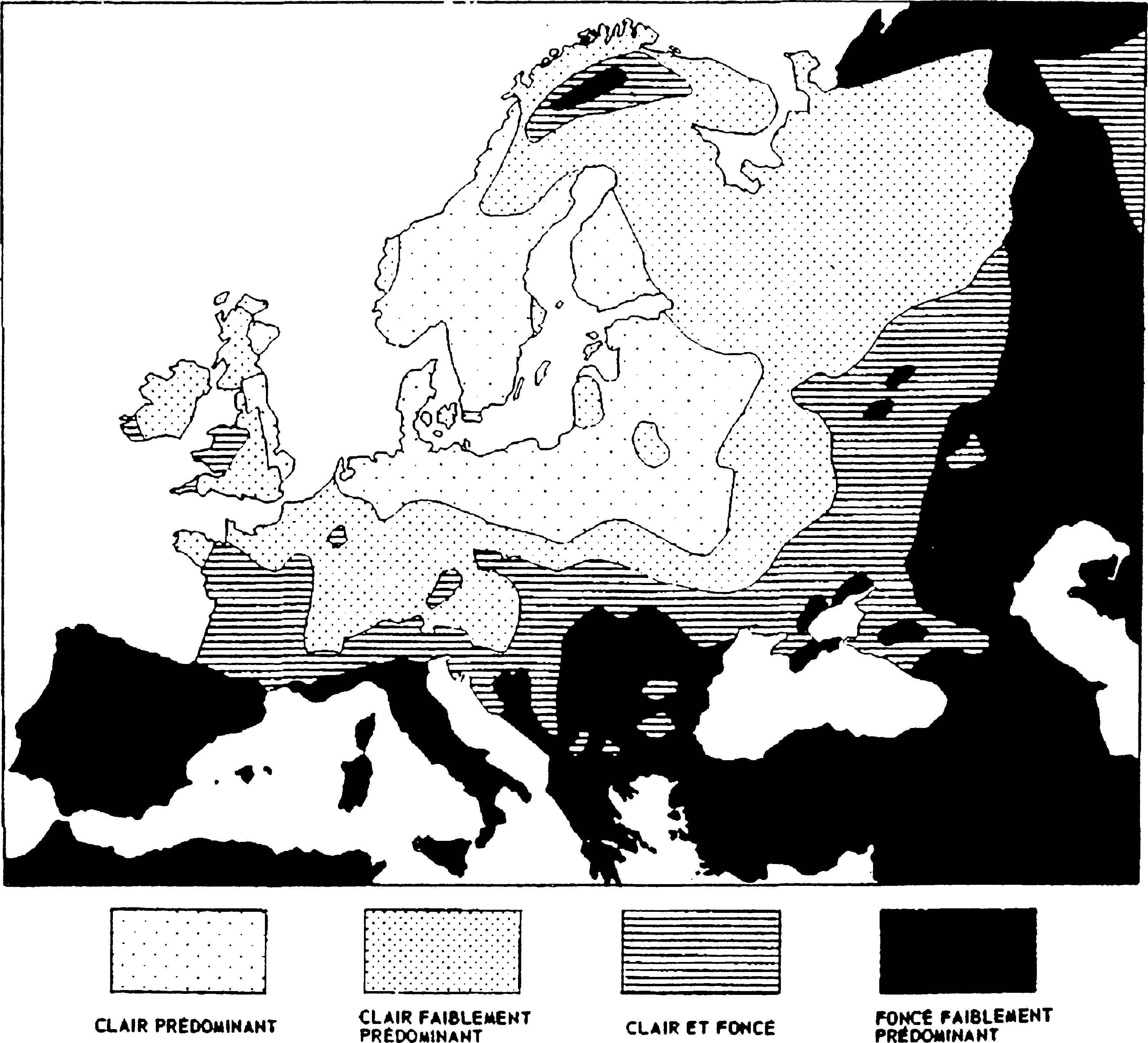}\hfill%
		\includegraphics[height=0.35\columnwidth]{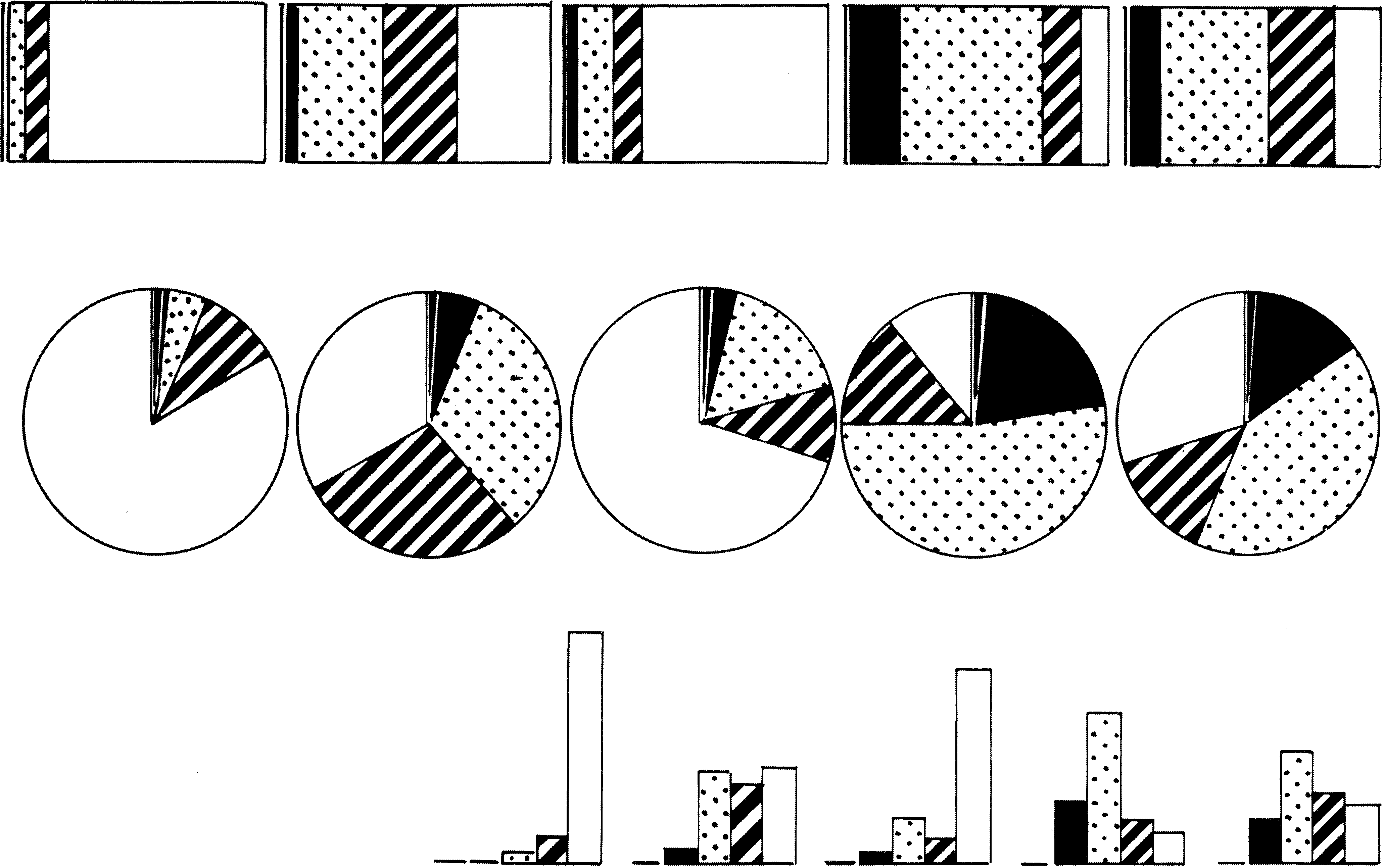}		
    \caption{Examples of categorical visualizations with black-and-white texture from Bertin \cite{bertin:1983:semiology,Bertin:1998:SG}; \textcopyright\ EHESS, used with permission.}%\vspace{-1ex}
    \label{fig:bertin_examples}
\end{figure}

\begin{figure}
    \centering
    \includegraphics[height=0.355\columnwidth]{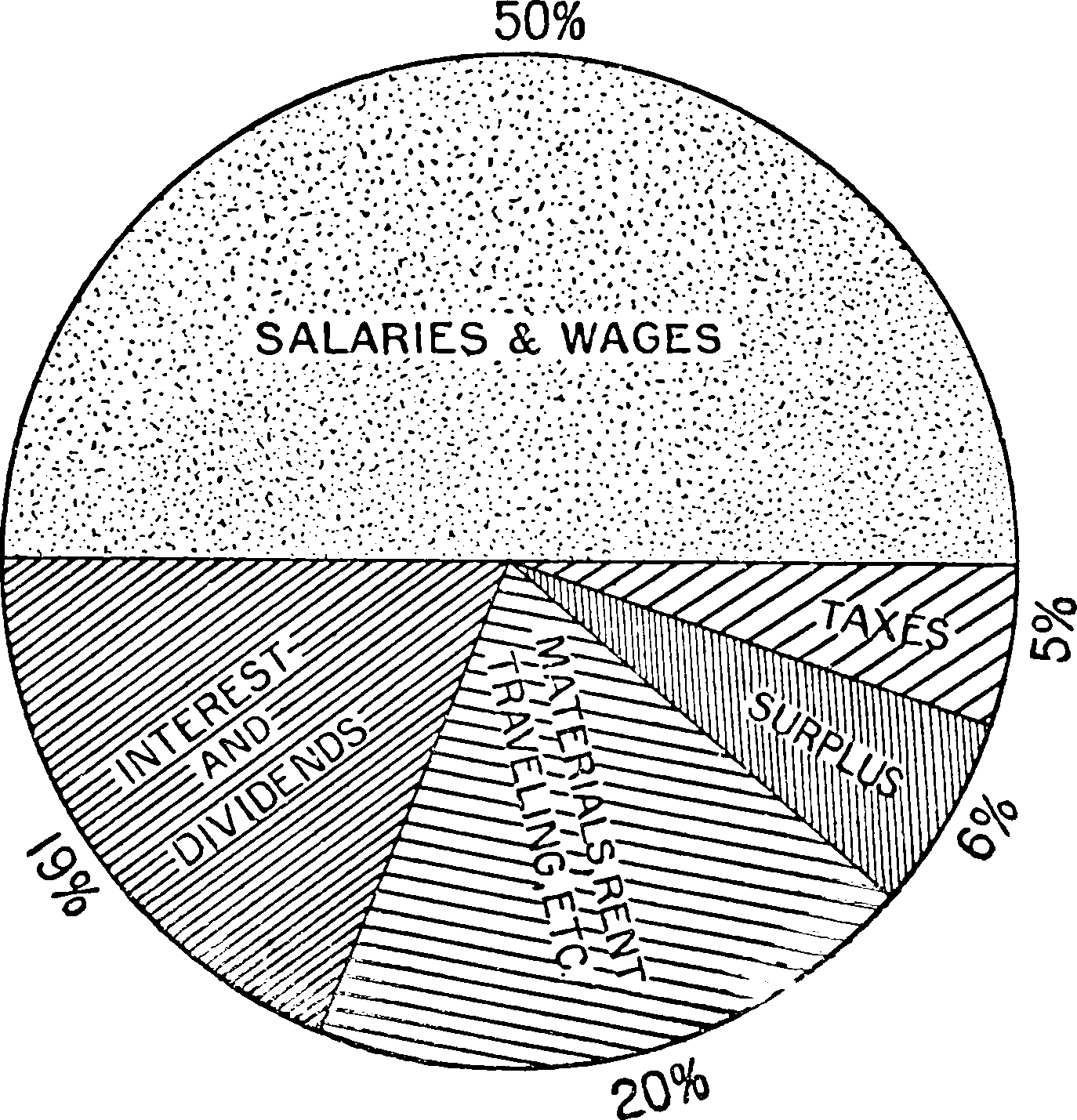}\hfill%
    \includegraphics[height=0.355\columnwidth]{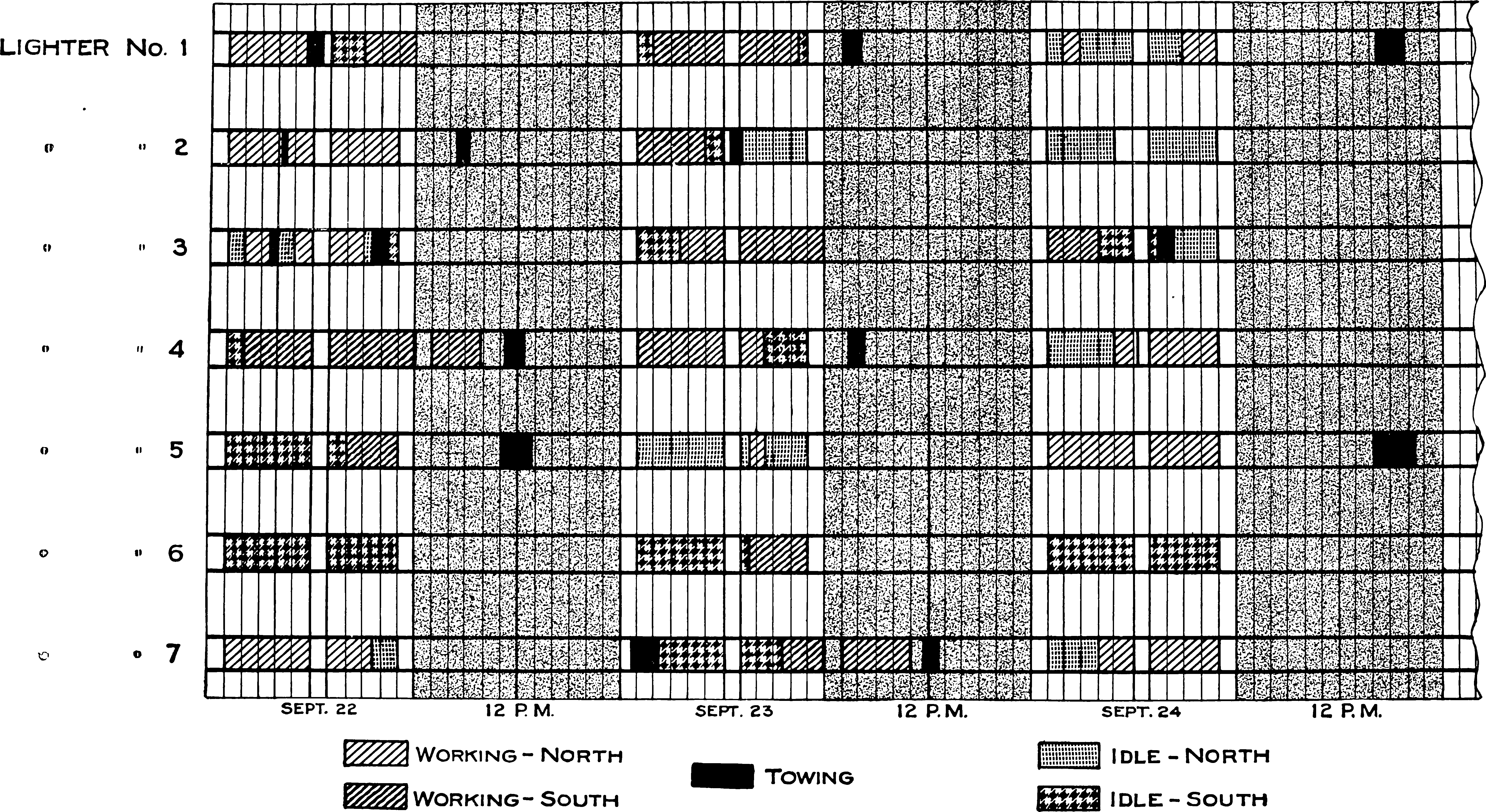}%
		% alternatives: page 36 (pie charts), 56 (table), 58 (table), 168 (table), 
    \caption{Examples of categorical visualizations with black-and-white texture from Brinton \cite{brinton:1919:graphic}; \ccPublicDomain\ the images are in the public domain.}%\vspace{-1ex}
    \label{fig:brinton_examples}
\end{figure}

Black-and-white textures continue to have many benefits in visualization today, and they were already in widespread use before color printing became affordable and common practice.
A century ago, texture was a prevalent visual channel for data mapping in news graphics \cite{brinton:1919:graphic,Brinton:1939:GP}, often featuring beautifully hand-crafted representations. Recreating this aesthetic is another benefit of using black-and-white textures today. In \autoref{fig:bertin_examples} we show some examples from Bertin's Semiology of Graphics \cite{bertin:1983:semiology,Bertin:1998:SG} and in \autoref{fig:brinton_examples} some examples from Brinton's book \cite{brinton:1919:graphic} that all served as an inspiration for us. Yet, surprisingly little design advice has persisted from this time and the possibility of automatically generating and parameterizing textures opens up new opportunities in computer-generated visualization.

 %(We might not need this? I also tried to remove the focus of "replacing" color to just using texture for its many potential benefits..)In previous research on the effectiveness of data encodings \cite{mackinlay:1986:automating}, color-related encodings were ranked highly for ordinal and nominal data, while texture is the channel followed color. Therefore, we focus on studying the replacement of color with 2D black-and-white texture to visualize nominal data in this work. 

When we use the term \emph{texture} in this paper, we follow a definition from computer graphics \cite{hawkins:1970:textural, haralick:1979:statistical} and 
%, who defined the notion of texture as ``some local `order' [\,\dots] repeated over a region which is large in comparison to the order's size, the order consists in the non-random arrangement of elementary parts, and the parts are roughly uniform entities having approximately the same dimensions everywhere within the textured region.'' Specifically, we 
consider a texture to be a repeated tiling pattern characterized by shapes that make up the pattern and the shape's attributes (\eg, density, size, etc.).
The shapes in a texture can be simple primitives such as lines or dots \inlinevis{-1pt}{1em}{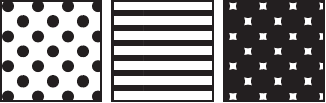} to form what we refer to as a \emph{geometric textures} . However, they may also be more figurative icons \inlinevis{-1pt}{1em}{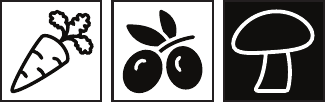} that represent objects the data may stand for, similar to the icons used in ISOTYPE visualizations \cite{neurath:2010:hieroglyphics}. We call these textures \emph{iconic textures} and investigate their use due to potential benefits shown for ISOTYPE representations in prior work \cite{haroz:2015:isotype}.

%This representation inspired us to investigate icons as shape attributes of textures to encode categorical data because it can add semantics to abstract textures. We refer to this type of texture as \emph{iconic texture}. We hypothesize that using iconic textures to represent categorical data can enhance people's performance in chart reading tasks, similar to using semantically-resonant color hues \cite{lin:2013:selecting}.

Despite the historical context and the potential benefits of textures as a qualitative visual channel, there has been little empirical research within the visualization community on how to use textures for visualization. Textures have rich attributes, such as shape type, density, size, or orientation, that can be varied to create new texture variations for additional categories. However, if used improperly, textures can bring negative effects such as vibratory effect \hty{(an optical illusion making patterns seem unstable, linked to the Moiré effect)}\cite{Bertin:1998:SG,bertin:1983:semiology,tufte:1985:visual} and visual clutter that may ultimately be distracting, lead to ineffective graphics, or simply lead to visualizations with an unappealing aesthetic.
%Yet, the existing guidelines on using textures in visualizations are limited. How these attributes work together in a single data display, how they affect the perception of data, and how they affect aesthetics are largely unknown. 

%TI: did we really develop a design space? I do not think that the iconic textures are new, are they really? I would focus this WHAT statement more on the actual things you did, the study! and then, for the study, you developed the tool, ran the experiments, etc.
%To fill this gap, we developed a design space for 2D black-and-white textures to summarize the attributes that can be used for encoding data. 
Ultimately, therefore, our fundamental research question is how to aesthetically and effectively use black-and-white textures for categorical data visualization.
%To learn more about the use of black and white textures for categorical data visualization, 
To answer this question, we derived a first design characterization that summarizes the attributes of 2D black-and-white textures that can be used for encoding data. Next, we conducted three experiments to explore the use of these attributes in visualization. As we conduct the first study in this area, we limited our research to three simple charts (bar charts, pie charts and maps) and two types of textures (geometric and iconic textures).
%We propose a new direction about black-and-white textures, which we call the iconic texture. We designed a tool for simple charts (bar chart, pie chart, and map) filled with geometric or iconic textures by adjusting the parameters of each texture attribute. 
%Then we conducted three experiments to explore the usage of these texture attributes in visualizations.
First, we invited 30 visualization experts to design geometric and iconic textured bar charts, pie charts and maps by adjusting parameters of each texture attribute. We collected 66 designs and experts' design strategies and opinions on using textures for visualizations. Then, we conducted a crowd-sourced experiment, in which we had 150 participants rate the designs we collected for their aesthetics. Finally, we conducted another crowd-sourced experiment with 150 participants to perceptually assess how quickly and accurately people can read the bar and pie charts filled with the top-rated geometric and iconic textures as well as a unicolor fill.

\hty{In summary, we contribute to the understanding of using 2D black-and-white textures for visualizations through (1) a summary of texture attributes, (2) an experiment on designing geometric and iconic textures for bar charts, pie charts, and maps, (3) an experiment examining the different textures' visual appearances, and (4) a perceptual experiment comparing geometric and iconic textures with respect to chart reading efficiency, accompanied by readability and aesthetics ratings.}
%\begin{itemize}[nosep]
%\item a summary of texture attributes,
%\item an experiment on designing geometric and iconic textures for bar charts, pie charts, and maps,
%\item an experiment examining texture's visual appearance, and
%\item a perceptual experiment comparing geometric and iconic textures for chart reading efficiency, accompanied by readability and aesthetics ratings.
%\end{itemize}

%% file: sections/02_related_work.tex
\section{Related Work}
\label{sec:related_work}
Texture is a topic of interest in numerous domains such as human visual perception, computer vision, and computer graphics, among others. We start this section by reviewing definitions of texture, followed by discussing prior research on texture dimensions. Subsequently, we examine previous work on the use of textures in visualizations, with a focus on traditional geometric textures. Finally, we describe research on pictographs, which inspired our use of iconic textures.

\subsection{Texture definition and discrimination}
\label{sub_sec:texture_definition}
Precise texture definitions can be elusive. In 1970, Hawkins \cite{hawkins:1970:textural} proposed a concept of texture based on three key components: \emph{``(1) some local `order' is repeated over a region which is large in comparison to the order's size, (2) the order consists in the non-random arrangement of elementary parts , and (3) the parts are roughly uniform entities having approximately the same dimensions everywhere within the textured region.''} Although this description of texture is widely accepted, it is important to emphasize that the role of randomness in texture properties was later acknowledged (as we mention in \autoref{sub-sec:texture_dimensions}) and is also considered as a factor in non-photorealistic rendering (NPR; \eg, \cite{Barla:2006:SPA,Martin:2017:SDS,salisbury:1994:interactive}). Subsequently, researchers have explored various texture analysis methods to describe the concept. Among them, the structural approach closely relates to the textures we study here. It defines texture as an organized area phenomenon formed by the combination of distinct texture elements such as evenly-spaced parallel lines. According to this approach, there are two fundamental dimensions for defining texture: (1) the properties of the primitives that form texture and (2) the placement rules and relationship between these primitives \cite{haralick:1979:statistical}.

The idea behind the structural approach originated from Julesz's research on texture discrimination \cite{julesz:1962:visual}, which refers to the ability of the human visual system to distinguish between different textures based on their specific structures. Initially, Julesz \cite{julesz:1962:visual,julesz:1975:experiments} conjectured that textures that can be differentiated by humans must have differences in mean luminance (\ie, first-order statistics) or texture orientation (\ie, second-order statistics), but this was later shown to be inadequate for explaining human texture discrimination \cite{julesz:1973:inability}. Julesz then introduced the texton theory \cite{julesz:1981:theory, julesz:1981:textons, julesz:1983:human} suggesting that textons are the fundamental micro-structures of pre-attentive human texture perception. This idea led to the development of the structural approach, which emphasizes extracting texture primitives for effective texture description.
%and only the first-order statistics of textons have perceptual significance. This theory gave rise to the structural approach, which involves extracting texture primitives as local features for texture description.

\subsection{Dimensions of textures}
\label{sub-sec:texture_dimensions}
%To be able to use texture for encoding data effectively, it is vital to understand how the human visual system perceives texture. Texture lacks a standardized description model like color  (\eg, CIE XYZ, CIE L*a*b*), but prior research has tried to identify essential perceptual dimensions for textures (also known as texture features). Tamura et al. \cite{tamura:1978:textural} identified six basic features, while Amadasun and King \cite{amadasun:1989:textural} suggested five perceptual attributes computationally. Ware and Knight \cite{ware:1992:orderable} introduced the orientation, size, contrast (OSC) texture model for data display. Rao and Lohse \cite{rao:1996:towards} proposed a Texture Naming System focusing on three key dimensions in natural texture perception. Liu and Picard \cite{liu:1996:periodicity} determined three crucial orthogonal dimensions, while Cho et al. \cite{cho:2000:reliability} reported four texture dimensions. Healey and Enns \cite{healey:1998:building} developed three-dimensional perceptual texture elements (pexels) for visualizing multidimensional datasets. Despite these contributions, the majority of this research \cite{tamura:1978:textural, amadasun:1989:textural, rao:1996:towards, liu:1996:periodicity, cho:2000:reliability} focused on natural textures, with no specific focus on the geometric or iconic textures explored in this paper.

To be able to use texture for encoding data effectively, it is vital to understand how the human visual system perceives textures. While there is no standardized model to describe texture like there is for color (\eg, CIE XYZ, CIE L*a*b*), previous studies have attempted to identify essential perceptual dimensions for textures (also known as texture features).
%previous studies have attempted to identify a limited set of perceptual dimensions (also known as texture features) that are essential for texture perception in the human visual system. 
Tamura et al. \cite{tamura:1978:textural} proposed six basic texture features, namely, coarseness, contrast, directionality, line-likeness, regularity, and roughness. Amadasun and King \cite{amadasun:1989:textural} approximated 5 perceptual texture attributes in computational form, namely coarseness, contrast, busyness, complexity, and strength of texture. Ware and Knight \cite{ware:1992:orderable} proposed the orientation, size, contrast (OSC) model of texture for data display. Rao and Lohse \cite{rao:1996:towards} identified a Texture Naming System with three most significant dimensions in natural texture perception: ``repetitive vs.\ non-repetitive; high-contrast and non-directional vs.\ low-contrast and directional; granular, coarse and low-complexity vs.\ non-granular, fine and high-complexity.'' Liu and Picard \cite{liu:1996:periodicity} identified three mutually orthogonal dimensions of texture that are important to human texture perception, namely periodicity, directionality and randomness. Cho et al. \cite{cho:2000:reliability} extended the perceptual research and reported four texture dimensions, namely coarseness, contrast, lightness and regularity. Healey and Enns \cite{healey:1998:building} built three-dimensional perceptual texture elements, or called pexels, for visualizing multidimensional datasets. Pexels can be varied in three separated texture dimensions, which are height, density, and regularity, and color of each pexel. 
Most of this work that examines texture dimensions \cite{tamura:1978:textural, amadasun:1989:textural, rao:1996:towards, liu:1996:periodicity, cho:2000:reliability}, however, has concentrated on natural textures, utilizing textures taken from Brodatz' album \cite{brodatz:1966:textures}, and none have specifically addressed the geometric or iconic textures that we investigate in this paper.

\subsection{Using texture for visualization}
\label{sub-sec:texture-for-vis}
In his seminal book, Bertin \cite{Bertin:1998:SG, bertin:1983:semiology} comprehensively discussed how to use 2D geometric textures for visualizations. Bertin referred to visual channels as \emph{retinal variables} and proposed 7 key ones, including \emph{planar position}, \emph{size}, \emph{value} (black/white ratio), \emph{texture}, \emph{color}, \emph{orientation}, and \emph{shape}. These visual channels are mainly used to manipulate 2D marks such as points, lines, and polygons in printed data graphics. We note here that Bertin's terminology differs from what we commonly use today. For instance, he used the term \emph{texture} to refer to the number of distinct marks in a given area, which is similar to what we call \emph{density} (\eg, \inlinevis{-3pt}{1.2em}{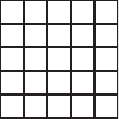} vs.\ \inlinevis{-3pt}{1.2em}{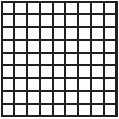}) . Meanwhile, Bertin used \emph{pattern} to denote variations in shapes applied to a mark, which is similar to our use of \emph{primitives} of textures (\eg, \inlinevis{-3pt}{1.2em}{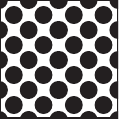} vs.\ \inlinevis{-3pt}{1.2em}{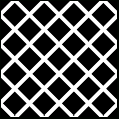}). Our notion of texture encompasses several visual variables mentioned by Bertin (texture (density), size, orientation, shape), so his ideas on the use of these variables are important for us. For instance, Bertin identified four perceptual qualities---\emph{associative}, \emph{selective}, \emph{ordered}, and \emph{quan\-ti\-ta\-tive}---to determine which visual channels are suitable for representing different types of data. Both associative and selective qualities are important for nominal data. Associative perception helps designers to balance variations and groupings across all categories of a given variable, while selective perception indicates that a variable has enough diversity for people to distinguish all the elements of this category from others. Bertin found that texture (density) as a visual variable is both selective and associative, making it ideal for encoding categorical data.

%\hty{Question: should we remove this paragraph? I checked Mackinlay's paper, he did not make it clear what ``texture'' they he was talking about, maybe it's just ``density''?}
More visual channels were proposed and evaluated after Bertin. 
Cleveland and McGill \cite{cleveland:1985:graphical} evaluated various channels for accuracy, but excluded texture. 
% position, length, direction, angle, area, volumne, curvature, shading (density of grid texture) and color
Mackinlay \cite{mackinlay:1986:automating} extended this research to 13 visual channels, ranking them based on their effectiveness in encoding quantitative, ordinal, and nominal data. For nominal data, texture ranked third, outperformed only by position and hue.

The use of texture in visualizations has not been extensively researched. There are several design guidelines on how to use texture in visualizations, but they are limited and mostly borrowed from the psychophysics field directly, without empirical research using visual data representations---which is what we provide. Some visualization design books \cite{ware:2019:information,kosslyn:2006:graph} recommend ensuring that visual properties are distinguishable when using texture. For instance, it is suggested that orientation varies by at least 30\textdegree\ and that the spacing of texture patterns with similar orientations varies by at least a ratio of 2 to 1. Both Tufte \cite{tufte:1985:visual} and Bertin \cite{Bertin:1998:SG, bertin:1983:semiology} mentioned that textures may produce vibratory effect. Bertin pointed out that this visual effect represents a remarkable selective possibility, so designers can make good use of it. Tufte \cite{tufte:1985:visual}, however, believed this effect should be avoided altogether. We empirically investigate this effect further in our own work.

\subsection{Reseach on pictographs}
Pictographs, or pictorial visual representations, use an icon-based language to represent data visually \cite{zhang:2020:dataquilt}. They have a long history and have been shown to have many positive effects. The ISOTYPE system, an icon-based visual language developed by Otto Neurath, Marie Neurath, and Gerd Arntz in the mid-1920s, is a well-known example of pictographic visualization. ISOTYPE visualizations feature rows or arrays of icons, using repetition rather than size of icons to represent quantitative data \cite{neurath:2010:hieroglyphics}. Chen and Floridi \cite{chen:2013:analysis} organized over 30 visual channels into a simple taxonomy consisting of four categories, namely geometric, optical, topological and semantic channels. Icons and ISOTYPE are classified as semantic channels in this taxonomy. 

Studies by Haroz et al. \cite{haroz:2015:isotype} on bar charts with ISOTYPE found that pictographs are beneficial for working memory and engagement, and do not significantly impact chart reading performance. Burns et al. \cite{burns:2021:designing} conducted a comparison between part-to-whole visualization using pictographs and found that pictographs made it easier for people to envision what was happening in the charts.

Since icons have lots of shape attributes, researchers also investigated how to support the design of pictographs. Borgo et al. \cite{borgo:2013:glyph} did a comprehensive survey of glyph-based visualization and proposed a set of design guidelines. Morais et al. \cite{morais:2020:showing} created a design space for anthropographics, a type of visualization that incorporates human-related information. One common approach in anthropographic design is to use pictographs in the shape of humans. Shi et al.\ \cite{shi:2022:supporting} explored the design patterns of pictorial visualizations that can be used to guide their generation. All these studies and the established pictograph qualities inspired us to investigate the use of icons as texture primitives.

Pictograph, however, can also be seen as a type of visual embellishment (or `chart junk'), which are extraneous elements in a chart or visualization that do not represent data \cite{bateman:2010:useful}. Tufte's \cite{tufte:1985:visual} design principles suggest to maximize the data-ink ratio and to avoid chart junk. To investigate this issue, Bateman et al.\ \cite{bateman:2010:useful} conducted a study comparing plain and embellished charts. They discovered that adding embellishments did not have any impact on interpretation accuracy, but it did improve long-term recall, made the topic and details of the chart more memorable (an effect later confirmed by Borkin et al.\ \cite{Borkin:2013:WMV,Borkin:2016:BMV}), and embellished charts were preferred by participants. These results also led us to investigate icon-based textures more closely.

%% file: sections/design_characterization.tex
\section{Design Characterization}
\label{sec:design_characterization}

We started our analysis by discussing the characteristics that we need to consider when designing black-and-white textures for visualizations. Extending our previous poster \cite{zhong:2020:BWT} on this topic, we adopt a structural approach that defines texture as recurring tiled primitives with specific properties (\autoref{sub_sec:texture_definition}). In \autoref{tab:design-space} we summarize the characteristics that we consider important \hty{for visualization researchers and designers}, and show examples for possible textures for points, lines, and grids. These are the most commonly employed primitive shapes in historical black-and-white texture-based visualization based on our review of illustrations from the books of Bertin \cite{bertin:1983:semiology,Bertin:1998:SG}, Tufte \cite{tufte:1985:visual}, Brinton \cite{brinton:1919:graphic,Brinton:1939:GP} and other historical visualization examples like works of Minard, 
%with additional inspiration, in particular, for the randomness category coming from the field of non-photorealistic rendering (NPR; \eg, \cite{Barla:2006:SPA,Martin:2017:SDS,salisbury:1994:interactive}).
with additional inspiration, especially for the randomness category, coming from the field of non-photorealistic rendering (NPR; \cite{Barla:2006:SPA,Martin:2017:SDS,salisbury:1994:interactive}).

%In this section, we extended our previous poster \cite{zhong:2020:BWT} by discussing the characteristics to consider when designing black-and-white textures for visualizations. As said in Section \ref{sub_sec:texture_definition}, we adopt a structural approach, defining texture as recurring tiled primitives with specific properties, therefore, we discussed design characteristics of textures for visualization from these two perspectives. We summarized all properties we discussed above with examples of points, lines, and grids (the most commonly employed primitive shapes in historical black-and-white texture visualization) textures in Table \ref{tab:design-space}. The examples we used are mostly based on texture used in Bertins' book\cite{bertin:1983:semiology,Bertin:1998:SG} , for the example of randomness properties we also included examples from NPR \cite{salisbury:1994:interactive}. We will explain them in detail below. 

\begin{table}[t]
\caption{Texture properties with examples.}
\label{tab:design-space}
\centering%
\footnotesize%
\begin{tabu}{lccc}
\toprule
properties \textbackslash\ primitives & point-based                           & line-based                          & grid-based                          \\
\midrule
\raisebox{1.5mm}{shape type}                           & \includegraphics[height=.5cm]{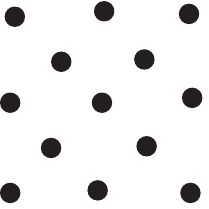} \includegraphics[height=.5cm]{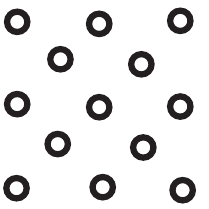}             & \includegraphics[height=.5cm]{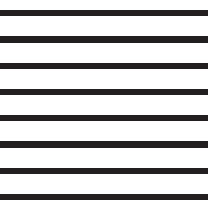} \includegraphics[height=.5cm]{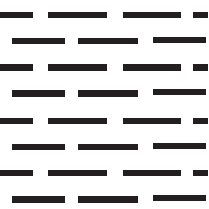}            & \includegraphics[height=.5cm]{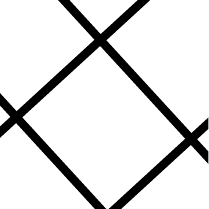} \includegraphics[height=.5cm]{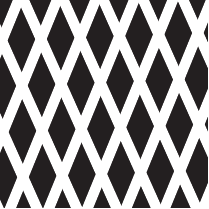}             \\
\raisebox{1.5mm}{density}                              & \includegraphics[height=.5cm]{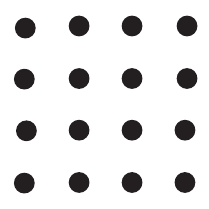} \includegraphics[height=.5cm]{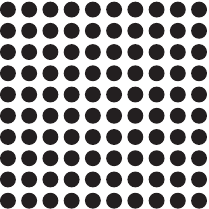}         & \includegraphics[height=.5cm]{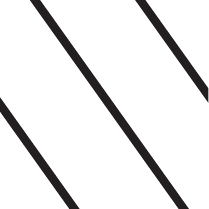} \includegraphics[height=.5cm]{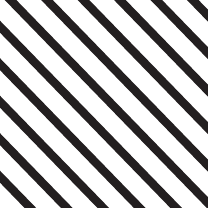}         & \includegraphics[height=.5cm]{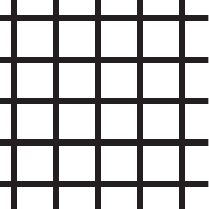} \includegraphics[height=.5cm]{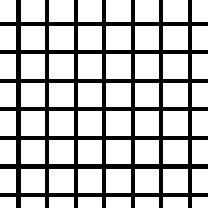}         \\
\raisebox{1.5mm}{shape size}                           & \includegraphics[height=.5cm]{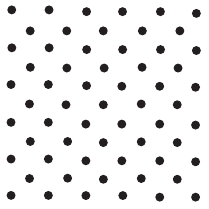} \includegraphics[height=.5cm]{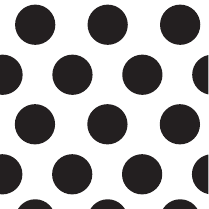}               & \includegraphics[height=.5cm]{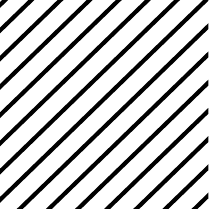} \includegraphics[height=.5cm]{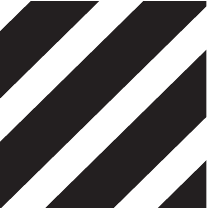}               & \includegraphics[height=.5cm]{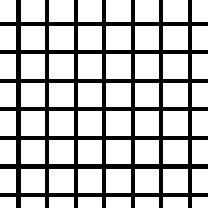} \includegraphics[height=.5cm]{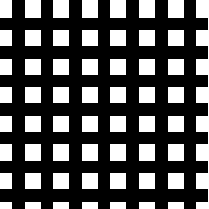}               \\
\raisebox{1.5mm}{orientation}                          & \includegraphics[height=.5cm]{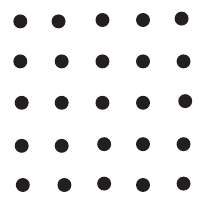} \includegraphics[height=.5cm]{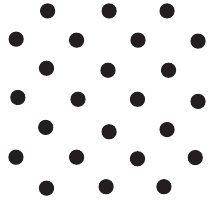} & \includegraphics[height=.5cm]{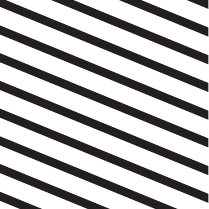} \includegraphics[height=.5cm]{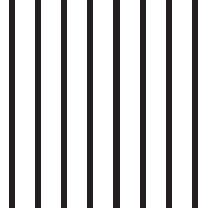} & \includegraphics[height=.5cm]{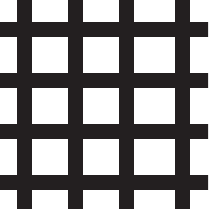} \includegraphics[height=.5cm]{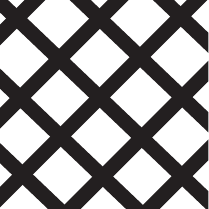} \\
\raisebox{1.5mm}{background color}                     & \includegraphics[height=.5cm]{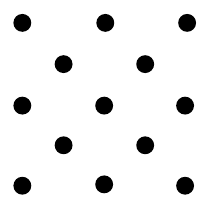} \includegraphics[height=.5cm]{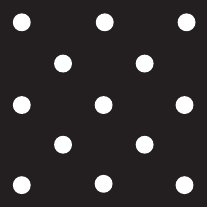}   & \includegraphics[height=.5cm]{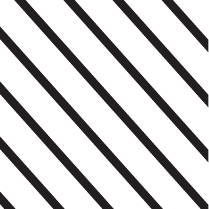} \includegraphics[height=.5cm]{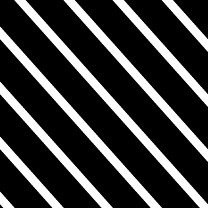}   & \includegraphics[height=.5cm]{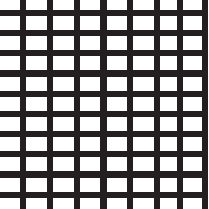} \includegraphics[height=.5cm]{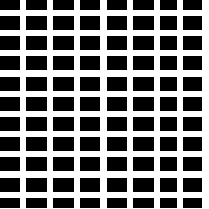}   \\
\raisebox{1.5mm}{randomness}                           & \includegraphics[height=.5cm]{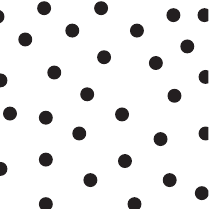}                          & \includegraphics[height=.5cm]{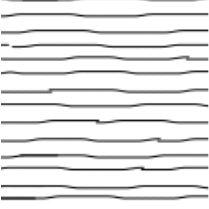}                         & \includegraphics[height=.5cm]{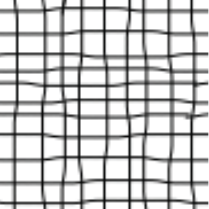}\\
\bottomrule
\end{tabu}
\end{table}

\subsection{Shape type of the primitives}
Bertin \cite{Bertin:1998:SG, bertin:1983:semiology} considered his \emph{pattern} variable to be applicable to points, lines, and areas. In most 2D vector graphics tools today, styles (including texture) can be applied to the fill and to the stroke of a 2D primitive. Yet for a reasonably flexible application of texture to strokes (beyond dashing patterns) we would need to make the stroke rather thick, effectively turning it into a 2D area---which also applies to textured points in Bertin's examples. Below we thus focus on the texturing of 2D areas, as they apply to many data representations such as bar or area graphs.

%Primitives can be applied to both fill and stroke. However, in this study, we focus on fill, as textures cannot be incorporated into strokes without making the stroke thicker (effectively turning it into a fill). Therefore, all the properties discussed in this study relate to a 2D area.

In a 2D area, textures can have a fundamental structure that is either point-based (0D) or line-based (1D). In pure geometry, a point is a dimensionless mark that indicates a specific position in space, with no width or height, and moving point primitives in any direction changes the texture's visual appearance. A line is a one-dimensional, straight connection between two points that further extends infinitely in both directions. Moving line primitives along their primary direction does not affect the texture's appearance. These basic structures can be grouped to form more complex structures. For example, a grid can be seen as a group of two line-based structures. If such grouping occurs, the relationship between the group can also become a property. In a grid, for instance, the angle between two lines can be a property.

For textures we relax the geometric definitions such that they can adopt any shape based on these structures, meaning that a ``point'' includes graphical elements such as circles or dots and a ``line'' does not have to be continuous, straight line but can take on various forms of linear structures. Partially because of this flexibility, point-based or line-based structures should be seen as a spectrum. Some textures, for example \inlinevis{-2pt}{1.2em}{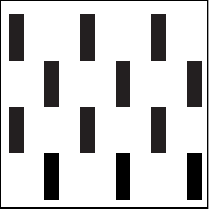}, can be seen as either point-based or line-based. But even without this notion we can have ambiguities: if we consider the negative space within a grid such as \inlinevis{-2pt}{1.2em}{bertin_def_texture0} as primitives then it could be seen as being composed of (white) point primitives.

The variety of shape types then also allows designers to establish semantic associations with textures by using iconic representations (short: icons) as points on a point-based structure, resulting in iconic textures. We hypothesize that integrating icons into the texture that represents a data category may have a similar effect to using semantically-resonant colors \cite{lin:2013:selecting} in improving chart reading task performance, and we examine this hypothesis in our experiments in \autoref{sec:experiment_design}--\ref{sec:experiment3}.

\subsection{Properties of the primitives}

%We have identified the following properties as significant for the textures we studied, based on review of historical texture visualizations, and texture propoerties proposed by prior reasearch as we mentioned in \autoref{sub_sec:texture_dimensions}.
%Based on a review of historical texture visualizations and of texture properties proposed in prior work (\autoref{sub-sec:texture_dimensions}), we identified the following properties as significant for the textures we want to study (\autoref{tab:design-space}). \hty{Please note that the parameter space varies for each dimension. For instance, the parameter space for dimensions like ``orientation'' or ``randomness'' is quite large, whereas the ``background color'' property is binary.}

Based on a review of historical texture visualizations and of texture properties proposed in prior work (\autoref{sub-sec:texture_dimensions}), we identified key properties for our target textures (\autoref{tab:design-space}). \hty{The parameter space differs across dimensions. For example, the ``randomness'' dimension encompasses a large parameter space, while ``background color'' is binary.}

\begin{description}[itemsep=-2pt]%, leftmargin=0pt, itemindent=0pt]
\item[Density:] The number of primitives per unit area.
\item[Primitive Size:] The dimensions of the primitives, such as the stroke width for lines or the radius for points. Changing size while main\-taining density alters the black/white ratio (value), while \textls[-5]{modifying size and density in sync preserves the black/white ratio.}
\item[Orientation:] The angular positioning or alignment of primitives within a given space, in relation to a given reference frame or axis. For point-based textures, both the whole texture as well as the individual primitives can be rotated, whereas for line-based textures only the textures as a whole can be turned.
\item[Background ``color'':] Either black or white.
\item[Randomness:] The irregularity in the distribution of primitives.
\end{description}

Furthermore, some properties relate not only to the texture itself but also to the chart and the overall visual design of the black-and-white texture visualization. These properties include:

\begin{description}[itemsep=-2pt]%, leftmargin=0pt, itemindent=0pt]
\item[Position of texture:] The placement or arrangement of patterns within chart components like bars, pie slices, or (map) areas.
\item[Outline stroke width of the chart \hty{components}:] The thickness or width of lines or borders defining the chart components.
\item[White halo width of the chart:] The white halo technique is widely employed in historical visualization to improve readability. It adds a white outline or glow around elements, enhancing contrast between chart elements and backgrounds. In our case, the white halo is between the textures and the outline of the chart element.
\end{description}

%% file: sections/experiment1_design.tex
\section{Experiment 1: Design}
\label{sec:experiment_design}
To better understand how to effectively combine these properties in designing textures for visualization, we conducted a series of experiments. In our first experiment we focused on the perspective of visualization professionals. We reached out to visualization experts with a design background and asked them to create designs using textures, to identify the characteristics of effective textured visualizations in their eyes and to study their approaches to parameter arrangement. To keep the workload manageable, we narrowed our scope to three basic chart types (bar charts, pie charts, maps) and two texture categories (geometric, iconic).

\subsection{Texture design interface as a technology probe}
\label{sub-sec:design-probe}

%\begin{figure}[t] % htbp are optional placement specifiers (here, top, bottom, page)
	%\centering % Centers the figure
	%\includegraphics[width=\columnwidth]{figures/experiment1/design_probe_screenshot.pdf}\
	%\caption{Technology probe for designing a pie chart with geometric textures.}
  %\label{fig:design_probe_screenshot}
%\end{figure}

To collect input from professionals, we developed a web-based technology probe \cite{Hutchinson:2003:TechProbe} (we show screenshots in \autoref{appendix:additional-exp1}). %(see the screenshot in \autoref{fig:design_probe_screenshot}). 
This tool allowed the experts to create chart designs using black-and-white textures by adjusting parameters. The probe comprised three main views: the visualization itself, the controllers, and the toolbar.

\subsubsection{Visualization view} 
In the central view we show the chart---a bar chart, a pie chart, or a map ``colored'' with black-and-white geometric or iconic textures---and its legend. The chart represents unspecified quantities for seven vegetable items (carrots, celery, corn, eggplant, mushrooms, olives, tomatoes).
%We generated the specific values following Lin et al.'s \cite{lin:2013:selecting} semantically-resonant color experiment.

When opening the interface, we showed the chart with default textures and dataset. For geometric textures, we provided five default texture sets, all sourced from Bertin's book \cite{bertin:1983:semiology, Bertin:1998:SG} (see \autoref{tab:bertin-textures} in \autoref{appendix:default-bertin}). Bertin used these textures to encode nominal and ordered data. We picked the default texture set from the Bertin set randomly per participant. For iconic textures we used a neutral style (with icons from \href{https://icons8.com/}{Icon8.com} \cite{icons8}, see below and \autoref{tab:fruit-icons} in \autoref{appendix:default-icon}) as a default with parameters that we deemed visually appealing (see \autoref{fig:iconic-default-texture} in \autoref{appendix:additional-exp1}).

The visualization experts could then edit any given vegetable's texture by clicking the corresponding section of the chart (\eg, the bar or pie piece) or the vegetable's legend entry. We showed a blue round dot next to the vegetable on the chart and the legend to indicate the vegetable currently being edited. The experts could also swap textures by dragging and dropping, such as dragging the texture from the carrot's bar and dropping it on the mushroom's bar. For iconic textures, naturally, we then switched only the parameters and not the vegetable icons themselves (\eg, the carrot bar always used carrot icons).

\subsubsection{Controls for adjusting the textures} 

Our interface relied on buttons and sliders. After selecting a vegetable's texture, the experts could modify the chart's textures using these controls by first choosing a primitive type and then adjusting its properties.

\textbf{Primitive shapes.}
For the geometric textures we provided three primitive types: lines, dots, and a grid. For the iconic textures, we selected two professionally designed, neutral, stylized icon sets from \href{https://icons8.com/}{Icon8.com} \cite{icons8} (ensuring we would have icons for all data items). We also provided two matching simplified texture sets, created by removing internal details and simplifying outlines from the original versions. In total, we offered four sets of icons (all shown in \autoref{tab:fruit-icons} in \autoref{appendix:default-icon}).

 %One set features light vegetable outlines, while the other contains dark-filled vegetables. Icons within the same set were consistently designed by professional designers. We also provide two corresponding simplified texture sets, created by removing internal details and simplifying outlines from the original version. In total, we offer four sets of icons. 

\textbf{Properties.}
The visualization experts could adjust all texture properties as described in \autoref{sec:design_characterization}, including primitive type, density, size, orientation, the texture position in the chart, and the chart outline width. In addition to these common properties, dot textures could be modified to display circles, while grid textures allowed angle adjustments between two lines. For icons, the entire texture could be rotated and the individual icons themselves  as well. For pie charts and maps with connected regions, we also added an optional white halo between the texture and the black outline and allowed the experts to adjust its width. 

We pilot-tested our technology probe within our research group and, based on this pilot, chose reasonable value ranges for density, size, outline, and the white halo width. For other parameters such as orientation we allowed the full spread of possibilities (\ie, a full 360\textdegree{} rotation). We also offered controls to quickly set certain properties to special values, such as rotating the texture in steps of 45\textdegree. We selected these special values because they are common in historical visualization examples. \hty{We also added a ``for all'' checkbox to the property controllers that, when checked, applies changes across all textures of the same type (for geometric textures, \eg, all grid textures; for iconic textures for the textures of all seven vegetables).}
%We also add a ``for all'' checkbox to all property controllers that ensured that any changes would be applied to all textures of the same primitive type (for geometric textures, \eg, all grid textures; for iconic textures for the textures of all seven vegetables).

\subsubsection{Toolbar}

At the top of the interface we offered a toolbar for managing operations on textures and datasets, loading default texture sets, as well as undo and redo functionality. A reset button allowed the visualization experts to revert all textures to their respective default settings. We also provided a button to load a new, random dataset or to return to the default dataset (that we use throughout this paper; \eg, Tables~\ref{tab:exp2-bar-geo}--\ref{tab:exp2-map-icon} or \autoref{appendix:all-designs}).

%The toolbar, located at the top of the interface, functions for managing operations on textures and datasets, as well as importing default texture sets. The undo and redo buttons allow designers to reverse or reapply texture parameter adjustments. The reset button enables designers to revert all textures to their default sets. There is a button for switching to different random data sets and another button for returning to the default data set, allowing designers to test their designs on various data sets. We also provide a drop-down box for importing default texture sets specifically for geometric textures.

\subsubsection{User feedback on the tool}

Although we did not specifically request participants to comment on our tool, four participants voluntarily commented in their free-text answers in Experiment 1 and nine participants provided voluntary, unprompted comments in response to the invitation e-mail. They said that they enjoyed using our tool (mentioned 10\texttimes) and found that the interface was well-designed (1\texttimes), that the controls made it easy to manipulate the textures (1\texttimes) and allowed them to create expressive textures (1\texttimes).

%\subsection{Participant recruiting}

\subsection{Method and procedure}

We used a mixed design with the between-subjects variable \emph{chart type} (bar, pie, map) and the within-subject variable \emph{texture type} (geometric, iconic). The experiment was pre-registered (\href{https://osf.io/r4z2p/}{\texttt{osf.io/r4z2p}}) and IRB-approved (Inria COERLE, avis \textnumero\ 2023-01).

We recruited participants by reaching out to visualization experts with design expertise within our network via e-mail. We also requested that these experts share the e-mail with their colleagues, friends, or students. Furthermore, we sent our experiment link to the design-related Slack channels of the Data Visualization Society \cite{datavissociety}.

We started the study by asking participants for their informed consent and background information. Following a tutorial to familiarize them with the interface, we assigned participants randomly to a chart type and instructed them to design two charts, one with geometric and one with iconic textures, in random order. We asked them to adjust the parameters to create effective visualizations. Subsequently, we asked them about their goals and design strategies, their opinions on the two texture types, and their thoughts on the transferability of their designed textures to the other two chart types by showing them their designs automatically applied to the respective other charts.
After completing the experiment once, we gave the participants the chance to continue. For any repetition they could select their preferred chart type to use, while we still randomly assigned the texture type order.

\subsection{Results}

We collected 66 designs from 30 experts (12 female, 18 male; ages: mean = 40.1, SD = 14.4; prior experience in visualization design: mean = 13.4 years, SD = 11.0 years). The designs consisted of 14 bar charts, 30 pie charts, and 22 maps. Six of the pie charts were from \hty{participants who had completed this experiment at least once before}. Half of the designs used geometric textures, while the other half used iconic textures. We show all created designs in \autoref{appendix:all-designs} and qualitatively coded the free-text answers using open coding. We discuss our observations and findings next.

\subsubsection{Design strategies}
\label{sub-sec:design-strategies}

We broadly categorized the primary objectives of the experts into two classes: those related to data readability (\eg, distinguishing between categories, ensuring clarity of the chart, creating semantic associations) and those focused on aesthetics (visual pleasure, balance). %Next, we explore these objectives and their corresponding design strategies in greater detail.

\textbf{Distinguishability.}
27 participants mentioned wanting to make the categories distinguishable (5\texttimes{} bar, 12\texttimes{} pie, 10\texttimes{} map). To achieve this goal, the use of varied visual channels was the most commonly used method (mentioned 12\texttimes). For differentiating geometric textures, most participants used background color, density, and size as the key visual channels. For iconic textures, background color, orientation, icon style, and density were generally considered helpful. In addition, participants mentioned that, for designing pie charts, outlines (1\texttimes) and white halos (3\texttimes) contributed to creating a more distinct separation. We indeed observed many designs with thick outlines (11\texttimes) and white halos (10\texttimes) in pie chart designs, which is not common in other chart types. For iconic textures, specifically, 5 participants found it important to show complete icons in the chart. One response also mentioned that upright icons were generally easier to recognize.

\textbf{Clarity.}
14 visualization experts tried to make the chart clear and readable (4\texttimes{} bar, 4\texttimes{} pie, 6\texttimes{} map). To be specific, 5 responses mentioned they participants focused on avoiding clutter and overwhelming elements. Fading icons into the background is considered as a way to avoid icons being overwhelming and distracting (2\texttimes). The white halo was also considered useful in preventing \hty{a perception of clutter} (1\texttimes).

\textbf{Semantic association.}
Five participants tried to create a semantic association between the textures and the vegetable items (2\texttimes{} bar, 2\texttimes{} pie, 1\texttimes{} map). With iconic textures, people applied the vegetables' relative size to the icons of their representative textures (4\texttimes), such as making the tomato icon larger than the olive icon due to their physical size difference. In addition, in 2 designs the visualization experts selected the textures for vegetables (dark vs.\ bright) based on their respective colors. Remarkably, 4\texttimes{} the experts sought to establish conceptual matches between the vegetable items and the geometric textures. One did this by considering the vegetables' color, while two others tried to elicit visual associations by incorporating various visual channels. Furthermore, one employed dots to signify vegetables typically planted in rows such as carrots, celery, and corn.

\textbf{Visual pleasure.} 
In 13 responses the participants tried to make the chart visually pleasing (4\texttimes{} bar, 5\texttimes{} pie, 4\texttimes{} map). One common strategy was to maintain a consistent visual style throughout all categories by applying a uniform orientation, line width, density, or icon style (9\texttimes). Other strategies included selecting an aesthetically pleasing default texture set (2\texttimes) and striving to create a visual experience that was harmonious (1\texttimes), clean and elegant (1\texttimes), or sketch-like (1\texttimes). 

\textbf{Visual balance.}
In 12 responses the visualization experts mentioned to attempt a visually balanced chart (2\texttimes{} bar, 7\texttimes{} pie, 3\texttimes{} map). To achieve this objective, a strategy was to use a consistent ink density across all categories such that the textures maintain a roughly equal visual weight, preventing one pattern from dominating or becoming too weak (8\texttimes). Notably, in 3 responses the participants emphasized that, since our designs were aimed at general datasets, textures should be effective for small areas without being overpowering in larger ones, indicating that the texture should remain recognizable even when a category represents a small data point.

\textbf{Other design strategies.}
Apart from these primary design goals, our participants employed several other noteworthy strategies.

\textit{Abstracting iconic textures:} One person aimed to create an abstract representation of iconic textures by making the icons overlap (BI4 in \autoref{tab:exp2-bar-icon} or \autoref{fig:BI4} in \autoref{appendix:all-designs}). This approach produced an interesting texture-like style, in which the vegetables are still distinguishable. 

\textit{Avoiding conflicts with chart outlines:} Another participant avoided using vertical line textures when designing bar charts with geometric textures, as these would conflict and compete with the vertical bars.

\textit{Using dense icons for iconic maps:} When designing iconic maps, people often used small and dense icons. Two participants mentioned to make an explicit effort to incorporate this design approach. We observed 8 out of 11 iconic maps with dense icons.

\textit{Connecting areas:} Two participants removed borders to allow the same patterns to connect between areas on a map. Unfortunately, \hty{this visual grouping of regions} in the maps (\autoref{fig:MI7} in \autoref{appendix:all-designs}; to some degree also MG4 in \autoref{tab:exp2-map-geo} or \autoref{fig:MG4} in \autoref{appendix:all-designs}) may be misleading, resembling a texture version of the rainbow color map \cite{Borland:2007:RCM}.

\textit{Avoiding negative texture effects:} Participants also employed various strategies to address the potential negative effects caused by textures. For example, one participant attempted to avoid the vibratory effect, while another was cautious not to incorporate too many patterns that could generate an aliasing effect. In addition, one participant adjusted the density of dot textures to minimize spatial density associations with adjacent textures, thereby reducing the adverse effect of densities altering the perception of grouping.

\subsubsection{Using geometric and iconic textures}
\label{sub-sec:compare-two-textures}
After designing textures for both geometric and iconic shapes, we asked participants to share their thoughts on using these texture types in data representation. The most notable difference that was mentioned by participants was the semantic association provided by iconic textures (10\texttimes), which made iconic textures self-explanatory. In addition, participants generally found geometric textures easier to handle (3\texttimes), had more variation (2\texttimes), and better for distinguishing bar chart columns (1\texttimes). Despite the novelty of iconic textures (2\texttimes), they were perceived as more cluttered and harder to read (7\texttimes).

\subsubsection{Application of textures to other charts}
We also applied the textures designed by participants to the two other chart types and asked the participants whether they thought that the textures still worked and to provide their reasoning. \hty{\autoref{tab:apply_to_another_charts} in \autoref{appendix:additional-exp1} summarizes the percentage of designs that participants considered to still work in each condition}. We can see that textures designed for bar and pie charts were rarely considered to work well on maps. Textures designed for maps, in contrast, were considered to be quite suitable for both bar and pie charts. The primary reason for this discrepancy is that the space available for filling textures in maps can be relatively small compared to bar and pie charts. Applying textures designed for bar and pie charts to maps can thus lead to visual clutter or generally bad readability. This observation highlights the significant impact that the available space in a chart has on the effective use of textures. Experts should therefore tailor their textures specifically to the target chart.

%\todo{Tobias is editing here.}

%% file: sections/experiment2_rate.tex
\section{Experiment 2: Rating}
\label{sec:experiment_rate}

After we collected a diverse set of texture designs, we wanted to know how the general public would experienced them in terms of their visual appeal.
%In our second experiment we thus looked for the visually most preferred texture design for each chart type as well as compared the visual appeal of geometric and iconic textures. 
We asked participants about the collected designs' aesthetics, vibratory effect, and overall preference. We included questions about the vibratory effect because it is a well-known negative effect that textures can produce. According to some experts \cite{Bertin:1998:SG,bertin:1983:semiology,tufte:1985:visual}, its negative impact makes the use of textures for visualization undesirable (see \autoref{sub-sec:texture-for-vis}). This experiment was also pre-registered (\href{https://osf.io/nyru7/}{\texttt{osf.io/nyru7}}) and IRB-approved (Inria COERLE, avis \textnumero\ 2023-01).

\subsection{Participants}
We recruited 150 valid participants (fluent English speakers, of legal age---18 years in most countries) through the Prolific platform. Participants received a compensation equivalent to \EUR10.20 per hour.

\subsection{Stimuli selection}
To avoid a lengthy experiment and given the similarity between some designs, we chose a subset of aesthetically appealing designs that represented a diverse range of aesthetic styles for our experiment.
%
 %As previously mentioned in Section \autoref{sec:experiment_design}, we collected 66 designs during Experiment 1. To avoid an overly lengthy experiment and given the similarity between some designs, our team discussed and aimed to select the most representative designs for inclusion in the second experiment. Our objective was to choose aesthetically appealing designs that represented a diverse range of aesthetic styles.
%
To facilitate this selection process, we first printed each of the 66 designs from Experiment~1 (\autoref{appendix:all-designs}) using the default dataset on individual A4 paper sheets. Subsequently, we classified these designs based on their distinguishing aesthetic characteristics. Some of these attributes included unique texture properties such as the use of a predominantly black background or overlapping icons. We also looked at the overall impression the design conveyed such as an appearance of regularity or a sense of calmness. While this classification process was inherently subjective, we made an effort to ensure a balanced representation of various aesthetic styles. After identifying different styles, we selected 24 design  we considered aesthetically pleasing; with four images representing each combination of chart type and texture type (see Tables~\ref{tab:exp2-bar-geo}--\ref{tab:exp2-map-icon}).

\subsection{Method}

We employed a mixed design using the between-subjects variable \emph{chart type} (bar, pie, map) and the within-subjects variable \emph{texture type} (geometric, iconic). We randomly assigned participants to one chart type.

We started the study by asking participants to complete a consent form and to provide their background information. We then gave them a brief explanation of the vibratory effect, \hty{and instructed them to focus only on the visual appearance of the charts.} Subsequently, we asked them to evaluate a total of eight images, presented in two separate blocks: one containing geometric and the other iconic textures. Each block contained four images, with the block order and the images within them randomized. For each block, we asked participants to rate the aesthetics of each visualization using a 7-point Likert scale via the 5 items of the BeauVis scale \cite{He:2023:BVS} and added 1 item to assess the degree to which they perceived a vibratory effect. We included one attention check question in this section. Following the rating section, we asked participants to rank the four visualizations they had just evaluated based on their overall preference. In addition, we asked them to provide a rationale for their selection of the highest-ranking visualization. 

\subsection{Data analysis and interpretation}
\label{sub-sec:exp2-data-analysis}
%\textbf{Data Analysis.}

For each design, we computed the BeauVis score as the mean of the five BeauVis Likert items. We then calculated the average BeauVis and vibratory scores for each design across all participants. We also counted the number of times a design was ranked first for overall preference.

For each chart type, we also computed the average BeauVis score for both the four geometric and the four iconic texture designs per participant. We report the sample means of BeauVis scores along with their 95\% Bootstrap confidence intervals (CIs; 10,000 bootstrap iterations, indicating that we have 95\% confidence that the calculated interval encompasses the population mean). We also first averaged the question on the vibratory effect across the four images per texture type, and then across all participants, and report the sample mean with its 95\% CI. We present the CIs of the mean differences between two texture types for each chart type for BeauVis score and vibratory score.
%We also present the CIs of the mean differences between two texture types for each chart type for BeauVis score and vibratory score, adjusted using the Bonferroni correction \cite{higgins:2004:introduction}.

%\textbf{Interpretation Methods.}
In our analysis, we derive inferences from the graphically presented point estimates and interval estimates, thus eliminating the need for conducting significance tests or reporting $p$-values. As suggested in the literature \cite{besanccon:2019:continued, cockburn:2020:threats, cumming:2013:understanding,dragicevic:2016:fair, blascheck:2023:studies}, we interpret CIs as providing different levels of evidence for the population mean. To compare different techniques, we examine the CIs of mean differences. When the CI bar of the mean difference between two techniques does not intersect with 0, we can conclude that there is evidence of a difference between these two techniques, which is equivalent to the results being statistically significant in traditional $p$-value tests.

\newlength{\tableimageheight}%
\setlength{\tableimageheight}{0.18\columnwidth}%
\begin{table}[t]
\centering%
\footnotesize%
\caption{BeauVis score with distribution, \# ranked first \hty{(total: 53)}, and vibratory score for geom.\ bars BG1--4 (left--right; larger in \autoref{appendix:all-designs}).}\vspace{-1ex}
\label{tab:exp2-bar-geo}
\begin{tabu}{lllll}
\toprule
              & BG1    & BG2    & BG3   & BG4   \\
							\midrule
BeauVis \hty{(1--7)}      & 4.70 \includegraphics[height= 5mm]{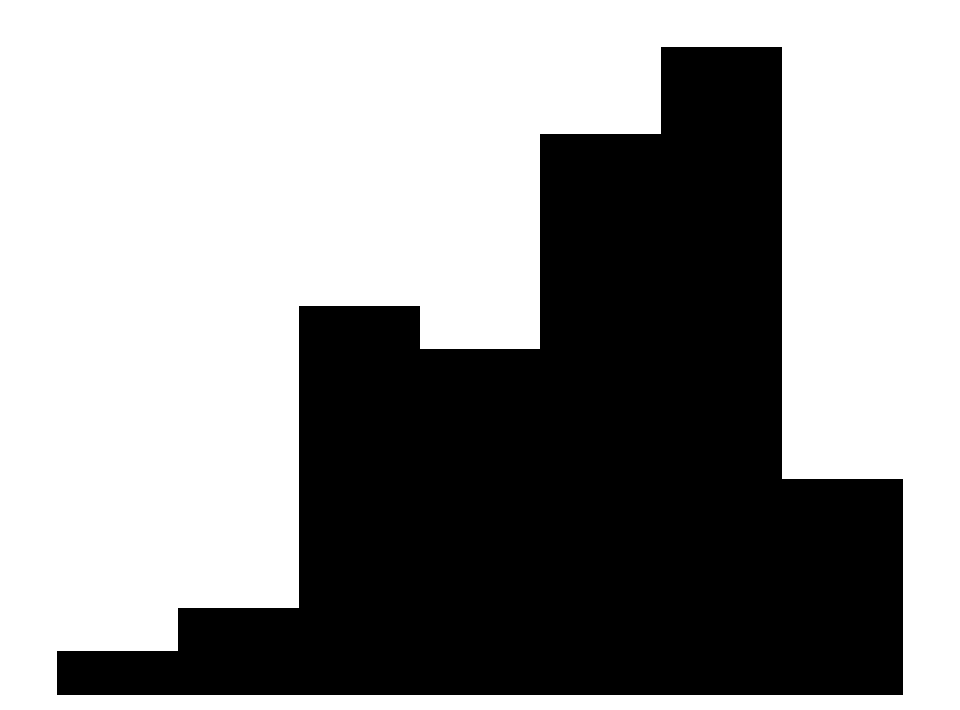}   & 4.45 \includegraphics[height= 5mm]{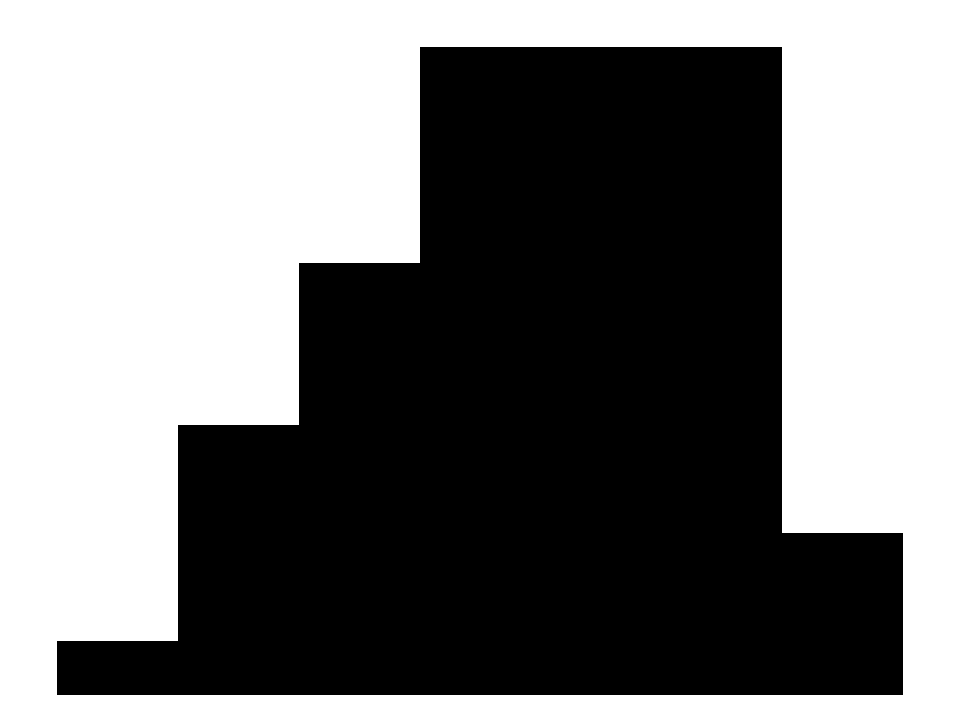}   & 3.92 \includegraphics[height= 5mm]{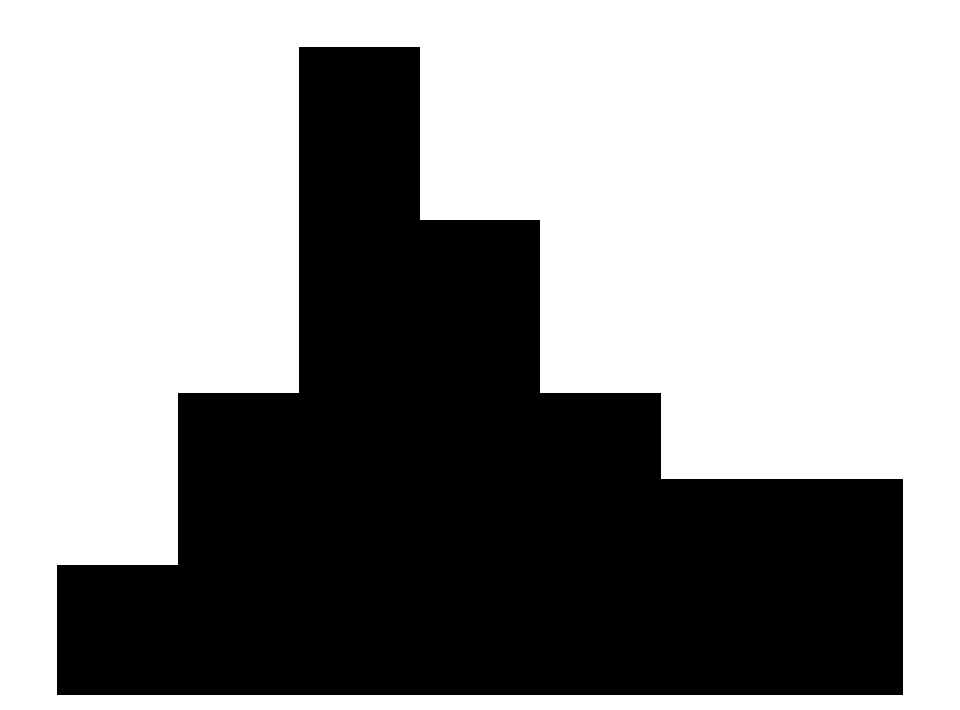}   & 3.84 \includegraphics[height= 5mm]{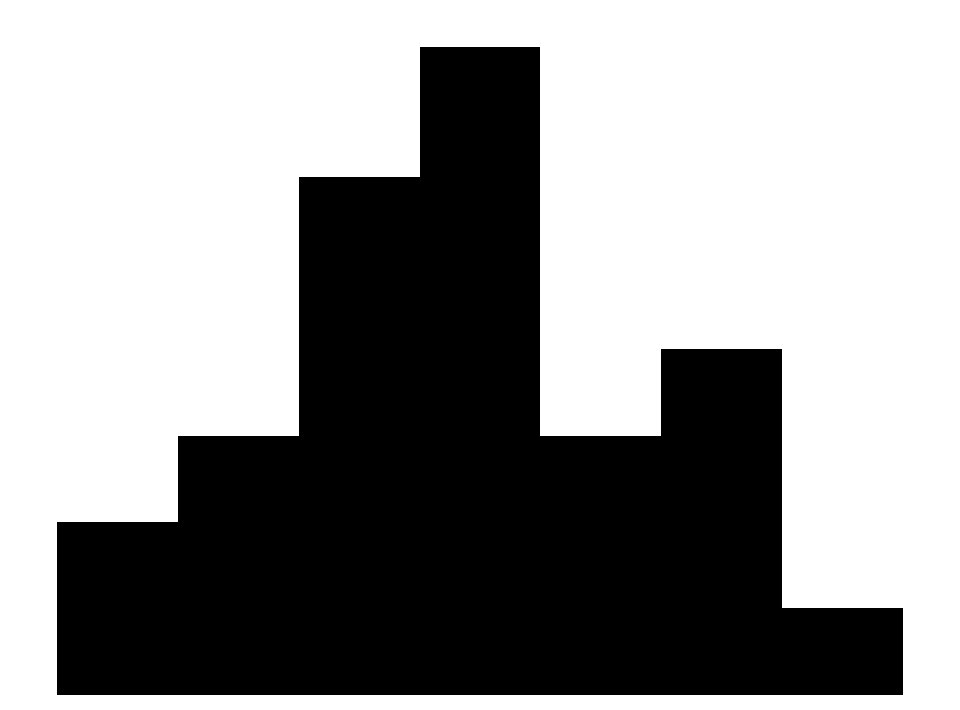}   \\
ranked first & ~16      & ~20      & ~13      & ~~~4       \\
vibratory \hty{(1--7)}    & 3.83    & 3.66    & 3.00    & 5.13   \\
\bottomrule
\end{tabu}\\[1ex]%
\includegraphics[height=\tableimageheight]{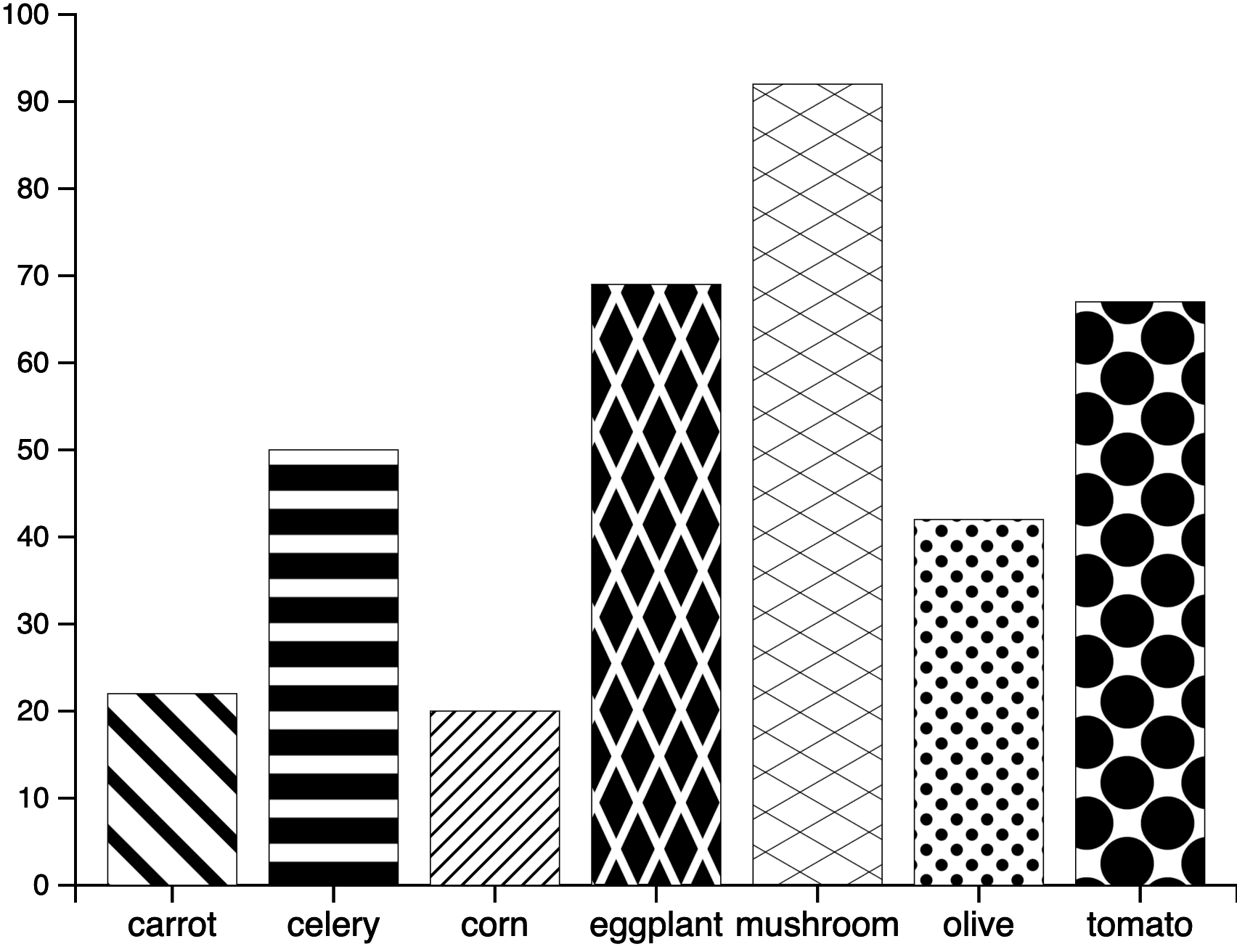}\hfill%BarGeo3
\includegraphics[height=\tableimageheight]{figures/experiment1/collected_designs/BG2.png}\hfill%BarGeo4
\includegraphics[height=\tableimageheight]{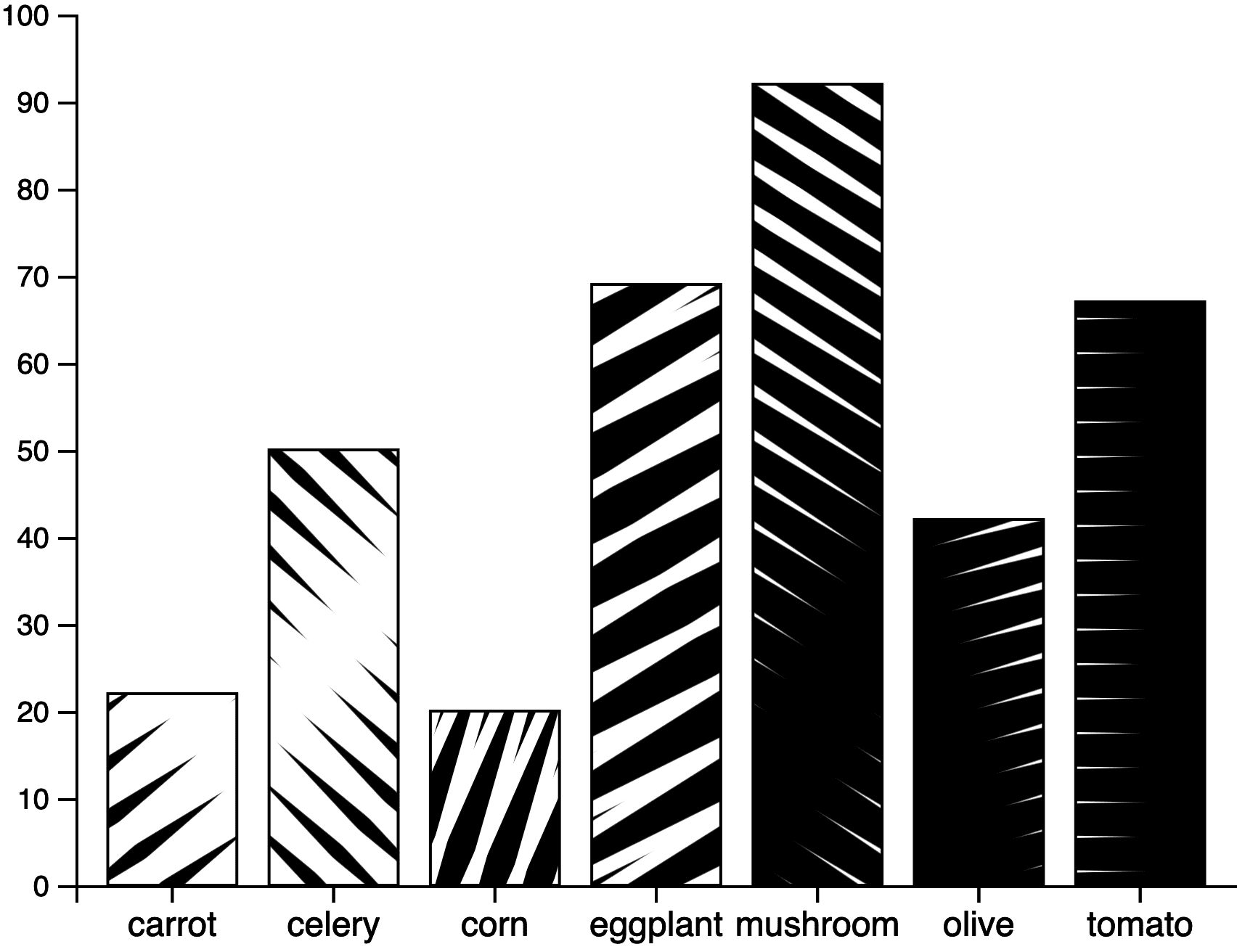}\hfill%BarGeo2
\includegraphics[height=\tableimageheight]{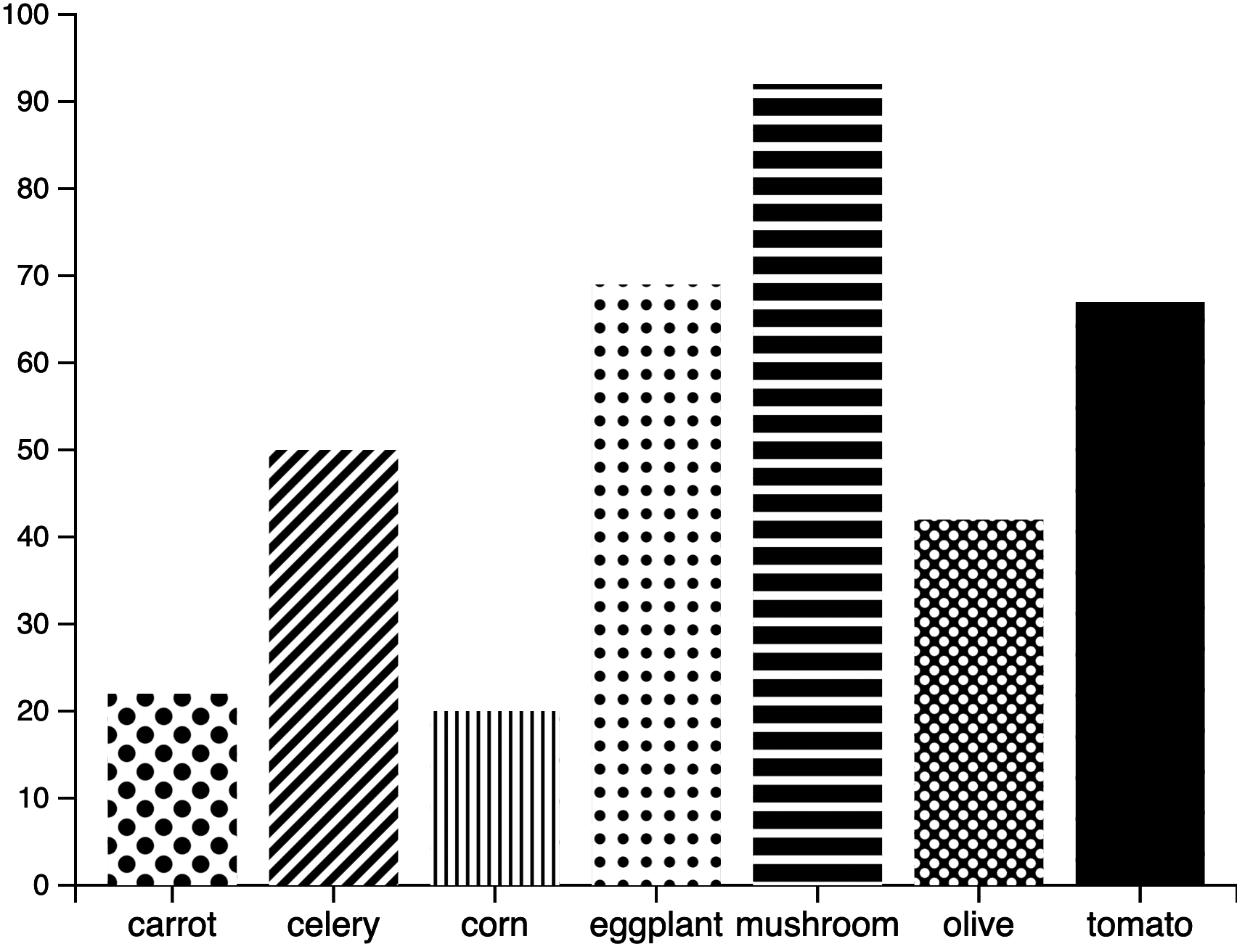}%BarGeo1
\vspace{0.5ex}
\end{table}

\begin{table}[t]
\centering%
\footnotesize%
\caption{BeauVis score with distribution, \# ranked first \hty{(total: 53)}, and vibratory score for iconic bars BI1--4 (left--right; larger in \autoref{appendix:all-designs}).}\vspace{-1ex}
\label{tab:exp2-bar-icon}
\begin{tabu}{lllll}
\toprule
              & BI1    & BI2    & BI3     & BI4    \\
              \midrule
BeauVis \hty{(1--7)}      & 5.07 \includegraphics[height= 5mm]{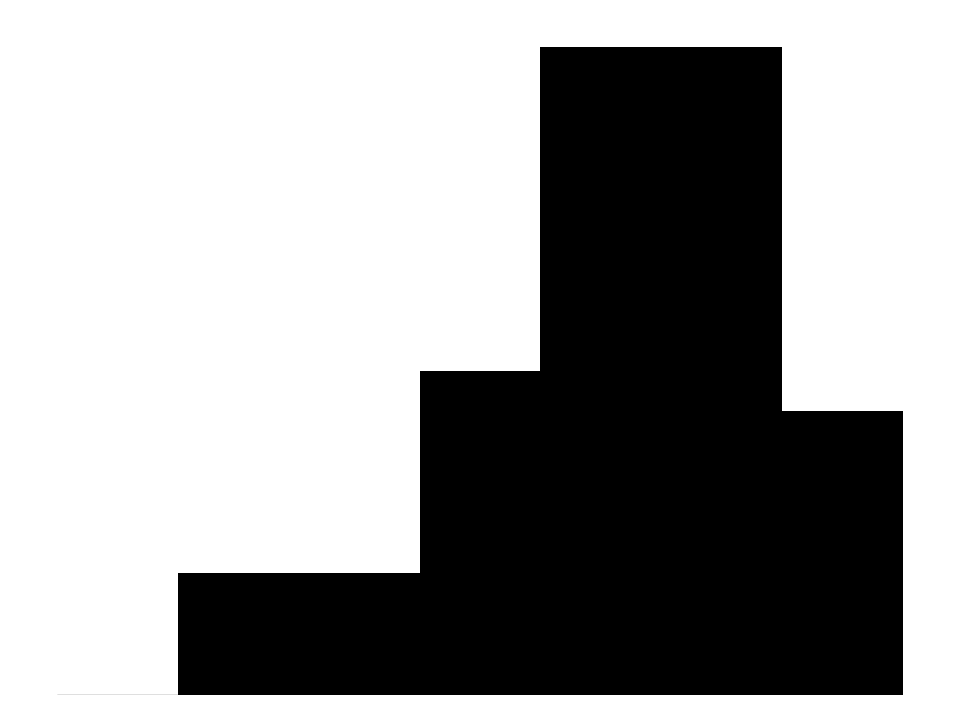}   & 4.71 \includegraphics[height= 5mm]{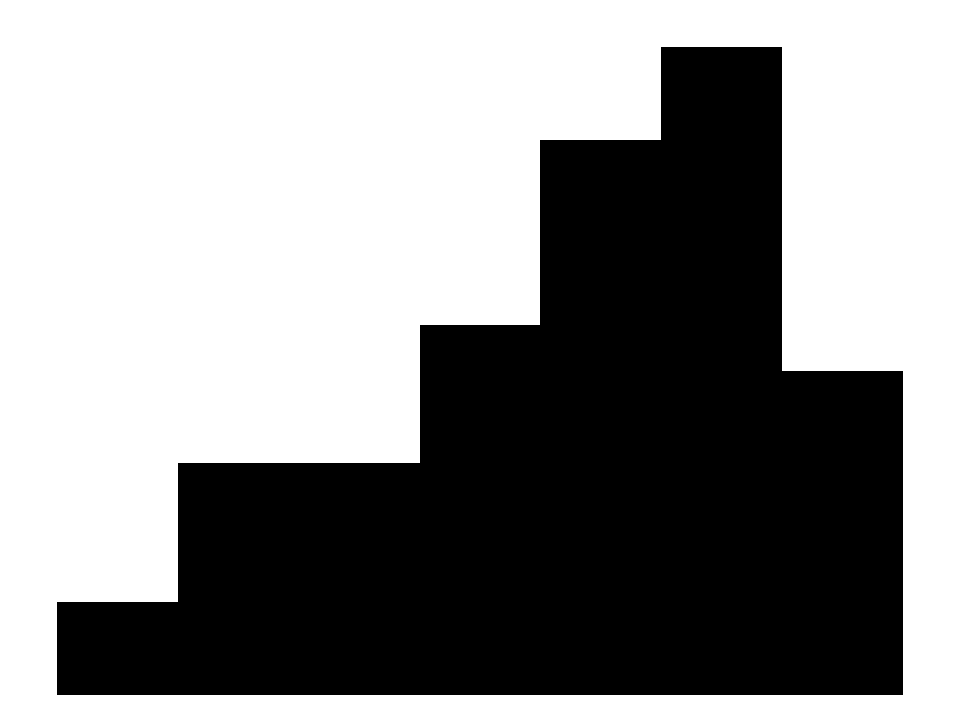}  & 4.29 \includegraphics[height= 5mm]{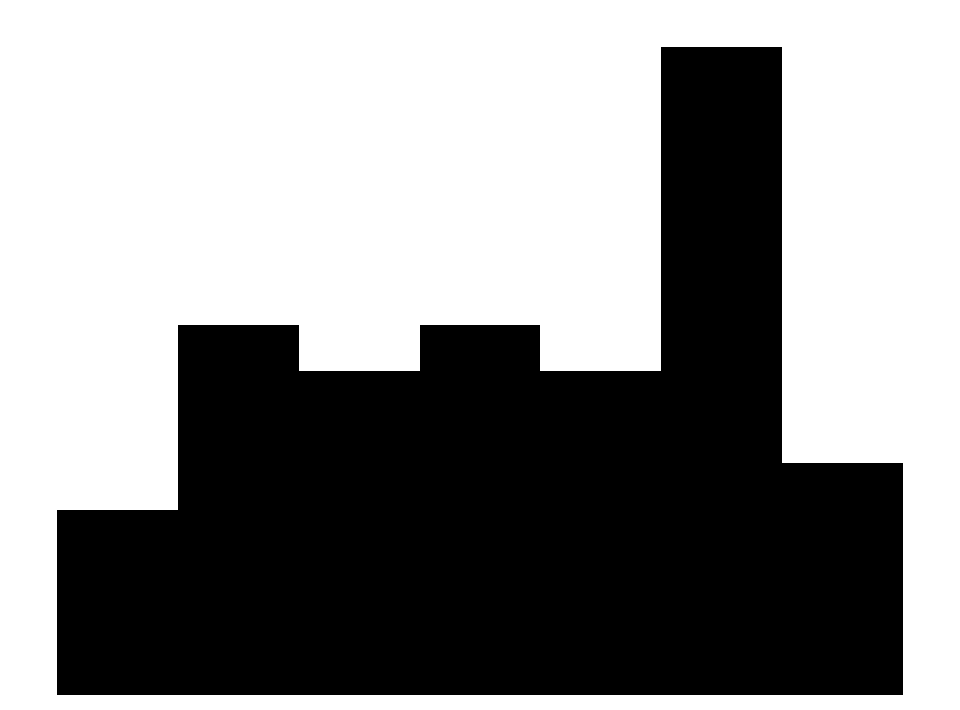}   & 3.79 \includegraphics[height= 5mm]{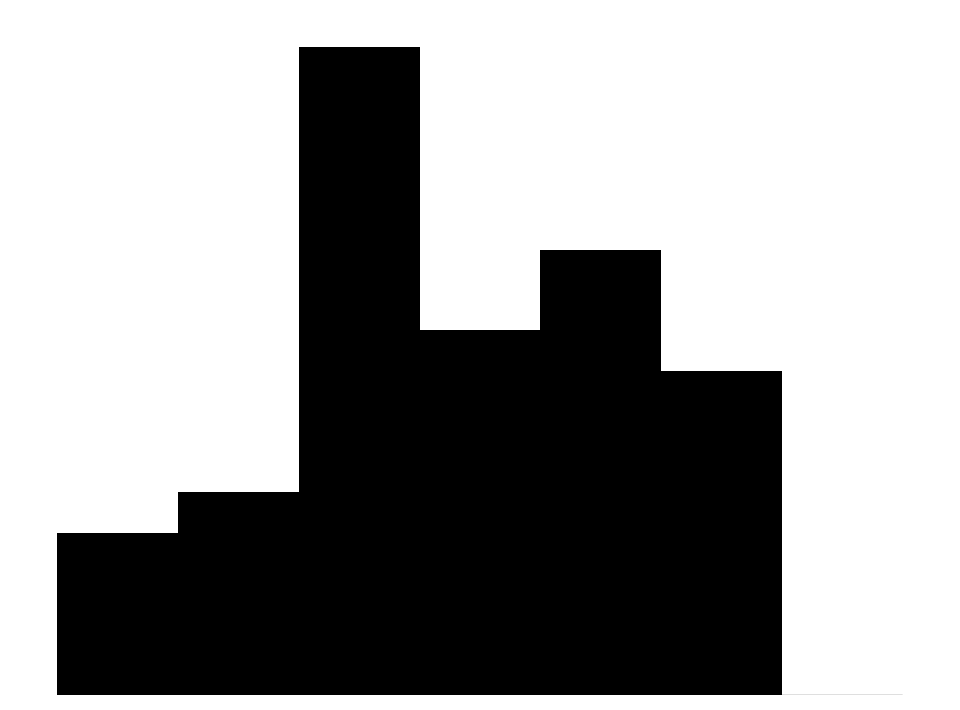}  \\
ranked first & ~16      & ~13      & ~19      & ~~~5       \\
vibratory \hty{(1--7)}    & 2.89   & 2.02    & 3.42    & 2.92  \\
\bottomrule
\end{tabu}\\[1ex]%
\includegraphics[height=\tableimageheight]{figures/experiment1/collected_designs/BI1.png}\hfill%BarIcon3
\includegraphics[height=\tableimageheight]{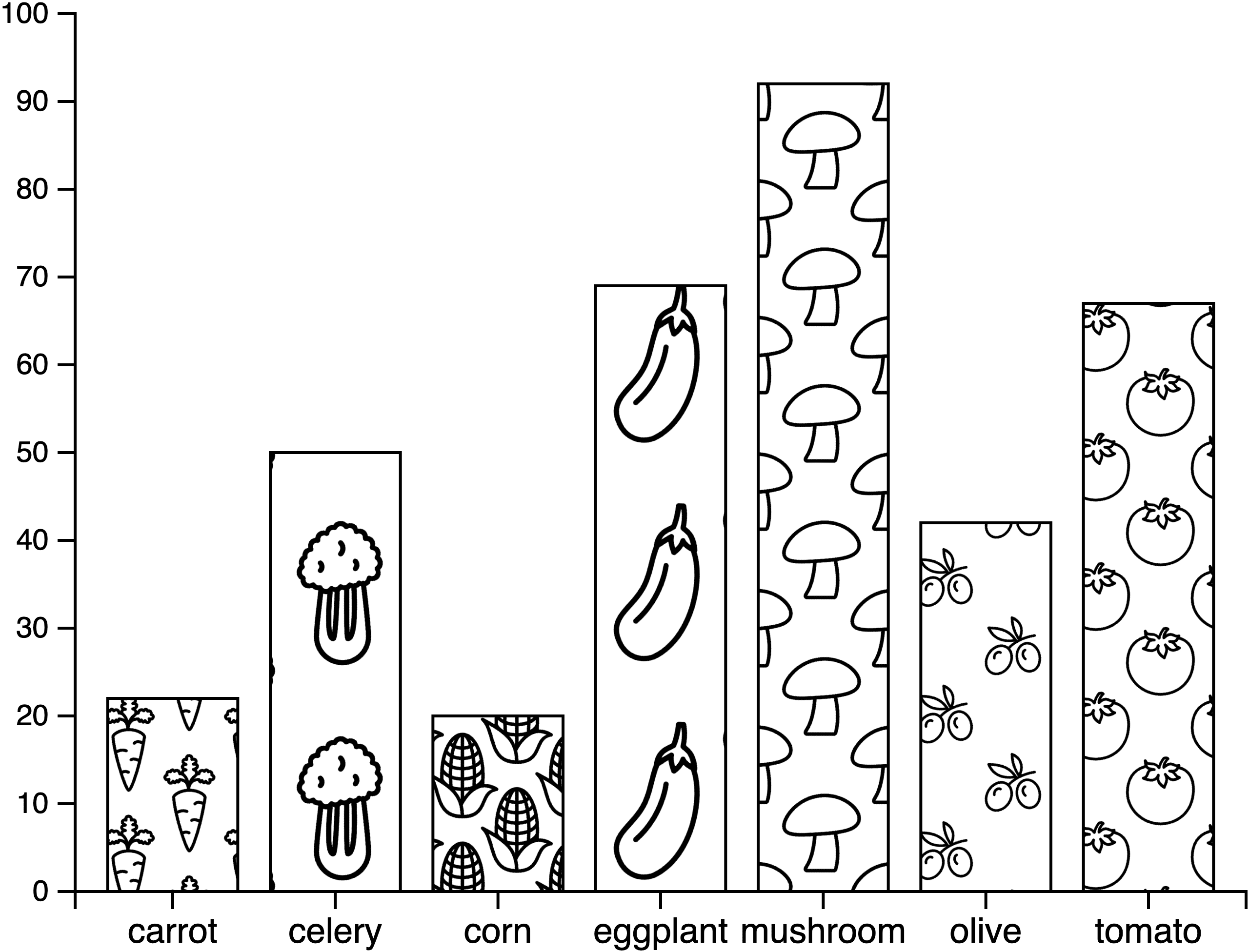}\hfill%BarIcon1
\includegraphics[height=\tableimageheight]{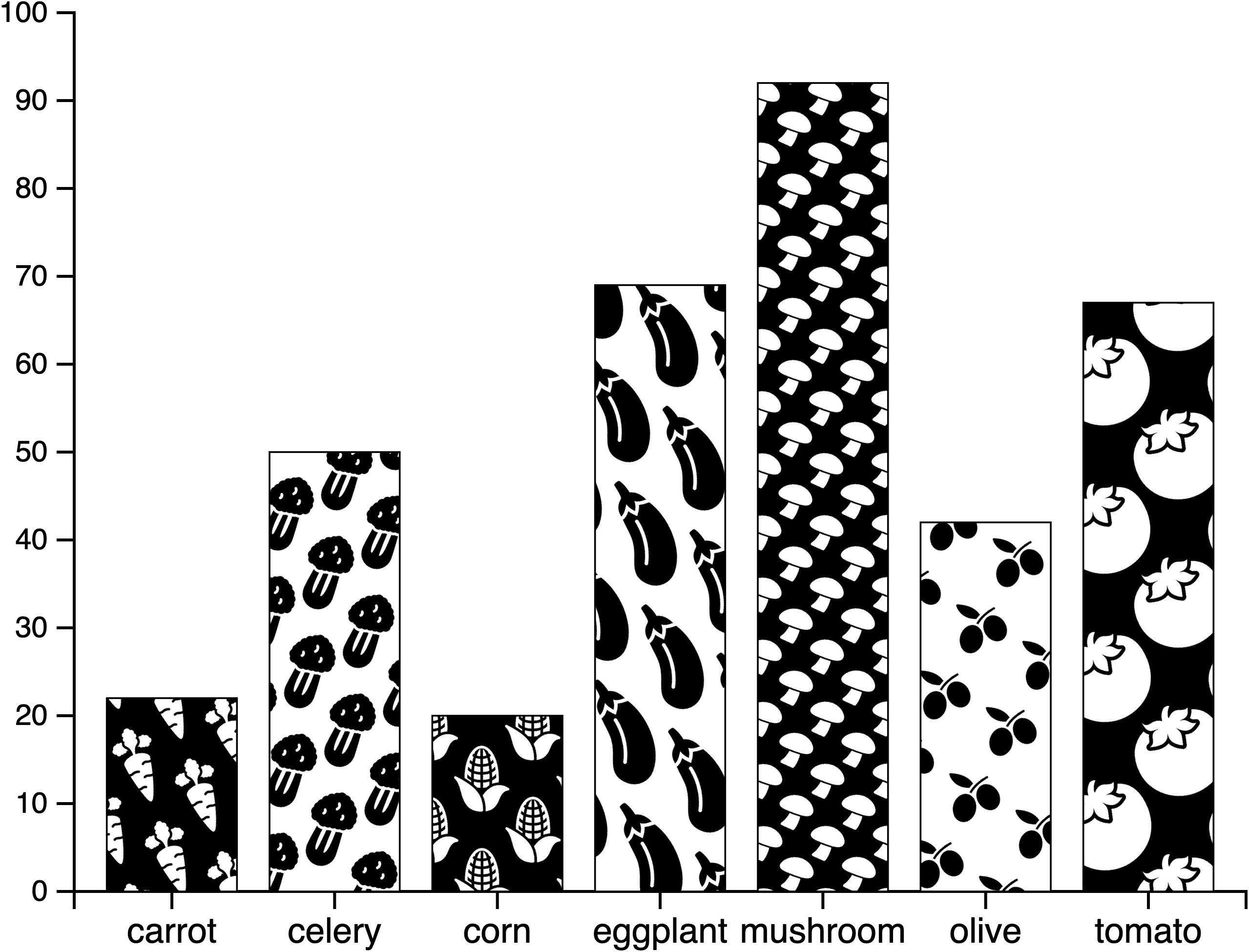}\hfill%BarIcon2
\includegraphics[height=\tableimageheight]{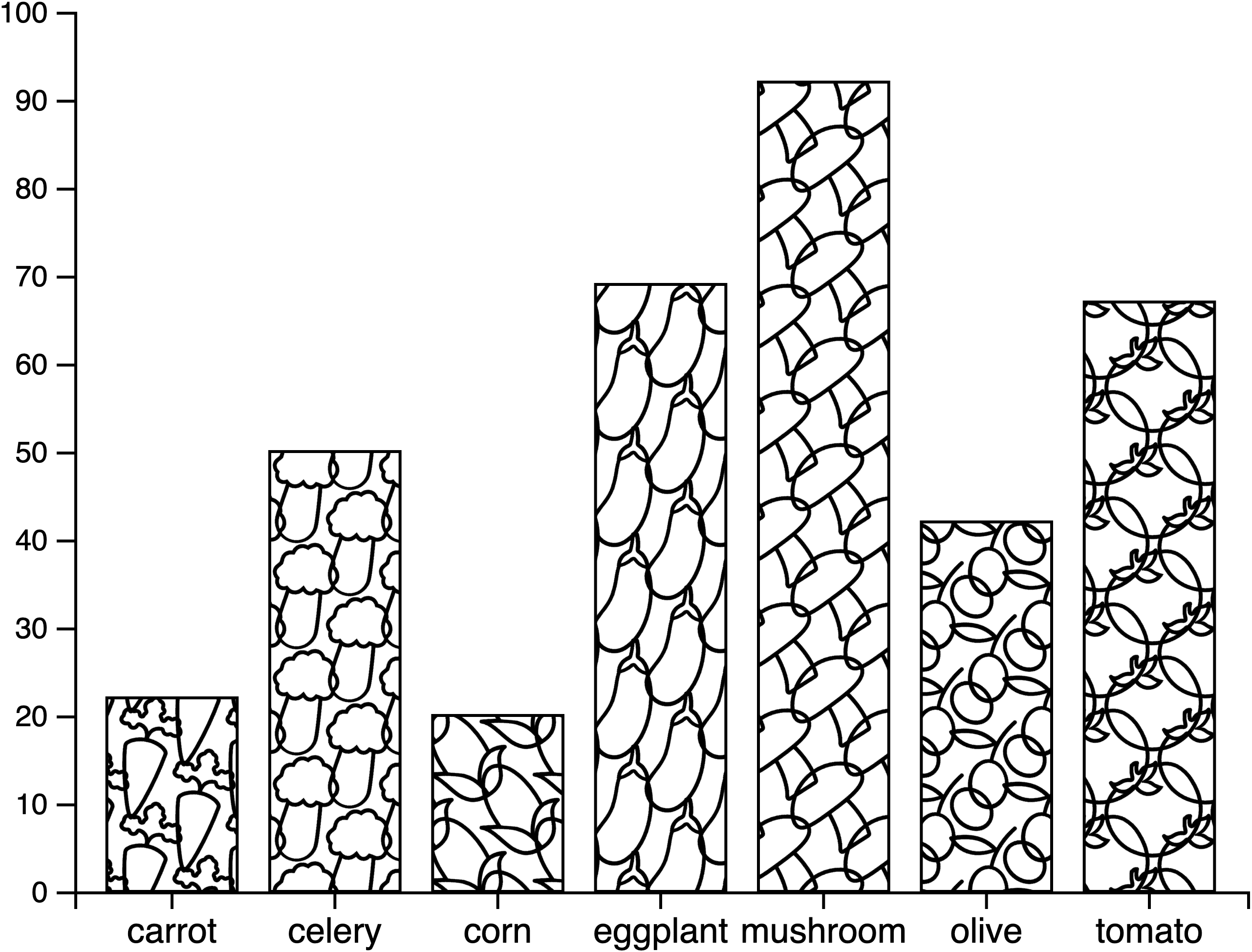}%BarIcon4
%\vspace{0.5ex}
\end{table}

\subsection{Results}

We received 170 responses from Prolific. After excluding those who failed our attention check question, we obtained 150 valid responses for our analysis (75 female, 75 male; ages: mean $=$ 28.2, SD $=$ 8.9; education: 87 Bachelor or equivalent, 27 Master's or equivalent, 3 PhD or equivalent, 33 other). Among them, 53 participated in the bar condition, 44 in the pie condition, and 53 in the map condition. 

Tables~\ref{tab:exp2-bar-geo}--\ref{tab:exp2-map-icon} show the BeauVis score (and its respective distribution), the number of times a design was ranked first, and the vibratory score for each design. While we calculated these scores primarily for selecting stimuli for our next experiment (\autoref{sec:experiment3}), they can also provide insight into the general public's opinion on each design. 

Looking at the pairwise differences of the two texture types for each chart (\autoref{fig:exp2-beauvis} and~\ref{fig:exp2-vibratory}), we only found evidence of a difference between iconic and geometric textures for maps where geometric textures were perceived as more aesthetically pleasing. Participants perceived iconic textures to have a lower vibratory effect than geometric textures across all three chart types. The average BeauVis scores were lowest for the iconic maps at just below average on the 7-point scale and hovered around or just above average for most other designs. The chart with the highest BeauVis score was an iconic bar chart with a rating of 5.07 on average.
This finding is particularly intriguing, prompting us to delve deeper into the data to examine the distribution of BeauVis scores for each design, which we included as word-scale histogram visualizations \cite{Goffin:2014:ETP} alongside the BeauVis scores in Tables~\ref{tab:exp2-bar-geo}--\ref{tab:exp2-map-icon}. From these we see that, in each condition except for iconic maps, the highest average score one (located on the leftmost side of the table) all have a normal-like BeauVis score distribution, which means \hty{that people's opinions are consistent}. This consistency gives us confidence in utilizing the BeauVis score as a reliable reference indicator for selecting the most suitable texture within each condition to serve as stimuli. But we can also see that opinions diverge for designs that received lower average scores, such as BI3 and BI4. Notably, the BeauVis score distributions for PI3, PI4, and MI1 are uniform or even bimodal, suggesting that participants hold varying views about these designs. \hty{Therefore, textures with lower average scores should not be directly counted as bad since they may appeal to certain individuals, as also demonstrated by the fact that each chart was ranked as the top choice by some participants.}

\setlength{\tableimageheight}{0.22\columnwidth}%
\begin{table}[t]
\centering%
\footnotesize%
\caption{BeauVis score  with distribution, \# ranked first \hty{(total: 44)}, and vibratory score for geometric pies PG1--4 (left--right; larger in \autoref{appendix:all-designs}).}\vspace{-1ex}
\label{tab:exp2-pie-geo}
\begin{tabu}{lllll}
\toprule
              & PG1    & PG2    & PG3     & PG4    \\
              \midrule
BeauVis \hty{(1--7)}      & 4.95 \includegraphics[height= 5mm]{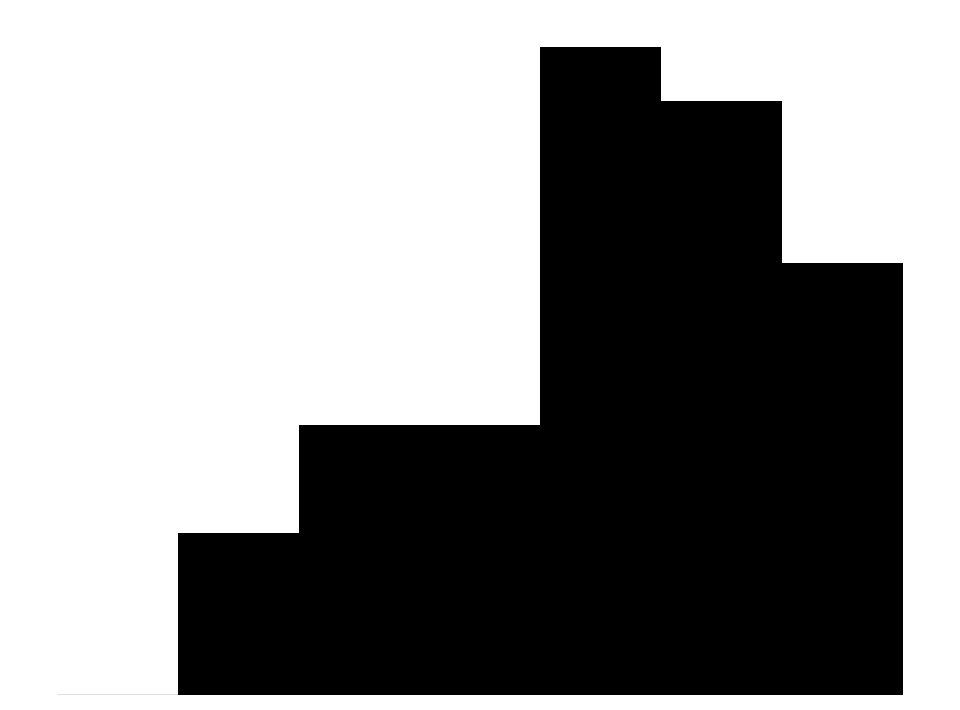}   & 4.40 \includegraphics[height= 5mm]{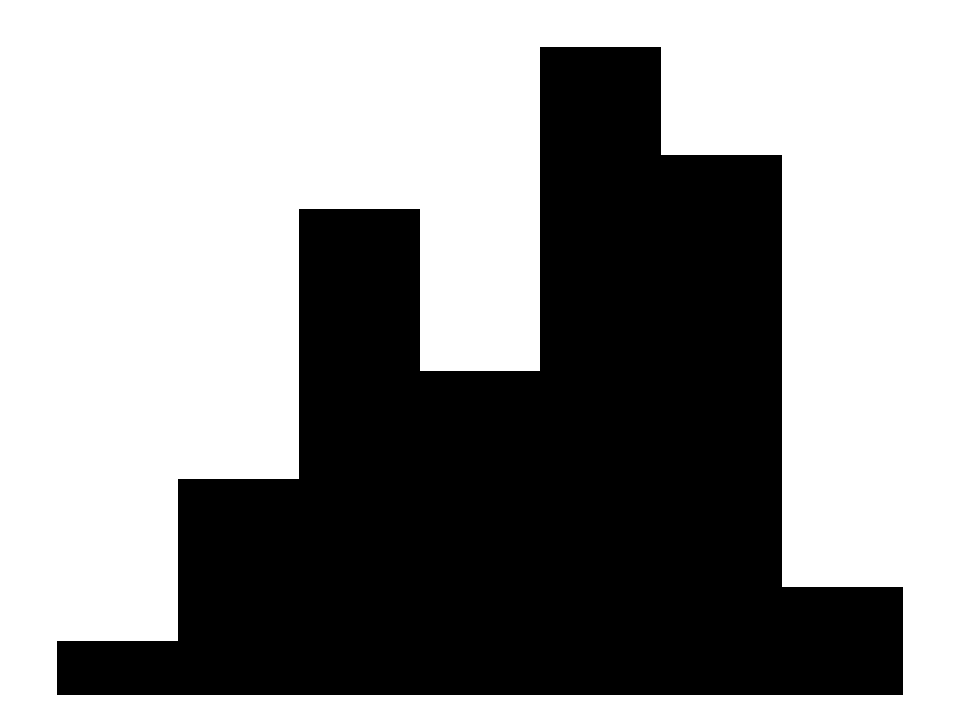}  & 4.37 \includegraphics[height= 5mm]{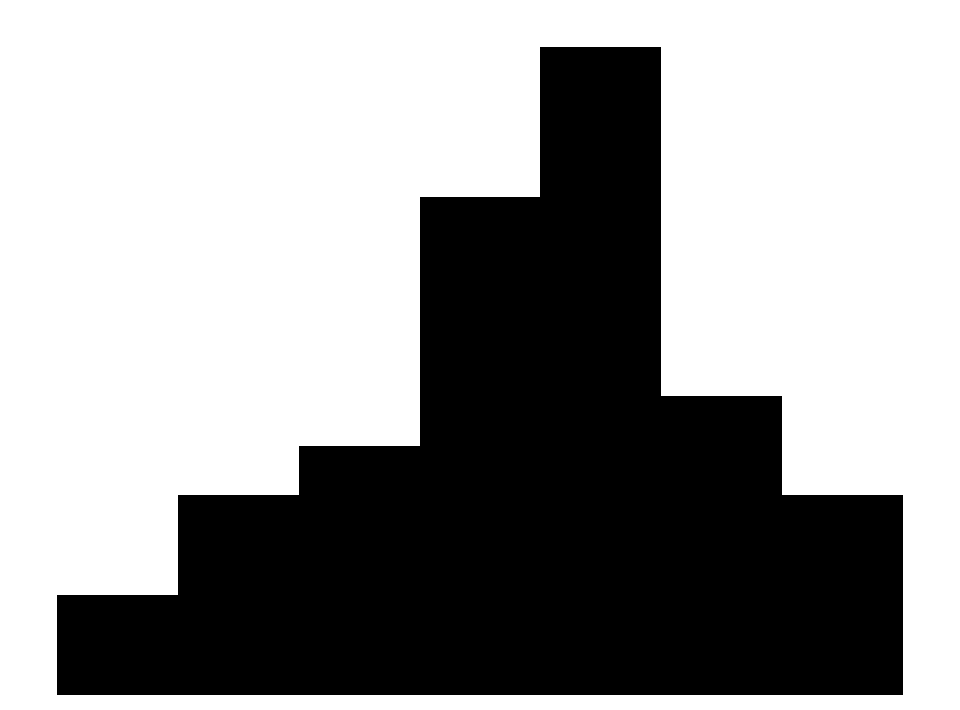}   & 4.33 \includegraphics[height= 5mm]{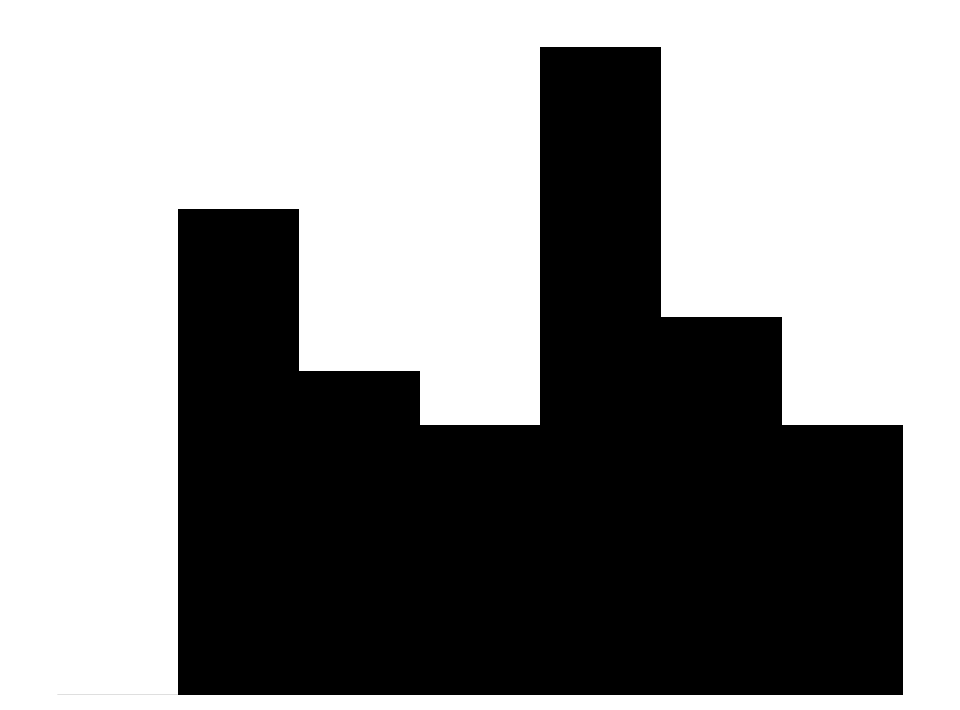}  \\
ranked first & ~17      & ~13      & ~~~4      & ~10      \\
vibratory \hty{(1--7)}    & 4.30   & 3.73    & 5.02    & 3.64 \\
\bottomrule
\end{tabu}\\[1ex]%
\includegraphics[height=\tableimageheight]{figures/experiment1/collected_designs/PG1_nolabel.png}\hfill%PieGeo2
\includegraphics[height=\tableimageheight]{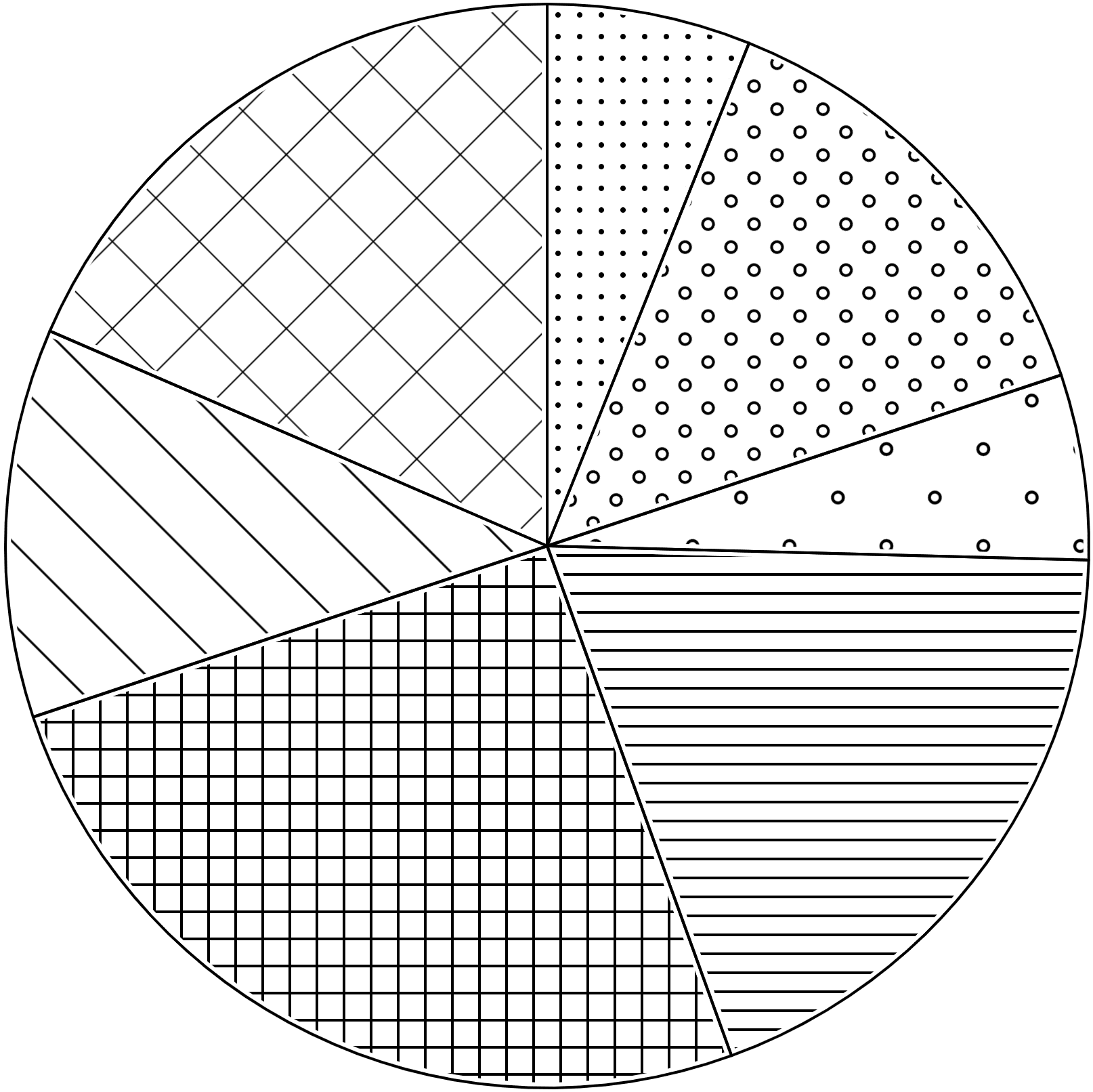}\hfill%PieGeo3
\includegraphics[height=\tableimageheight]{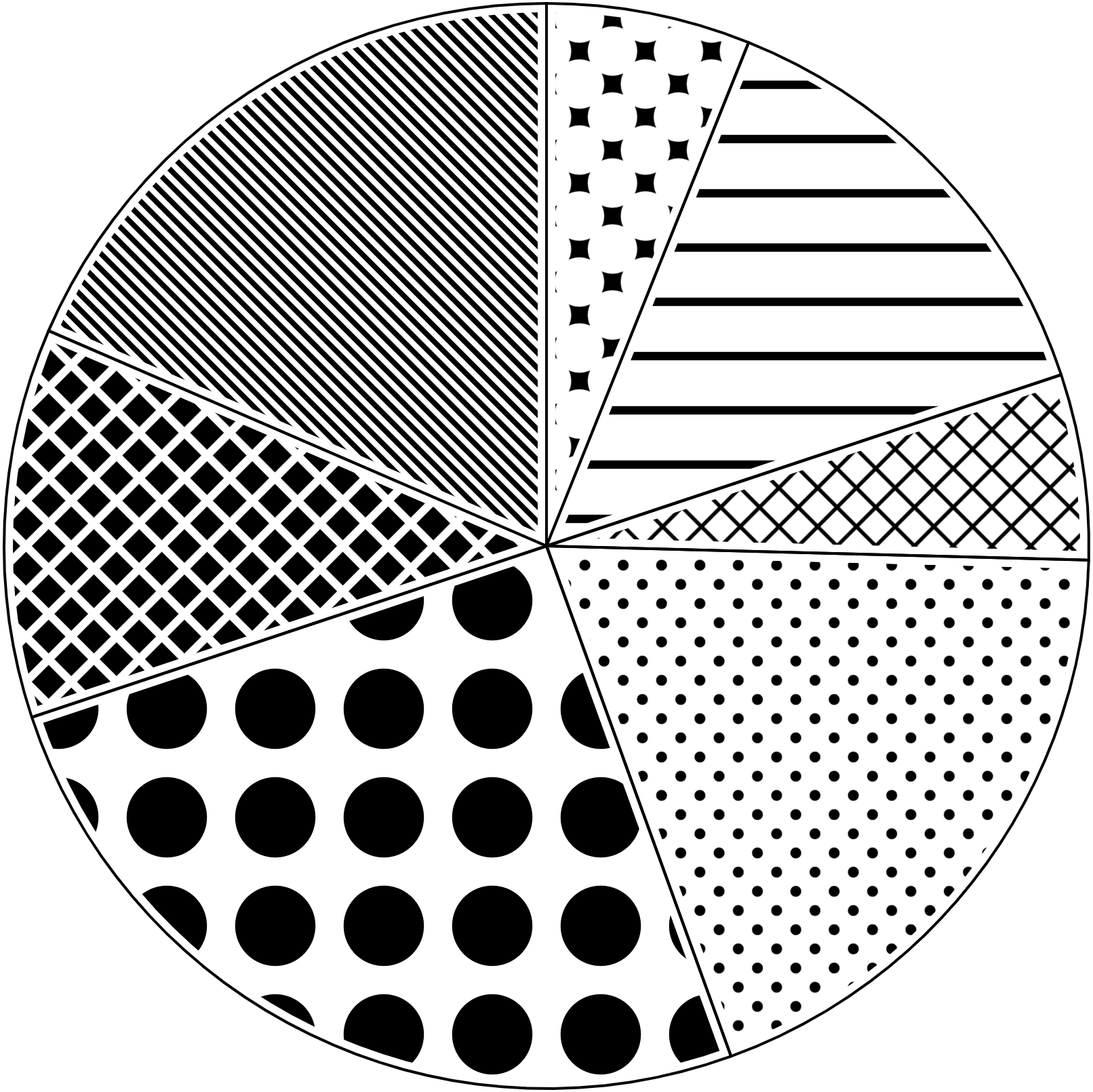}\hfill%PieGeo1
\includegraphics[height=\tableimageheight]{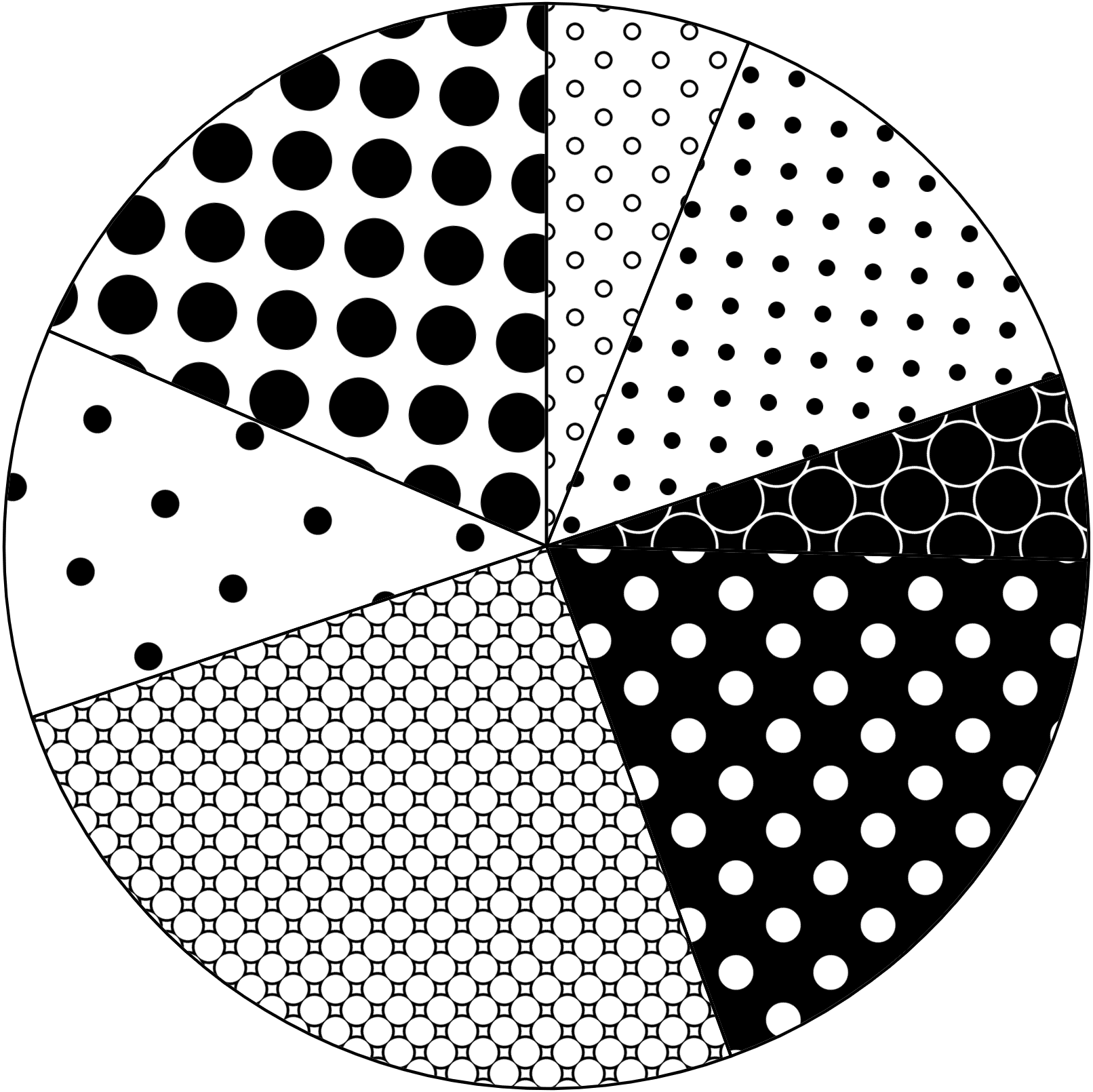}%PieGeo4
\vspace{0.5ex}
\end{table}

\begin{table}[t]
\centering%
\footnotesize%
\caption{BeauVis score with distribution, \# ranked first \hty{(total: 44)}, and vibratory score for iconic pies PI1--4 (left--right; larger in \autoref{appendix:all-designs}).}\vspace{-1ex}
\label{tab:exp2-pie-icon}
\begin{tabu}{lllll}
\toprule
              & PI1    & PI2    & PI3     & PI4    \\
              \midrule
BeauVis \hty{(1--7)}      & 4.81 \includegraphics[height= 5mm]{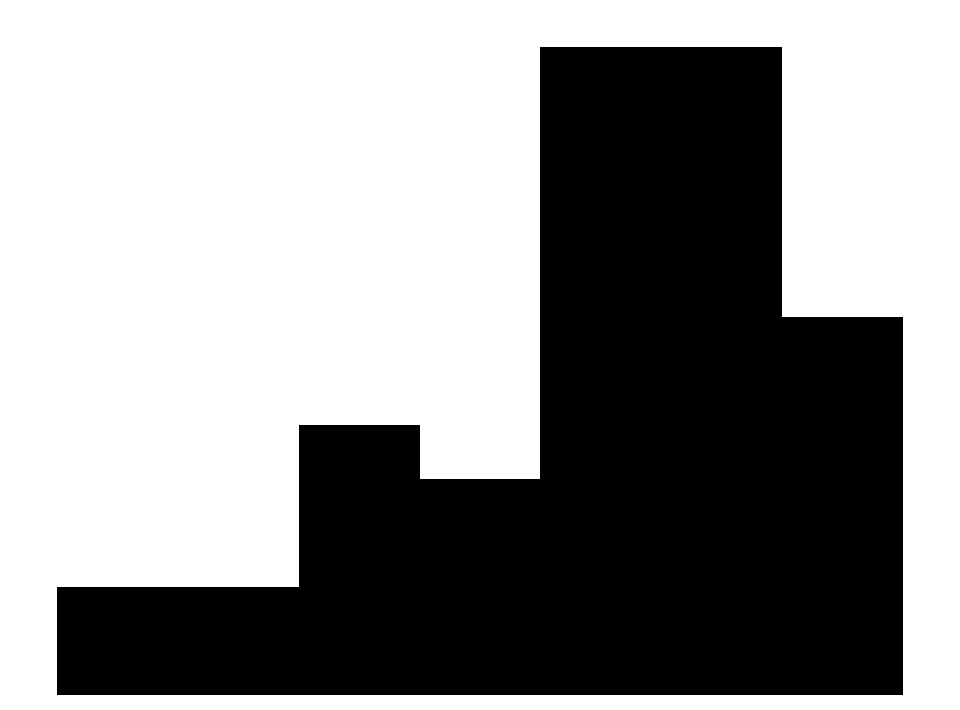}   & 4.69 \includegraphics[height= 5mm]{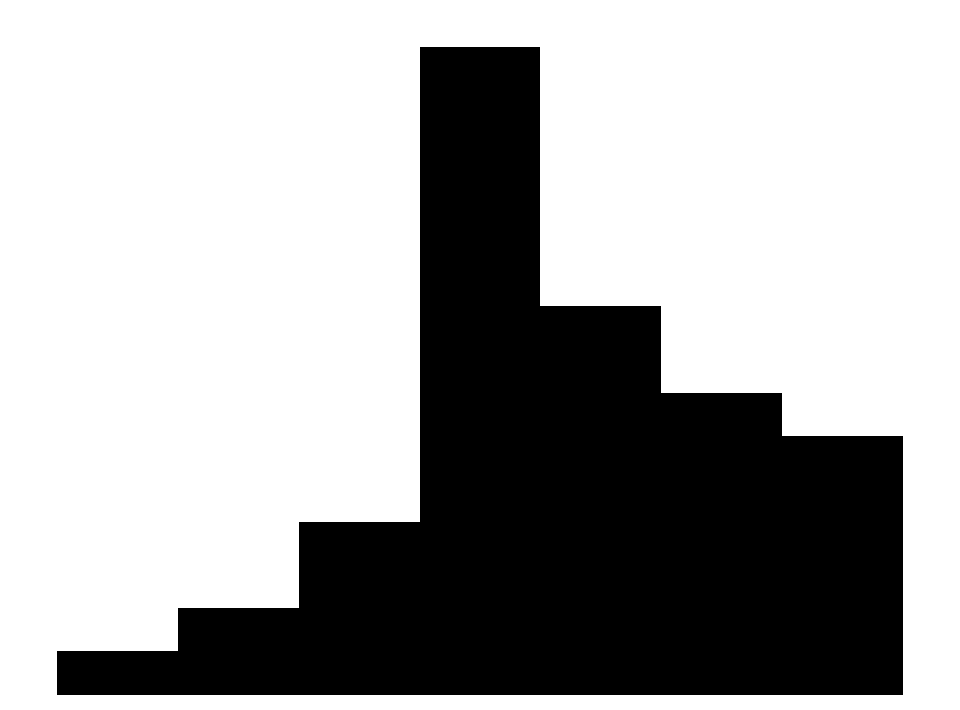}   & 4.60 \includegraphics[height= 5mm]{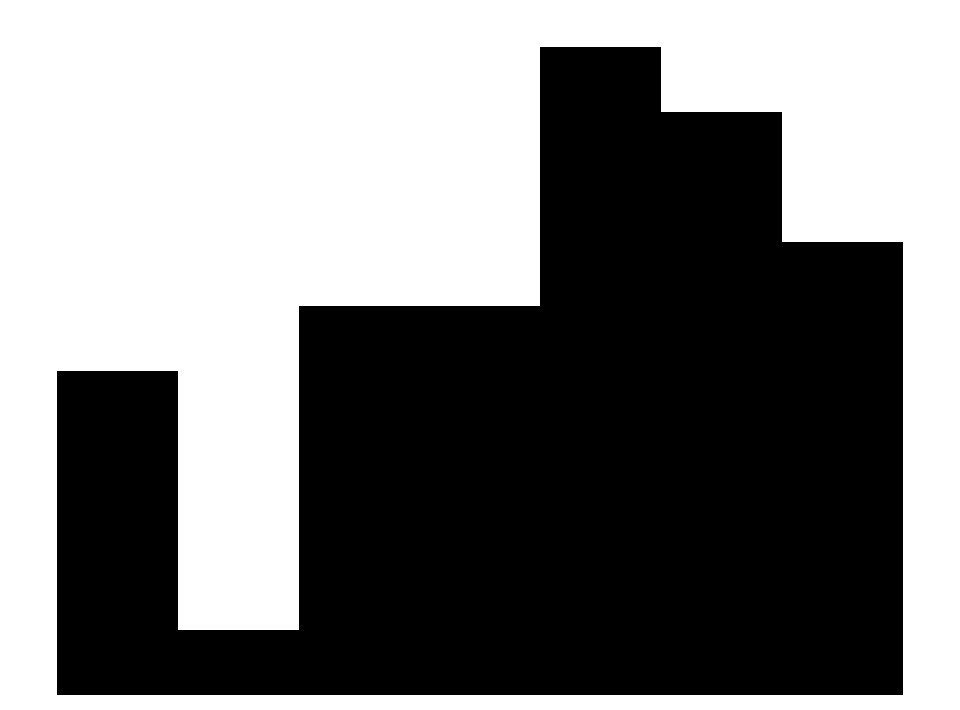}   & 4.48 \includegraphics[height= 5mm]{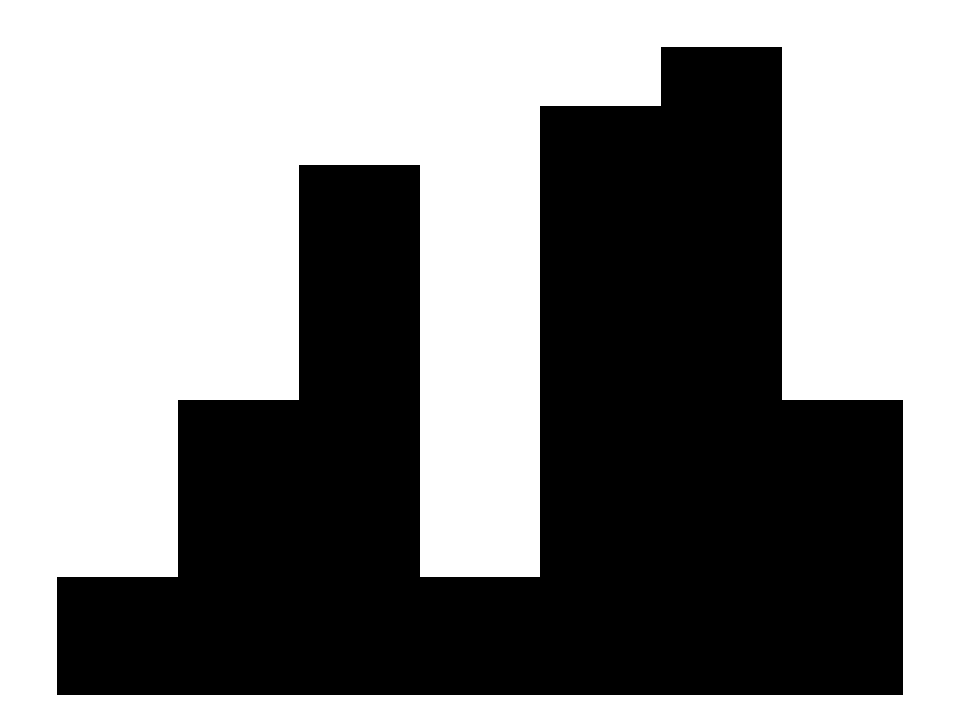}   \\
ranked first & ~13      & ~~~9      & ~10      & ~12       \\
vibratory \hty{(1--7)}    & 2.55    & 2.95    & 2.59    & 3.57   \\
\bottomrule
\end{tabu}\\[1ex]%
\includegraphics[height=\tableimageheight]{figures/experiment1/collected_designs/PI1_nolabel.png}\hfill%PieIcon1
\includegraphics[height=\tableimageheight]{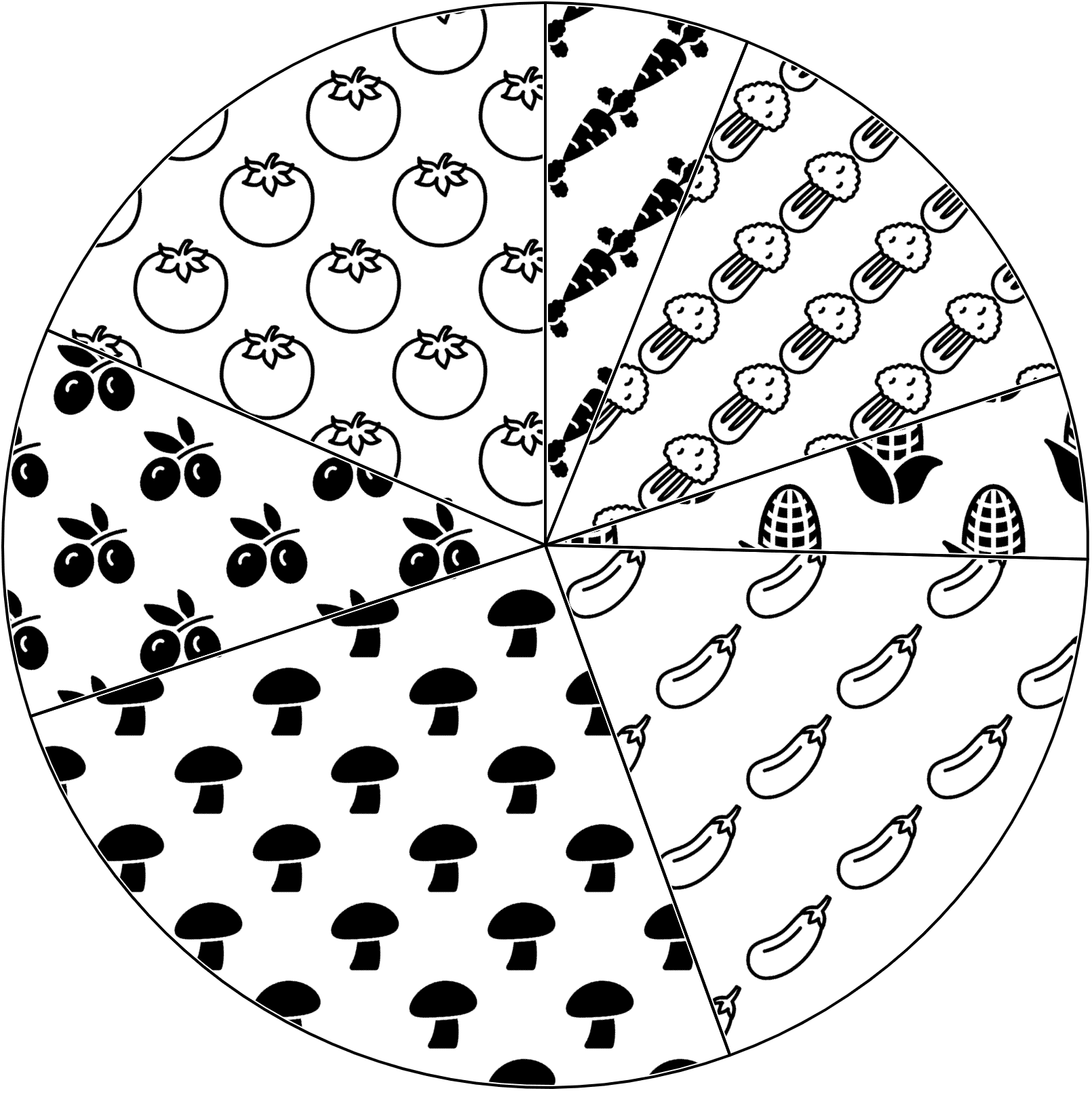}\hfill%PieIcon3
\includegraphics[height=\tableimageheight]{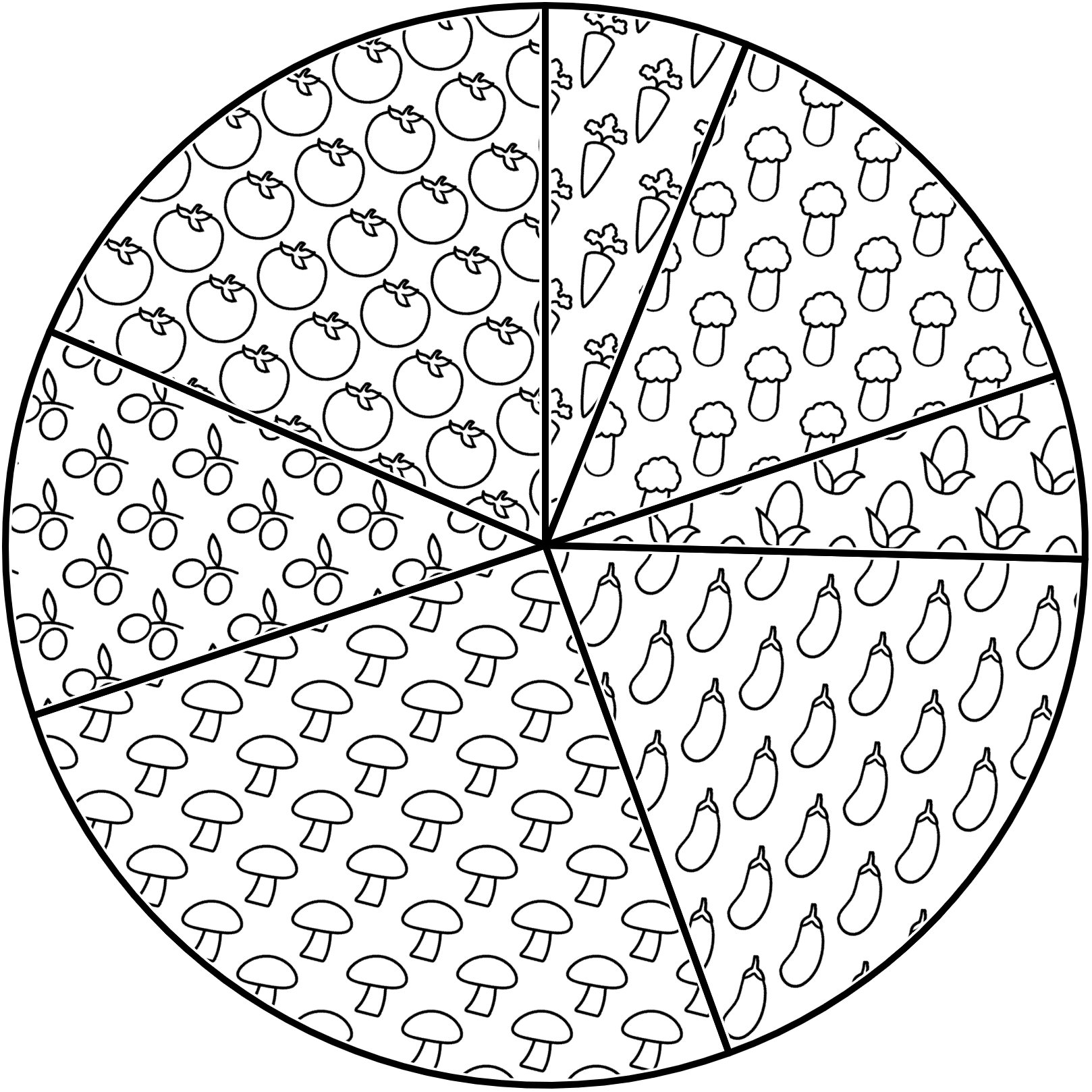}\hfill%PieIcon2
\includegraphics[height=\tableimageheight]{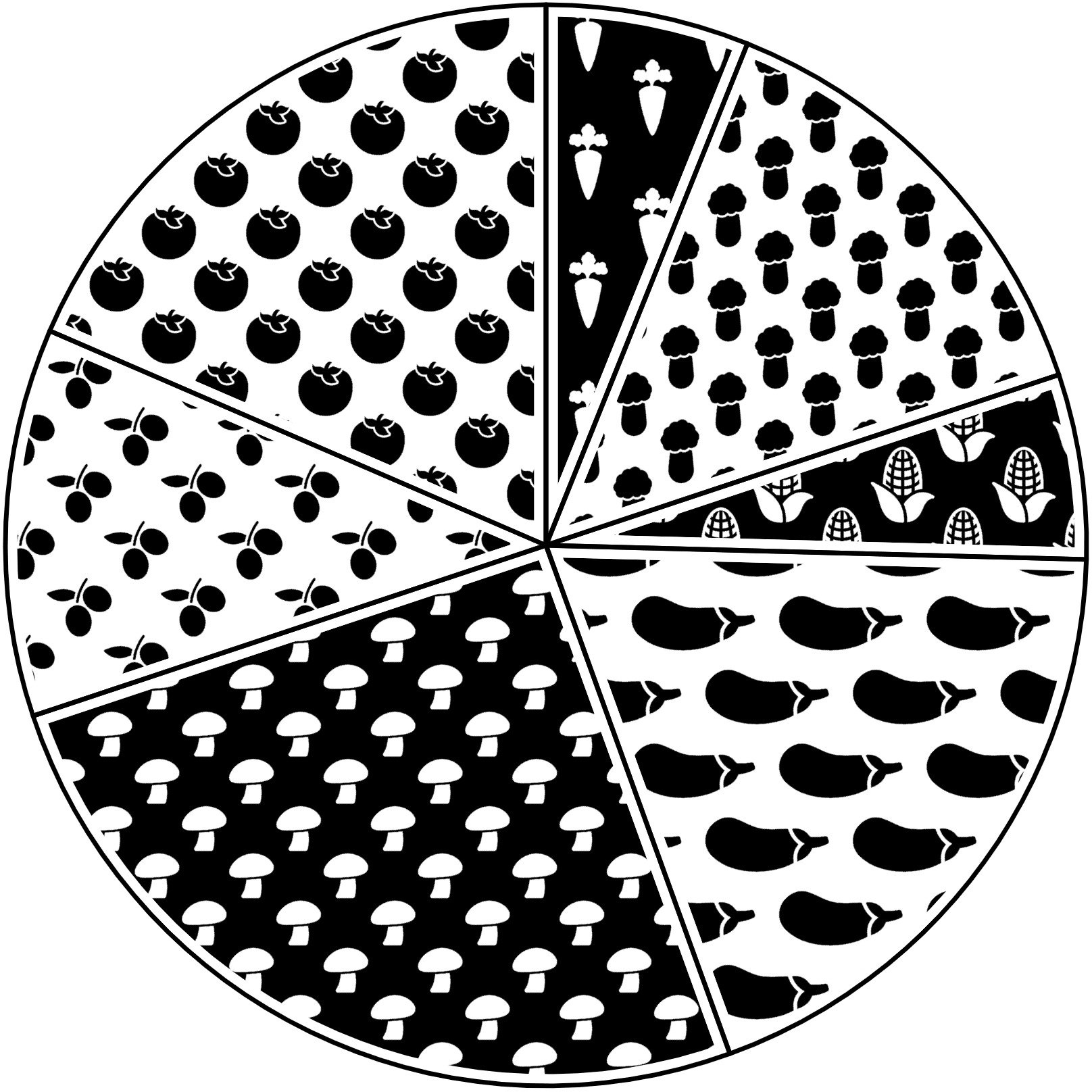}%PieIcon4
\vspace{0.5ex}
\end{table}

\begin{table}[t]
\centering%
\footnotesize%
\caption{\textls[-20]{BeauVis score with distribution, \# ranked first \hty{(total: 53)}, and vibratory score for geometric maps MG1--4 (left--right; larger in \autoref{appendix:all-designs})}.}\vspace{-1ex}
\label{tab:exp2-map-geo}
\begin{tabu}{lllll}
\toprule
              & MG1    & MG2    & MG3     & MG4    \\
              \midrule
BeauVis \hty{(1--7)}      & 4.27 \includegraphics[height= 5mm]{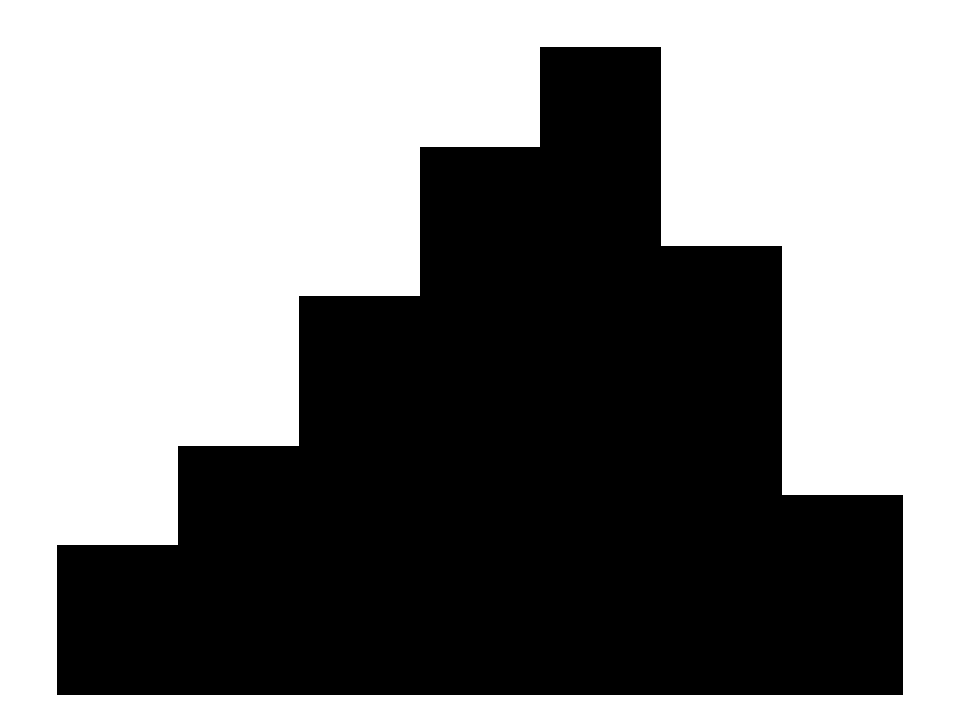}    & 4.25 \includegraphics[height= 5mm]{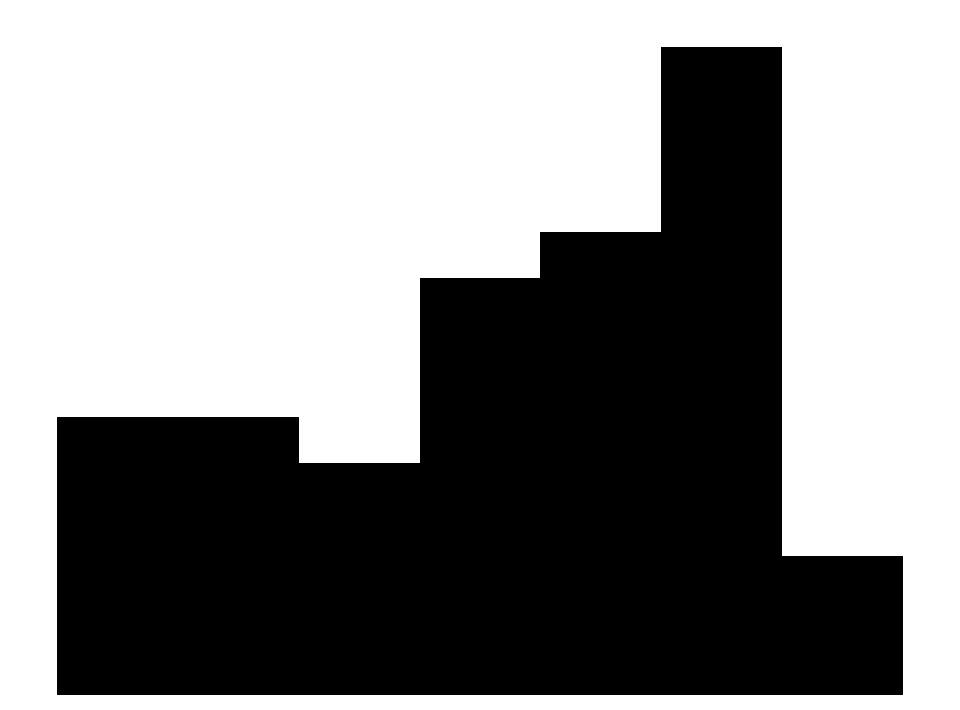}    & 3.57 \includegraphics[height= 5mm]{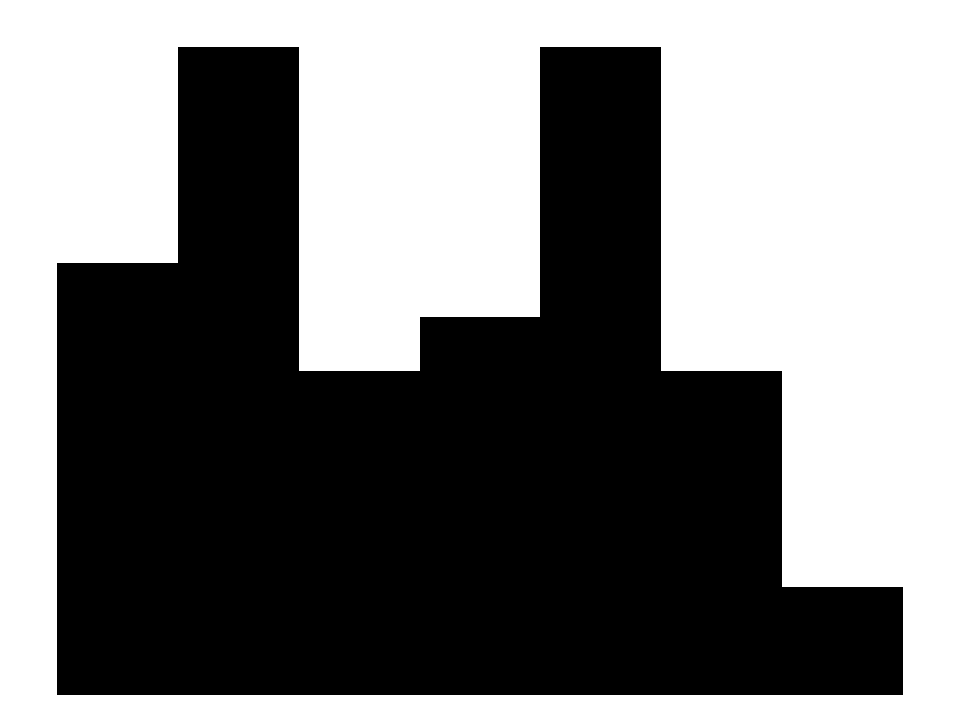}   & 3.15 \includegraphics[height= 5mm]{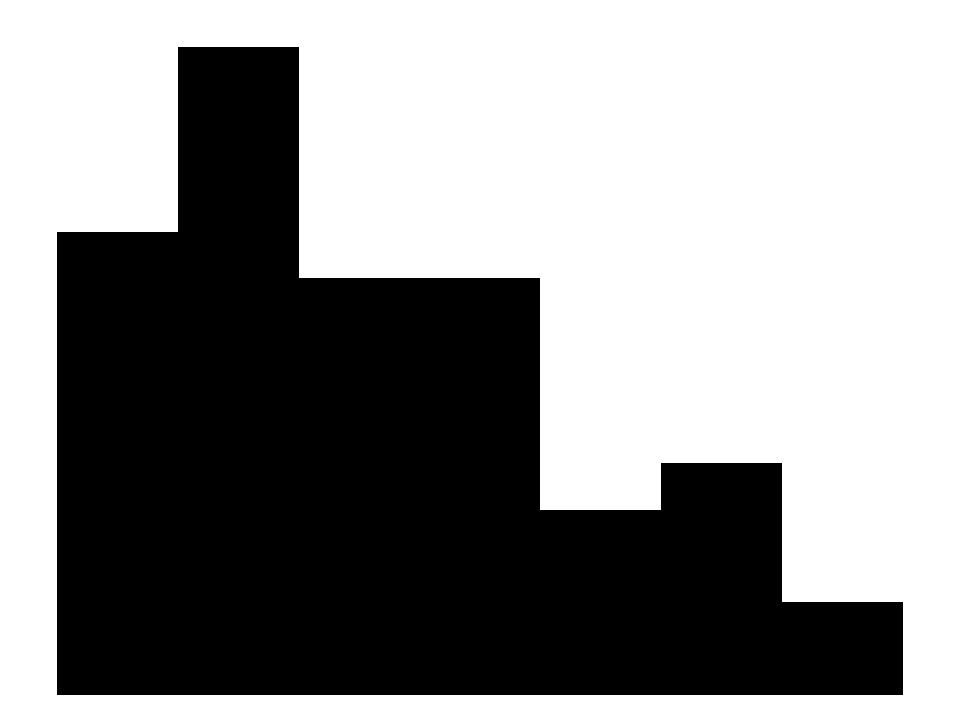}   \\
ranked first & ~21      & ~18      & ~~~6      & ~~~8       \\
vibratory \hty{(1--7)}    & 3.42    & 4.43    & 4.38    & 3.08   \\
\bottomrule
\end{tabu}\\[1ex]%
\includegraphics[height=\tableimageheight]{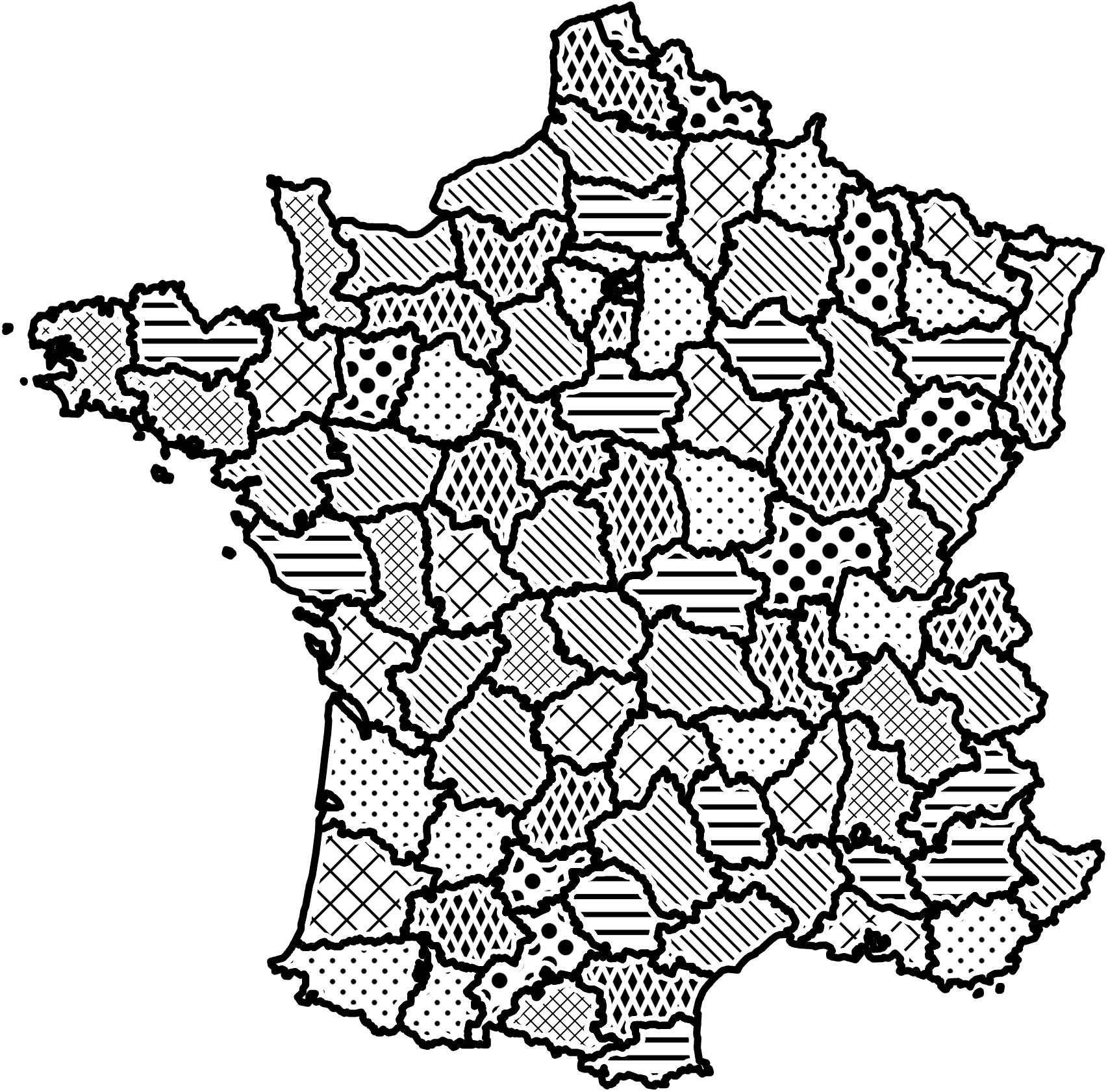}\hfill%MapGeo4
\includegraphics[height=\tableimageheight]{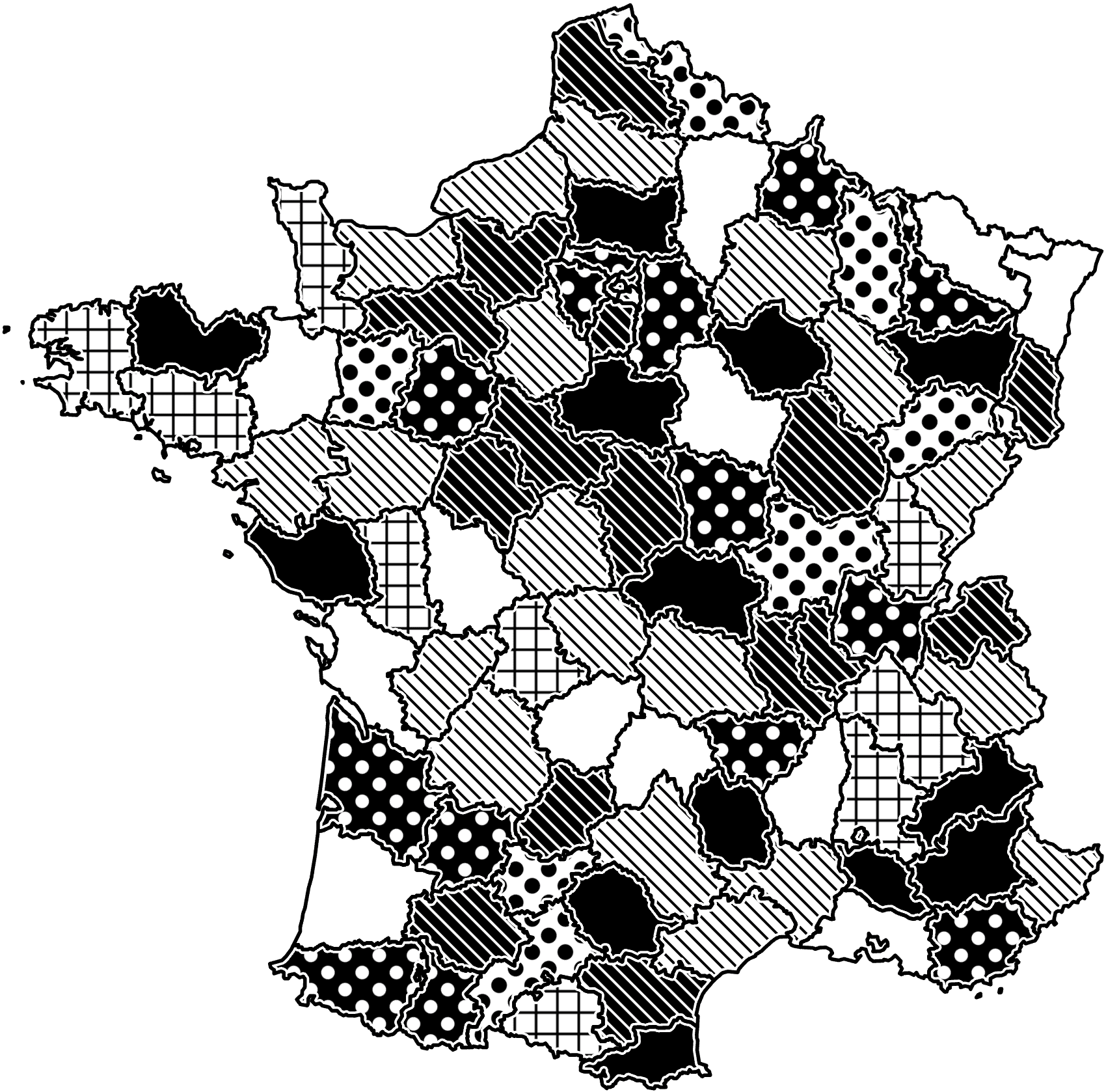}\hfill%MapGeo3
\includegraphics[height=\tableimageheight]{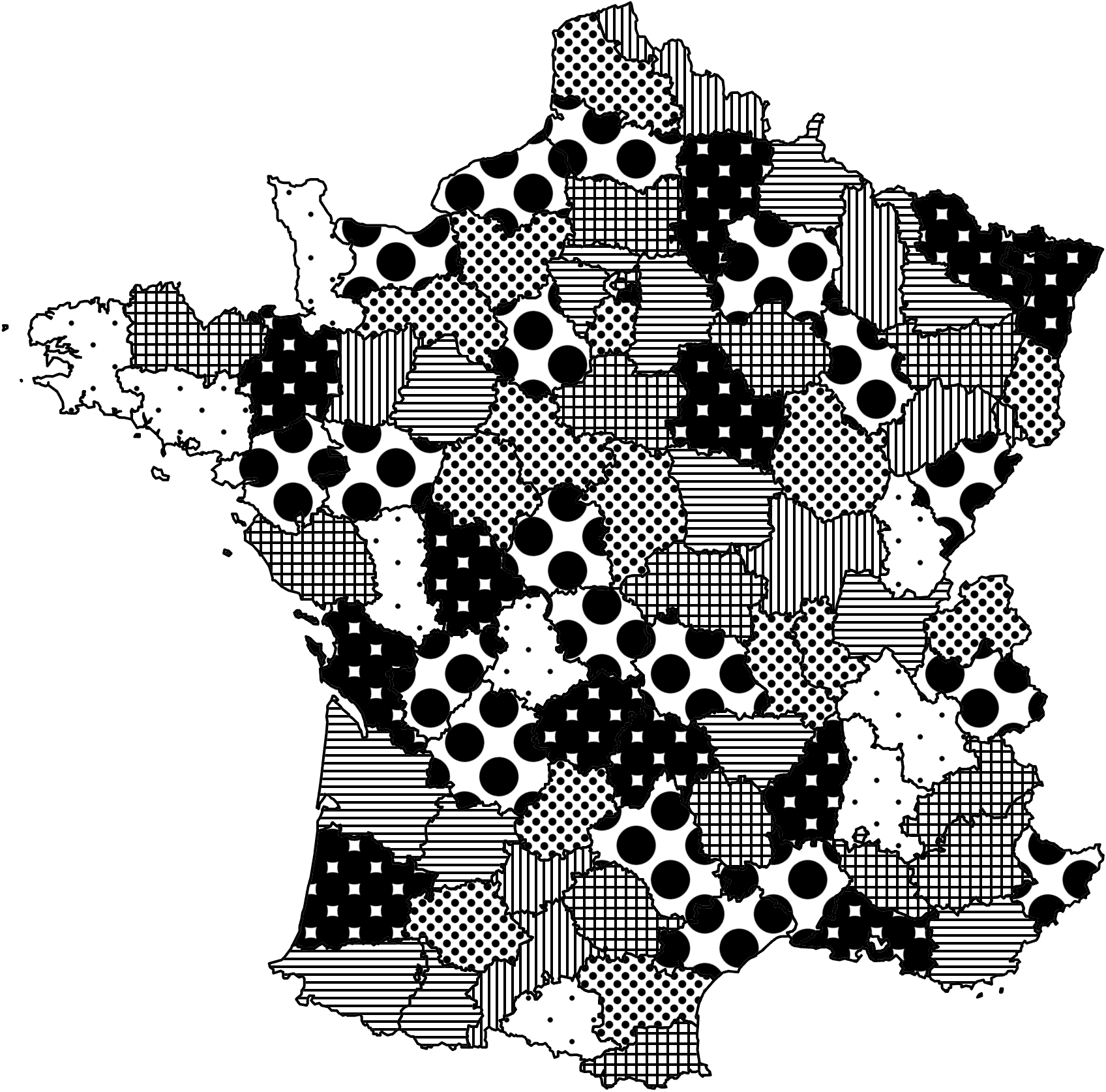}\hfill%MapGeo1
\includegraphics[height=\tableimageheight]{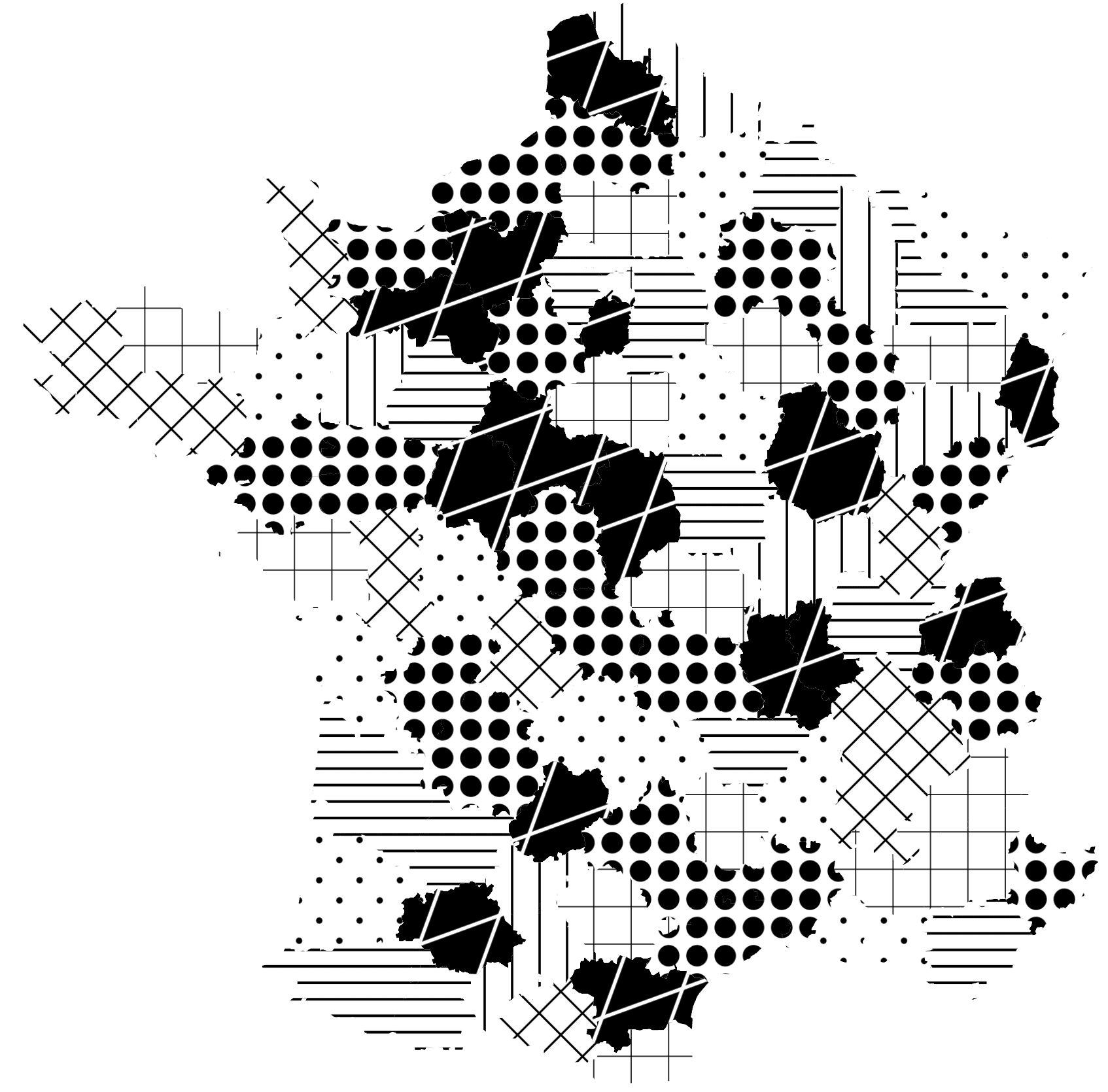}%MapGeo2
\vspace{0.5ex}
\end{table}

\begin{table}[t]
\centering%
\footnotesize%
\caption{BeauVis score with distribution, \# ranked first  \hty{(total: 53)}, and vibratory score for iconic maps MI1--4 (left--right; larger in \autoref{appendix:all-designs}).}\vspace{-1ex}
\label{tab:exp2-map-icon}
\begin{tabu}{lllll}
\toprule
              & MI1    & MI2    & MI3     & MI4    \\
              \midrule
BeauVis \hty{(1--7)}      & 3.58 \includegraphics[height= 5mm]{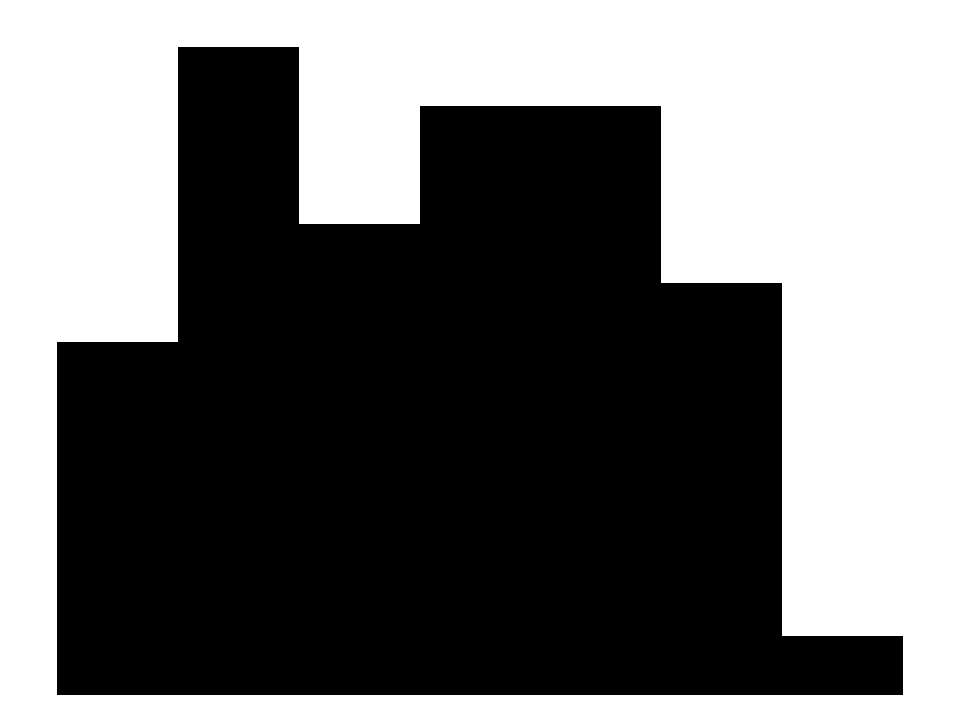}   & 3.55 \includegraphics[height= 5mm]{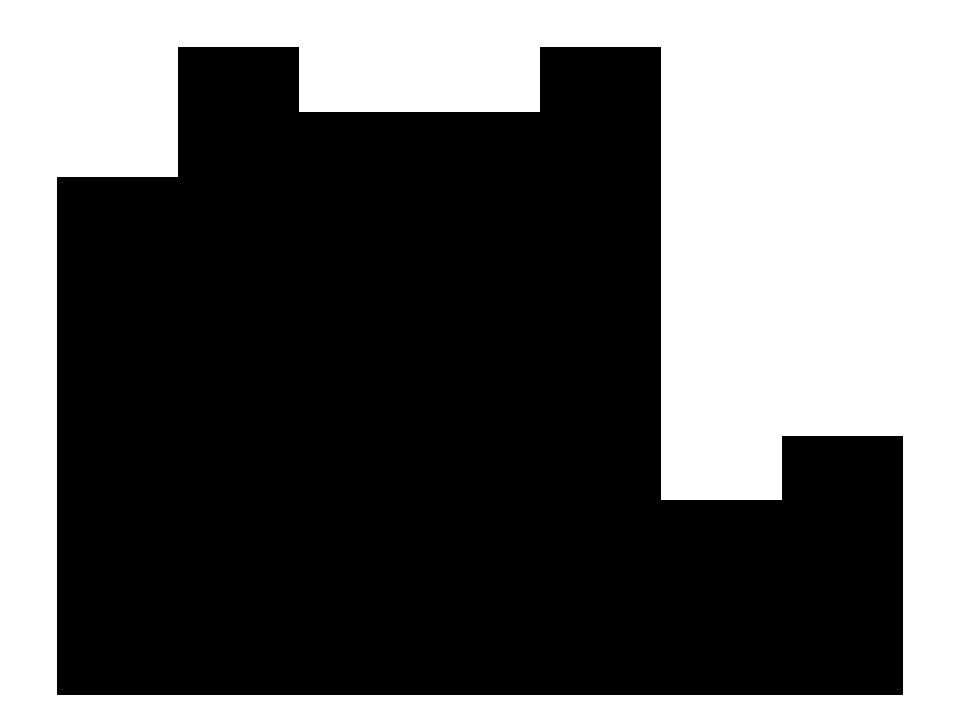}   & 3.32 \includegraphics[height= 5mm]{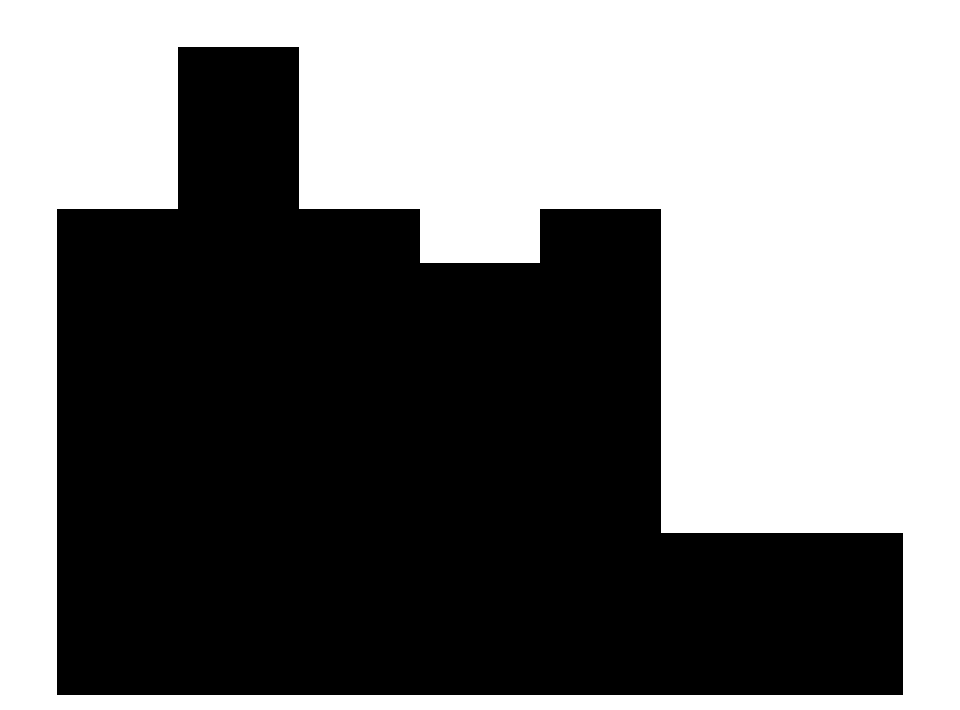}   & 2.66 \includegraphics[height= 5mm]{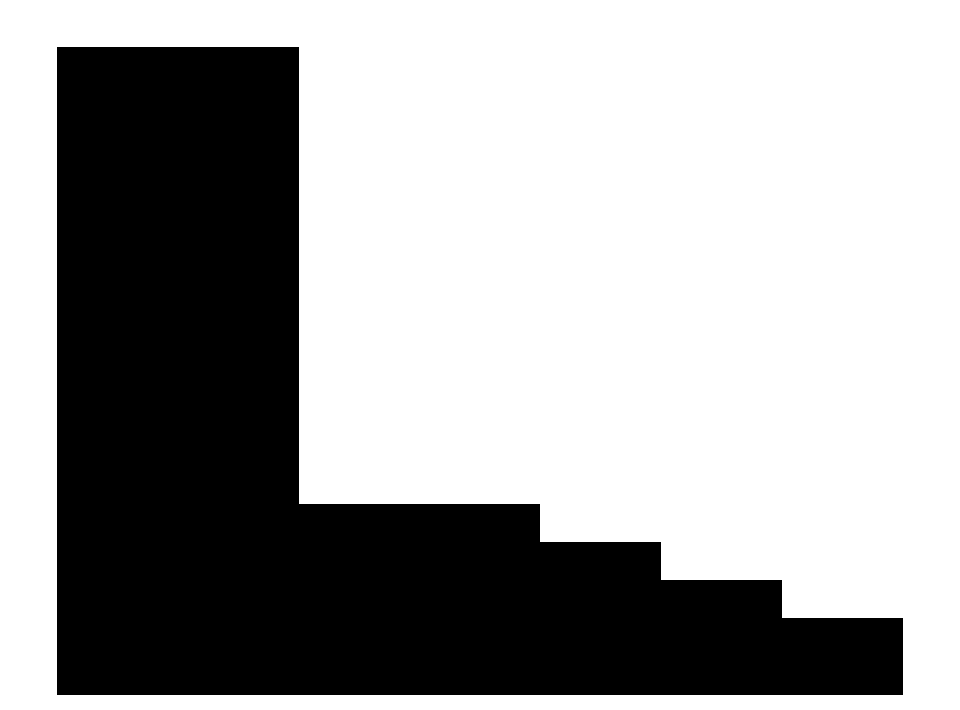}   \\
ranked first & ~17      & ~18      & ~16      & ~~~2       \\
vibratory \hty{(1--7)}    & 2.81    & 3.68    & 2.32    & 3.55   \\
\bottomrule
\end{tabu}\\[1ex]%
\includegraphics[height=\tableimageheight]{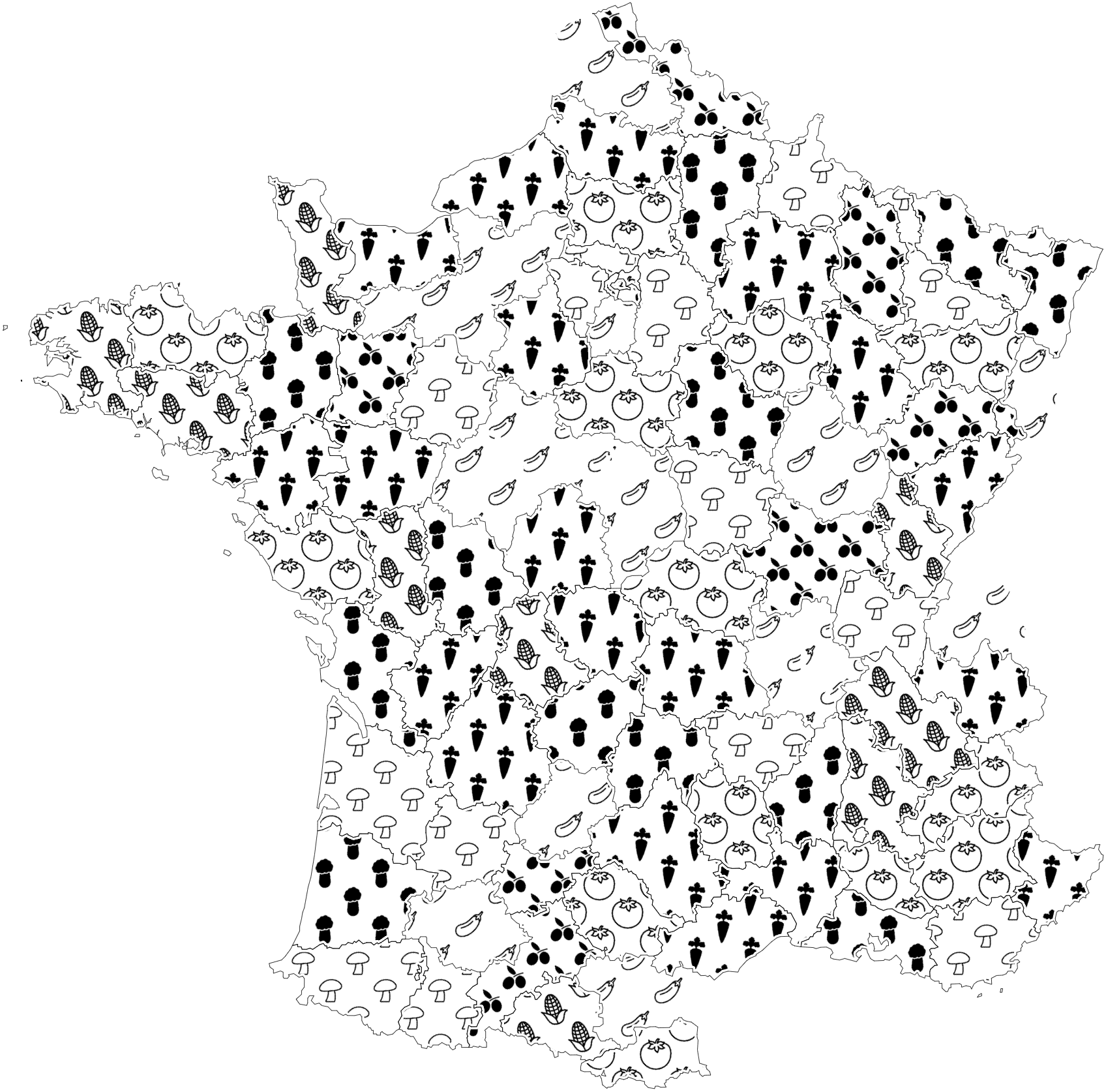}\hfill%MapIcon4
\includegraphics[height=\tableimageheight]{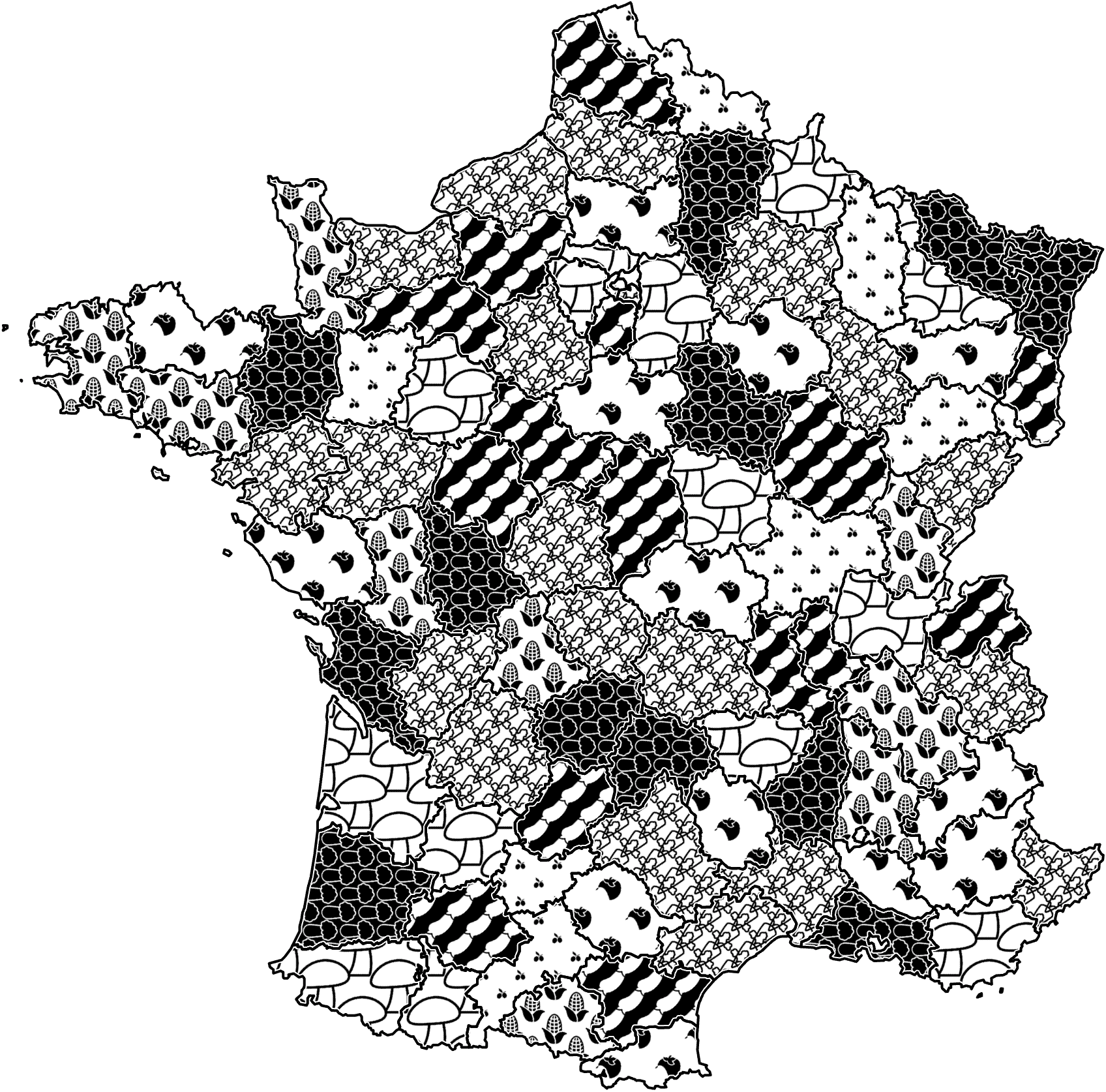}\hfill%MapIcon1
\includegraphics[height=\tableimageheight]{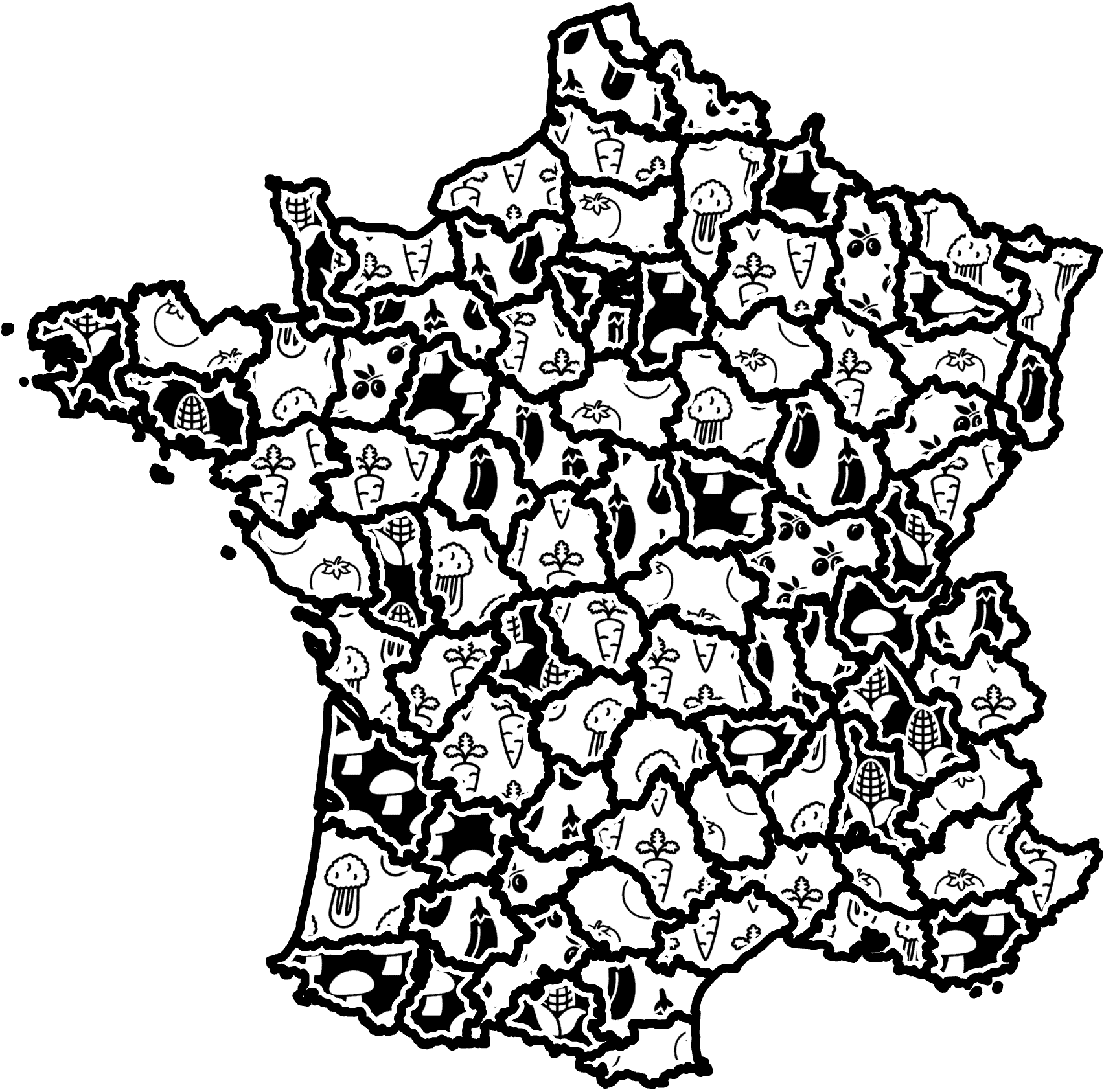}\hfill%MapIcon2
\includegraphics[height=\tableimageheight]{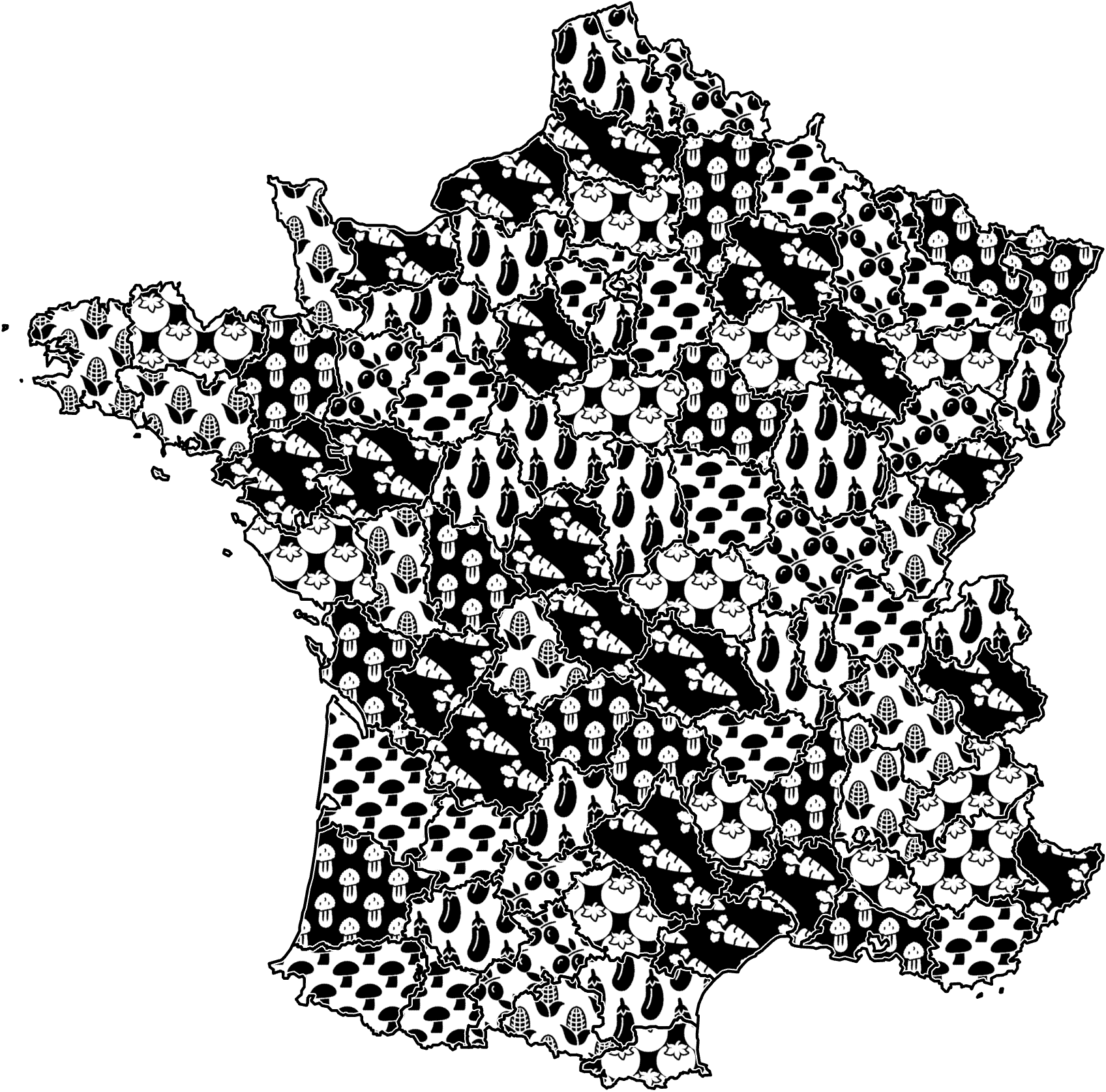}%MapIcon3
%\vspace{0.5ex}
\end{table}

\setlength{\figraisecaptionoffset}{-12pt}%
\setlength{\figaftercaptionoffset}{-6pt}%
\setlength{\figaftercaptionextraoffset}{-10pt}%
\setlength{\figraisecaptionoffsetwithextra}{\figraisecaptionoffset}%
\addtolength{\figraisecaptionoffsetwithextra}{\figaftercaptionextraoffset}%
\begin{figure}[t]
    \centering
    \subcaptionbox{~\hspace{\columnwidth}~}
    {\hspace{.05\columnwidth}\includegraphics[width=0.95\columnwidth,trim={0 23pt 0 0},clip]{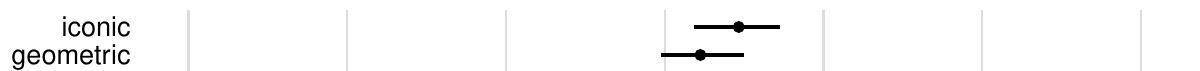}\vspace{\figraisecaptionoffset}}\\[\figaftercaptionoffset]
    \subcaptionbox{~\hspace{\columnwidth}~}
    {\hspace{.05\columnwidth}\includegraphics[width=0.95\columnwidth,trim={0 23pt 0 0},clip]{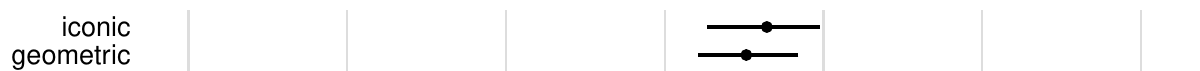}\vspace{\figraisecaptionoffset}}\\[\figaftercaptionoffset]
    \subcaptionbox{~\hspace{\columnwidth}~}
    {\hspace{.05\columnwidth}\includegraphics[width=0.95\columnwidth,trim={0 0 0 0},clip]{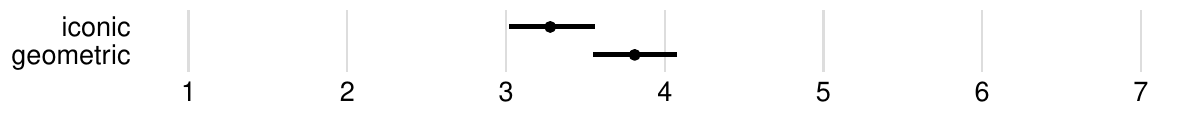}\vspace{\figraisecaptionoffsetwithextra}}\\[\figaftercaptionoffset]
\vspace{6pt}%
\setlength{\figraisecaptionoffset}{-11pt}%
\setlength{\figaftercaptionoffset}{-9pt}%
\setlength{\figaftercaptionextraoffset}{-10pt}%
\setlength{\figraisecaptionoffsetwithextra}{\figraisecaptionoffset}%
\addtolength{\figraisecaptionoffsetwithextra}{\figaftercaptionextraoffset}%
    \subcaptionbox{~\hspace{\columnwidth}~}
    {\hspace{.05\columnwidth}\includegraphics[width=0.95\columnwidth,trim={0 21pt 0 0},clip]{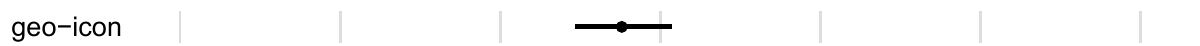}\vspace{\figraisecaptionoffset}}\\[\figaftercaptionoffset]
    \subcaptionbox{~\hspace{\columnwidth}~}
    {\hspace{.05\columnwidth}\includegraphics[width=0.95\columnwidth,trim={0 21pt 0 0},clip]{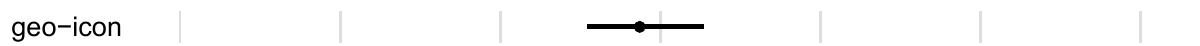}\vspace{\figraisecaptionoffset}}\\[\figaftercaptionoffset]
    \subcaptionbox{~\hspace{\columnwidth}~}
    {\hspace{.05\columnwidth}\includegraphics[width=0.95\columnwidth,trim={0 0 0 0},clip]{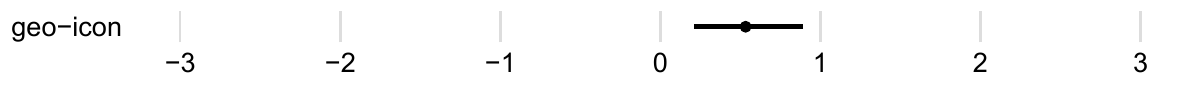}\vspace{\figraisecaptionoffsetwithextra}}
		\vspace{-\figaftercaptionextraoffset}%
		\vspace{-2.5ex}%
    \caption{Aesthetics analysis: BeauVis score for each fill type for (a) bar charts, (b) pie charts, and (c) maps; (d)--(f) corresponding pairwise comparisons between the two fill types. Error bars are 95\% Bootstrap confidence intervals (CIs).}\vspace{-1ex}
    \label{fig:exp2-beauvis}
\end{figure}

\setlength{\figraisecaptionoffset}{-12pt}%
\setlength{\figaftercaptionoffset}{-6pt}%
\setlength{\figaftercaptionextraoffset}{-10pt}%
\setlength{\figraisecaptionoffsetwithextra}{\figraisecaptionoffset}%
\addtolength{\figraisecaptionoffsetwithextra}{\figaftercaptionextraoffset}%
\begin{figure}[t]
    \centering
    \subcaptionbox{~\hspace{\columnwidth}~}
    {\hspace{.05\columnwidth}\includegraphics[width=0.95\columnwidth,trim={0 23pt 0 0},clip]{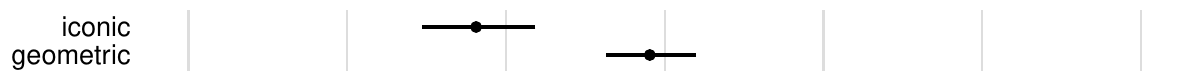}\vspace{\figraisecaptionoffset}}\\[\figaftercaptionoffset]
    \subcaptionbox{~\hspace{\columnwidth}~}
    {\hspace{.05\columnwidth}\includegraphics[width=0.95\columnwidth,trim={0 23pt 0 0},clip]{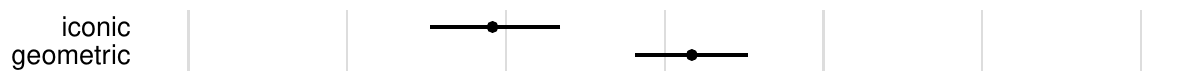}\vspace{\figraisecaptionoffset}}\\[\figaftercaptionoffset]
    \subcaptionbox{~\hspace{\columnwidth}~}
    {\hspace{.05\columnwidth}\includegraphics[width=0.95\columnwidth,trim={0 0 0 0},clip]{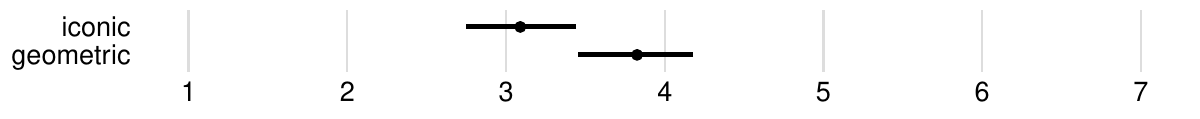}\vspace{\figraisecaptionoffsetwithextra}}\\[\figaftercaptionoffset]%
\vspace{6pt}%
\setlength{\figraisecaptionoffset}{-11pt}%
\setlength{\figaftercaptionoffset}{-9pt}%
\setlength{\figaftercaptionextraoffset}{-10pt}%
\setlength{\figraisecaptionoffsetwithextra}{\figraisecaptionoffset}%
\addtolength{\figraisecaptionoffsetwithextra}{\figaftercaptionextraoffset}%
    \subcaptionbox{~\hspace{\columnwidth}~}
    {\hspace{.05\columnwidth}\includegraphics[width=0.95\columnwidth,trim={0 21pt 0 0},clip]{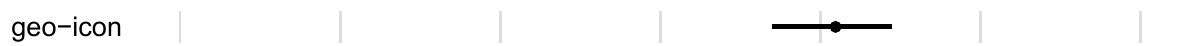}\vspace{\figraisecaptionoffset}}\\[\figaftercaptionoffset]
    \subcaptionbox{~\hspace{\columnwidth}~}
    {\hspace{.05\columnwidth}\includegraphics[width=0.95\columnwidth,trim={0 21pt 0 0},clip]{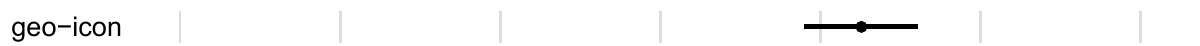}\vspace{\figraisecaptionoffset}}\\[\figaftercaptionoffset]
    \subcaptionbox{~\hspace{\columnwidth}~}
    {\hspace{.05\columnwidth}\includegraphics[width=0.95\columnwidth,trim={0 0 0 0},clip]{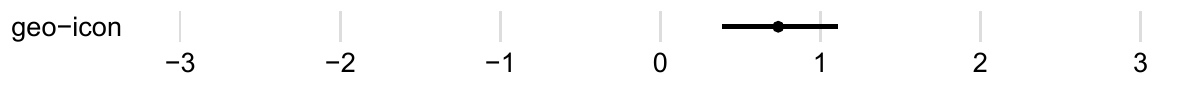}\vspace{\figraisecaptionoffsetwithextra}}
		\vspace{-\figaftercaptionextraoffset}%
		\vspace{-2.5ex}%
    \caption{Vibratory effect analysis: vibratory score for each fill type for (a) bar charts, (b) pie charts, and (c) maps; (d)--(f) corresponding pairwise comparisons between the two fill types. Error bars: 95\% CIs.}\vspace{-1ex}
    \label{fig:exp2-vibratory}
\end{figure}

%% file: sections/experiment3_perception.tex
\section{Experiment 3: Chart Reading}
\label{sec:experiment3}

Beyond this feedback on visual appearance, however, it is also important to understand how the use of textures influences chart reading.
%
%In this paper, we aim to investigate the overall impact of using texture in visualizations and address the lack of empirical work in this field. As discussed in Section, the most important qualities of texture visualization, as perceived by experts, are aesthetics and readability. We have already gathered general public preferences on different texture designs. To use texture effectively, it is essential to understand how it influences chart reading.
%

Lin et al. \cite{lin:2013:selecting} found that employing semantically-resonant colors can improve performance in chart reading tasks\hty{, while Haroz et al. \cite{haroz:2015:isotype} found that pictographs do not significantly affect chart reading time. When we initially pre-registered this experiment, we assumed that iconic textures, like pictographs, would have no impact on chart reading speed. However, given the emphasis by experts in our Experiment 1 on the semantic association characteristic of iconic textures, we revised our hypothesis before beginning the experiment to that iconic textures may also enhance chart reading speed.}
% Previous work \cite{lin:2013:selecting} had shown that employing semantically-re\-so\-nant colors can improve performance in chart reading tasks. Semantic association is a characteristic of iconic textures, which was emphasized by experts in our Experiment 1 (\autoref{sub-sec:compare-two-textures}). We thus hypothesized that iconic textures may also enhance chart reading speed. 
Bertin \cite{bertin:1983:semiology,Bertin:1998:SG} proposed that geometric textures possess selective qualities, enabling them to assist viewers in distinguishing between categories. Consequently, we hypothesized that these textures may also have a positive influence on chart reading speed. With this in mind, we hypothesized (\textbf{H1}) that \emph{both iconic and geometric textures can lead to faster chart reading}.
Earlier studies, however, demonstrated that pictographs can improve engagement \cite{haroz:2015:isotype} and that people tend to find embellished charts more attractive than those without \cite{bateman:2010:useful}. Interestingly, our previous experiment (\autoref{sec:experiment_rate}) showed no evidence of a difference between geometric and iconic textures in terms of aesthetics for bar and pie charts. This contrast led us to questions whether the focus on participants' first impressions in our prior study may be a factor. We thus decided to investigate if aesthetic preferences changed after actually using the visualizations and formulated our second hypothesis \textbf{H2} that \emph{iconic textures will be perceived as more aesthetically pleasing compared to geometric textures, after people have completed chart reading tasks.}

To test our hypotheses, we conducted a third experiment to compare the most preferred geometric and iconic textures with respect to effectiveness, aesthetics, and readability. We limited the chart types to bar and pie charts as they are suitable for the chart reading tasks studied in previous research \cite{haroz:2015:isotype}, and maps overall received a lower BeauVis score in our Experiment 2. Specifically, participants answered which one of two specific data values represented more or fewer items. This experiment was again pre-registered (\href{https://osf.io/8cy62/}{\texttt{osf.io/8cy62}}) and IRB-approved (Inria COERLE, avis \textnumero\ 2023-01).

\subsection{Participants}
Following the sample size used in previous experiments\cite{haroz:2015:isotype, lin:2013:selecting}, we recruited 150 English-fluent, legal-age participants and compensated them at a rate of \EUR 10.20 per hour.

%\subsection{Stimuli}
\subsection{Texture selection}
To select the best textures for bar and pie charts in this experiment, we considered their BeauVis scores and the number of times each was ranked first in Experiment~2. In instances where the BeauVis scores and ranking counts did not align, we took into account the vibratory effect scores for each image and the open responses to the strategy question. Only if the result remained inconclusive we prioritized the BeauVis score (for more details about the selection see \autoref{appendix:decision-e3}). This process resulted in the four designs we show in \autoref{fig:teaser}, and we added a light gray fill for bar and pie charts as a baseline.

%In summary, the selected textures for stimuli in this experiment are the second design in \autoref{tab:exp2-bar-geo} and the first designs in \autoref{tab:exp2-bar-icon, tab:exp2-pie-geo, tab:exp2-pie-icon, tab:exp2-map-geo, tab:exp2-map-icon}. We also used a unicolor (light gray) as a baseline for filling bar charts and pie charts.

\setlength{\figraisecaptionoffset}{-16pt}%
\setlength{\figaftercaptionoffset}{0pt}%
\setlength{\figaftercaptionextraoffset}{-10pt}%
\setlength{\figraisecaptionoffsetwithextra}{\figraisecaptionoffset}%
\addtolength{\figraisecaptionoffsetwithextra}{\figaftercaptionextraoffset}%
\begin{figure}
    \centering
    \subcaptionbox{~\hspace{\columnwidth}~}
    {\hspace{.05\columnwidth}\includegraphics[width=0.95\columnwidth,trim={0 23pt 0 0},clip]{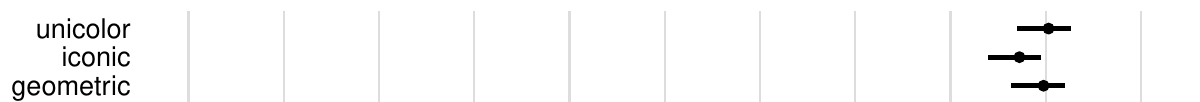}\vspace{\figraisecaptionoffset}}\\[\figaftercaptionoffset]
    \subcaptionbox{~\hspace{\columnwidth}~}
    {\hspace{.05\columnwidth}\includegraphics[width=0.95\columnwidth,trim={0 0 0 0},clip]{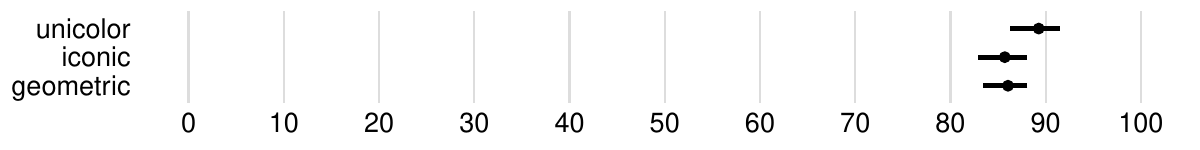}\vspace{\figraisecaptionoffsetwithextra}}\\[\figaftercaptionoffset]%
\vspace{8pt}%
\setlength{\figraisecaptionoffset}{-17pt}%
\setlength{\figaftercaptionoffset}{-2pt}%
\setlength{\figaftercaptionextraoffset}{-10pt}%
\setlength{\figraisecaptionoffsetwithextra}{\figraisecaptionoffset}%
\addtolength{\figraisecaptionoffsetwithextra}{\figaftercaptionextraoffset}%
    \subcaptionbox{~\hspace{\columnwidth}~}
    {\hspace{.05\columnwidth}\includegraphics[width=0.95\columnwidth,trim={0 20pt 0 0},clip]{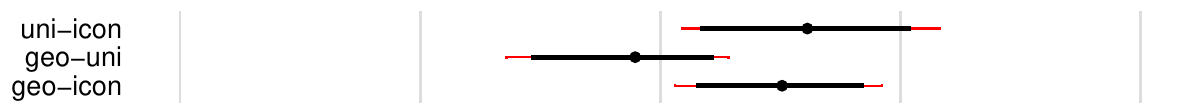}\vspace{\figraisecaptionoffset}}\\[\figaftercaptionoffset]
    \subcaptionbox{~\hspace{\columnwidth}~}
    {\hspace{.05\columnwidth}\includegraphics[width=0.95\columnwidth,trim={0 0 0 0},clip]{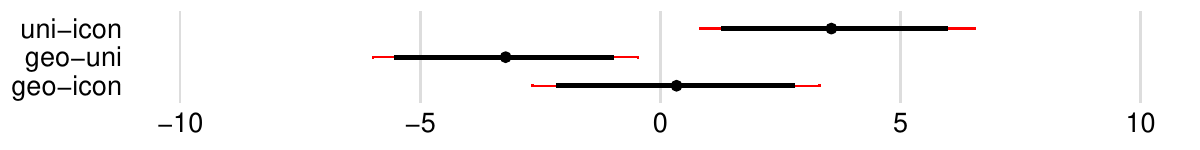}\vspace{\figraisecaptionoffsetwithextra}}
		\vspace{-\figaftercaptionextraoffset}%
		\vspace{-1.5ex}%
    \caption{Correct answer rates in \% for (a) bar and (b) pie charts; (c), (d) corresponding pairwise comparisons between the fill types. Error bars: 95\% CIs. Red bars: CIs for Bonferroni-corrected pairwise comparison.}%\vspace{-1ex}
    \label{fig:exp3-correct-rate}
\end{figure}

\setlength{\figraisecaptionoffset}{-16pt}%
\setlength{\figaftercaptionoffset}{0pt}%
\setlength{\figaftercaptionextraoffset}{-10pt}%
\setlength{\figraisecaptionoffsetwithextra}{\figraisecaptionoffset}%
\addtolength{\figraisecaptionoffsetwithextra}{\figaftercaptionextraoffset}%
\begin{figure}
    \centering
    \subcaptionbox{~\hspace{\columnwidth}~}
    {\hspace{.05\columnwidth}\includegraphics[width=0.95\columnwidth,trim={0 23pt 0 0},clip]{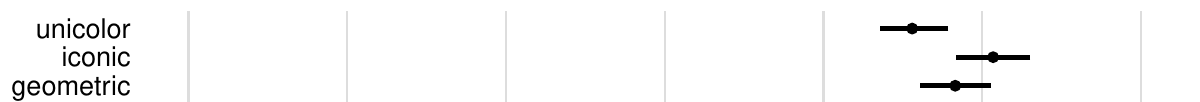}\vspace{\figraisecaptionoffset}}\\[\figaftercaptionoffset]
    \subcaptionbox{~\hspace{\columnwidth}~}
    {\hspace{.05\columnwidth}\includegraphics[width=0.95\columnwidth,trim={0 0 0 0},clip]{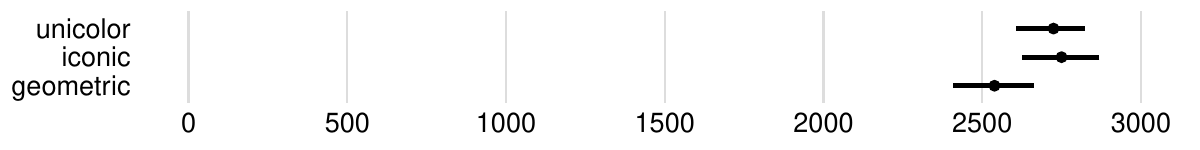}\vspace{\figraisecaptionoffsetwithextra}}\\[\figaftercaptionoffset]%
\vspace{5pt}%
\setlength{\figraisecaptionoffset}{-17pt}%
\setlength{\figaftercaptionoffset}{-2pt}%
\setlength{\figaftercaptionextraoffset}{-10pt}%
\setlength{\figraisecaptionoffsetwithextra}{\figraisecaptionoffset}%
\addtolength{\figraisecaptionoffsetwithextra}{\figaftercaptionextraoffset}%
    \subcaptionbox{~\hspace{\columnwidth}~}
    {\hspace{.05\columnwidth}\includegraphics[width=0.95\columnwidth,trim={0 20pt 0 0},clip]{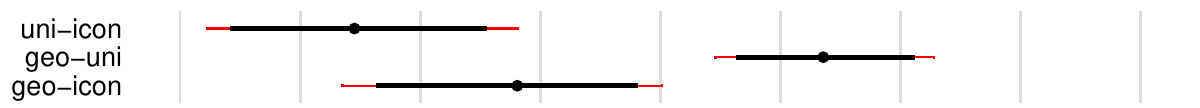}\vspace{\figraisecaptionoffset}}\\[\figaftercaptionoffset]
    \subcaptionbox{~\hspace{\columnwidth}~}
    {\hspace{.05\columnwidth}\includegraphics[width=0.95\columnwidth,trim={0 0 0 0},clip]{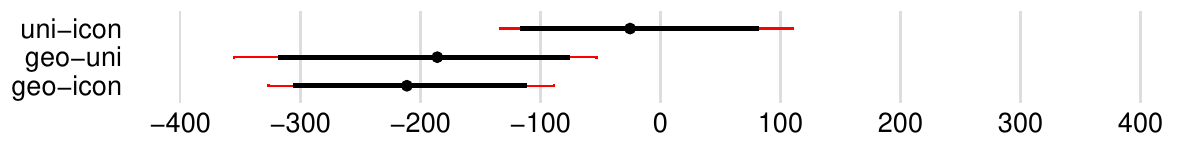}\vspace{\figraisecaptionoffsetwithextra}}
\vspace{-\figaftercaptionextraoffset}%
		\vspace{-1.5ex}%
    \caption{Response times in ms for (a) bar and (b) pie charts; (c), (d) corresponding pairwise comparisons between the fill types. Error bars: 95\% CIs. Red bars: CIs for Bonferroni-corrected pairwise comparison.}%\vspace{-1ex}
    \label{fig:exp3-response-time}
\end{figure}

\subsection{Method}
We used a mixed design with a between-subjects variable \emph{chart type} (bar, pie) and a within-subjects variable \emph{fill type} (geometric, iconic, unicolor). We also used two question types (more, fewer). At the beginning of the experiment, we asked participants to complete a consent form and to provide background information.

% \autoref{tab:experiment3_variables}.
% \begin{table}[]
% \caption{Independent variables in Experiment 3}
% \label{tab:experiment3_variables}
% \begin{tabular}{ll}
% \hline
% \textbf{Variables} & \textbf{Levels}                                                              \\ \hline
% Chart Type (between-subjects)         & bar chart or pie chart                                                       \\
% Encoding Type  (within-subjects)        & geometric texture, iconic texture or Gray                                    \\
% Question Type (within-subjects)     & \begin{tabular}[c]{@{}l@{}}"Who has FEWER?" or\\ "Who has MORER?"\end{tabular} \\ \hline
% \end{tabular}
% \end{table}

Inspired by the studies of Haroz et al. \cite{haroz:2015:isotype} and Lin et al. \cite{lin:2013:selecting} who measured chart reading speed and accuracy, we asked each participant to complete 60 trials, consisting of 2 question types \texttimes{} 3 fill types \texttimes{} 10 repetitions. We grouped the trials by question type and sub-grouped by fill type. At the beginning of each block, we presented participants with instructions that explained the task and instructed participants to complete the tasks as quickly and accurately as possible. In each fill type block, we asked participants to first examine a chart with the texture type to familiarize themselves with the chart fill and then to proceed to the training. We required the participants to complete three correct training trials, before they could advance to the real experiment. We randomized the chart type, the order of each block, and the stimuli.

During each trial, we presented two targets (\eg, \emph{olive} and \emph{corn}) and one of two questions: ``Which has MORE?'' or ``Which has FEWER?'' We represented the targets as a vegetable name and an image, the latter being a geometric texture, an icon, or a blank light gray square depending on the chart fill condition. Participants needed to press the space bar to initiate the trial, reveal the chart, and start the timer. We instructed them to press the left or right arrow key to indicate which target answered the question. We ensured that, for both bar and pie charts, the item designated by the left arrow key consistently appeared on the left side relative to the item identified by the right arrow key. After 5 seconds, the question and chart disappeared, and we showed participants the result of their response (correct, incorrect, or timed out). We conducted a pilot within our research group and determined that 5 seconds was a reasonable time to be able to give an answer. 

Finally, we showed participants three charts with default data values, each featuring a different fill type, in random order. We asked the them to rate each chart using the 5 items of the BeauVis scale and an additional readability item on a 7-point Likert scale.

\subsection{Datasets generation}
Following the approach used by Lin et al. \cite{lin:2013:selecting, cleveland:1985:graphical}, we generated 10 datasets for our experiment, with seven data values each. We randomly selected these seven values from a range of 5 to 95 on a 0--100 scale, ensuring that the values of two targets for comparison were separated by at least 5 points on the scale. 
With the 10 datasets, we generated images for each fill type \texttimes{} chart type condition, resulting in 60 images in total. 
In the experiment, we used the 30 images of bar or pie charts twice due to the two question types.
For each image, we randomly shuffled the order of the seven vegetable items on the chart (\eg, in a bar chart, the carrot bar could appear at any position). We also generated additional stimuli for training trials, following the same rules. 

\subsection{Data analysis and interpretation}
We calculated average correct rates, response times, readability, and BeauVis scores for each fill and chart type combination (\eg, geometric bars) across participants. We report sample means and pairwise mean differences of our three fill types with 95\% CIs. We adjusted the CIs of pairwise differences using the Bonferroni correction to reduce the risk of type I errors when doing multiple comparisons simultaneously \cite{higgins:2004:introduction}. 
% We report sample means with 95\% CIs and pairwise mean differences between each of our three fill types with 95\% CIs adjusted using the Bonferroni correction \cite{higgins:2004:introduction}.
%For response time, we only counted the trials in which participants gave an answer (both correct trials and incorrect trials). 
%We removed the few timed-out trials ($<$\,1.5\%) as we could not estimate whether a person was distracted or how much more time they would have needed. 
We interpret the results in the same way as in Experiment~2. (\autoref{sub-sec:exp2-data-analysis}).

\subsection{Results}
We received 150 valid responses (67\texttimes{} bar, 83\texttimes{} pie), which we used for our analysis (74 female, 76 male; ages: mean = 28.0, SD = 8.1;  education: 99 Bachelor or equivalent, 23 Master's or equivalent, 1 PhD or equivalent, 27 other). We should have received 9000 valid experiment trials, but we lost data from 12 trials due to log file issues. Among the remaining 8988 trials, there were 125 timed-out trials. \hty{\autoref{tab:exp3-timeout} in \autoref{appendix:additional-exp3} shows the distribution of time-out trials.} %There might be a trend that iconic textures make chart harder to read.  

\textbf{Accuracy rate.} 
\autoref{fig:exp3-correct-rate} shows the mean values and pairwise comparisons of the accuracy rates for all fill types in bar and pie charts. \hty{All conditions yielded high average accuracy rates, exceeding 85\%, much higher than the 50\% correct rate for random guessing.} 
Pairwise comparisons, shown in \autoref{fig:exp3-correct-rate}(c, d), reveal that for bar charts, unicolor and geometric textures outperform iconic textures, while for pie charts, unicolor surpasses both textures. We note, however, that the difference is quite small in practice ($<$\,3.6\%). \harozissue{After examining individual correct rates, we revised our pre-registrated analysis plan of including all participants (see \autoref{appendix:original-analysis-e3}) to only include the 86 participants who achieved $\geq$\,90\% overall accuracy (45\texttimes{} bar, 41\texttimes{} pie) for the following analysis, minimizing the effect of random guesses. We counted 2 trials recorded with correct answers but durations slightly over 5s as correct.}

%An examination of the pairwise differences in accuracy rates, as depicted in \autoref{fig:exp3-correct-rate}(c, d), shows that, for bar charts, unicolor and geometric textures perform better than iconic textures. For pies charts, unicolor is better than both iconic and geometric textures. We note, however, that the difference is quite small in practice ($<$\,3.6\%).
 % \todo{Petra: I think it says For bars: uni \& geo are better than icon. For pies: uni is better than icon and geo. Please double check. I would also mention that the difference is quite small in practice (around 2.5 percentage points)}
 % reveals some evidence that the unicolor fill type results in a higher accuracy rate compared to both geometric and iconic textures for both bar and pie charts. However, we see no evidence of a difference between geometric and iconic textures.

\textbf{Response time.} 
\label{sec:exp3-response-time}
% \harozissue{We had initially counted all participants' response times (\autoref{fig:exp3-response-time-original} in \autoref{appendix:original-analysis-e3}) as pre-registered. But we decided to adjust to only count correct trials from those 86 participants who achieved the 90\% overall accuracy threshold (45\texttimes{} bar, 41\texttimes{} pie), to ensure the interpretability of our results and to minimize the influence of chance.} 
\harozissue{We only counted the response times of correct trials from the 86 participants to ensure the interpretability of our results.} 
\autoref{fig:exp3-response-time} shows the mean values and pairwise comparisons of response times for all fill types in bar and pie charts. The analysis of the pairwise differences shows that, for bar charts, we have evidence that 
\harozissue{both textures have a longer response time than unicolor.}
% iconic textures have a longer response time than the other two.  
For pie charts, we see evidence that geometric textures have shorter response times than the other two fill types. There was no evidence of a difference for any other combination of fill types. Again, we note that the differences are minimal, within a range of 
\harozissue{$<$\,255\,ms.}
% $<230$\,ms.

\textbf{Readability.} 
\autoref{fig:exp3-readable} presents the mean values and pairwise comparisons of readability scores for all fill types for bar charts and pie charts, which we measured using a 7-point Likert item. For bar charts, the pairwise differences in \autoref{fig:exp3-readable}(c) indicate that unicolor filling was considered more readable than the other two types; however, we have no evidence for a difference between the two textures. We observe a consistent trend across all three analyses (correct rate, response time, and readability) for bar charts, showing that unicolor outperforms geometric textures, which in turn outperform iconic textures. This trend aligns with the distribution of the number of timed-out trials. Regarding pie charts, the pairwise differences in \autoref{fig:exp3-readable}(d) reveal no evidence of differences in readability among the three fill types.

\setlength{\figraisecaptionoffset}{-18pt}%
\setlength{\figaftercaptionoffset}{1pt}%
\setlength{\figaftercaptionextraoffset}{-10pt}%
\setlength{\figraisecaptionoffsetwithextra}{\figraisecaptionoffset}%
\addtolength{\figraisecaptionoffsetwithextra}{\figaftercaptionextraoffset}%
\begin{figure}
    \centering
    \subcaptionbox{~\hspace{\columnwidth}~}
    {\hspace{.05\columnwidth}\includegraphics[width=0.95\columnwidth,trim={0 23pt 0 0},clip]{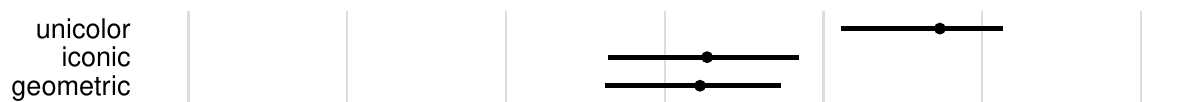}\vspace{\figraisecaptionoffset}}\\[\figaftercaptionoffset]
    \subcaptionbox{~\hspace{\columnwidth}~}
    {\hspace{.05\columnwidth}\includegraphics[width=0.95\columnwidth,trim={0 0 0 0},clip]{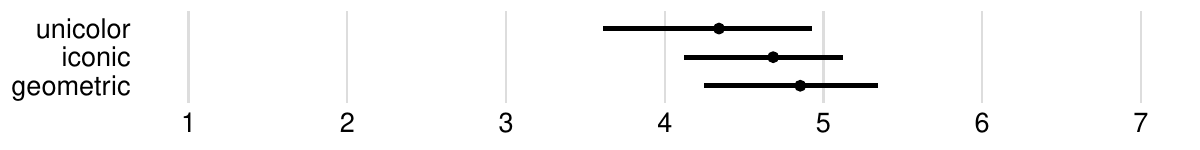}\vspace{\figraisecaptionoffsetwithextra}}\\[\figaftercaptionoffset]%
\vspace{8pt}%
\setlength{\figraisecaptionoffset}{-18pt}%
\setlength{\figaftercaptionoffset}{-1pt}%
\setlength{\figaftercaptionextraoffset}{-10pt}%
\setlength{\figraisecaptionoffsetwithextra}{\figraisecaptionoffset}%
\addtolength{\figraisecaptionoffsetwithextra}{\figaftercaptionextraoffset}%
    \subcaptionbox{~\hspace{\columnwidth}~}
    {\hspace{.05\columnwidth}\includegraphics[width=0.95\columnwidth,trim={0 20pt 0 0},clip]{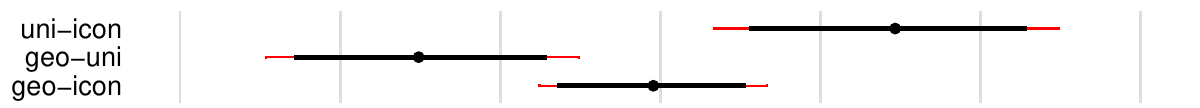}\vspace{\figraisecaptionoffset}}\\[\figaftercaptionoffset]
    \subcaptionbox{~\hspace{\columnwidth}~}
    {\hspace{.05\columnwidth}\includegraphics[width=0.95\columnwidth,trim={0 0 0 0},clip]{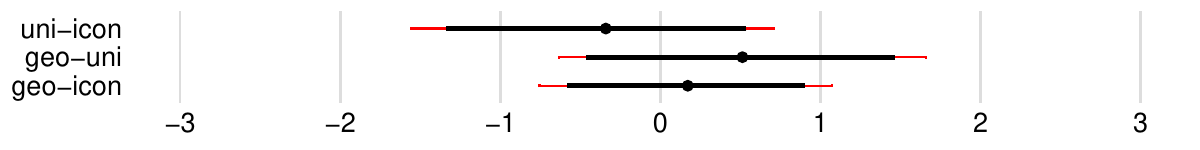}\vspace{\figraisecaptionoffsetwithextra}}
\vspace{-\figaftercaptionextraoffset}%
		\vspace{-1.5ex}%
    \caption{Readability scores for (a) bar and (b) pie charts; (c), (d) corresponding pairwise comparisons between the fill types. Error bars: 95\% CIs. Red bars: CIs for Bonferroni-corrected pairwise comparison.}%\vspace{-1ex}
    \label{fig:exp3-readable}
\end{figure}

\textbf{Aesthetics.} 
\autoref{fig:exp3-beauvis} displays the mean values and pairwise comparisons of the BeauVis score for all fill types separated by bar and pie charts. 
%In the case of bar charts, the pairwise differences shown in \autoref{fig:exp3-beauvis}(c) reveal no evidence of a difference between geometric or iconic textures and unicolor, although iconic textures were considered more aesthetically pleasing than geometric textures. 
\harozissue{For bar charts, the pairwise differences in \autoref{fig:exp3-beauvis}(c) reveal no evidence of a difference between either geometric or iconic textures and unicolor,}
although iconic textures were considered more aesthetically pleasing than geometric textures. 
For pie charts there was evidence suggests that both geometric and iconic textures were perceived as more aesthetically pleasing than unicolor; no evidence, however, supports a difference between geometric and iconic textures in terms of aesthetics.

\setlength{\figraisecaptionoffset}{-18pt}%
\setlength{\figaftercaptionoffset}{1pt}%
\setlength{\figaftercaptionextraoffset}{-10pt}%
\setlength{\figraisecaptionoffsetwithextra}{\figraisecaptionoffset}%
\addtolength{\figraisecaptionoffsetwithextra}{\figaftercaptionextraoffset}%
\begin{figure}
    \centering
    \subcaptionbox{~\hspace{\columnwidth}~}
    {\hspace{.05\columnwidth}\includegraphics[width=0.95\columnwidth,trim={0 23pt 0 0},clip]{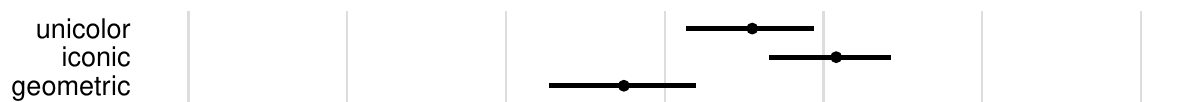}\vspace{\figraisecaptionoffset}}\\[\figaftercaptionoffset]
    \subcaptionbox{~\hspace{\columnwidth}~}
    {\hspace{.05\columnwidth}\includegraphics[width=0.95\columnwidth,trim={0 0 0 0},clip]{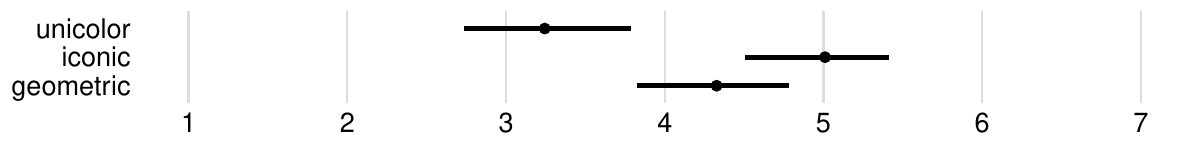}\vspace{\figraisecaptionoffsetwithextra}}\\[\figaftercaptionoffset]%
\vspace{8pt}%
\setlength{\figraisecaptionoffset}{-18pt}%
\setlength{\figaftercaptionoffset}{-1pt}%
\setlength{\figaftercaptionextraoffset}{-10pt}%
\setlength{\figraisecaptionoffsetwithextra}{\figraisecaptionoffset}%
\addtolength{\figraisecaptionoffsetwithextra}{\figaftercaptionextraoffset}%
    \subcaptionbox{~\hspace{\columnwidth}~}
    {\hspace{.05\columnwidth}\includegraphics[width=0.95\columnwidth,trim={0 20pt 0 0},clip]{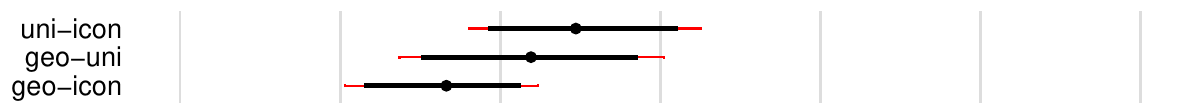}\vspace{\figraisecaptionoffset}}\\[\figaftercaptionoffset]
    \subcaptionbox{~\hspace{\columnwidth}~}
    {\hspace{.05\columnwidth}\includegraphics[width=0.95\columnwidth,trim={0 0 0 0},clip]{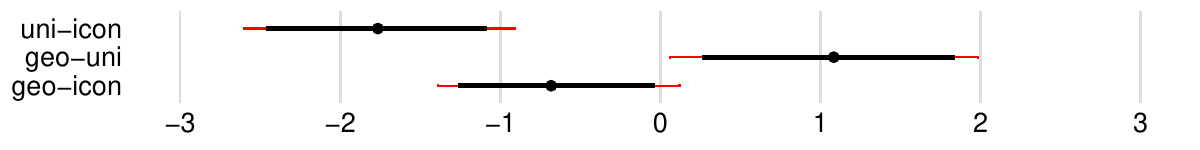}\vspace{\figraisecaptionoffsetwithextra}}
\vspace{-\figaftercaptionextraoffset}%
		\vspace{-1.5ex}%
    \caption{BeauVis scores for (a) bar and (b) pie charts; (c), (d) corresponding pairwise comparisons between the fill types. Error bars: 95\% CIs. Red bars: CIs for Bonferroni-corrected pairwise comparison.}%\vspace{-1ex}
    \label{fig:exp3-beauvis}
\end{figure}

\textbf{Summary.}
Our results show that for, bar charts, iconic textures performed worse than the other two types, resulting in more errors and slower responses. 
\harozissue{While geometric textures did not reduce accuracy, they did slow down response times.}
%Although geometric textures did not slow down response time, they were perceived as less readable than the unicolor ones. 
For bar charts our hypothesis H1 is thus incorrect, but, since geometric textures were perceived as less aesthetically pleasing than iconic textures, H2 is supported. For pie charts, the situation is reversed; geometric textures performed well, demonstrating faster response times and being considered more visually appealing than unicolor textures. There was also a trend towards higher readability for geometric textures. For pie charts, however, the iconic textures did not have a positive effect on chart reading effectiveness, supporting H1 only partially. Since there is no significant difference in aesthetics between geometric and iconic textures, H2 is not supported.

% Data: 
% Chart data were consistently distributed between 5\% and 95\% on a scale of 0 100\% following Cleveland and McGill's methodology\cite{cleveland:1985:graphical}. There was at least a 5\% difference between the two values to compare. In any two charts, at least 6 of the 7 values are different.
% \cite{lin:2013:selecting}

%% file: sections/discussion.tex
\section{Discussion and Limitations}
\label{sec:discussion}
% \marginpar{\todo{make sure that the hypothesis with semantic association for icons, like for colors, is discussed.}} 
\hty{The results of Experiments 2 and 3 slightly deviated from our expectations, but the overall differences were marginal. In Experiment 2, the average BeauVis scores hovered around `neutral,' with a range of opinions causing this median result. Experiment 3 saw the unicolored bar chart surpassing the two textures in terms of readability, time, and accuracy, but the differences were relatively minor (less than 3.6\% in accuracy, and under \harozissue{255ms} in response time). 
%The unicolored chart was also rated slightly more attractive than the geometric bar chart, with a less than 1 point difference on a 7-point Likert scale. 
Practically speaking, these differences may be too small to be substantial. In addition, the results from this simple test of Experiment 3 should \textbf{not} be over-generalized to broad conclusions that ``textures reduce accuracy.'' 
\harozissue{Since textures are considered to be as aesthetically pleasing as unicolor in bar charts, and even more aesthetically pleasing than unicolor in pie charts,} the use of textures could be recommended for those who have a strong preference for aesthetics or specific needs to incorporate textures into their charts.}

%For the pie chart both textures were considered more beautiful than the unicolored pie chart and the response time was faster for the geometric textures than the unicolored chart. Still, people made fewer errors with the unicolored pie chart than with the textured ones. 

Our hypothesis in Experiment 3 about the effects of semantic association on textures, although failed, is still intriguing. Ex\-pe\-ri\-ment~1 demonstrated that semantic association is a quality valued by experts, as they sought to achieve semantic association, not only for iconic textures but also for geometric ones. Interestingly, despite the evident semantic association of iconic textures, previous research on pictographs \cite{haroz:2015:isotype, burns:2021:designing} and our own experiment did not reveal any positive effects on chart reading speed like those observed with semantically resonant colors \cite{lin:2013:selecting}. This may potentially be because icons can be distracting and thus increase reading difficulty. One visualization expert's approach to using the overlapping of icons to abstract them is highly insightful, as it balances other expert strategies of retaining complete icons for their semantic association while simultaneously fading them into the background to prevent visual overload. This method reminded us of Escher's tessellations with recognizable figures and suggests that exploring a middle ground between iconic and geometric textures may be a promising direction in texture design.

%The past research on semantically resonant colors \cite{lin:2013:selecting} can serve as inspiration to explore other tasks where iconic textures might be beneficial; beyond the simple task we chose to replicate in Experiment 3.  

Our observed equal or slightly better performance of unicolor charts compared to textures may also be due to the fact that \emph{\textbf{all}} charts we showed to participants were labeled. The associative quality of textures \cite{bertin:1983:semiology, Bertin:1998:SG} may have enticed participants to use pattern or icon association for finding the right items, while for unicolor charts the lack of any pattern forced participants to read the labels. This was possible in a fast way, in particular for the short, one-word items we used and the lack of ``distraction'' from textures. In situations where the labels are longer or where there is no possibility to have labels in the first place the situation may thus be more favorable for textured charts. 

Finally, we want to acknowledge some limitations of our work. Especially Experiments 2 and 3 are based on specific instantiations of iconic and geometric textures and, as such, it is important not to make general claims for all possible textures of these two types. We also only focused on three basic chart types as we already noted. Textures have a much larger parameter space than color and are, as such, difficult to analyze comprehensively. We hope that our work will spark some interest in the community and that efforts in this space will continue.%
%\hty{Textures' associatve quality \cite{bertin:1983:semiology, Bertin:1998:SG} can help us to quickly identify elements within a group, can also offer potential benefits for data visualization, but is under explored. By applying textures to grouped bar charts and maps, elements within a group can be swiftly recognized, eliminating the need for labels. This advantage can also extends across charts, for example in \autoref{fig:bertin_examples} (right), with texture we can easily identify those parts representing the same category across different chart types.}

\section{\hty{Future Work}}
\hty{Establishing general design guidelines for textures would be extremely useful but will require more work in the community. Several factors influencing the utility of textures in visualizations warrant further exploration:
\textbf{Texture types:} Testing an expanded range of texture variations, such as textures characterized by randomness or freeform polygons, could yield insightful findings. A tool that allows designers complete freedom to add textures to a prepared  chart could also prove to be useful.
\textbf{Representation types:} Texture application significantly varies based on the representation type, as mentioned by our expert participants. Despite numerous examples from Bertin, textured maps underperformed in Experiment 2 according to their lower BeauVis ratings. Thus, the investigation of distinct textures for various chart types---maps in particular---presents an exciting opportunity for future research.
\textbf{Texture discrimination:} It would be interesting to study how many different categories can textures support. Examining the readability of textures at diverse sizes is also crucial. An automated tool warning users when their visualization design is incompatible with their data, especially when textures are indistinguishable, would be particularly beneficial.
\textbf{Task types:} The effect of textures on a wider range of tasks, including those involving long-term memory or multiple category differentiation, should be analyzed.
\textbf{Specific user groups:} Certain groups may benefit significantly from specific texture designs.}

\hty{Moreover, several additional texture application scenarios merit exploration:
\textbf{Accessible visualization: } Black-and-white textures can be easily printed physically or embossed and can be used in other contexts such as embroidery. Their impact on visually impaired individuals could yield valuable insights.
\textbf{Texture with color:} Studying optimal use of black-and-white textures with colors for multivariate data representation is necessary.
\textbf{Data emphasis: } Visual differences across textures may emphasize certain data points. To avoid that this happens unintentionally, we need to incorporate a design feedback loop. We can also potentially use this effect to our advantage. Could we craft textures specifically for pattern detection, or even to introduce bias?}

\section{Conclusion}
\label{sec:conclusion}

So, where does this leave us now? On the one hand, we could not show substantial benefits of textures as a means of associating data representations to data items---akin to a null result. On the other hand, we also learned a lot and the textures did not really fare worse than the baseline. So in situations where color and/or labels are not available for some reason they are valid options for the design of visual representations. What particularly encourages us to continue is the enthusiasm expressed by some of the visualization experts we had approached. One, \eg, stated: ``it's been a fun morning for me. I wish all my mornings could start like this.'' Another said ``nice to see someone doing work on textures''  and many expressed interest in the results. Some also saw the potential for visualization on alternative displays: ``Please make e-ink visualization displays a thing! My tired eyes will thank you.'' 
%\hty{So, ultimately, the appreciation of textures in visualization may be subjective---it depends on the individual beholder.}
So, ultimately, the use of textures in visualization may be in the eye of the beholder---both textures specifically and visualization in general are not “just” a science but also an art.

%In this study, we investigated the design characterization of black-and-white textures in visualization, resulting in a comprehensive design characterization of textures, a collection of 66 visualization designs from expert designers, corresponding design strategies, and two empirical experiments. Our findings elaborated in the previous sections, but the most important findings can be summarized as follows: 
% (1) On
% Iconic textures are perceived as more aesthetically pleasing than geometric textures in bar and pie charts, while the opposite is true for maps. Across all chart types examined, iconic charts were associated with lower vibratory effects compared to geometric textures. (2) Textures do not have a substantial impact on overall chart readability. However, the use of textures can enhance readability for specific chart types, such as employing geometric textures in pie charts.

%% file: sections/appendix.tex
\clearpage

\begin{strip}
\noindent\begin{minipage}{\textwidth}
\makeatletter
\centering%
\sffamily\bfseries\fontsize{15}{16.5}\selectfont
\vgtc@title\\[.5em]
\large Appendix\\[.75em]
\makeatother
\normalfont\rmfamily\normalsize\noindent\raggedright In this appendix we provide additional tables, plots, and charts that show data beyond the material that we could include in the main paper due to space limitations or because it was not essential for explaining our approach.%\vspace{-.5em}
\end{minipage}
\end{strip}

\appendix

\section{Default Geometric Texture Sets from Bertin}
\label{appendix:default-bertin}
In Experiment~1 we included five default texture sets for geometric textures in our texture design interface (see \autoref{sub-sec:design-probe} and \autoref{appendix:additional-exp1}). We derived these sets, which we show in \autoref{tab:bertin-textures} below, from visualizations that we found in Bertin's book \cite{bertin:1983:semiology, Bertin:1998:SG}. Bertin had originally employed these textures to represent nominal or ordinal data.

\setlength{\pictureheight}{0.087\columnwidth}
\begin{figure}[h]
	\centering
    \begin{tabular}{lllllll}
     \includegraphics[height=\pictureheight]{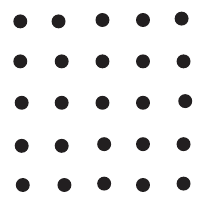}& \includegraphics[height=\pictureheight]{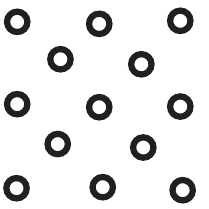}& \includegraphics[height=\pictureheight]{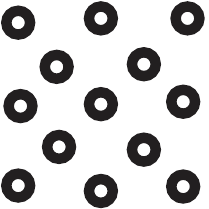}& \includegraphics[height=\pictureheight]{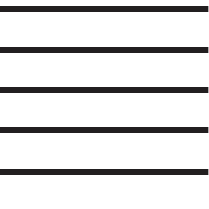}& \includegraphics[height=\pictureheight]{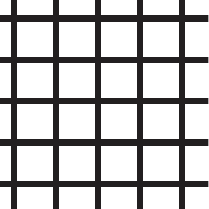}& \includegraphics[height=\pictureheight]{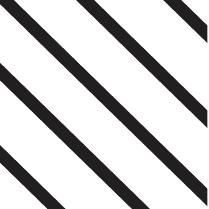}& \includegraphics[height=\pictureheight]{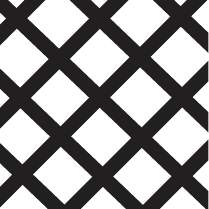}\\[1ex]
     \includegraphics[height=\pictureheight]{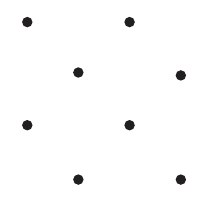}& \includegraphics[height=\pictureheight]{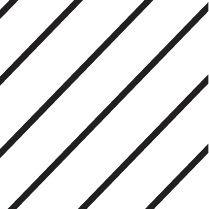}& \includegraphics[height=\pictureheight]{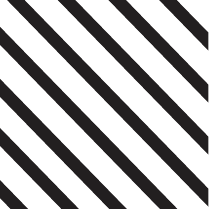}& \includegraphics[height=\pictureheight]{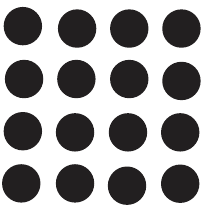}& \includegraphics[height=\pictureheight]{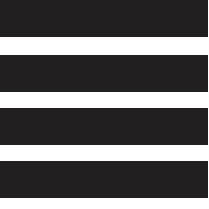}& \includegraphics[height=\pictureheight]{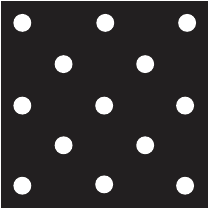}& \includegraphics[height=\pictureheight]{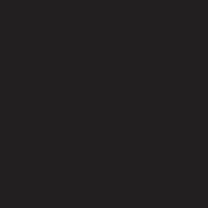}\\[1ex]
     \includegraphics[height=\pictureheight]{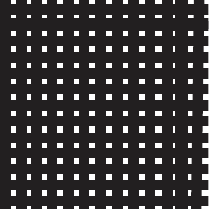}& \includegraphics[height=\pictureheight]{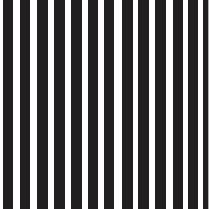}& \includegraphics[height=\pictureheight]{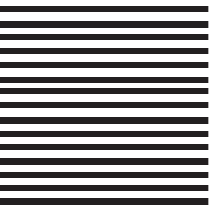}& \includegraphics[height=\pictureheight]{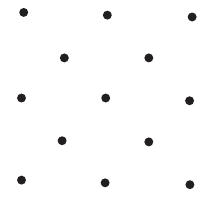}& \includegraphics[height=\pictureheight]{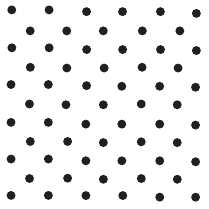}& \includegraphics[height=\pictureheight]{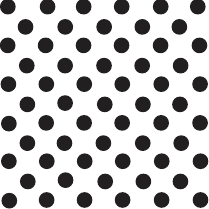}& \includegraphics[height=\pictureheight]{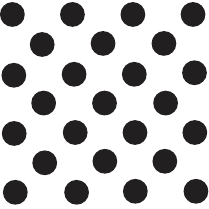}\\[1ex]
     \includegraphics[height=\pictureheight]{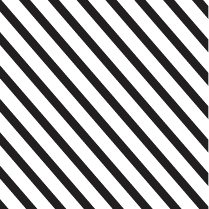}& \includegraphics[height=\pictureheight]{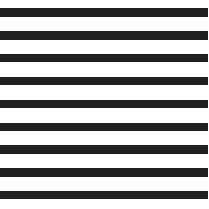}& \includegraphics[height=\pictureheight]{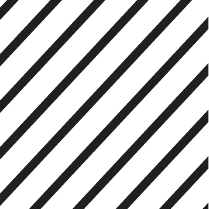}& \includegraphics[height=\pictureheight]{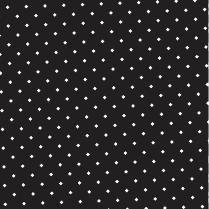}& \includegraphics[height=\pictureheight]{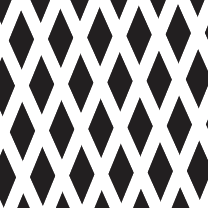}& \includegraphics[height=\pictureheight]{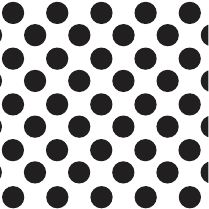}& \includegraphics[height=\pictureheight]{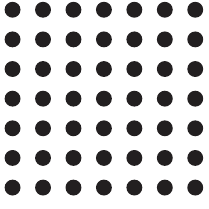}\\[1ex]
      \includegraphics[height=\pictureheight]{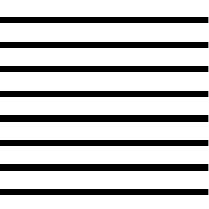}& \includegraphics[height=\pictureheight]{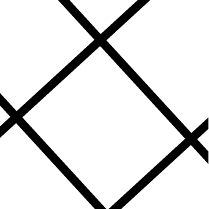}& \includegraphics[height=\pictureheight]{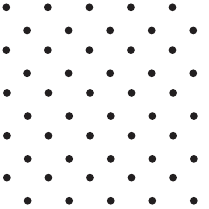}& \includegraphics[height=\pictureheight]{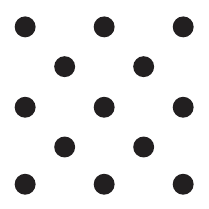}& \includegraphics[height=\pictureheight]{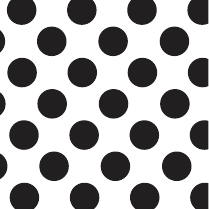}& \includegraphics[height=\pictureheight]{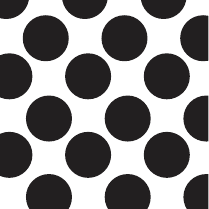}& \includegraphics[height=\pictureheight]{figures/experiment1/bertin_textures/black.pdf} 
    \end{tabular}
    \caption{The five texture sets (rows) we included in Experiment~1 as defaults, inspired by visualizations from Bertin's book \cite{bertin:1983:semiology, Bertin:1998:SG}.}
    \label{tab:bertin-textures}
\end{figure}

\section{Used Icon Sets from and Inspired by \href{https://icons8.com/}{Icon8.com}}
\label{appendix:default-icon}

For Experiment~1 we chose two professionally designed, neutral, and stylized icon sets from \href{https://icons8.com/}{Icon8.com} \cite{icons8} for our iconic textures to represent the vegetable items, as we show in the top two rows in \autoref{tab:fruit-icons} (one light one and the corresponding dark variant). In addition, we wanted to provided the participants with two corresponding simplified texture sets. As we did not find complete, matching sets on \href{https://icons8.com/}{Icon8.com}, we ourselves used the original, detailed icons and eliminated details and streamlined the outlines, as we show in the bottom two rows in \autoref{tab:fruit-icons}. The only icon that we did not change is that for the mushroom, as there was no detail that we could reasonably remove. In total, in Experiment~1 we thus provided participants with four distinct icon sets (\autoref{tab:fruit-icons}).

\setlength{\pictureheight}{0.087\columnwidth}
\begin{figure}[h]
\centering
    \begin{tabular}{lllllll}
     \includegraphics[height=\pictureheight]{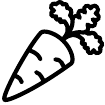}&\includegraphics[height=\pictureheight]{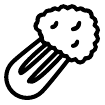}  &\includegraphics[height=\pictureheight]{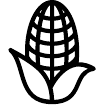}  &\includegraphics[height=\pictureheight]{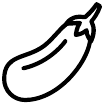} &\includegraphics[height=\pictureheight]{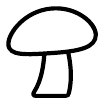}  & \includegraphics[height=\pictureheight]{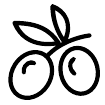} & \includegraphics[height=\pictureheight]{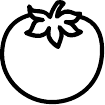}\\[1ex]
     \includegraphics[height=\pictureheight]{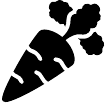}&\includegraphics[height=\pictureheight]{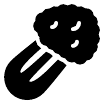}  &\includegraphics[height=\pictureheight]{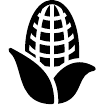}  &\includegraphics[height=\pictureheight]{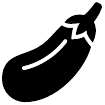}  &\includegraphics[height=\pictureheight]{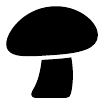}  & \includegraphics[height=\pictureheight]{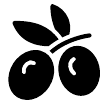} & \includegraphics[height=\pictureheight]{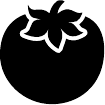}\\[1ex]
    \includegraphics[height=\pictureheight]{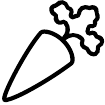}&\includegraphics[height=\pictureheight]{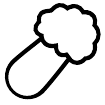}  &\includegraphics[height=\pictureheight]{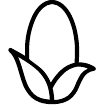}  &\includegraphics[height=\pictureheight]{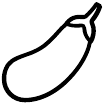}  &\includegraphics[height=\pictureheight]{figures/experiment1/fruit_icons/stroke-mushroom.pdf}  & \includegraphics[height=\pictureheight]{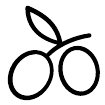} & \includegraphics[height=\pictureheight]{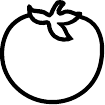}\\[1ex]
    \includegraphics[height=\pictureheight]{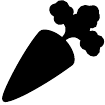}&\includegraphics[height=\pictureheight]{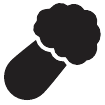}  &\includegraphics[height=\pictureheight]{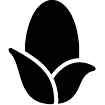}  &\includegraphics[height=\pictureheight]{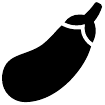}  &\includegraphics[height=\pictureheight]{figures/experiment1/fruit_icons/fill-mushroom.pdf}  & \includegraphics[height=\pictureheight]{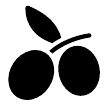} & \includegraphics[height=\pictureheight]{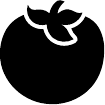} 
    \end{tabular}
    \caption{Icon sets included in Experiment~1. The first and second rows of icons are collected from \href{https://icons8.com/}{Icon8.com}, the third and fourth rows of icons are simplified versions we created outselves. The icons in the top two rows are \textcopyright{} \href{https://icons8.com/}{Icon8.com}, used with permission.}
    \label{tab:fruit-icons}
\end{figure}

\section{Additional Information on Experiment 1}
\label{appendix:additional-exp1}

\begin{figure}[b] % htbp are optional placement specifiers (here, top, bottom, page)
	\centering % Centers the figure
	\includegraphics[width=\columnwidth]{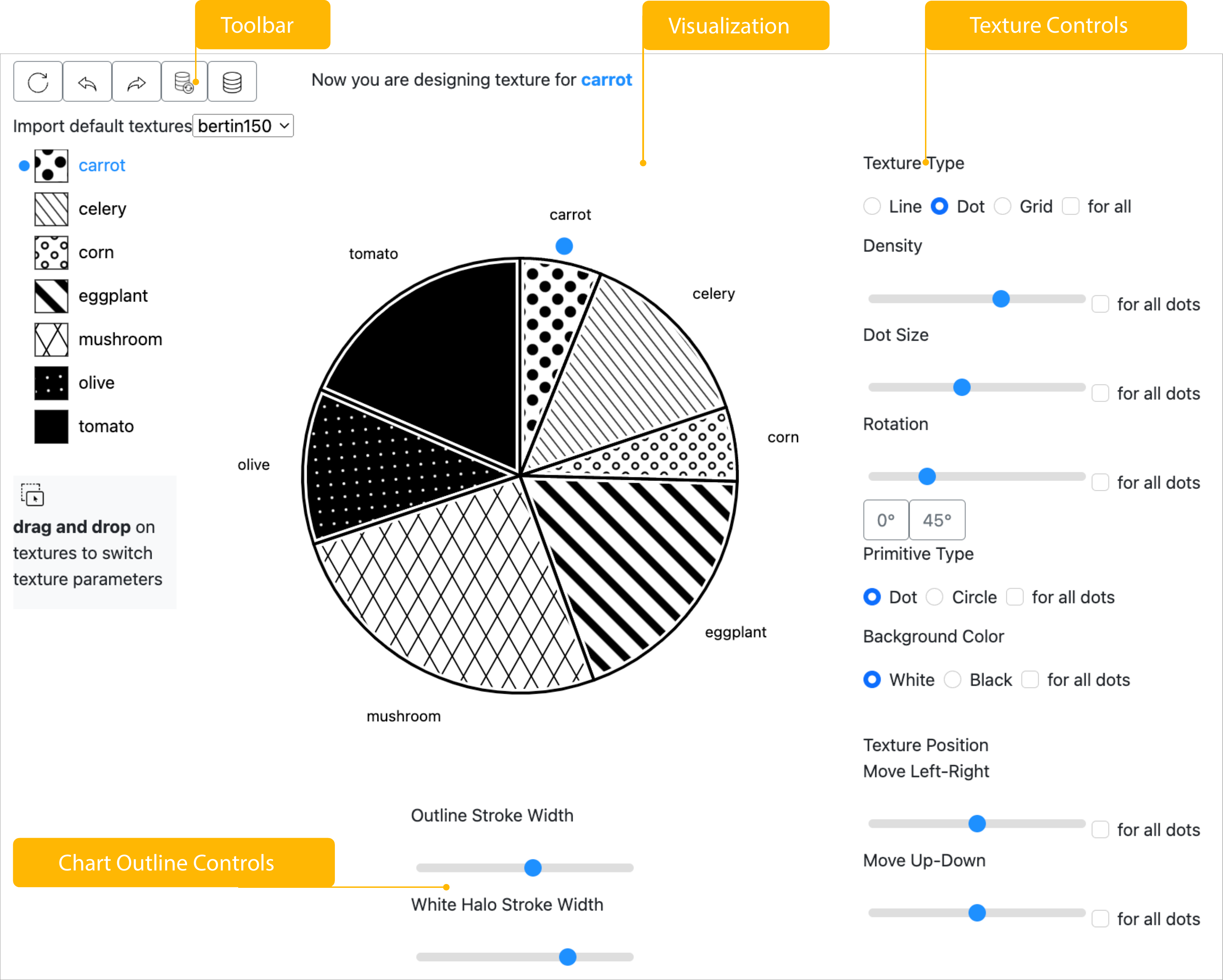}\
	\caption{Technology probe for designing a textures for used in charts, here shown for pie charts with geometric textures. The annotations point out the elements we discuss in \autoref{sub-sec:design-probe}.}
  \label{fig:design_probe_screenshot}
\end{figure}

\autoref{fig:design_probe_screenshot} shows a screenshot of our web-based technology probe for creating visualizations using black-and-white textures by adjusting parameters via buttons and sliders. \autoref{fig:iconic-default-texture} shows an iconic pie chart with the default iconic texture on our web-based technology probe.

\setlength{\picturewidth}{\columnwidth}
\setlength{\fboxsep}{2pt}
\addtolength{\picturewidth}{-2\fboxrule}
\addtolength{\picturewidth}{-2\fboxsep}
\begin{figure}[b] % htbp are optional placement specifiers (here, top, bottom, page)
	\centering % Centers the figure
	\fbox{%
	\includegraphics[width=\picturewidth]{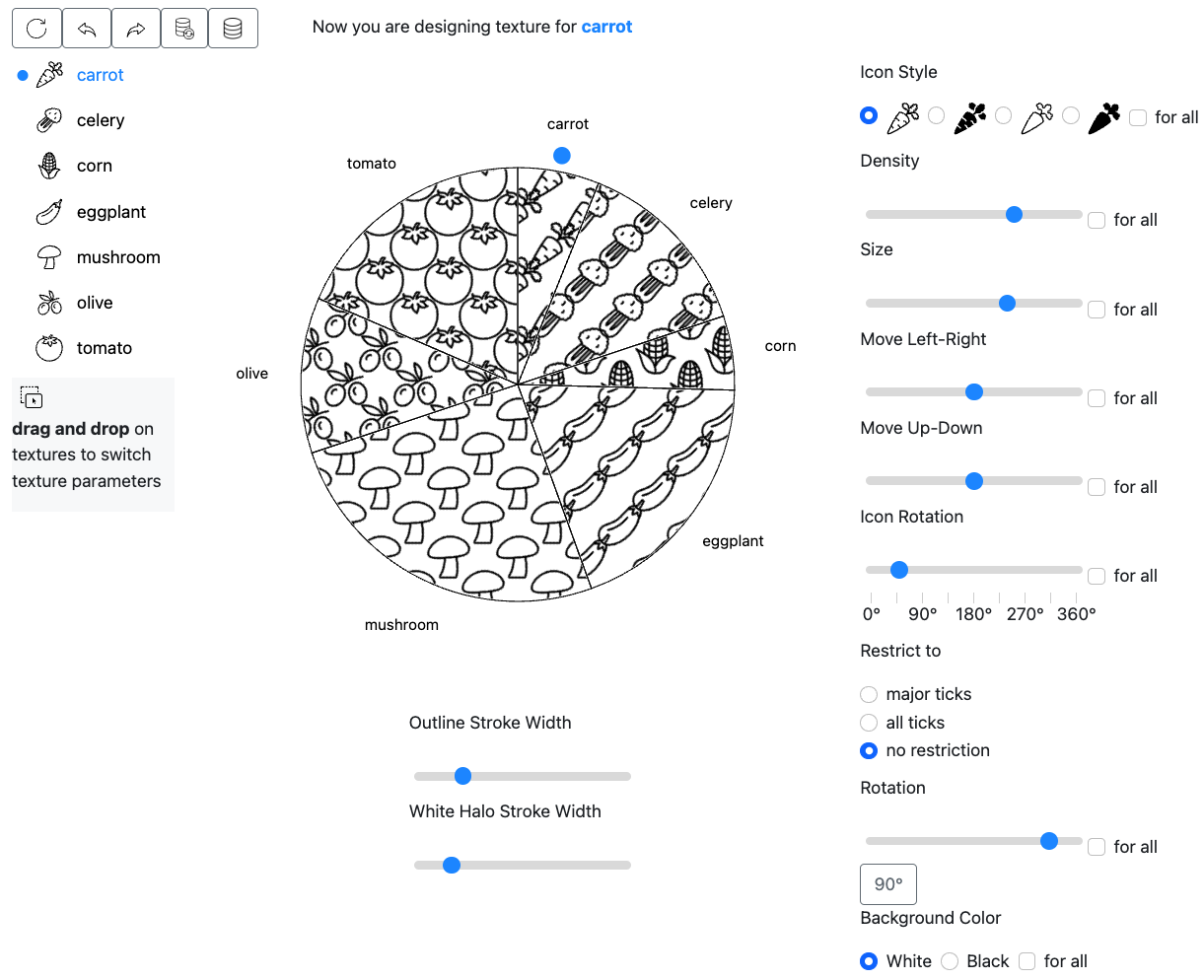}}
	\caption{An iconic pie chart with the default iconic texture, being edited in the technology probe. Notice the slightly changed interface elements compared to \autoref{fig:design_probe_screenshot}; these changes were necessary to allow users to control the different texture type.}
  \label{fig:iconic-default-texture}
\end{figure}

\hty{In Experiment 1, we applied the textures designed by participants for a specific chart type to two other chart types. We then asked the participants if they believed the textures still worked well with the other chart types. \autoref{tab:apply_to_another_charts} shows the percentage of designs that participants considered to still work in each condition.}

\begin{table}[t]
  \centering
\footnotesize%
  \caption{Percent of designs that still worked for another chart type.}\vspace{-1ex}
  \label{tab:apply_to_another_charts}
    \begin{tabular}{lccc}
    \toprule
    \cellcolor{white}texture design \textbackslash{} applied to & \cellcolor{white}bar & \cellcolor{white}pie & \cellcolor{white}map \\
    \midrule
    % \hline
    \cellcolor{white}geometric bar & \cellcolor{white}/ & \cellcolor{antiquefuchsia!25}57.1\% & \cellcolor{antiquefuchsia!20}28.6\% \\
    % \hline
    \cellcolor{white}iconic bar & \cellcolor{white}/ & \cellcolor{antiquefuchsia!50}100\% & \cellcolor{antiquefuchsia!10}28.6\% \\
    % \hline
    \cellcolor{white}geometric pie & \cellcolor{antiquefuchsia!25}53.3\% & \cellcolor{white}/ & \cellcolor{antiquefuchsia!10}26.7\% \\
    % \hline
    \cellcolor{white}iconic pie & \cellcolor{antiquefuchsia!35}73.3\% & \cellcolor{white}/ & \cellcolor{antiquefuchsia!5}13.3\% \\
    % \hline
    \cellcolor{white}geometric map & \cellcolor{antiquefuchsia!45}90.9\% & \cellcolor{antiquefuchsia!45}90.9\% & \cellcolor{white}/ \\
    % \hline
    \cellcolor{white}iconic map & \cellcolor{antiquefuchsia!35}72.7\% & \cellcolor{antiquefuchsia!40}81.8\% & \cellcolor{white}/ \\
    % \hline
    \bottomrule
    \end{tabular}%
\end{table}%

\section{Detailed Description of the Stimuli Selection Process for Experiment 3}
\label{appendix:decision-e3}
%\todo{I suggest to cut this and the following 2 paragraphs and to move them to the appendix. Just refer to the selected designs is enough here with the description above.}
For both geometric and iconic textures for pie charts, the texture with the highest BeauVis score and the highest number of being ranked first were consistent. We thus confidently selected these top 2 designs (PG1 in \autoref{tab:exp2-pie-geo} resp.\ \autoref{fig:PG1} and PI1 in \autoref{tab:exp2-pie-icon} resp.\ \autoref{fig:PI1}).
For iconic textures for bar charts, despite the BeauVis score and top-ranking frequency being inconsistent, the choice was clear. The most aesthetic design (BI1 in \autoref{tab:exp2-bar-icon} resp.\ \autoref{fig:BI1}) had a much higher BeauVis score than the most frequently ranked first design (BI3, shown in \autoref{fig:BI3}), and their top-ranking frequencies were similar. In addition, the most aesthetic design received a significantly lower vibratory score, leading us to choose it as the best iconic texture for bar charts.

Selecting the geometric texture for bar charts, however, was challenging as we had to decide between the first (BG1, shown in \autoref{fig:BG1}) and second (BG2, shown in \autoref{fig:BG2}) designs (comparison in \autoref{tab:exp2-bar-geo}). BG1 had the highest BeauVis score (4.70), ranked first 16\texttimes, and a vibratory score of 3.83. BG2 had a BeauVis score of 4.45, ranked first 20\texttimes{} (the highest), and had a better vibratory score (3.66) than BG1. To make a decision, we conducted a qualitative coding analysis of the reasons participants provided for ranking these textures as their top choice. For BG1, the most frequently mentioned reasons were aesthetics (7\texttimes) and ease of distinction (5\texttimes), while for BG2 they were aesthetics (9\texttimes) and visual comfort (6\texttimes). Considering these reasons collectively and factoring in the lower vibratory score of BG2, we decided on BG2. Notably, due to technical issues, the texture in BG2 was slightly shifted in our previous experiment, suggesting that the original version might have received even higher ratings. We thus chose to use the originally designed version in this experiment.    

\section{Additional Information on Experiment 3}
\label{appendix:additional-exp3}

\setlength{\picturewidth}{\columnwidth}
\setlength{\fboxsep}{2pt}
\addtolength{\picturewidth}{-2\fboxrule}
\addtolength{\picturewidth}{-2\fboxsep}
\begin{figure}[b] % htbp are optional placement specifiers (here, top, bottom, page)
	\centering % Centers the figure
	\fbox{%
	\includegraphics[width=\picturewidth]{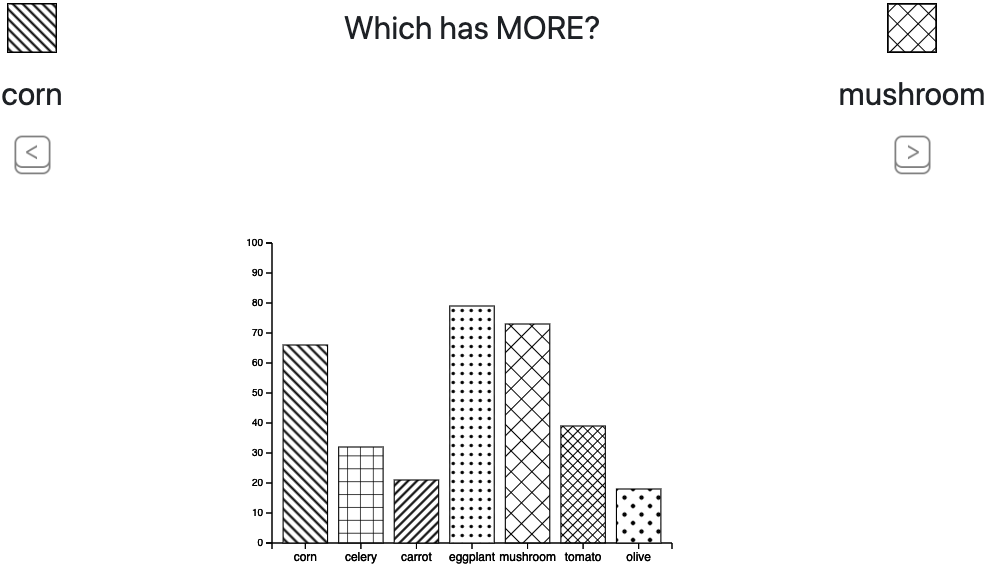}}
	\caption{One trial in Experiment 3 with bar charts, geometric textures, and asking to identify the item with a higher value (``MORE'').}
  \label{fig:exp3-screenshot-geo}
\end{figure}

\setlength{\picturewidth}{\columnwidth}
\setlength{\fboxsep}{2pt}
\addtolength{\picturewidth}{-2\fboxrule}
\addtolength{\picturewidth}{-2\fboxsep}
\begin{figure}[t] % htbp are optional placement specifiers (here, top, bottom, page)
	\centering % Centers the figure
	\fbox{%
	\includegraphics[width=\picturewidth]{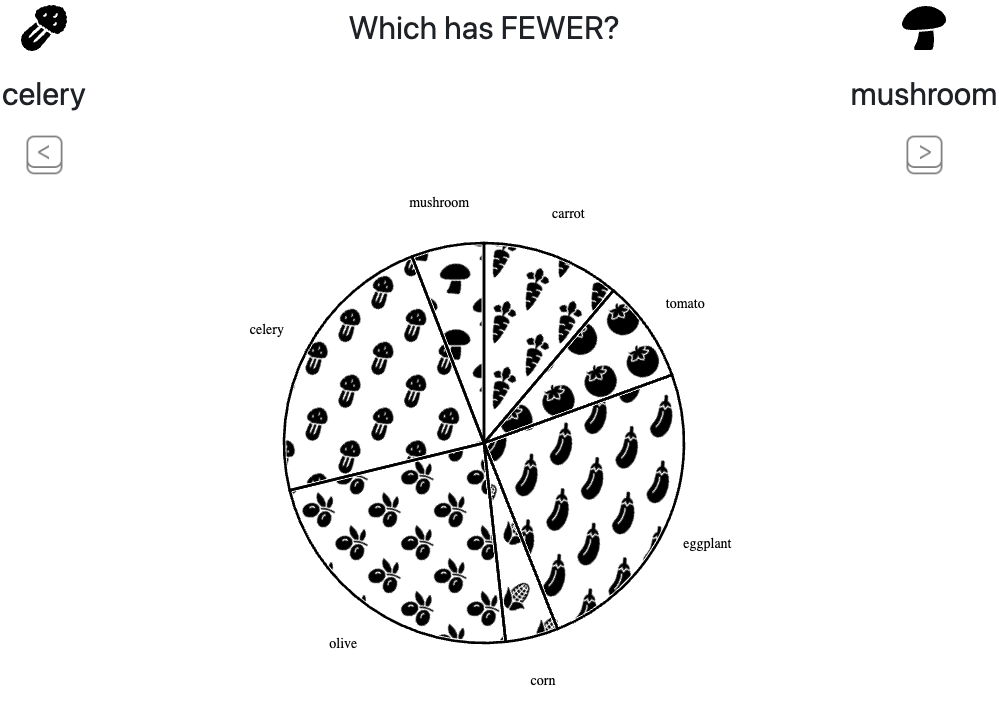}}
	\caption{One trial in Experiment 3 with pie charts, iconic textures, and asking to identify the item with a lower value (``FEWER'').}
  \label{fig:exp3-screenshot-icon}
\end{figure}

\autoref{fig:exp3-screenshot-geo} through \ref{fig:exp3-screenshot-gray} show three screenshots taken during a trial in Experiment 3, representing varying chart types, fill styles, and question categories. The target items are shown on the top left and top right, respectively, with their associated texture samples and with the question to be answered on the top in the middle.

\setlength{\picturewidth}{\columnwidth}
\setlength{\fboxsep}{2pt}
\addtolength{\picturewidth}{-2\fboxrule}
\addtolength{\picturewidth}{-2\fboxsep}
\begin{figure}[t] % htbp are optional placement specifiers (here, top, bottom, page)
	\centering % Centers the figure
	\fbox{%
	\includegraphics[width=\picturewidth]{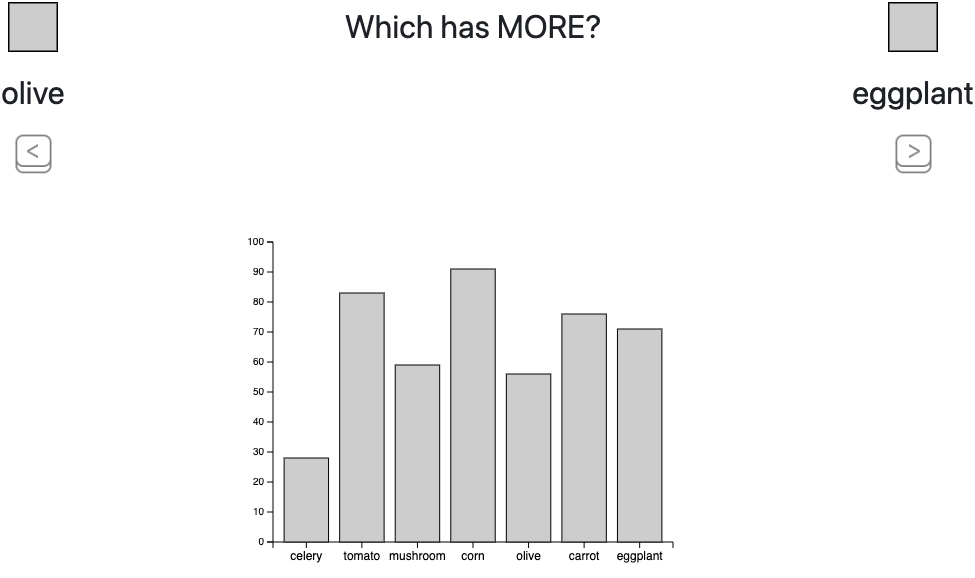}}
	\caption{One trial in Experiment 3 with bar charts, no textures, and asking to identify the item with a higher value (``MORE'').}
  \label{fig:exp3-screenshot-gray}
\end{figure}

\hty{\autoref{tab:exp3-timeout} shows the distribution of time-out trials in the different conditions in Experiment 3.}

\begin{table}[]
\caption{Number of trials per condition that timed out in Experiment 3.}
\label{tab:exp3-timeout}
\centering%
\footnotesize%
\begin{tabu}{lccc}
\toprule
chart \textbackslash{} texture & \multicolumn{1}{l}{geometric} & \multicolumn{1}{l}{iconic} & \multicolumn{1}{l}{unicolor} \\
\midrule
bar                          & 12                            & 26                         & 8  \\
pie                          & 23                            & 31                         & 25 \\
\bottomrule
\end{tabu}
\end{table}

\harozissue{\section{Original Analysis in Experiment 3}}
\label{appendix:original-analysis-e3}

\noindent\harozissue{Initially, we included all 150 responses in the analysis of response time, readability, and aesthetics, as pre-registered on the OSF platform. Later, in addition to our pre-registered analysis plan, we examined the individual accuracy rate per participant. We found 64 participants whose overall accuracy rate was below 90\%. We decided to adjust our approach by excluding these low-accuracy participants to minimize the influence of chance performance (\ie, random guessing), because these low-accuracy participants may have largely guessed randomly. This is an additional exclusion criterion, in addition to the exclusion criteria outlined in our pre-registration. We now only counted the 86 participants who achieved a 90\% overall accuracy threshold (45\texttimes{} bar, 41\texttimes{} pie). Note that there are two trials in our data where we recorded that the participants gave the correct answers, but their recorded response times were slightly above 5 seconds. The times for these two trials were 5.002 and 5.006 seconds, respectively, which should be timed out. We speculate that this situation occurred because, due to network latency, the page did not redirect in time, allowing the participants the opportunity to input their answers, which were then recorded. Given that we know that these two trials are correct, and that the differences between their duration and the 5-second threshold were minimal, we still counted them as correct trials when calculating the accuracy rate.}

\harozissue{In addition, we note that in our pre-registration we decided to remove ``incomplete responses'' from our analysis in Experiment 3. With this wording we intended to refer to those participants who did not complete our experiment; \ie, those who quit the experiment midway and did not reach the last page (and we indeed excluded those participants). Another interpretation of our wording could have been to refer to participants with missing trials due to the log file issue (we lost 12 trials out of 9,000 trials), which is what we did not intend to mean. So ultimately we did not remove the 6 participants with missing trials (4 missed 1 trial, 1 missed 2 trials, and 1 missed 6 trials), because these comparatively few missing trials do not affect other trials.}

\harozissue{Below we present \autoref{fig:exp3-response-time-original}--\ref{fig:exp3-beauvis-original}, which show the results of our original analysis (\ie, as pre-registered). We also discuss the difference observed in the refined analysis as compared to the original analysis.}

\harozissue{\subsection{Response time}}
\noindent\harozissue{We initially included all participants and counted both their correct and incorrect trials. We removed the few timed-out trials ($<$\,1.5\%) as we could not estimate whether a person was distracted or how much more time they would have needed. \autoref{fig:exp3-response-time-original} presents mean response times and pairwise comparisons for all fill types in bar and pie charts from the original analysis (as pre-registered). The pairwise differences indicate that, for bar charts, we have evidence that iconic textures have a longer response time than the other two fill types. For pie charts, we have evidence that geometric textures have shorter response times than the other two fill types. No other combination of fill types showed an evident difference. In addition, all these differences were minimal, within a range of $<$\,230\,ms.}

\harozissue{Later, we improved our analysis approach, as previously mentioned. In addition, for response time analysis specifically, we only counted the correct trials. This exclusion is necessary due to the difficulty in interpreting the speed of incorrect responses, and because averaging the response times of both correct and incorrect trials does not logically make sense. So, in our adjusted approach we now analyze the response times of correct trials from the 86 high-accuracy participants, excluding both incorrect and timed-out trials.}

\harozissue{\autoref{fig:exp3-response-time} presents the mean response times and pairwise comparisons for all fill types, as represented in both bar and pie charts, from our refined analysis. A detailed explanation of these results can be found in \autoref{sec:exp3-response-time} in the main paper. In summary, the only change in our findings is the observed evidence of longer response times for geometric textures compared to unicolor fill in bar charts, a difference that was not evident in our original analysis. All other results remained consistent, and the outcomes for both of our hypotheses were unaffected. The differences in response times across the three fill types remained minimal, within a range of $<$\,255\,ms, thereby also maintaining our overall conclusion.}

\setlength{\figraisecaptionoffset}{-16pt}%
\setlength{\figaftercaptionoffset}{0pt}%
\setlength{\figaftercaptionextraoffset}{-10pt}%
\setlength{\figraisecaptionoffsetwithextra}{\figraisecaptionoffset}%
\addtolength{\figraisecaptionoffsetwithextra}{\figaftercaptionextraoffset}%
\begin{figure}
    \centering
    \subcaptionbox{~\hspace{\columnwidth}~}
    {\hspace{.05\columnwidth}\includegraphics[width=0.95\columnwidth,trim={0 23pt 0 0},clip]{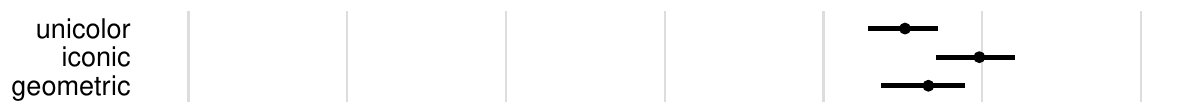}\vspace{\figraisecaptionoffset}}\\[\figaftercaptionoffset]
    \subcaptionbox{~\hspace{\columnwidth}~}
    {\hspace{.05\columnwidth}\includegraphics[width=0.95\columnwidth,trim={0 0 0 0},clip]{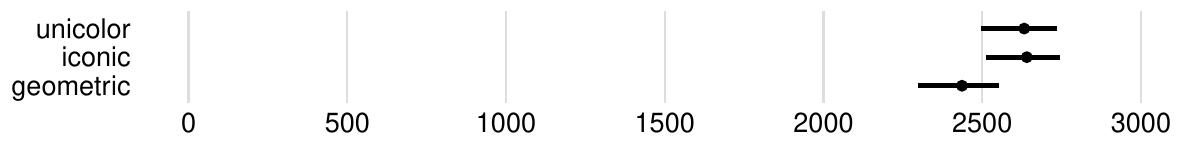}\vspace{\figraisecaptionoffsetwithextra}}\\[\figaftercaptionoffset]%
\vspace{5pt}%
\setlength{\figraisecaptionoffset}{-17pt}%
\setlength{\figaftercaptionoffset}{-2pt}%
\setlength{\figaftercaptionextraoffset}{-10pt}%
\setlength{\figraisecaptionoffsetwithextra}{\figraisecaptionoffset}%
\addtolength{\figraisecaptionoffsetwithextra}{\figaftercaptionextraoffset}%
    \subcaptionbox{~\hspace{\columnwidth}~}
    {\hspace{.05\columnwidth}\includegraphics[width=0.95\columnwidth,trim={0 20pt 0 0},clip]{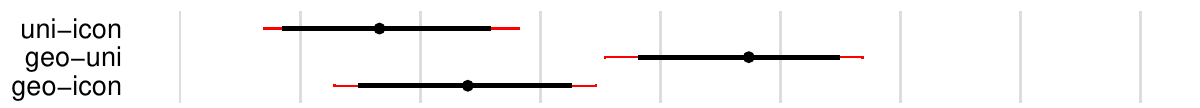}\vspace{\figraisecaptionoffset}}\\[\figaftercaptionoffset]
    \subcaptionbox{~\hspace{\columnwidth}~}
    {\hspace{.05\columnwidth}\includegraphics[width=0.95\columnwidth,trim={0 0 0 0},clip]{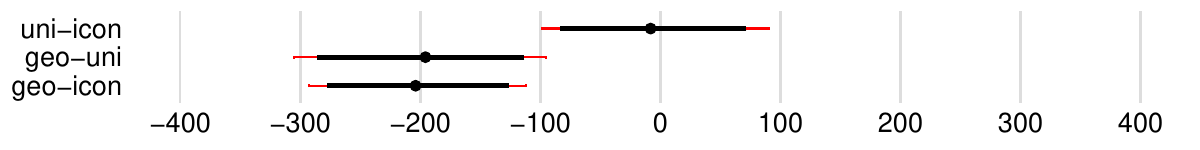}\vspace{\figraisecaptionoffsetwithextra}}
\vspace{-\figaftercaptionextraoffset}%
		\vspace{-1.5ex}%
    \caption{Results of our \harozissue{original} analysis for response times \harozissue{(as preregistered)}. Response times in ms for (a) bar and (b) pie charts; (c), (d) corresponding pairwise comparisons between the fill types. Error bars: 95\% CIs. Red bars: CIs for Bonferroni-corrected pairwise comparison.}%\vspace{-1ex}
    \label{fig:exp3-response-time-original}
\end{figure}

\harozissue{\subsection{Readability}}
\noindent\harozissue{In \autoref{fig:exp3-readable-original} we present the mean readability scores, along with pairwise comparisons, for all fill types in both bar and pie charts from the original analysis (as pre-registered). There is no change in the results between the original (pre-registered) and the refined analysis.}

\setlength{\figraisecaptionoffset}{-18pt}%
\setlength{\figaftercaptionoffset}{1pt}%
\setlength{\figaftercaptionextraoffset}{-10pt}%
\setlength{\figraisecaptionoffsetwithextra}{\figraisecaptionoffset}%
\addtolength{\figraisecaptionoffsetwithextra}{\figaftercaptionextraoffset}%
\begin{figure}
    \centering
    \subcaptionbox{~\hspace{\columnwidth}~}
    {\hspace{.05\columnwidth}\includegraphics[width=0.95\columnwidth,trim={0 23pt 0 0},clip]{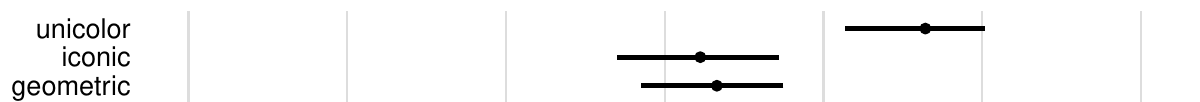}\vspace{\figraisecaptionoffset}}\\[\figaftercaptionoffset]
    \subcaptionbox{~\hspace{\columnwidth}~}
    {\hspace{.05\columnwidth}\includegraphics[width=0.95\columnwidth,trim={0 0 0 0},clip]{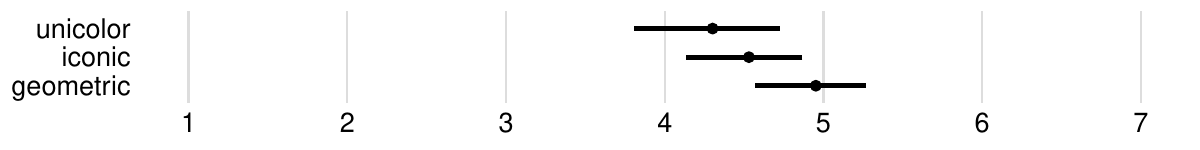}\vspace{\figraisecaptionoffsetwithextra}}\\[\figaftercaptionoffset]%
\vspace{8pt}%
\setlength{\figraisecaptionoffset}{-18pt}%
\setlength{\figaftercaptionoffset}{-1pt}%
\setlength{\figaftercaptionextraoffset}{-10pt}%
\setlength{\figraisecaptionoffsetwithextra}{\figraisecaptionoffset}%
\addtolength{\figraisecaptionoffsetwithextra}{\figaftercaptionextraoffset}%
    \subcaptionbox{~\hspace{\columnwidth}~}
    {\hspace{.05\columnwidth}\includegraphics[width=0.95\columnwidth,trim={0 20pt 0 0},clip]{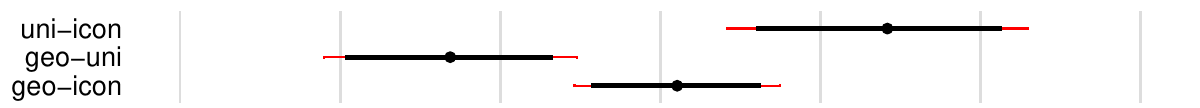}\vspace{\figraisecaptionoffset}}\\[\figaftercaptionoffset]
    \subcaptionbox{~\hspace{\columnwidth}~}
    {\hspace{.05\columnwidth}\includegraphics[width=0.95\columnwidth,trim={0 0 0 0},clip]{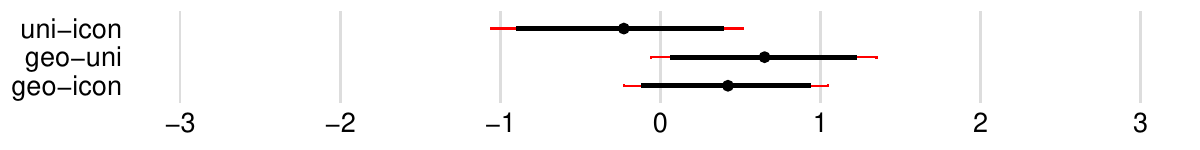}\vspace{\figraisecaptionoffsetwithextra}}
\vspace{-\figaftercaptionextraoffset}%
		\vspace{-1.5ex}%
    \caption{Results of our \harozissue{original} analysis for readability scores \harozissue{(as preregistered)}. Readability scores for (a) bar and (b) pie charts; (c), (d) corresponding pairwise comparisons between the fill types. Error bars: 95\% CIs. Red bars: CIs for Bonferroni-corrected pairwise comparison.}%\vspace{-1ex}
    \label{fig:exp3-readable-original}
\end{figure}

\harozissue{\subsection{Aesthetics}}
\noindent\harozissue{\autoref{fig:exp3-beauvis-original} presents mean BeauVis scores and pairwise comparisons for all fill types in bar and pie charts from the original analysis (as pre-registered). The only difference in the results is that in the original analysis, from pairwise differences (\autoref{fig:exp3-beauvis-original}(c)), we see evidence suggesting that geometric textures are considered to be less aesthetically pleasing than unicolor fill for bar charts. Pairwise differences of bar charts from the refined analysis (\autoref{fig:exp3-beauvis}(c)), however, reveal no evidence of difference between geometric textures and unicolor fill with respect to whether they are considered to be aesthetically pleasing or not. Because geometric textures are still considered by participants to be less aesthetically pleasing than iconic textures for bar charts, however, the refined analysis does not change our result of the related hypothesis (H2), nor does it affect our overall conclusion.}%\vspace{1em}

\setlength{\figraisecaptionoffset}{-18pt}%
\setlength{\figaftercaptionoffset}{1pt}%
\setlength{\figaftercaptionextraoffset}{-10pt}%
\setlength{\figraisecaptionoffsetwithextra}{\figraisecaptionoffset}%
\addtolength{\figraisecaptionoffsetwithextra}{\figaftercaptionextraoffset}%
\begin{figure}
    \centering
    \subcaptionbox{~\hspace{\columnwidth}~}
    {\hspace{.05\columnwidth}\includegraphics[width=0.95\columnwidth,trim={0 23pt 0 0},clip]{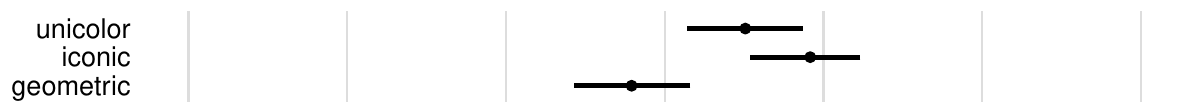}\vspace{\figraisecaptionoffset}}\\[\figaftercaptionoffset]
    \subcaptionbox{~\hspace{\columnwidth}~}
    {\hspace{.05\columnwidth}\includegraphics[width=0.95\columnwidth,trim={0 0 0 0},clip]{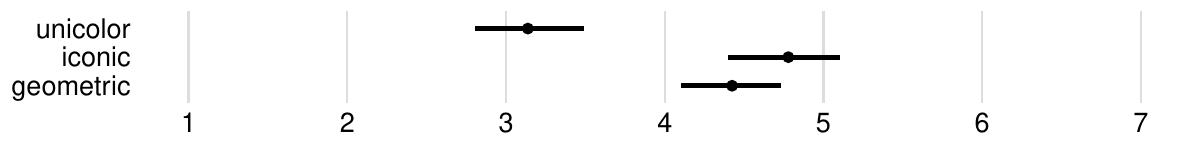}\vspace{\figraisecaptionoffsetwithextra}}\\[\figaftercaptionoffset]%
\vspace{8pt}%
\setlength{\figraisecaptionoffset}{-18pt}%
\setlength{\figaftercaptionoffset}{-1pt}%
\setlength{\figaftercaptionextraoffset}{-10pt}%
\setlength{\figraisecaptionoffsetwithextra}{\figraisecaptionoffset}%
\addtolength{\figraisecaptionoffsetwithextra}{\figaftercaptionextraoffset}%
    \subcaptionbox{~\hspace{\columnwidth}~}
    {\hspace{.05\columnwidth}\includegraphics[width=0.95\columnwidth,trim={0 20pt 0 0},clip]{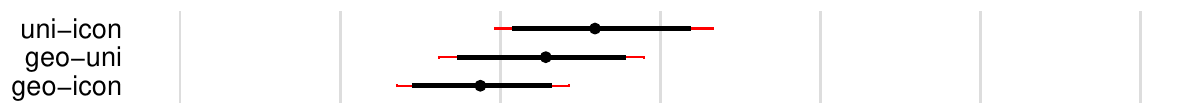}\vspace{\figraisecaptionoffset}}\\[\figaftercaptionoffset]
    \subcaptionbox{~\hspace{\columnwidth}~}
    {\hspace{.05\columnwidth}\includegraphics[width=0.95\columnwidth,trim={0 0 0 0},clip]{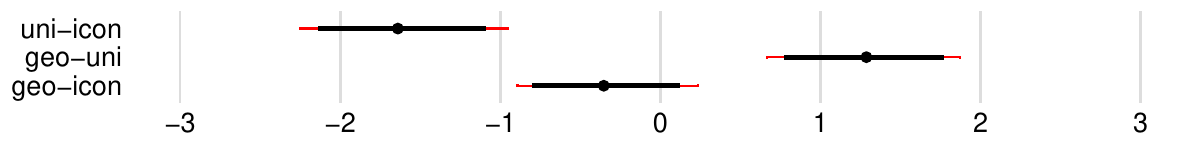}\vspace{\figraisecaptionoffsetwithextra}}
\vspace{-\figaftercaptionextraoffset}%
		\vspace{-1.5ex}%
    \caption{Results of our \harozissue{original} analysis for BeauVis scores \harozissue{(as preregistered)}. BeauVis scores for (a) bar and (b) pie charts; (c), (d) corresponding pairwise comparisons between the fill types. Error bars: 95\% CIs. Red bars: CIs for Bonferroni-corrected pairwise comparison.}%\vspace{-1ex}
    \label{fig:exp3-beauvis-original}
\end{figure}

\section{All Designs Generated by the Visualization Experts in Experiment 1}
\label{appendix:all-designs}

In \autoref{fig:BG1}--\ref{fig:MI11} we show the 66 designs we collected from 30 visualization designers in Experiment ~1. The collection comprises 14 bar charts (\autoref{fig:BG1}--\ref{fig:BI7}), 30 pie charts (\autoref{fig:PG1}--\ref{fig:PI15}), and 22 maps (\autoref{fig:MG1}--\ref{fig:MI11}).

We include all these images (here and also the images in the main paper such as the teaser (\autoref{fig:teaser}) and Tables \ref{tab:exp2-bar-geo}--\ref{tab:exp2-map-icon}) as pixel images on purpose because the SVG vector version relies on tiled texture samples, which---when converted to PDF for the inclusion in the paper---lead to unfortunate errors in the display in all PDF readers we tested. Likely this effect is due to numeric issues that affect the exact positions where the texture tiles meet. Nonetheless, you can find the original SVG images in our OSF repository at \href{https://osf.io/n5zut/}{\texttt{osf.io/n5zut}} and you can look at them with a browser such as Chrome, Microsoft Edge, or Firefox.

\section*{Images/graphs/plots/tables/data license/copyright}
We as authors state that all of our own figures, graphs, plots, and data tables in this appendix (\ie, those for which we did not cite a specific copyright in the caption) are and remain under our own personal copyright, with the permission to be used here. We also make them available under the \href{https://creativecommons.org/licenses/by/4.0/}{Creative Commons At\-tri\-bu\-tion 4.0 International (\ccLogo\,\ccAttribution\ \mbox{CC BY 4.0})} license and share them at \href{https://osf.io/n5zut/}{\texttt{osf.io/n5zut}}. 

\clearpage

\newlength{\appendixfigurewidth}
\setlength{\appendixfigurewidth}{.8\columnwidth}

\begin{figure}[t] % htbp are optional placement specifiers (here, top, bottom, page)
	\centering % Centers the figure
	\includegraphics[width=\appendixfigurewidth]{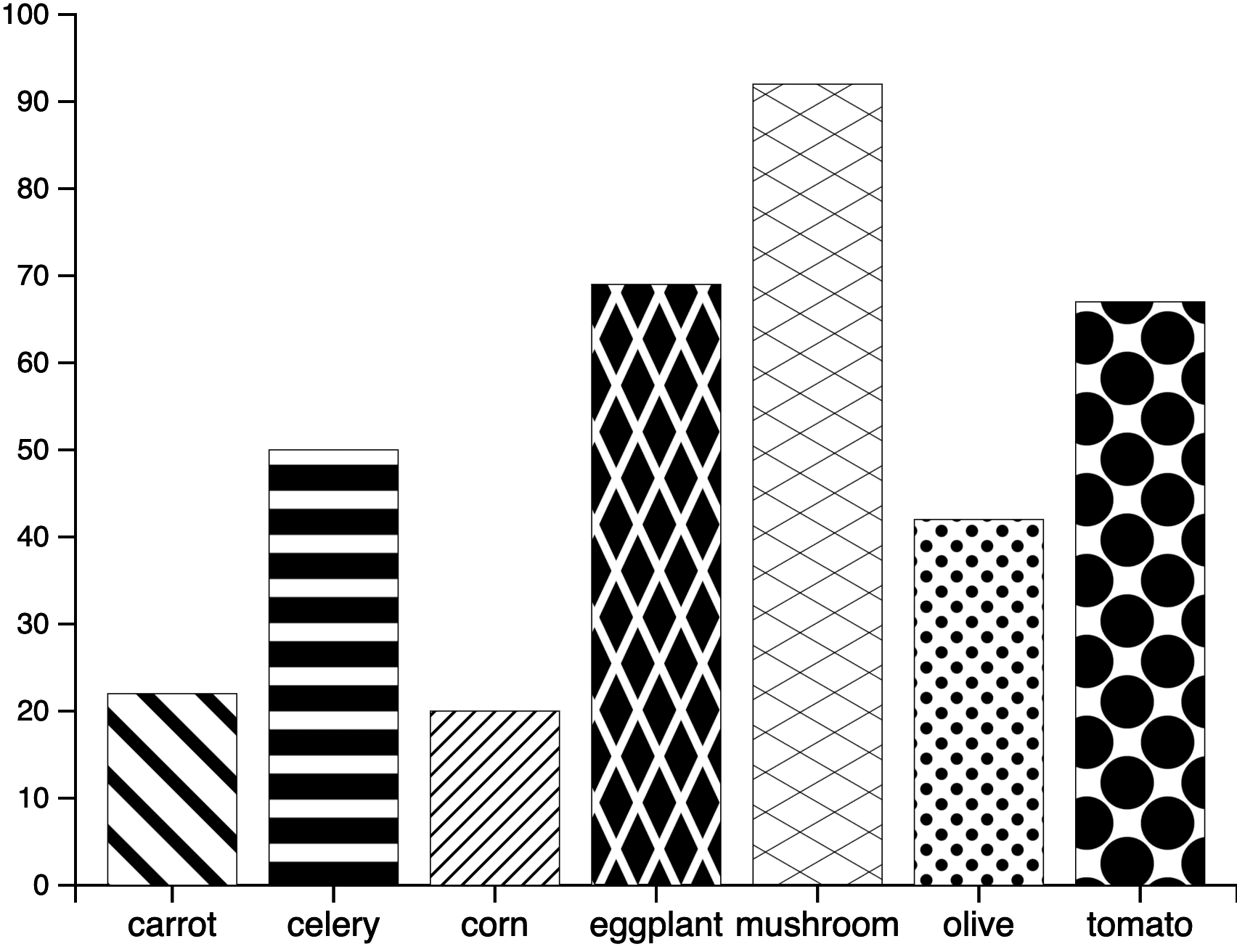}\\
	\caption{A geometric textured bar chart design (BG1) collected in our Experiment 1. This is a larger version of the first image in \autoref{tab:exp2-bar-geo}.}
  \label{fig:BG1}
\end{figure}

\begin{figure}[t] % htbp are optional placement specifiers (here, top, bottom, page)
	\centering % Centers the figure
	\includegraphics[width=\appendixfigurewidth]{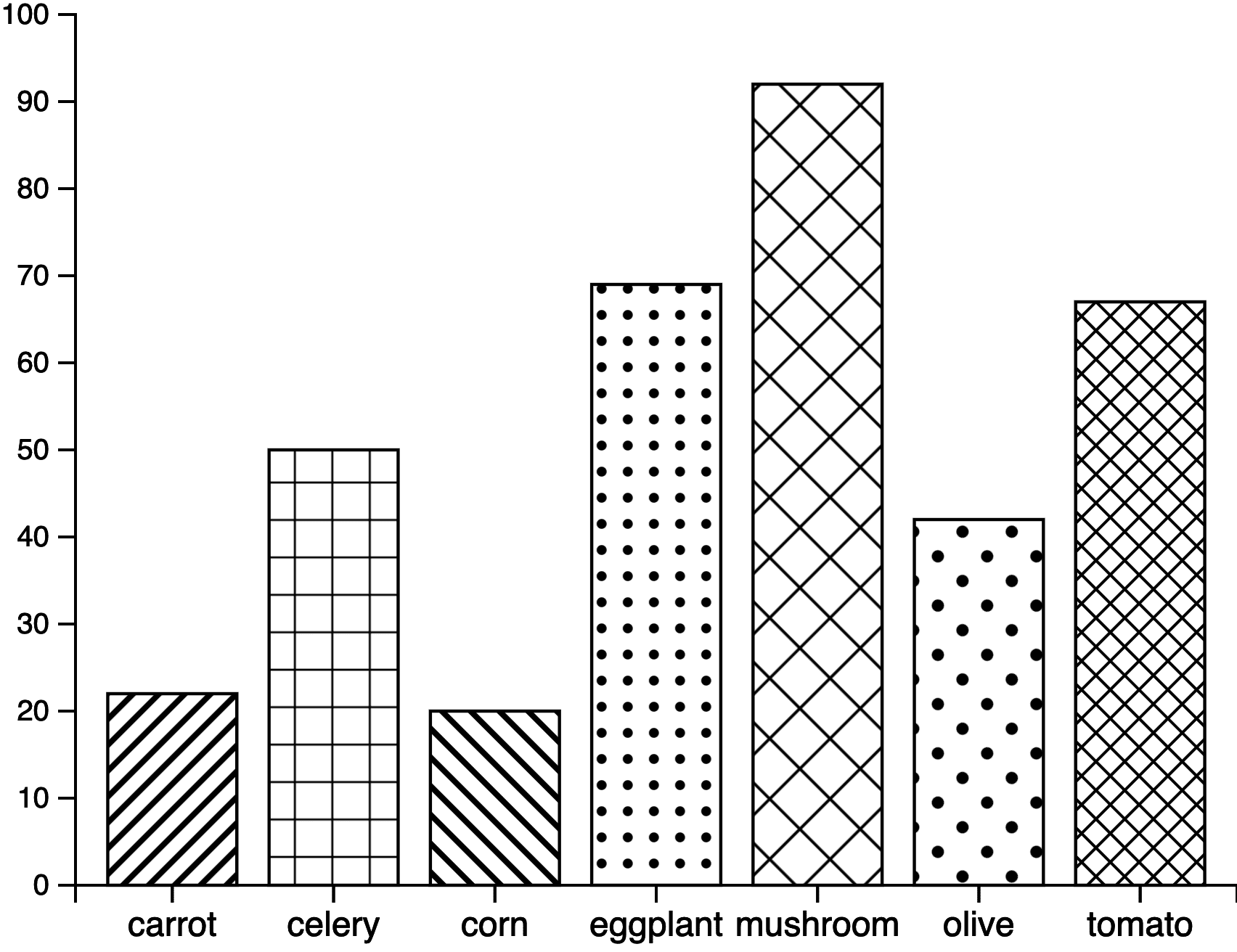}\\
	\caption{A geometric textured bar chart design (BG2) collected in our Experiment 1. This is a larger version of the second image in \autoref{tab:exp2-bar-geo}.}
  \label{fig:BG2}
\end{figure}

\begin{figure}[t] % htbp are optional placement specifiers (here, top, bottom, page)
	\centering % Centers the figure
	\includegraphics[width=\appendixfigurewidth]{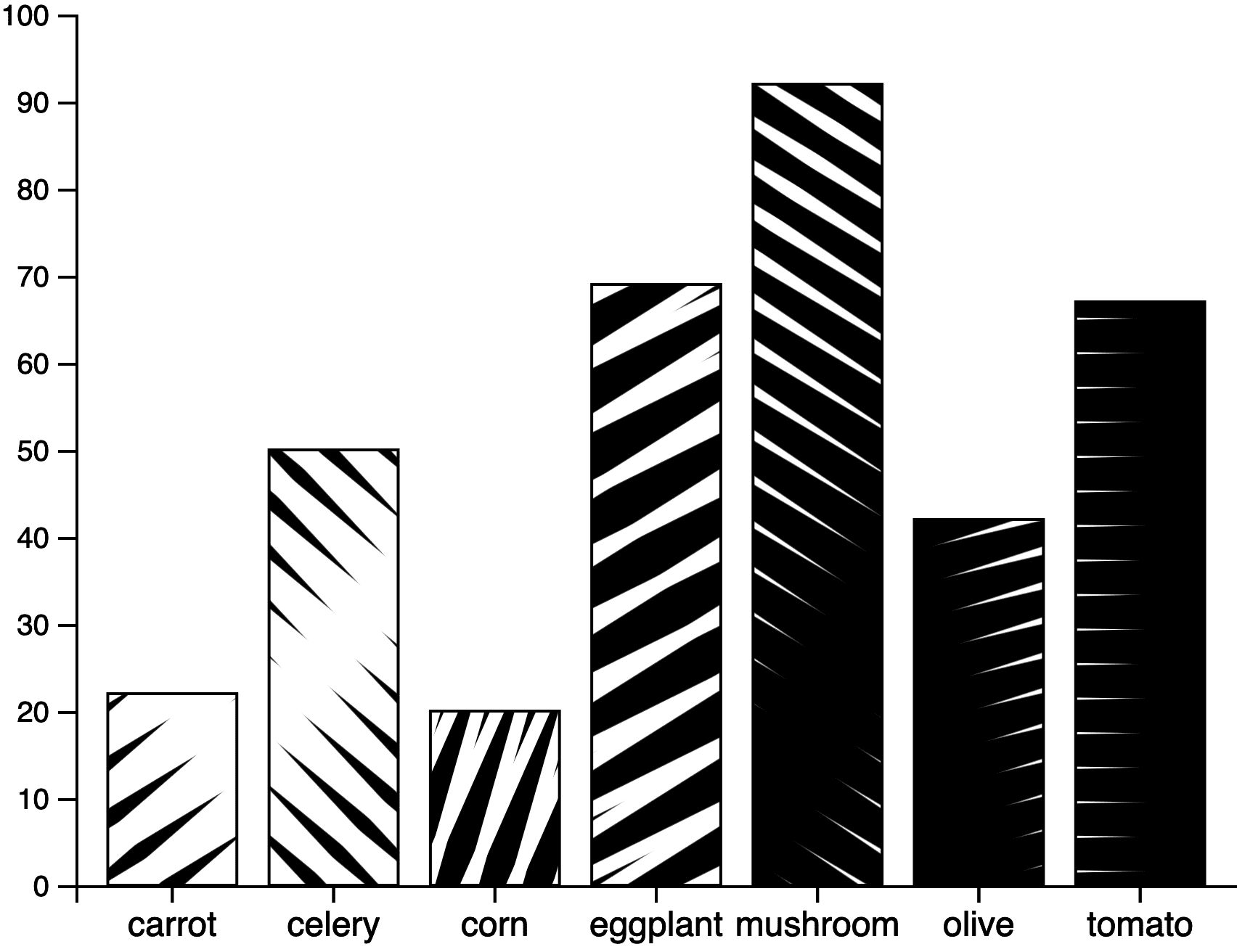}\\
	\caption{A geometric textured bar chart design (BG3) collected in our Experiment 1. This is a larger version of the third image in \autoref{tab:exp2-bar-geo}.}
  \label{fig:BG3}
\end{figure}

\begin{figure}[t] % htbp are optional placement specifiers (here, top, bottom, page)
	\centering % Centers the figure
	\includegraphics[width=\appendixfigurewidth]{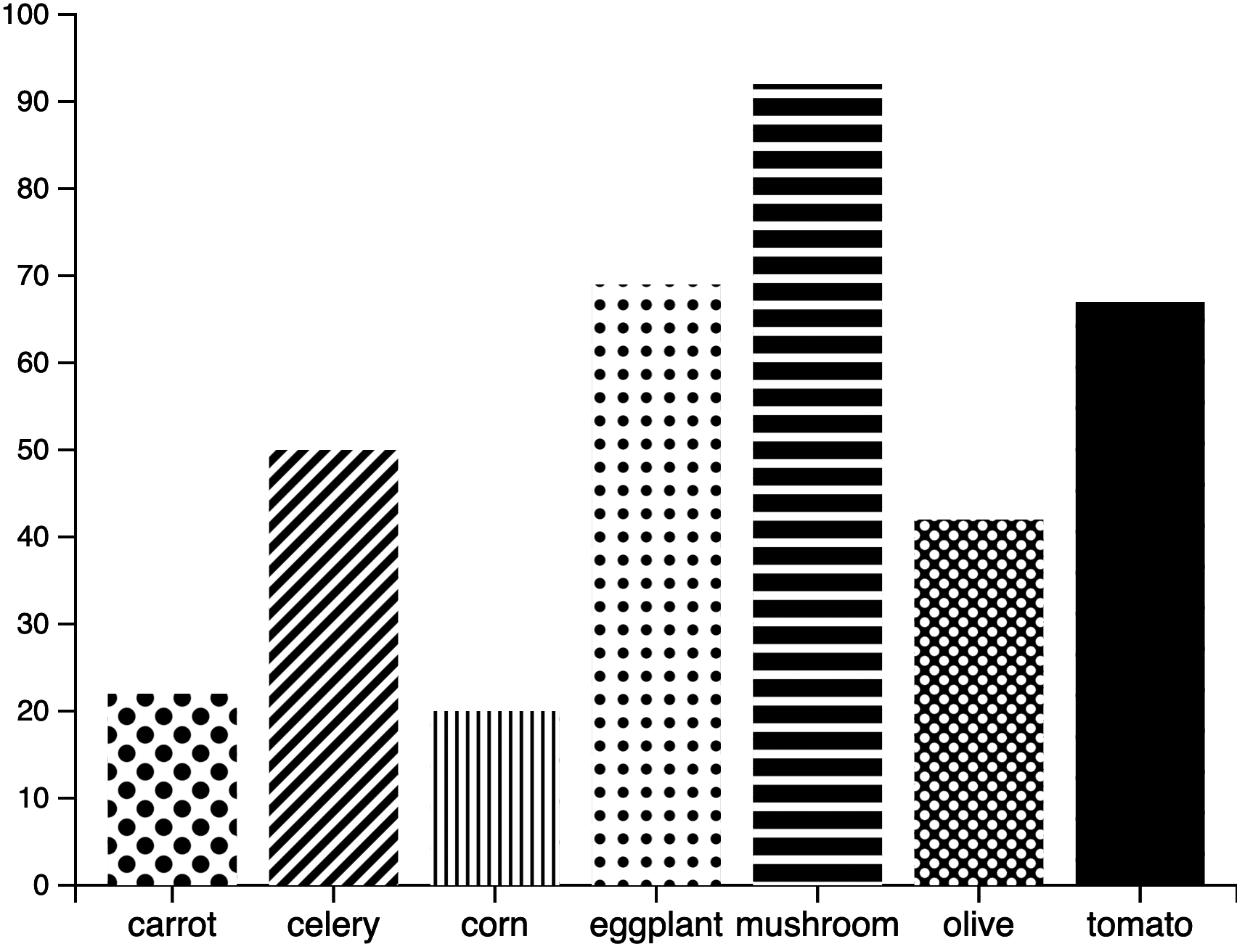}\\
	\caption{A geometric textured bar chart design (BG4) collected in our Experiment 1. This is a larger version of the fourth image in \autoref{tab:exp2-bar-geo}.}
  \label{fig:BG4}
\end{figure}

\begin{figure}[t] % htbp are optional placement specifiers (here, top, bottom, page)
	\centering % Centers the figure
	\includegraphics[width=\appendixfigurewidth]{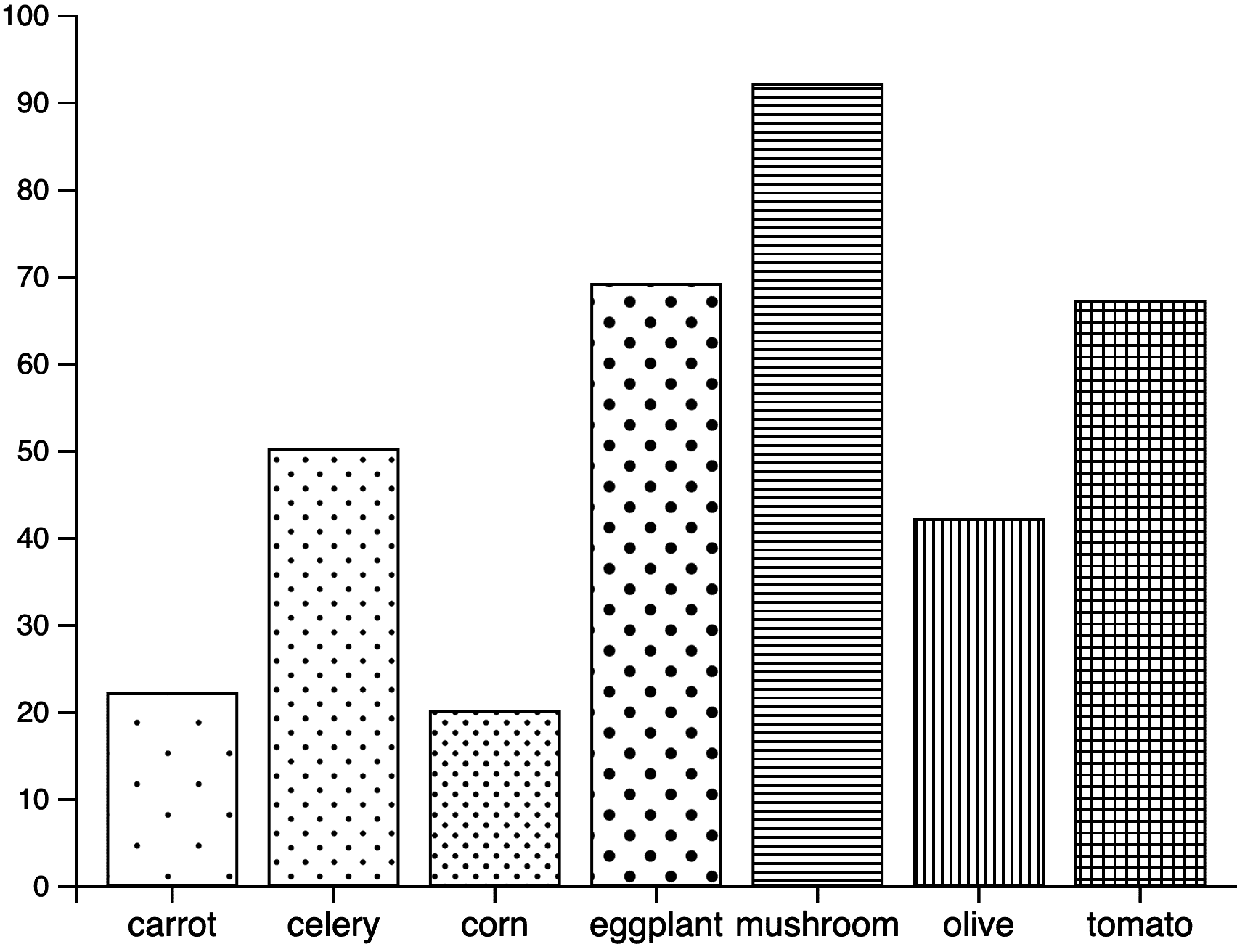}\\
	\caption{A geometric textured bar chart design (BG5) collected in our Experiment 1.}
  \label{fig:BG5}
\end{figure}

\begin{figure}[t] % htbp are optional placement specifiers (here, top, bottom, page)
	\centering % Centers the figure
	\includegraphics[width=\appendixfigurewidth]{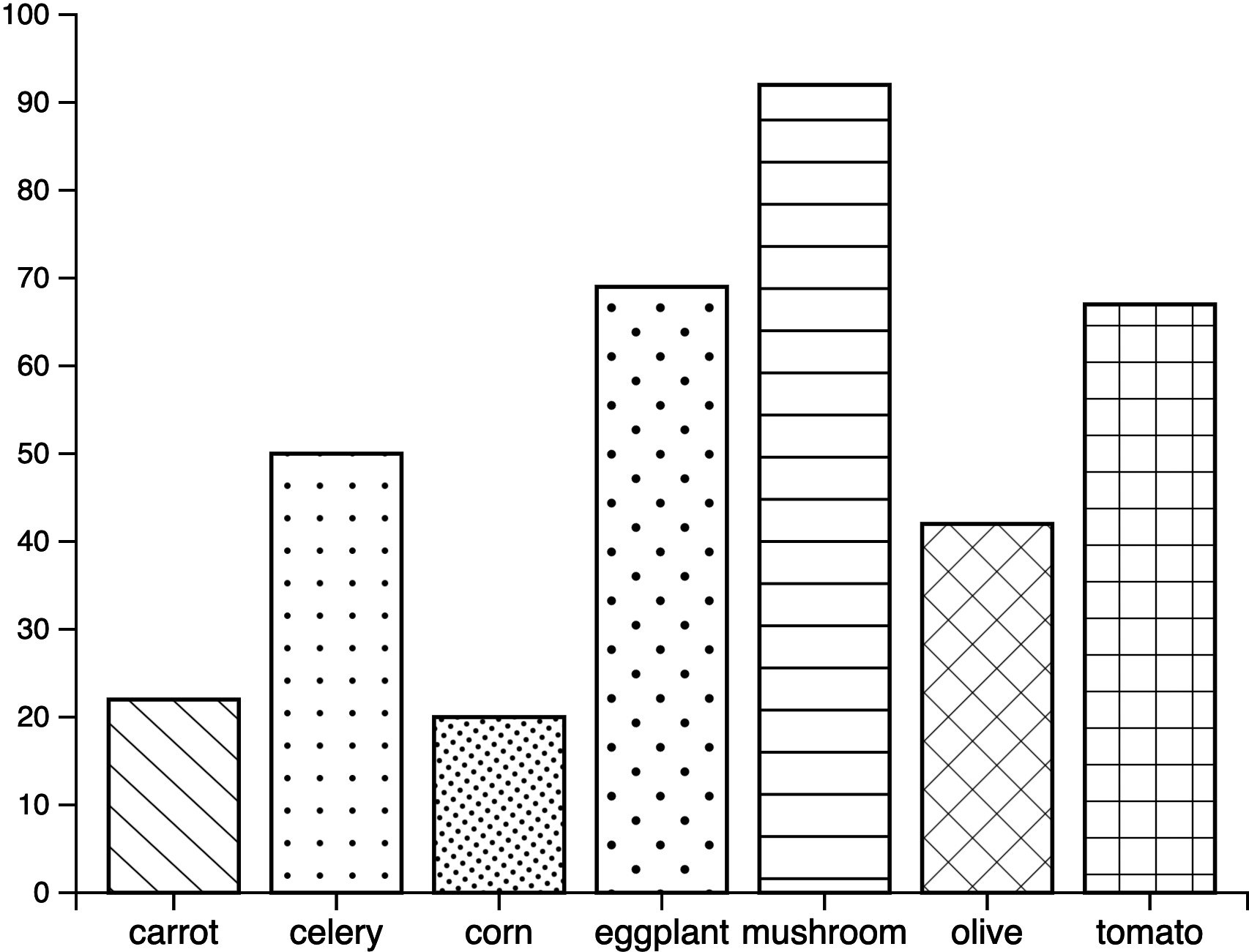}\\
	\caption{A geometric textured bar chart design (BG6) collected in our Experiment 1.}
  \label{fig:BG6}
\end{figure}

\clearpage

\begin{figure}[t] % htbp are optional placement specifiers (here, top, bottom, page)
	\centering % Centers the figure
	\includegraphics[width=\appendixfigurewidth]{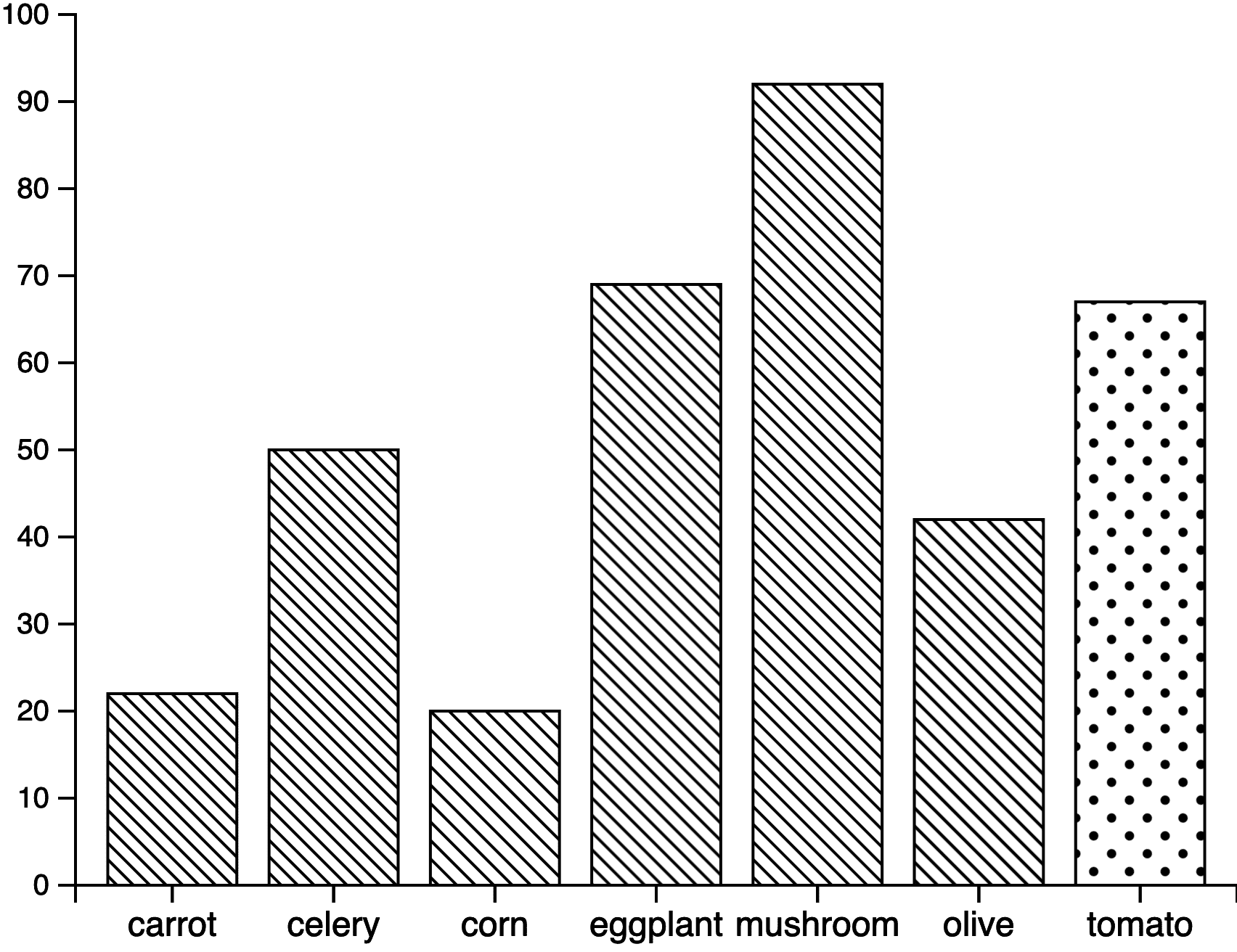}\\
	\caption{A geometric textured bar chart design (BG7) collected in our Experiment 1.}
  \label{fig:BG7}
\end{figure}

\begin{figure}[t] % htbp are optional placement specifiers (here, top, bottom, page)
	\centering % Centers the figure
	\includegraphics[width=\appendixfigurewidth]{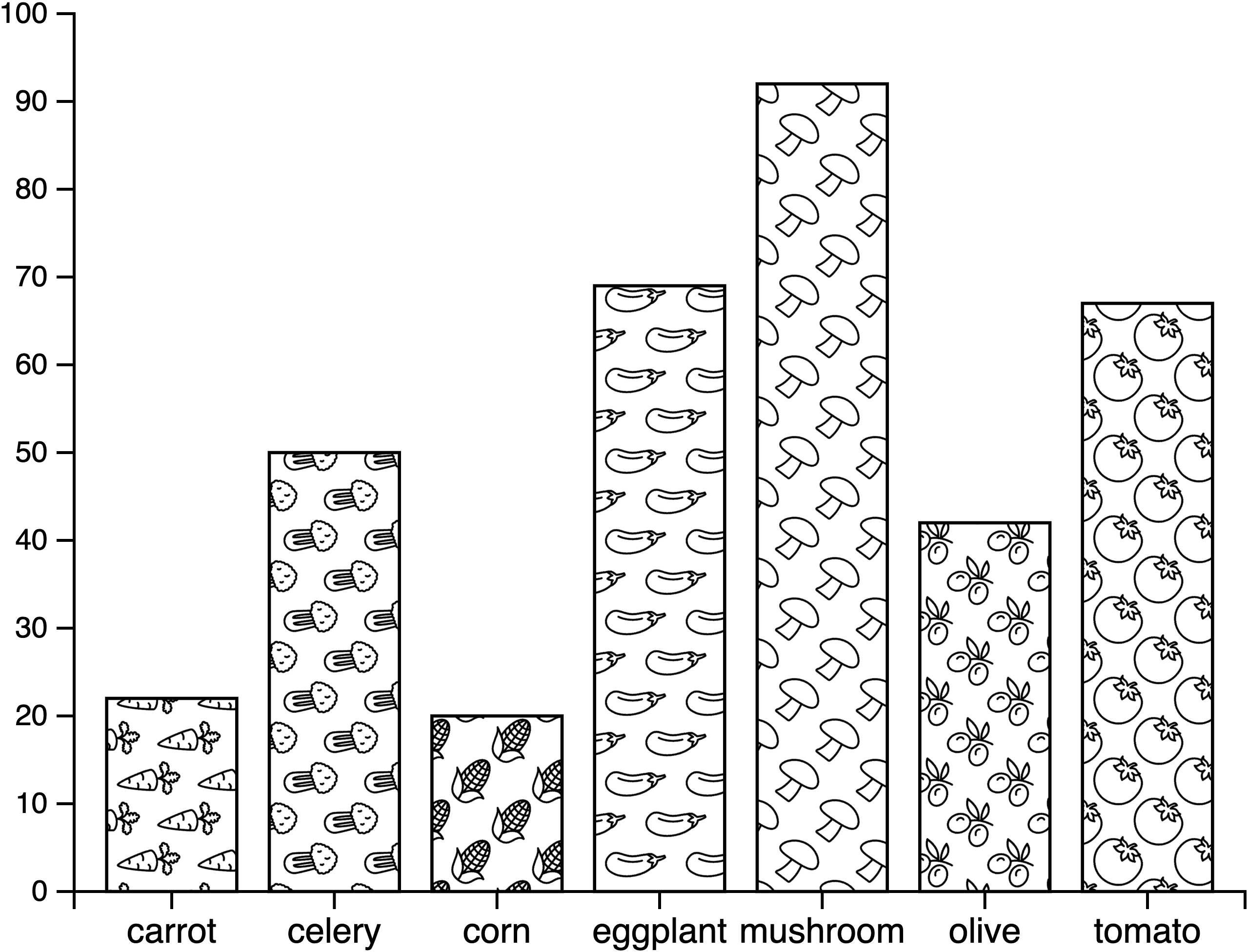}\\
	\caption{An iconic textured bar chart design (BI1) collected in our Experiment 1. This is a larger version of the first image in \autoref{tab:exp2-bar-icon}.}
  \label{fig:BI1}
\end{figure}

\begin{figure}[t] % htbp are optional placement specifiers (here, top, bottom, page)
	\centering % Centers the figure
	\includegraphics[width=\appendixfigurewidth]{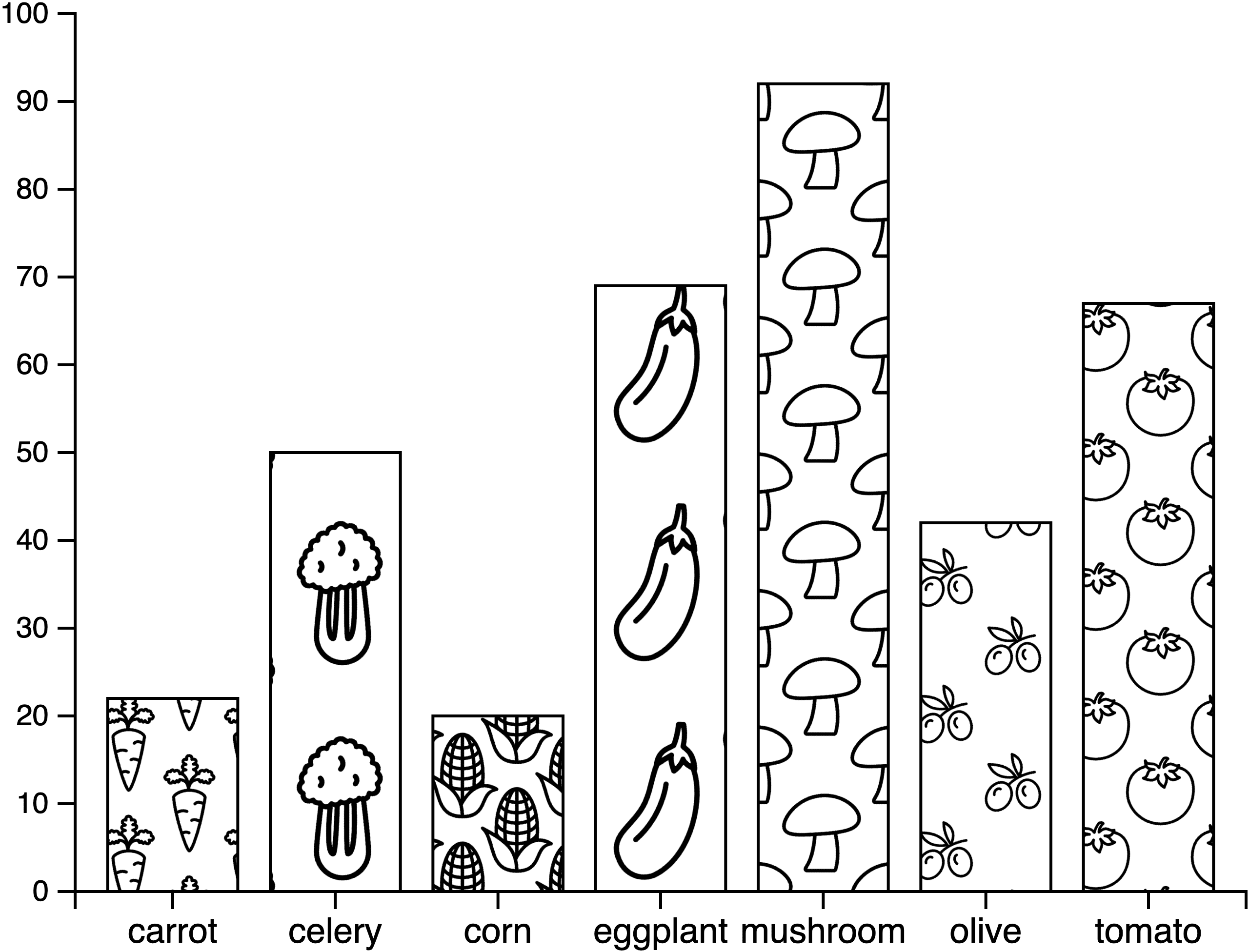}\\
	\caption{An iconic textured bar chart design (BI2) collected in our Experiment 1. This is a larger version of the second image in \autoref{tab:exp2-bar-icon}.}
  \label{fig:BI2}
\end{figure}

\begin{figure}[t] % htbp are optional placement specifiers (here, top, bottom, page)
	\centering % Centers the figure
	\includegraphics[width=\appendixfigurewidth]{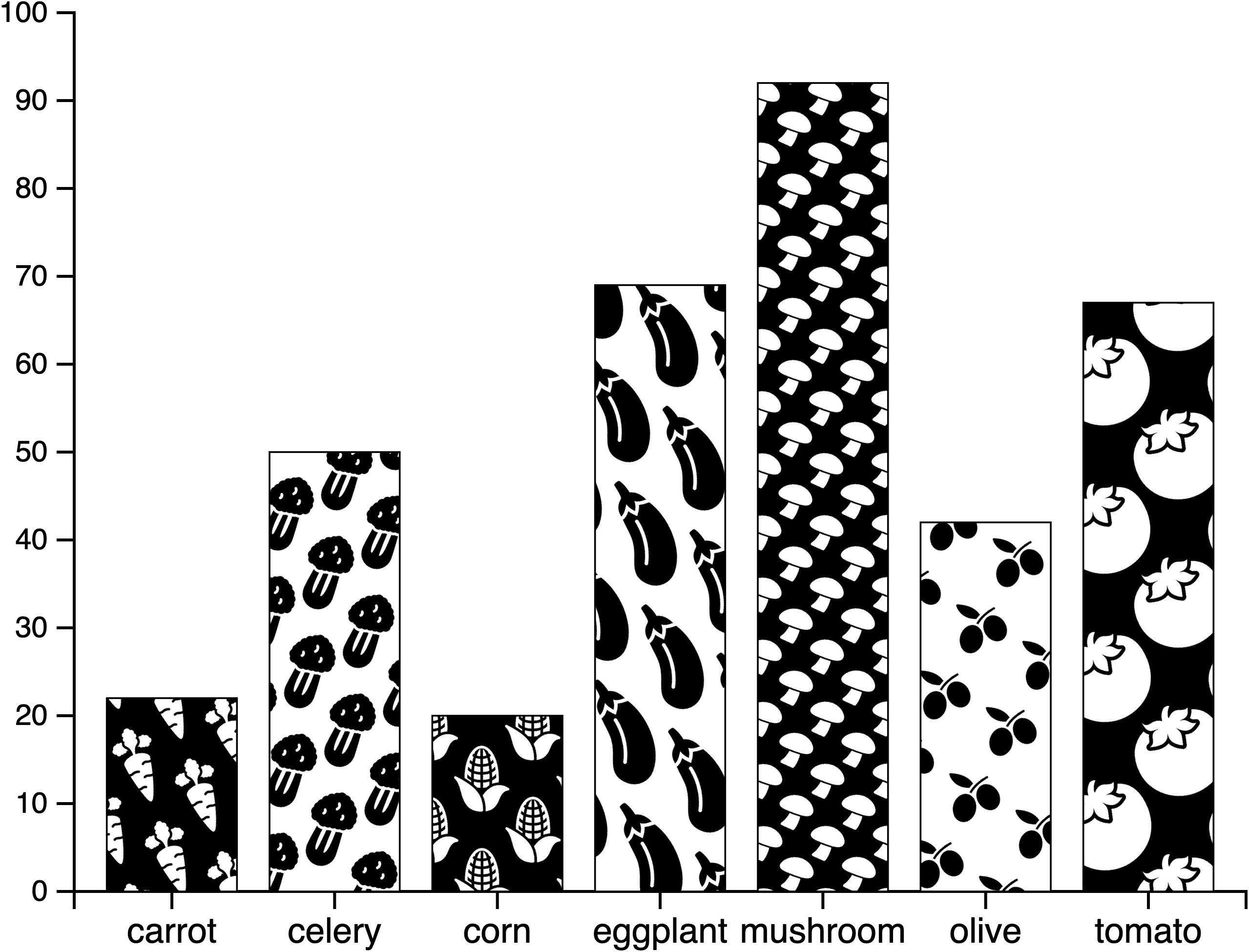}\\
	\caption{An iconic textured bar chart design (BI3) collected in our Experiment 1. This is a larger version of the third image in \autoref{tab:exp2-bar-icon}.}
  \label{fig:BI3}
\end{figure}

\begin{figure}[t] % htbp are optional placement specifiers (here, top, bottom, page)
	\centering % Centers the figure
	\includegraphics[width=\appendixfigurewidth]{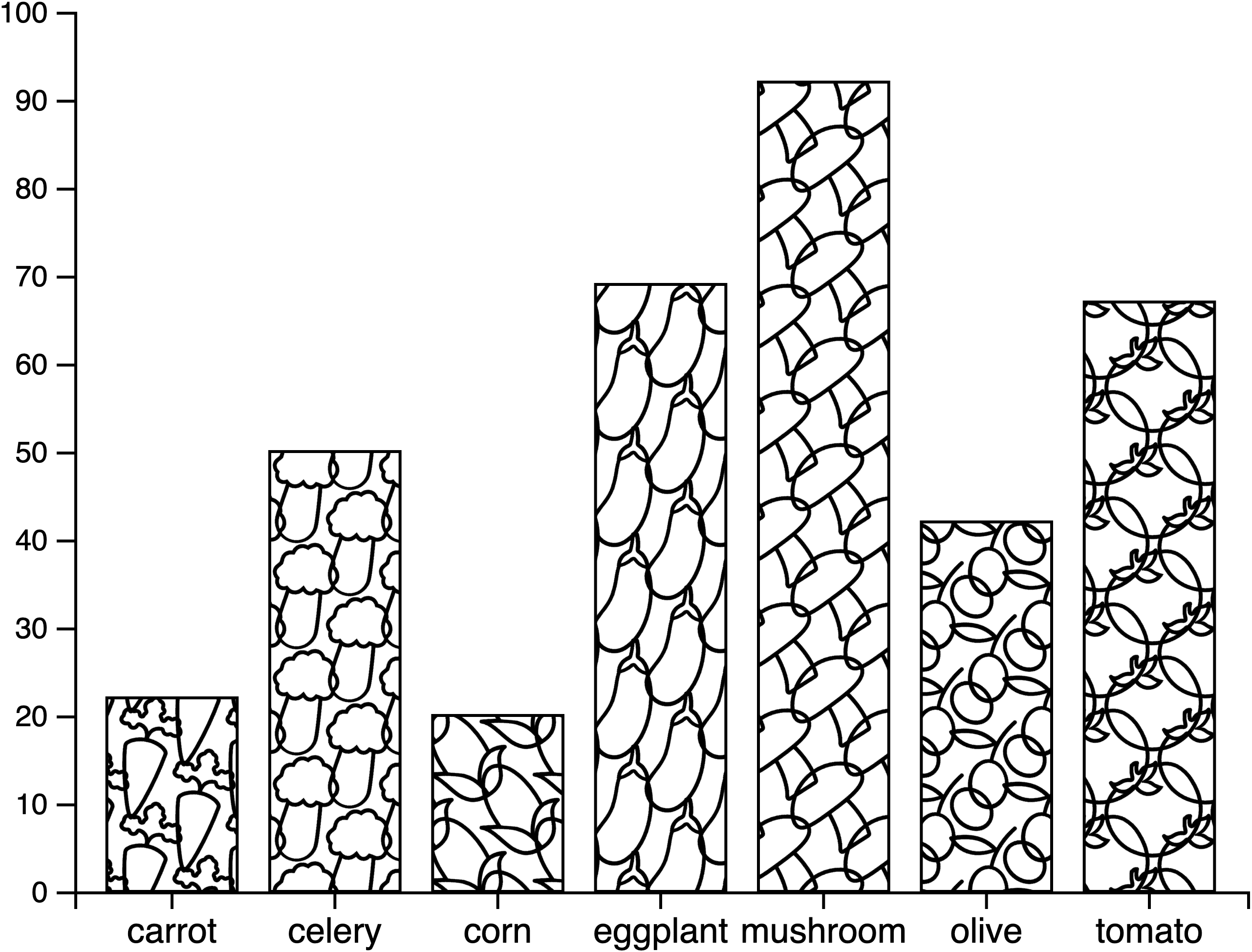}\\
	\caption{An iconic textured bar chart design (BI4) collected in our Experiment 1. This is a larger version of the fourth image in \autoref{tab:exp2-bar-icon}.}
  \label{fig:BI4}
\end{figure}

\begin{figure}[t] % htbp are optional placement specifiers (here, top, bottom, page)
	\centering % Centers the figure
	\includegraphics[width=\appendixfigurewidth]{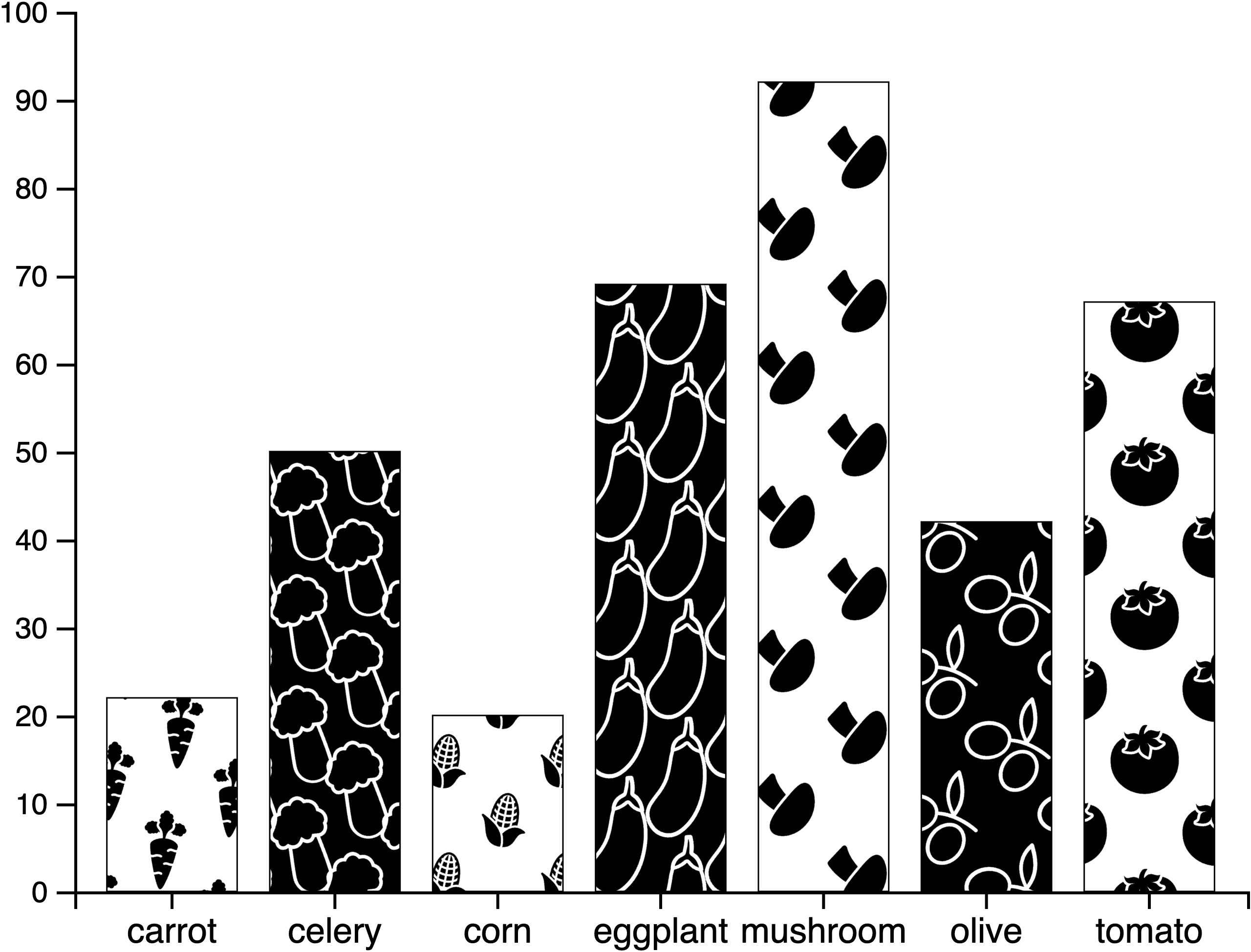}\\
	\caption{An iconic textured bar chart design (BI5) collected in our Experiment 1.}
  \label{fig:BI5}
\end{figure}

\clearpage

\begin{figure}[t] % htbp are optional placement specifiers (here, top, bottom, page)
	\centering % Centers the figure
	\includegraphics[width=\appendixfigurewidth]{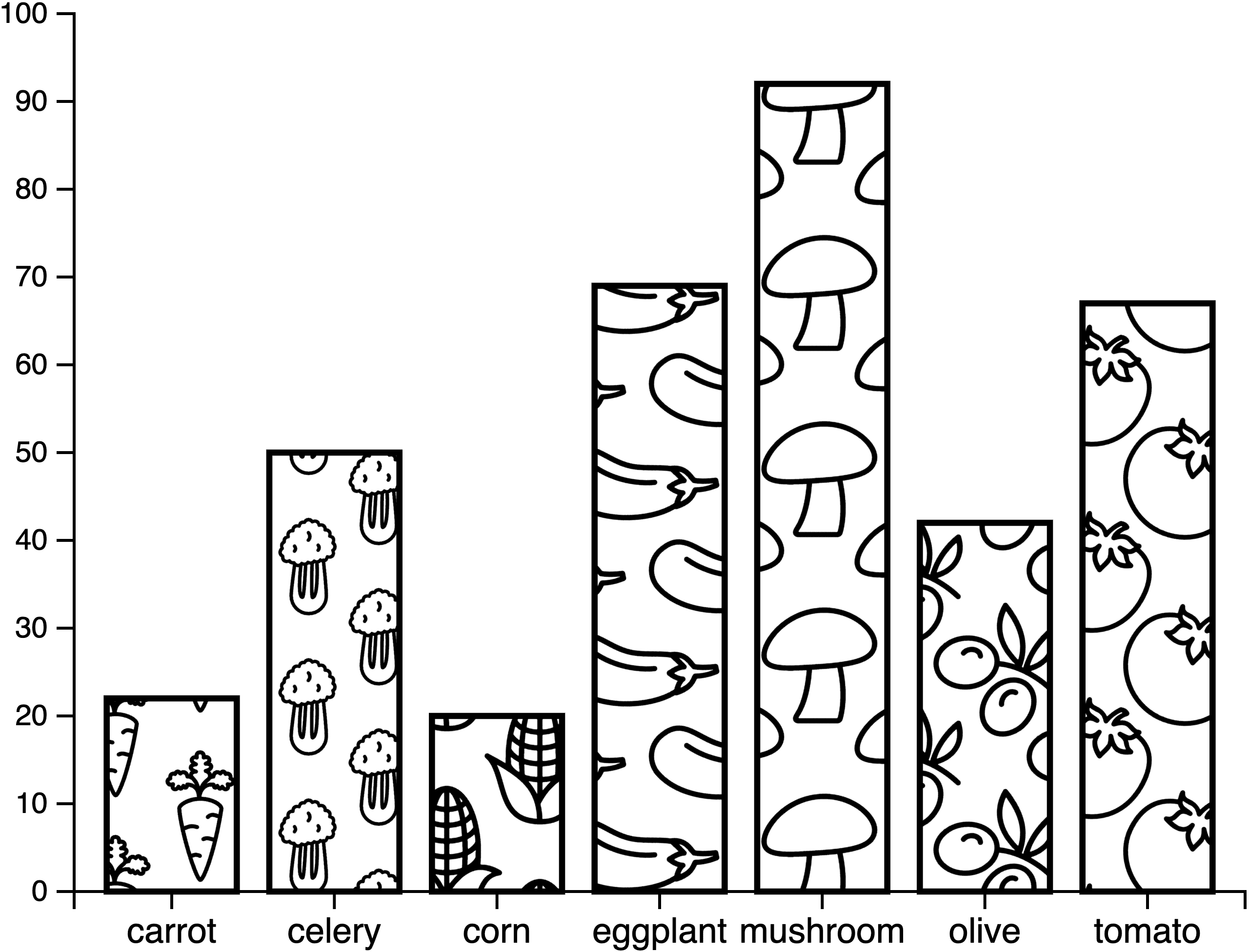}\\
	\caption{An iconic textured bar chart design (BI6) collected in our Experiment 1.}
  \label{fig:BI6}
\end{figure}

\begin{figure}[t] % htbp are optional placement specifiers (here, top, bottom, page)
	\centering % Centers the figure
	\includegraphics[width=\appendixfigurewidth]{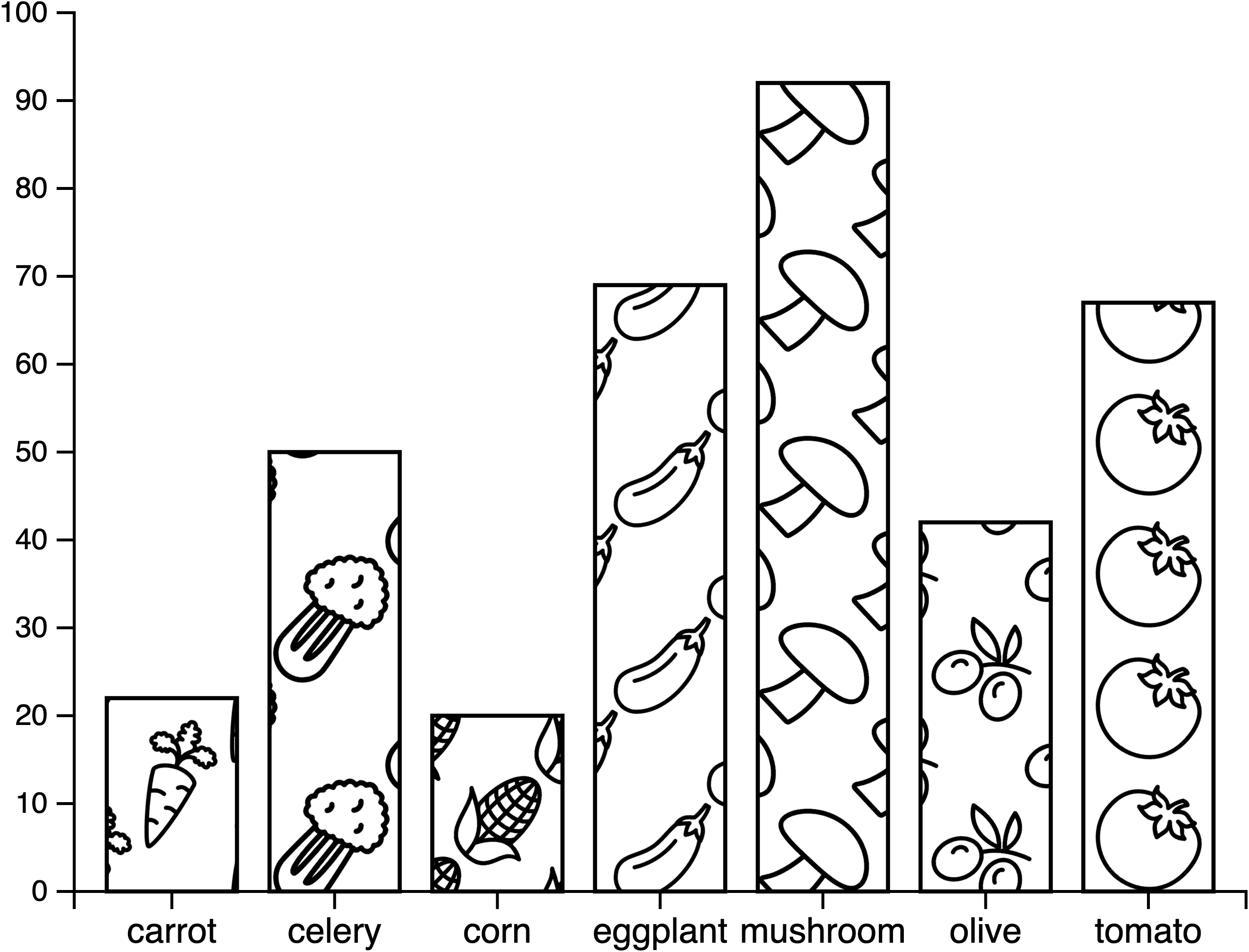}\\
	\caption{An iconic textured bar chart design (BI7) collected in our Experiment 1.}
  \label{fig:BI7}
\end{figure}
\begin{figure}[t] % htbp are optional placement specifiers (here, top, bottom, page)
	\centering % Centers the figure
	\includegraphics[width=\appendixfigurewidth]{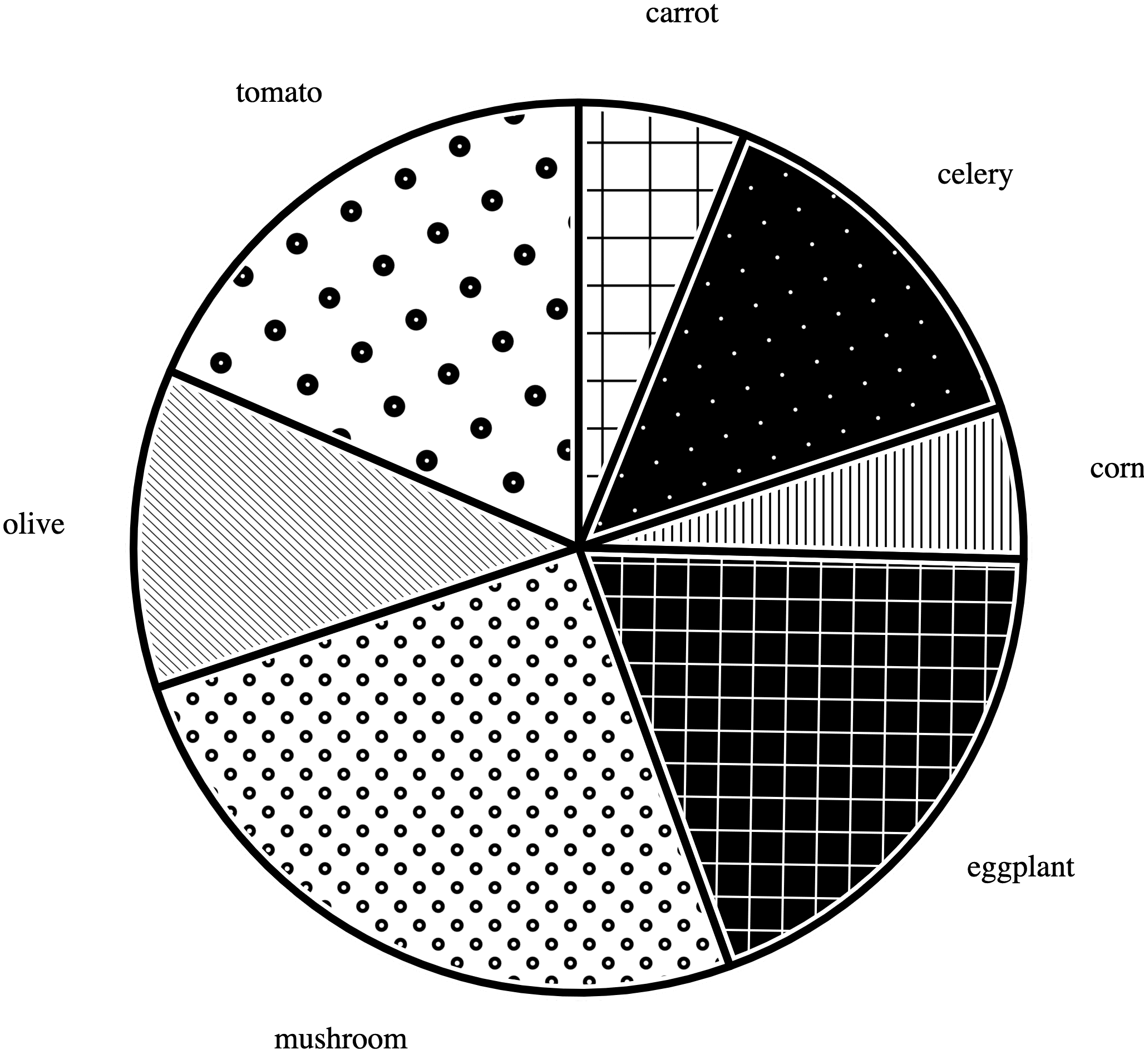}\\
	\caption{An geometric textured pie chart design (PG1) collected in our Experiment 1. This is a larger version of the first image in \autoref{tab:exp2-pie-geo}.}
  \label{fig:PG1}
\end{figure}

\begin{figure}[t] % htbp are optional placement specifiers (here, top, bottom, page)
	\centering % Centers the figure
	\includegraphics[width=\appendixfigurewidth]{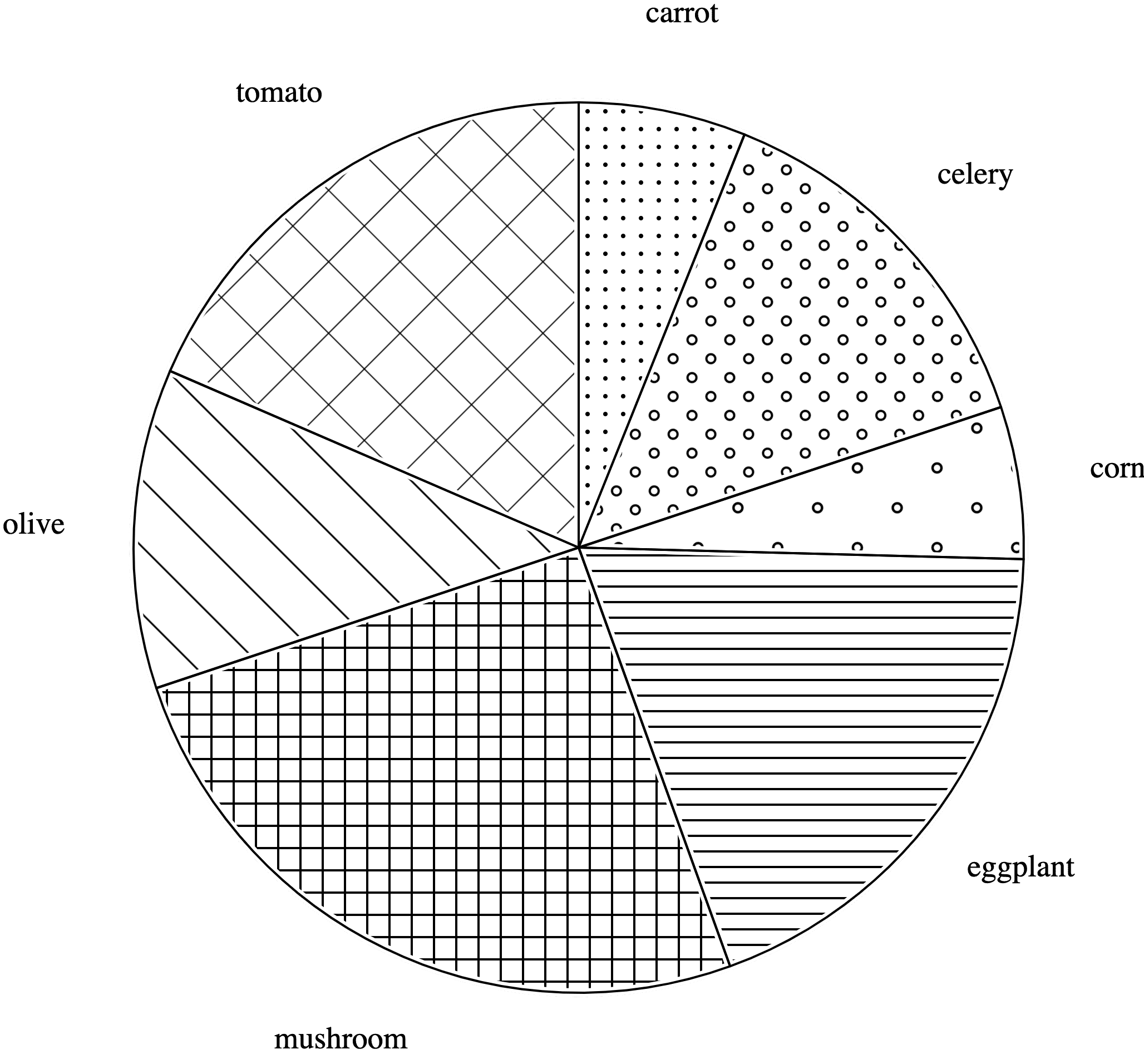}\\
	\caption{An geometric textured pie chart design (PG2) collected in our Experiment 1. This is a larger version of the second image in \autoref{tab:exp2-pie-geo}.}
  \label{fig:PG2}
\end{figure}

\begin{figure}[t] % htbp are optional placement specifiers (here, top, bottom, page)
	\centering % Centers the figure
	\includegraphics[width=\appendixfigurewidth]{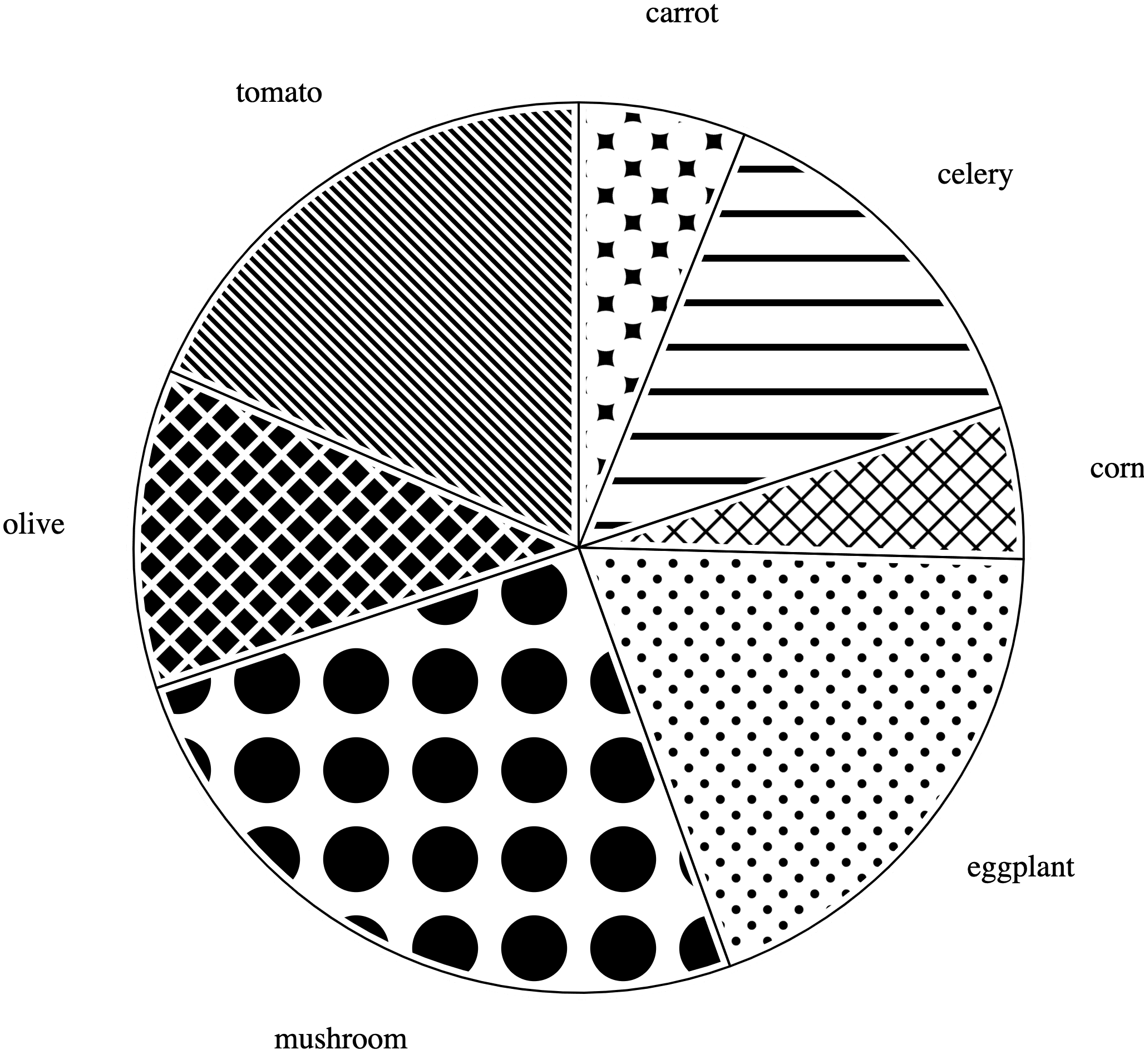}\\
	\caption{An geometric textured pie chart design (PG3) collected in our Experiment 1. This is a larger version of the third image in \autoref{tab:exp2-pie-geo}.}
  \label{fig:PG3}
\end{figure}

\begin{figure}[t] % htbp are optional placement specifiers (here, top, bottom, page)
	\centering % Centers the figure
	\includegraphics[width=\appendixfigurewidth]{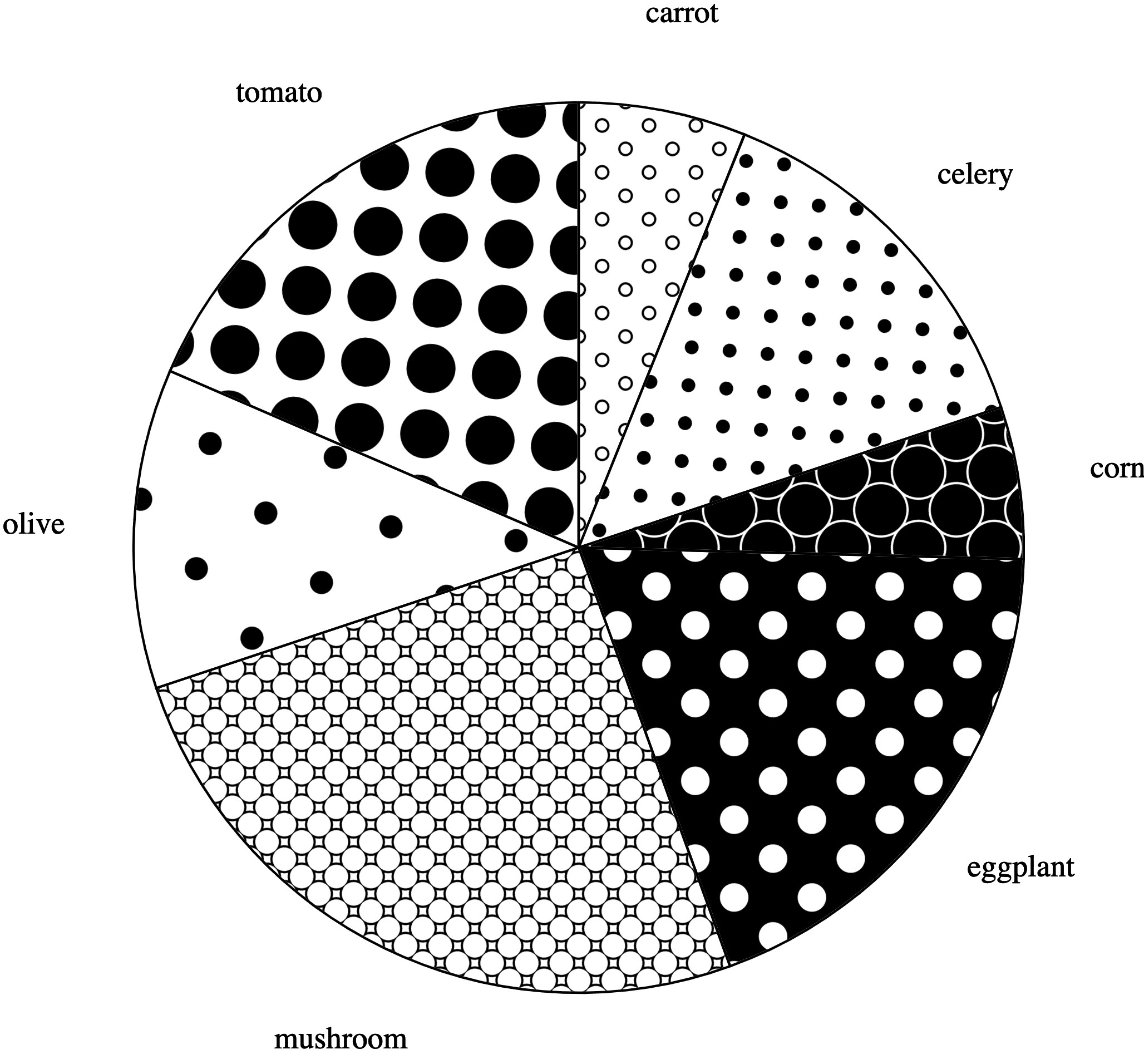}\\
	\caption{An geometric textured pie chart design (PG4) collected in our Experiment 1. This is a larger version of the fourth image in \autoref{tab:exp2-pie-geo}.}
  \label{fig:PG4}
\end{figure}

\clearpage

\begin{figure}[t] % htbp are optional placement specifiers (here, top, bottom, page)
	\centering % Centers the figure
	\includegraphics[width=\appendixfigurewidth]{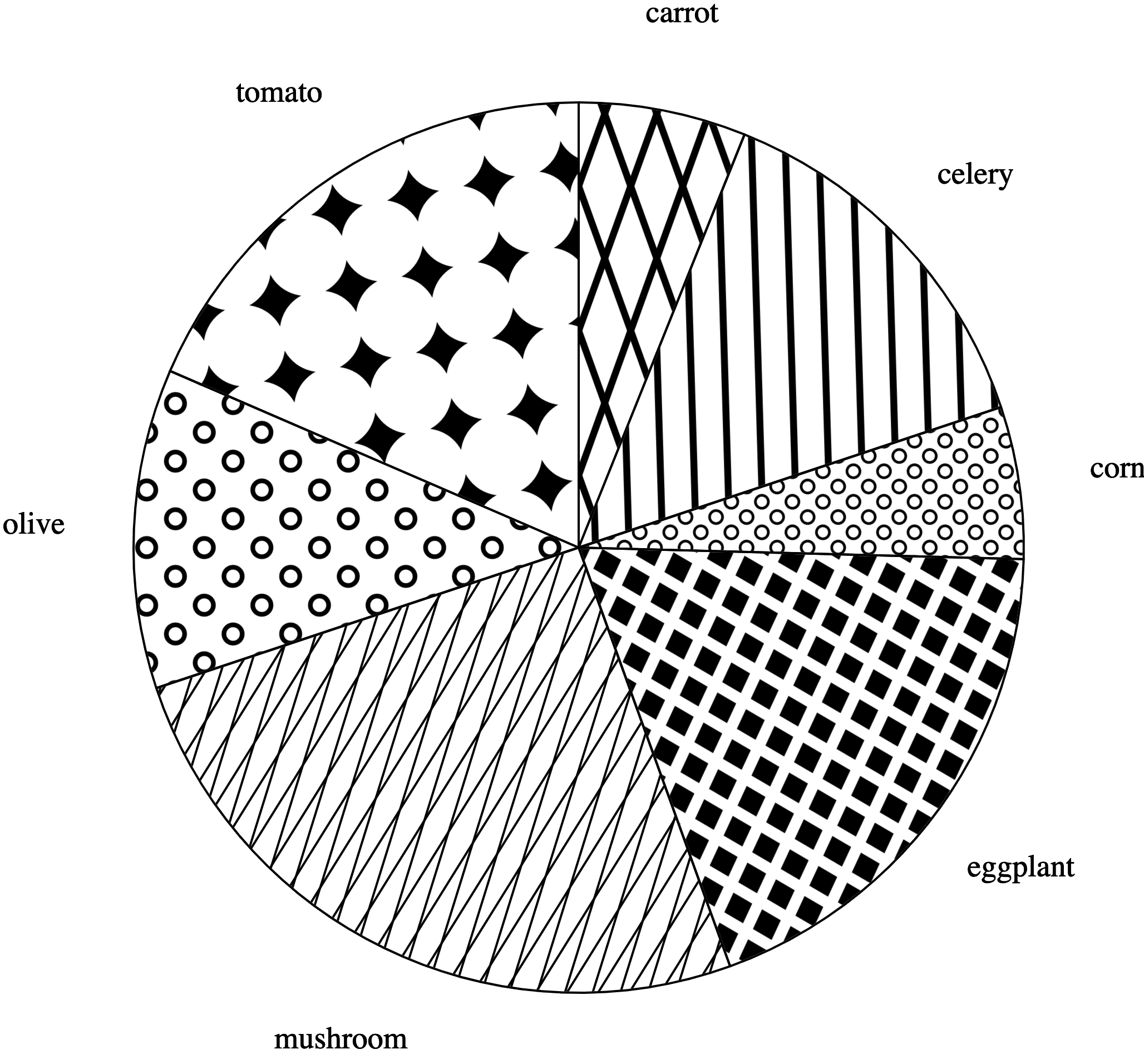}\\
	\caption{An geometric textured pie chart design (PG5) collected in our Experiment 1.}
  \label{fig:PG5}
\end{figure}

\begin{figure}[t] % htbp are optional placement specifiers (here, top, bottom, page)
	\centering % Centers the figure
	\includegraphics[width=\appendixfigurewidth]{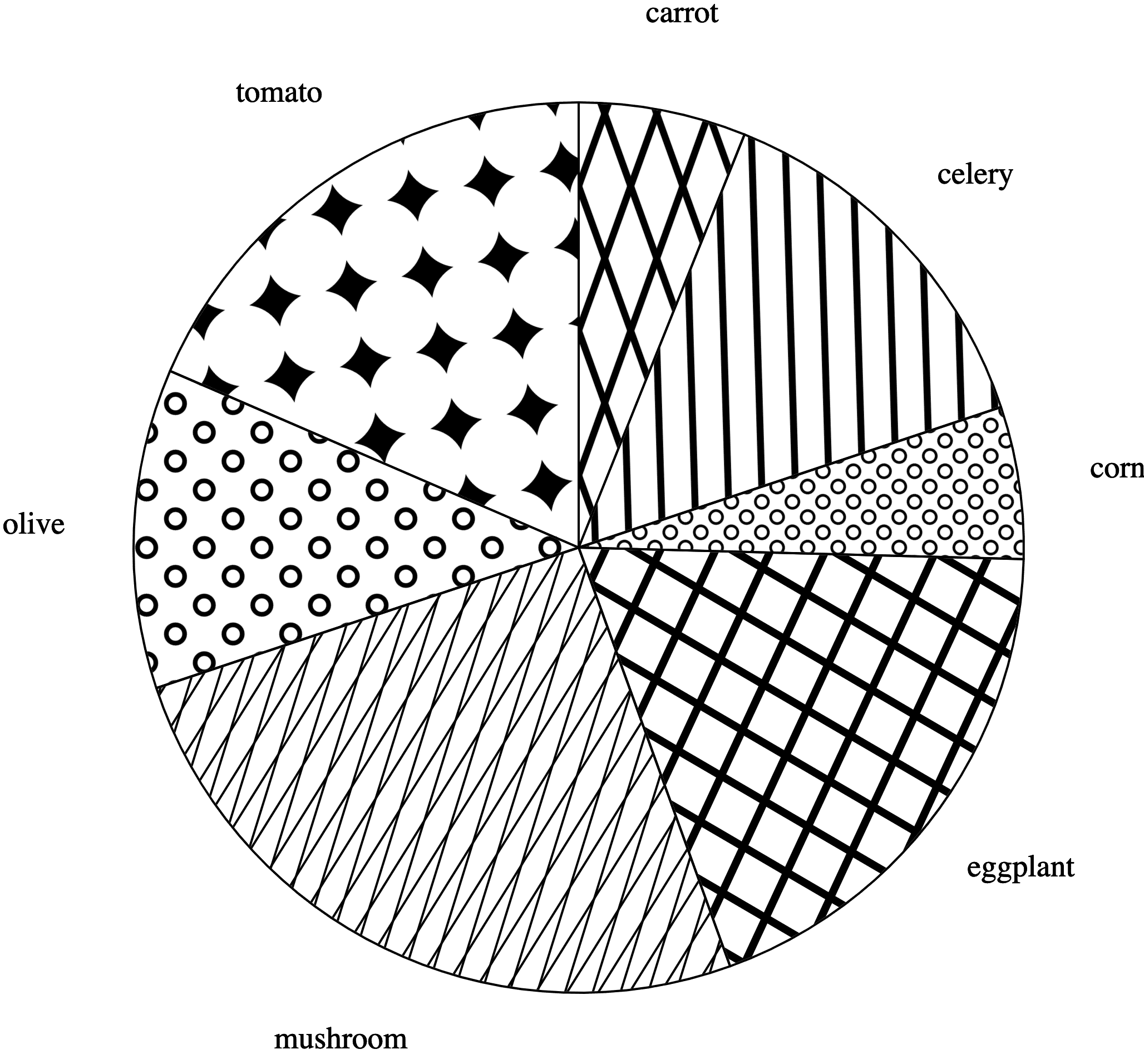}\\
	\caption{An geometric textured pie chart design (PG6) collected in our Experiment 1.}
  \label{fig:PG6}
\end{figure}

\begin{figure}[t] % htbp are optional placement specifiers (here, top, bottom, page)
	\centering % Centers the figure
	\includegraphics[width=\appendixfigurewidth]{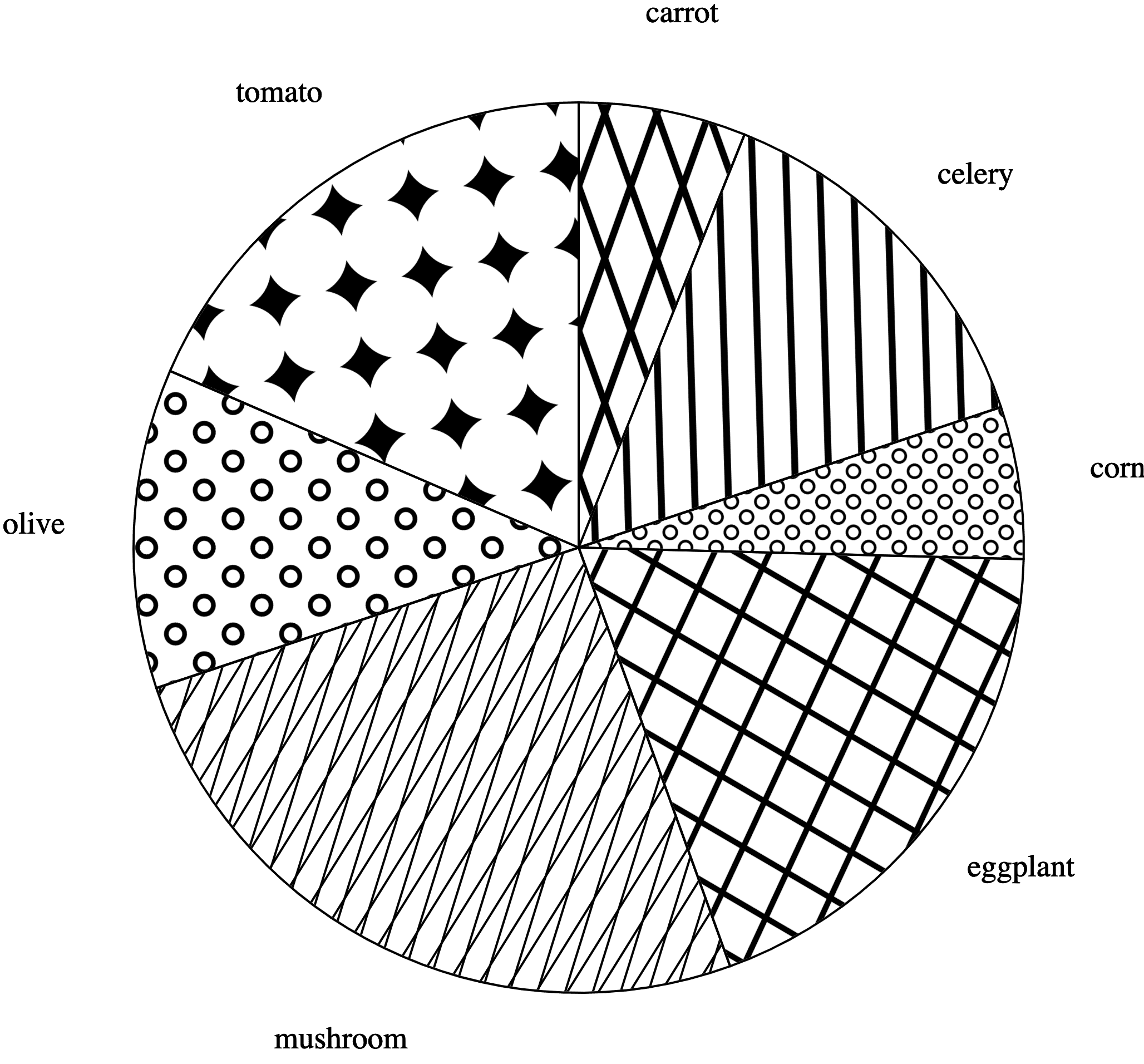}\\
	\caption{An geometric textured pie chart design (PG7) collected in our Experiment 1.}
  \label{fig:PG7}
\end{figure}

\begin{figure}[t] % htbp are optional placement specifiers (here, top, bottom, page)
	\centering % Centers the figure
	\includegraphics[width=\appendixfigurewidth]{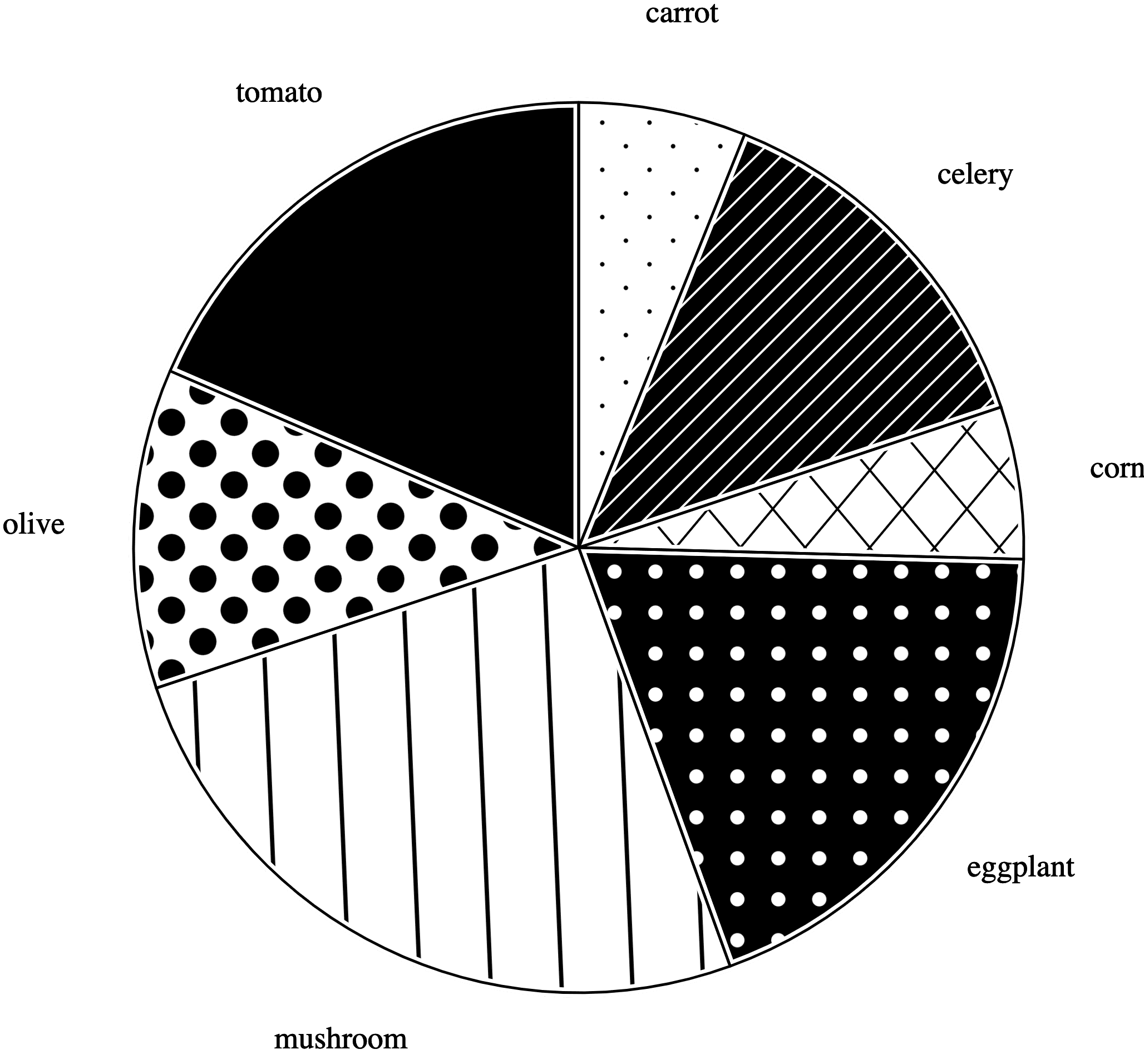}\\
	\caption{An geometric textured pie chart design (PG8) collected in our Experiment 1.}
  \label{fig:PG8}
\end{figure}

\begin{figure}[t] % htbp are optional placement specifiers (here, top, bottom, page)
	\centering % Centers the figure
	\includegraphics[width=\appendixfigurewidth]{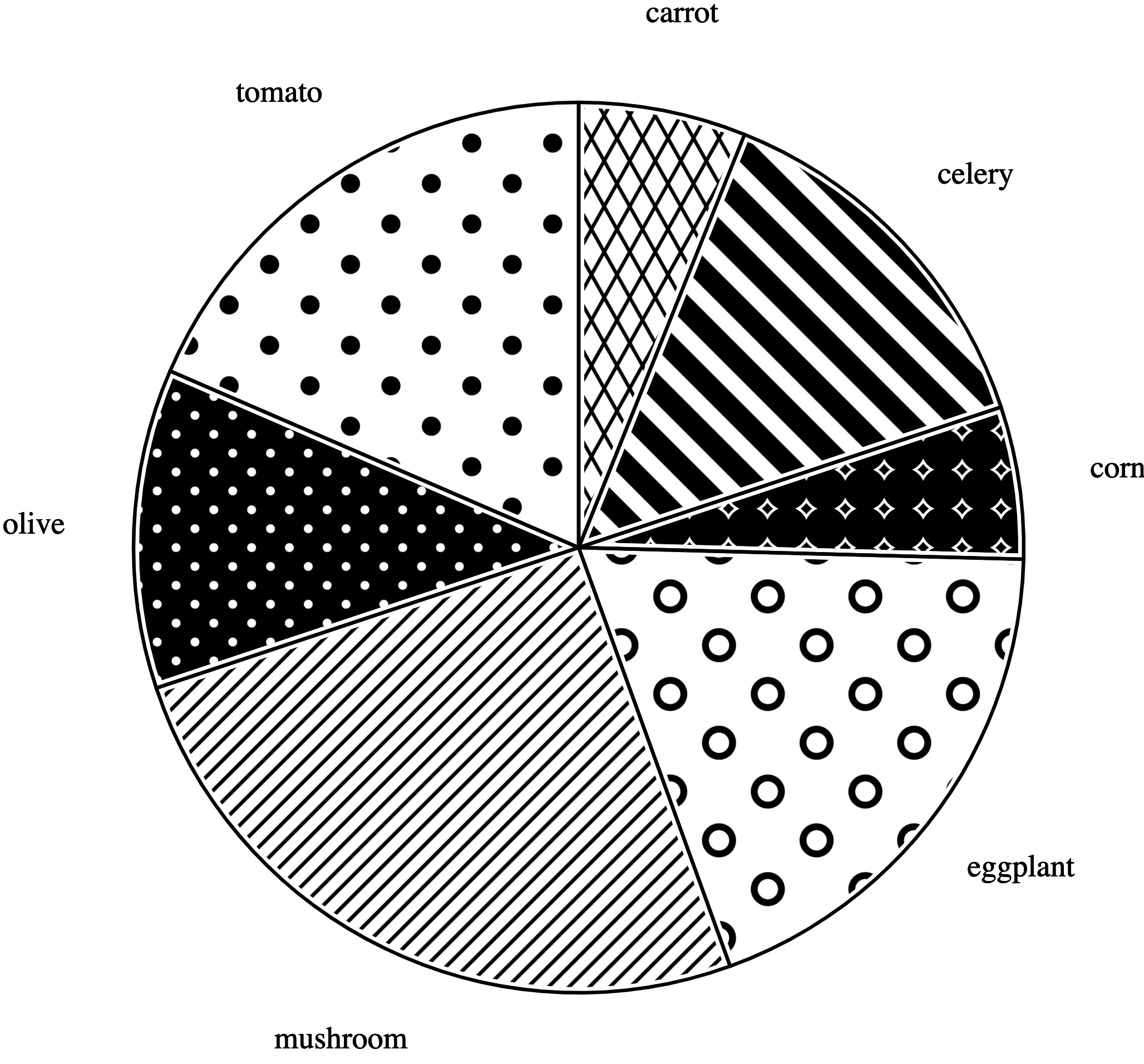}\\
	\caption{An geometric textured pie chart design (PG9) collected in our Experiment 1.}
  \label{fig:PG9}
\end{figure}

\begin{figure}[t] % htbp are optional placement specifiers (here, top, bottom, page)
	\centering % Centers the figure
	\includegraphics[width=\appendixfigurewidth]{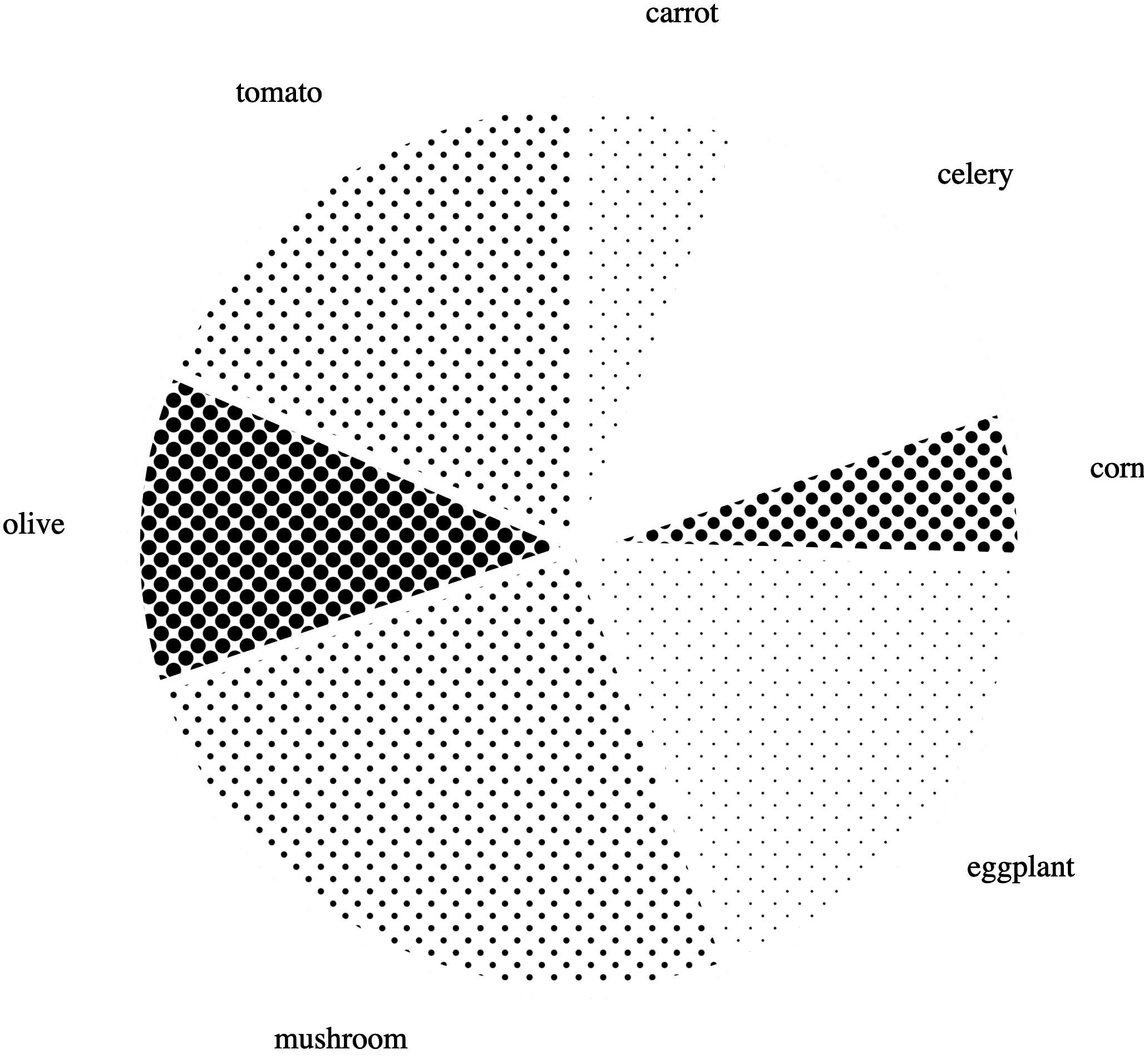}\\
	\caption{An geometric textured pie chart design (PG10) collected in our Experiment 1.}
  \label{fig:PG10}
\end{figure}

\clearpage

\begin{figure}[t] % htbp are optional placement specifiers (here, top, bottom, page)
	\centering % Centers the figure
	\includegraphics[width=\appendixfigurewidth]{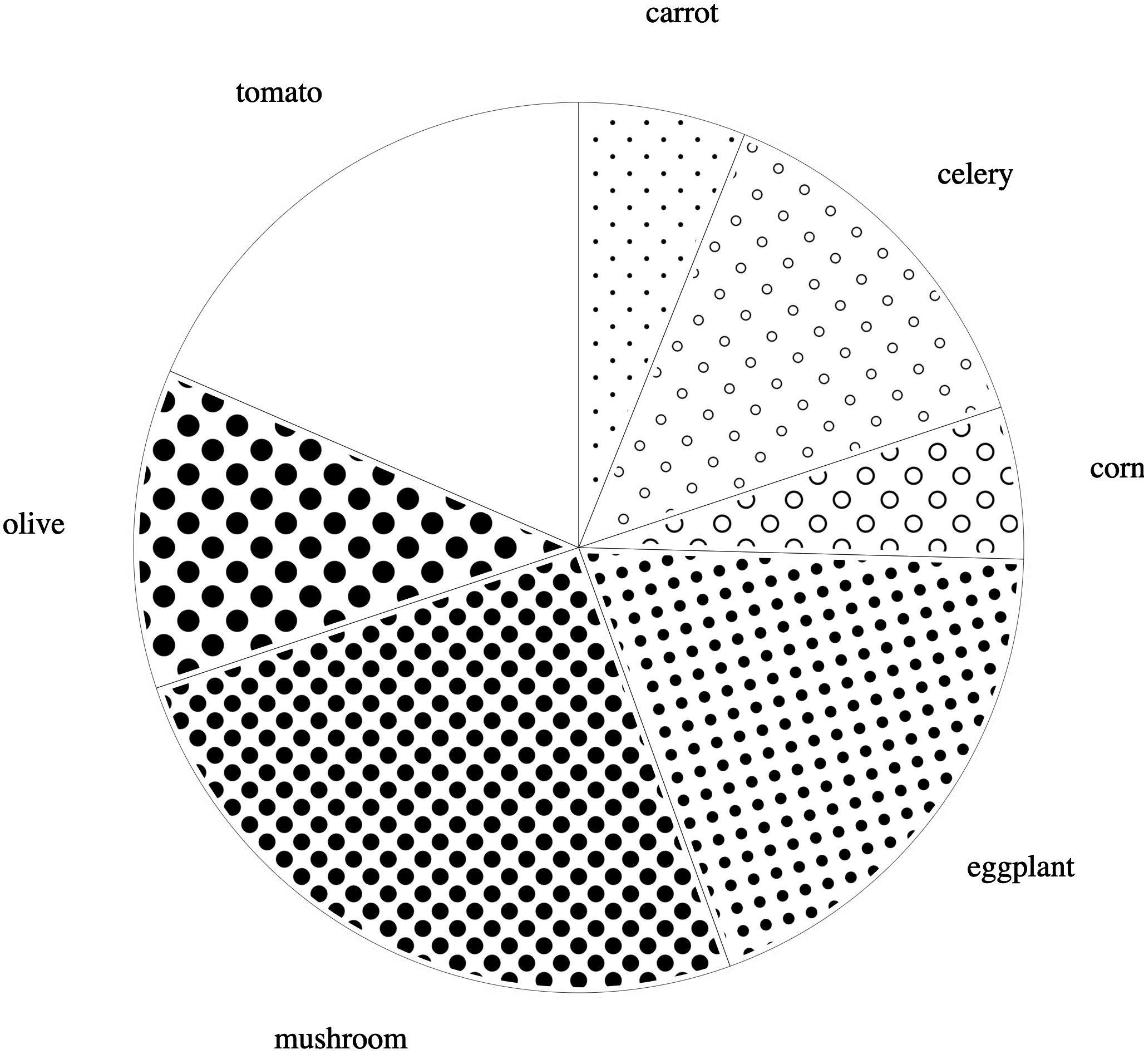}\\
	\caption{An geometric textured pie chart design (PG11) collected in our Experiment 1.}
  \label{fig:PG11}
\end{figure}

\begin{figure}[t] % htbp are optional placement specifiers (here, top, bottom, page)
	\centering % Centers the figure
	\includegraphics[width=\appendixfigurewidth]{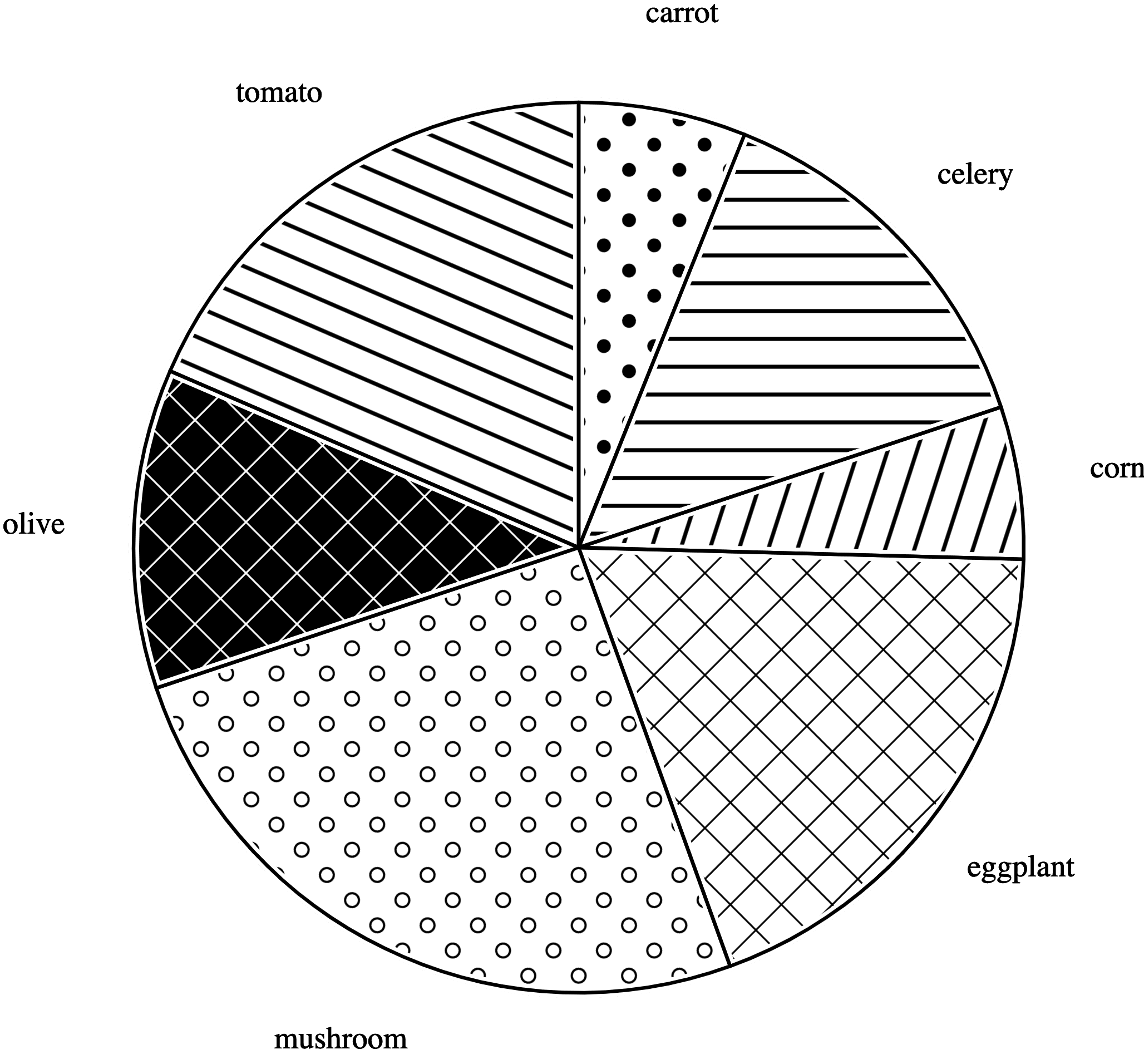}\\
	\caption{An geometric textured pie chart design (PG12) collected in our Experiment 1.}
  \label{fig:PG12}
\end{figure}

\begin{figure}[t] % htbp are optional placement specifiers (here, top, bottom, page)
	\centering % Centers the figure
	\includegraphics[width=\appendixfigurewidth]{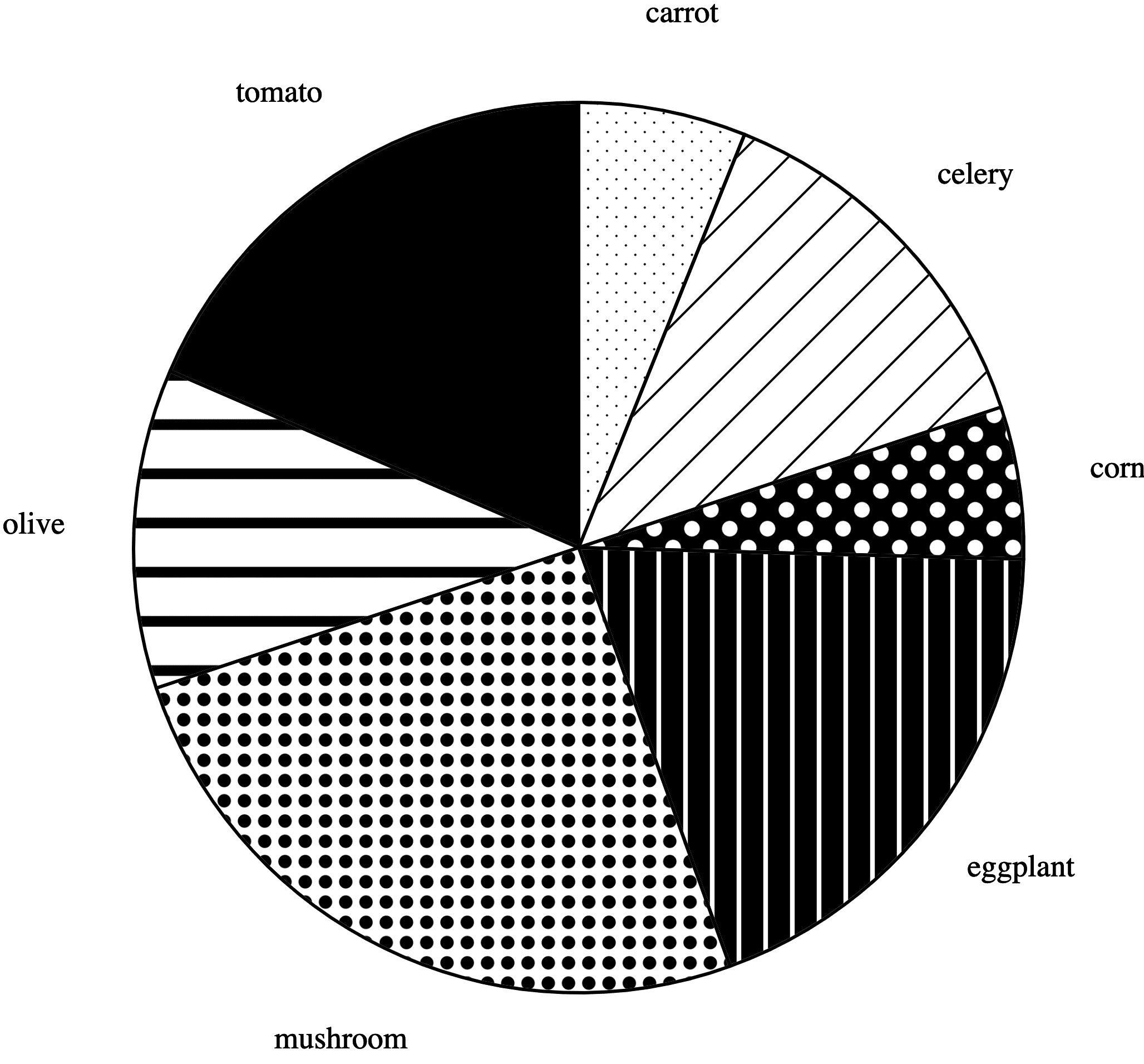}\\
	\caption{An geometric textured pie chart design (PG13) collected in our Experiment 1.}
  \label{fig:PG13}
\end{figure}

\begin{figure}[t] % htbp are optional placement specifiers (here, top, bottom, page)
	\centering % Centers the figure
	\includegraphics[width=\appendixfigurewidth]{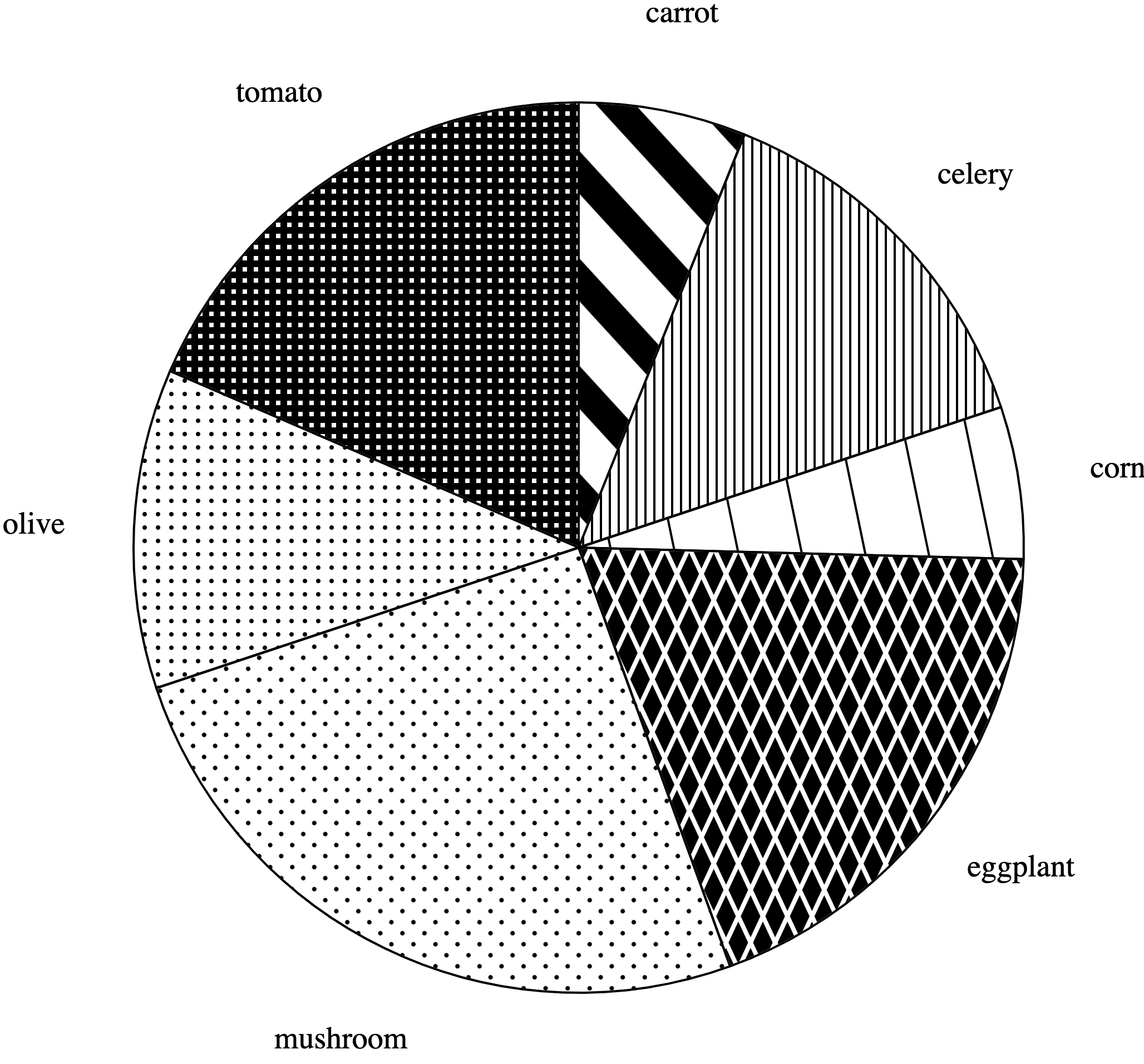}\\
	\caption{An geometric textured pie chart design (PG14) collected in our Experiment 1.}
  \label{fig:PG14}
\end{figure}

\begin{figure}[t] % htbp are optional placement specifiers (here, top, bottom, page)
	\centering % Centers the figure
	\includegraphics[width=\appendixfigurewidth]{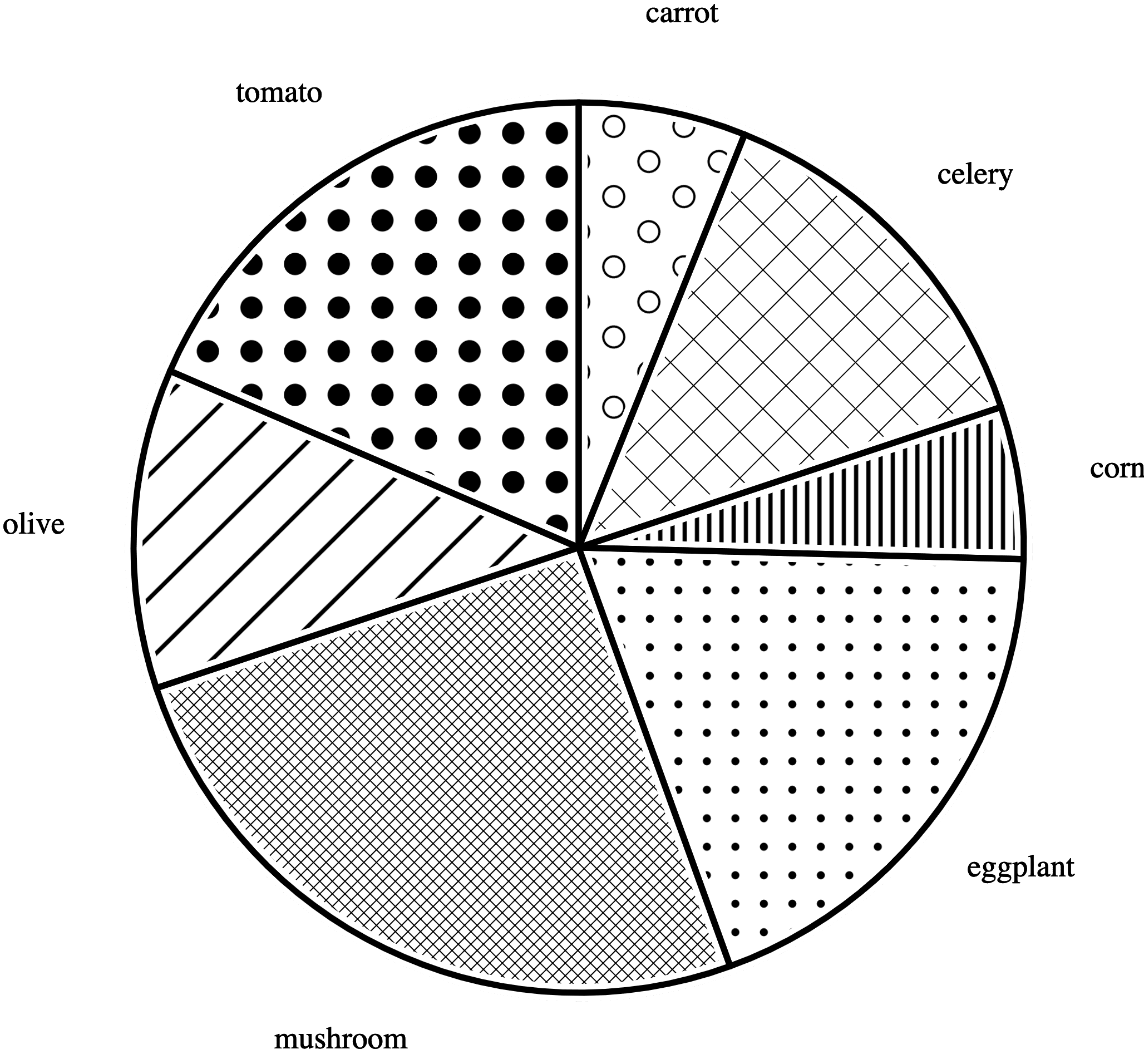}\\
	\caption{An geometric textured pie chart design (PG15) collected in our Experiment 1.}
  \label{fig:PG15}
\end{figure}

\begin{figure}[t] % htbp are optional placement specifiers (here, top, bottom, page)
	\centering % Centers the figure
	\includegraphics[width=\appendixfigurewidth]{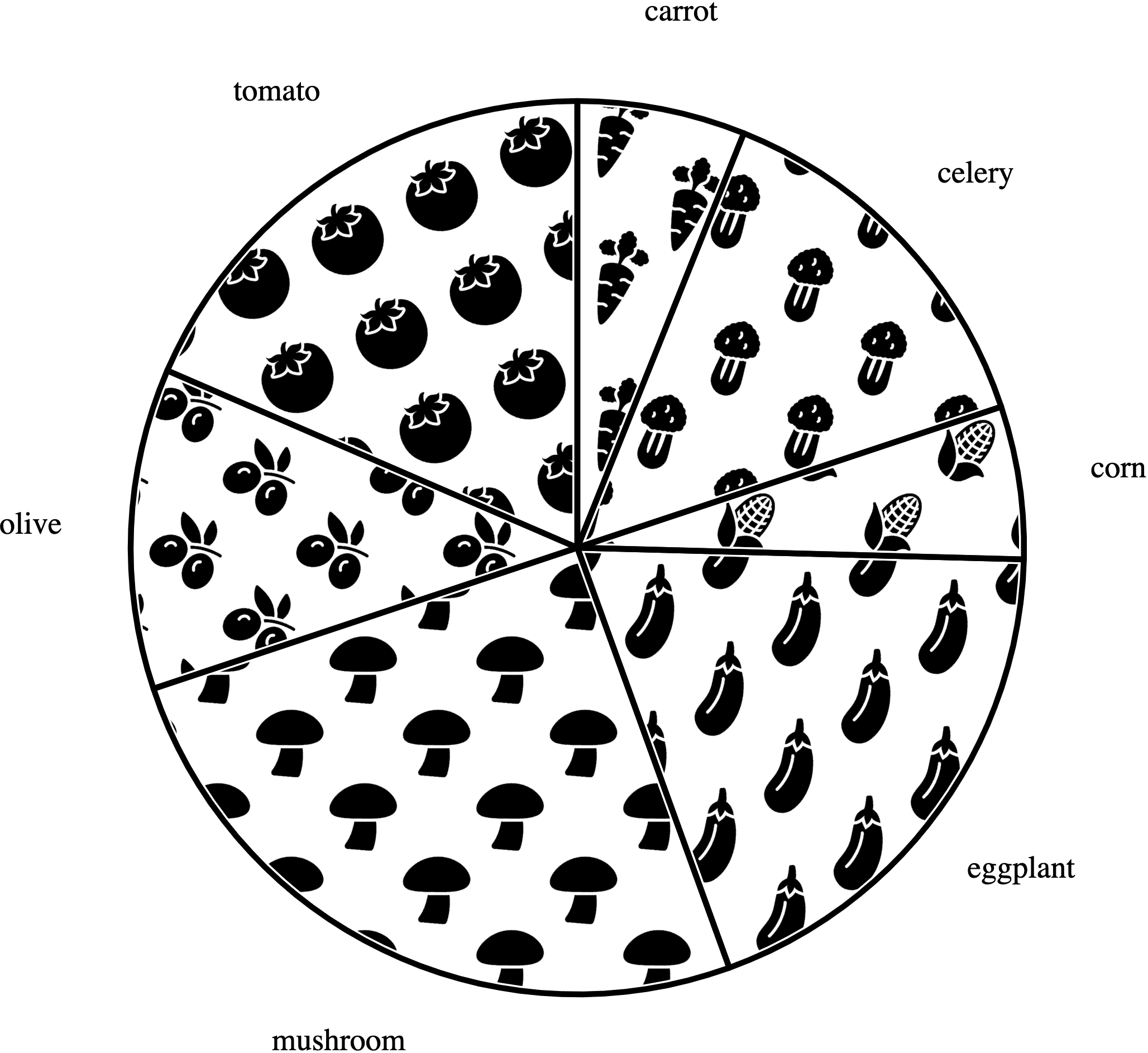}\\
	\caption{An iconic textured pie chart design (PI1) collected in our Experiment 1. This is a larger version of the first image in \autoref{tab:exp2-pie-icon}.}
  \label{fig:PI1}
\end{figure}

\clearpage

\begin{figure}[t] % htbp are optional placement specifiers (here, top, bottom, page)
	\centering % Centers the figure
	\includegraphics[width=\appendixfigurewidth]{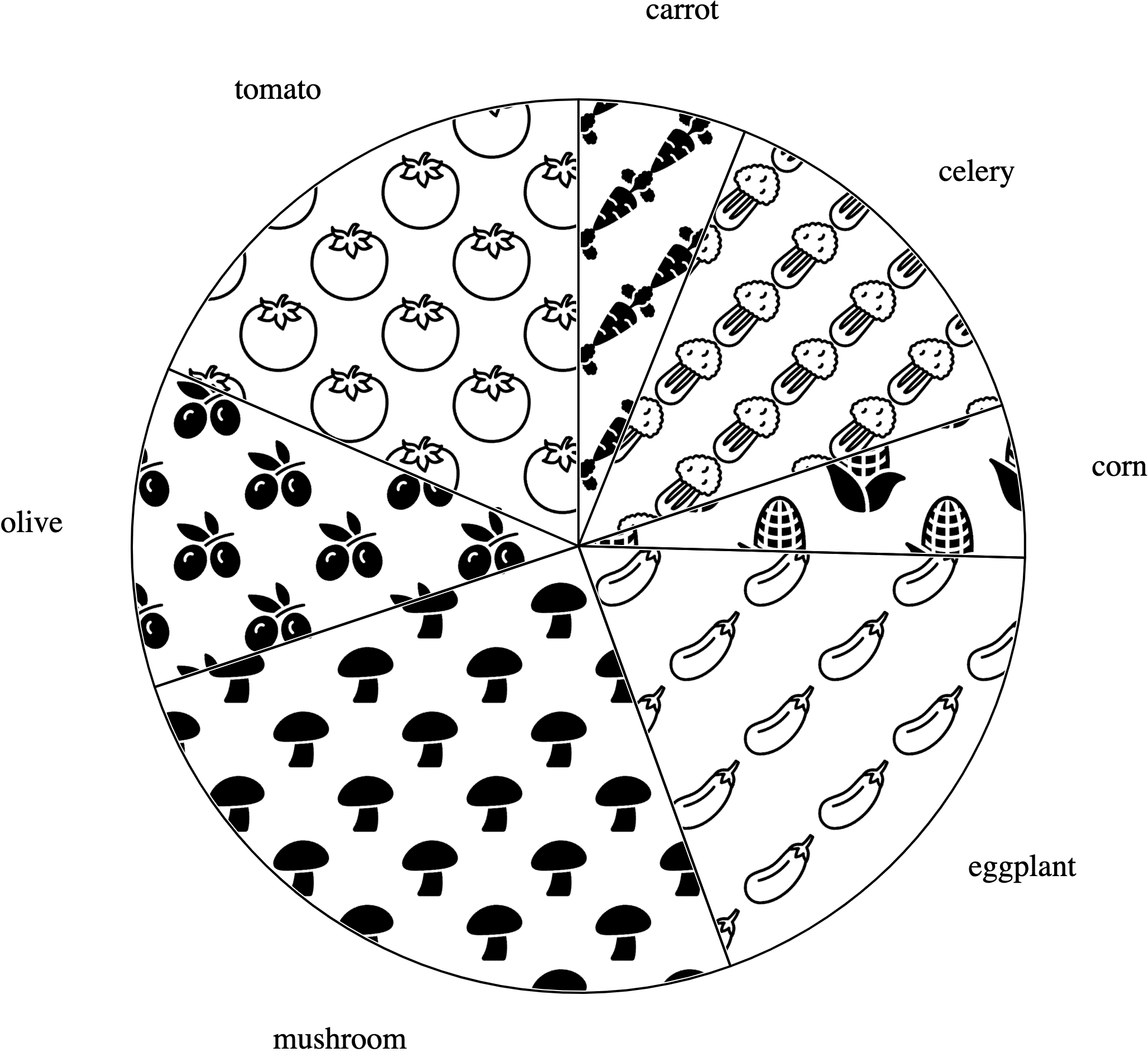}\\
	\caption{An iconic textured pie chart design (PI2) collected in our Experiment 1. This is a larger version of the second image in \autoref{tab:exp2-pie-icon}.}
  \label{fig:PI2}
\end{figure}

\begin{figure}[t] % htbp are optional placement specifiers (here, top, bottom, page)
	\centering % Centers the figure
	\includegraphics[width=\appendixfigurewidth]{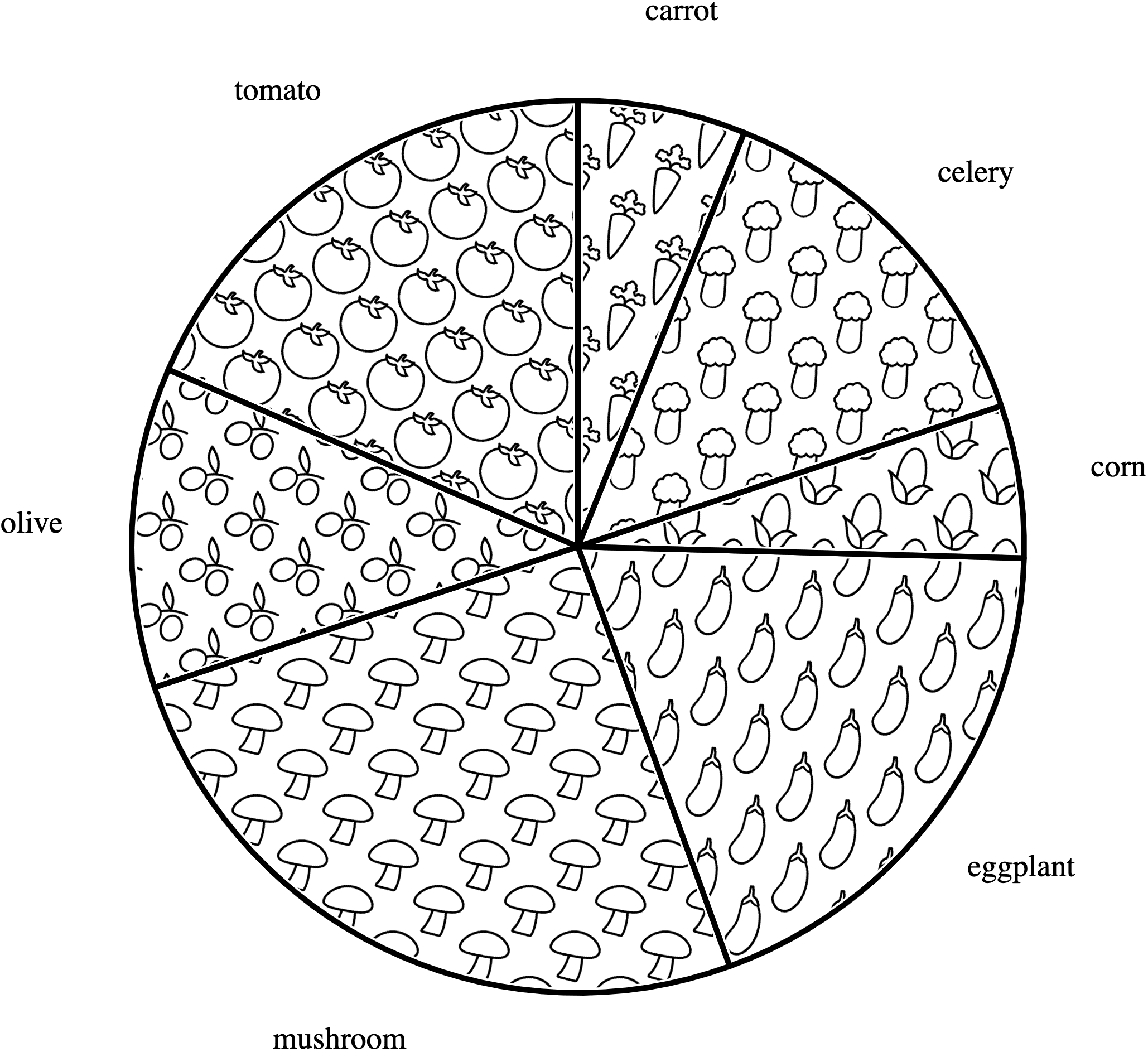}\\
	\caption{An iconic textured pie chart design (PI3) collected in our Experiment 1. This is a larger version of the third image in \autoref{tab:exp2-pie-icon}.}
  \label{fig:PI3}
\end{figure}

\begin{figure}[t] % htbp are optional placement specifiers (here, top, bottom, page)
	\centering % Centers the figure
	\includegraphics[width=\appendixfigurewidth]{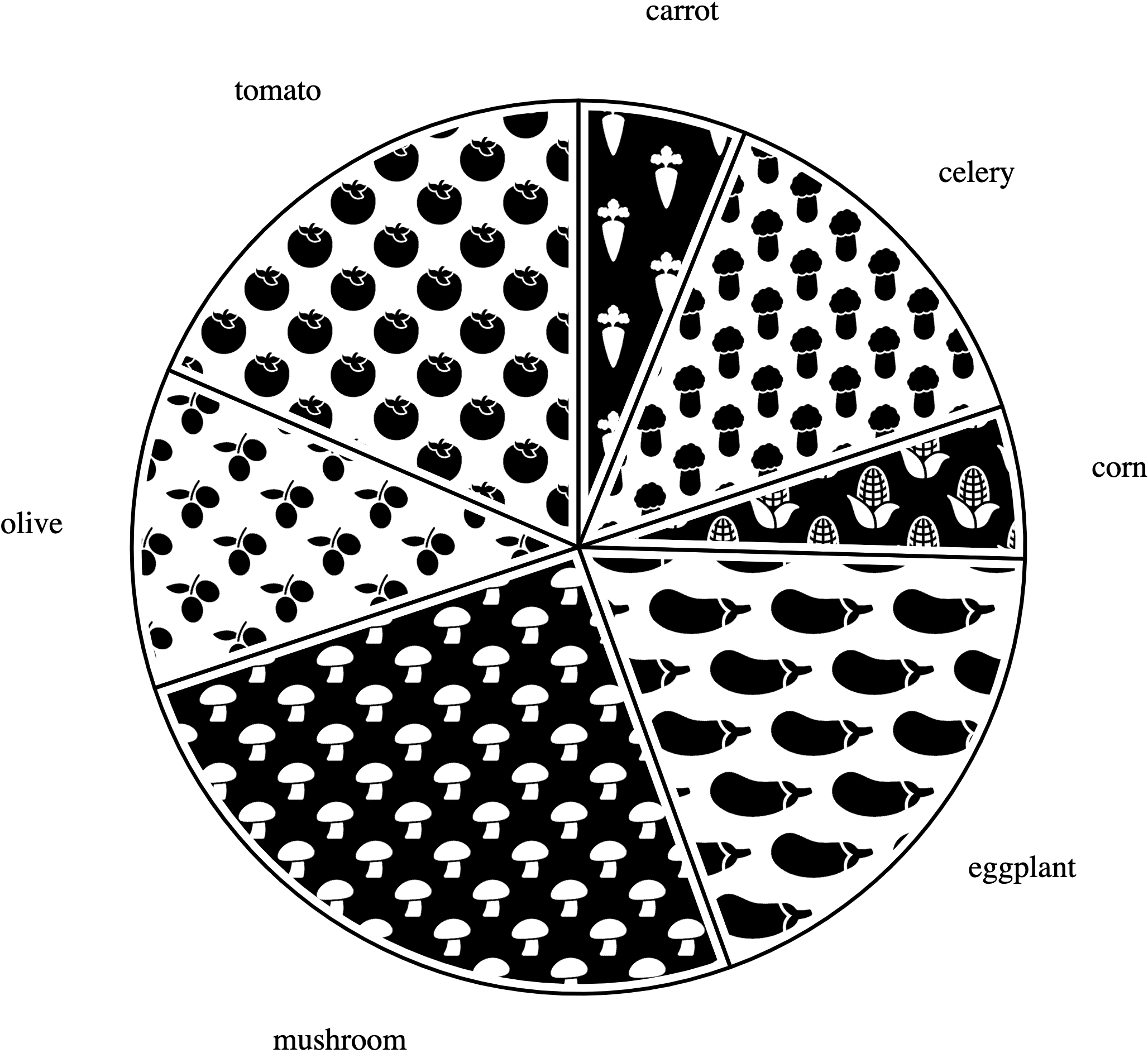}\\
	\caption{An iconic textured pie chart design (PI4) collected in our Experiment 1. This is a larger version of the fourth image in \autoref{tab:exp2-pie-icon}.}
  \label{fig:PI4}
\end{figure}

\begin{figure}[t] % htbp are optional placement specifiers (here, top, bottom, page)
	\centering % Centers the figure
	\includegraphics[width=\appendixfigurewidth]{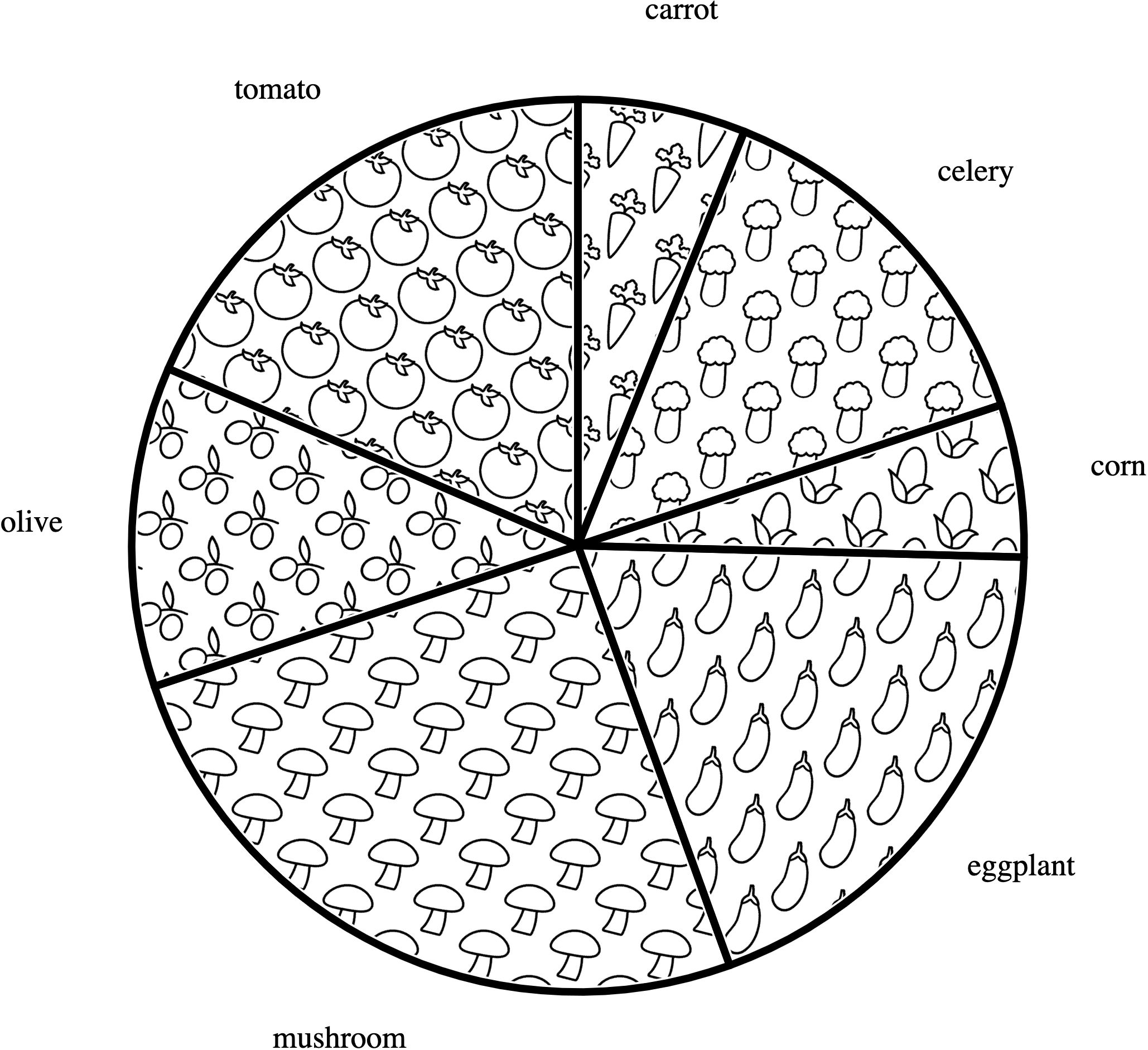}\\
	\caption{An iconic textured pie chart design (PI5) collected in our Experiment 1.}
  \label{fig:PI5}
\end{figure}

\begin{figure}[t] % htbp are optional placement specifiers (here, top, bottom, page)
	\centering % Centers the figure
	\includegraphics[width=\appendixfigurewidth]{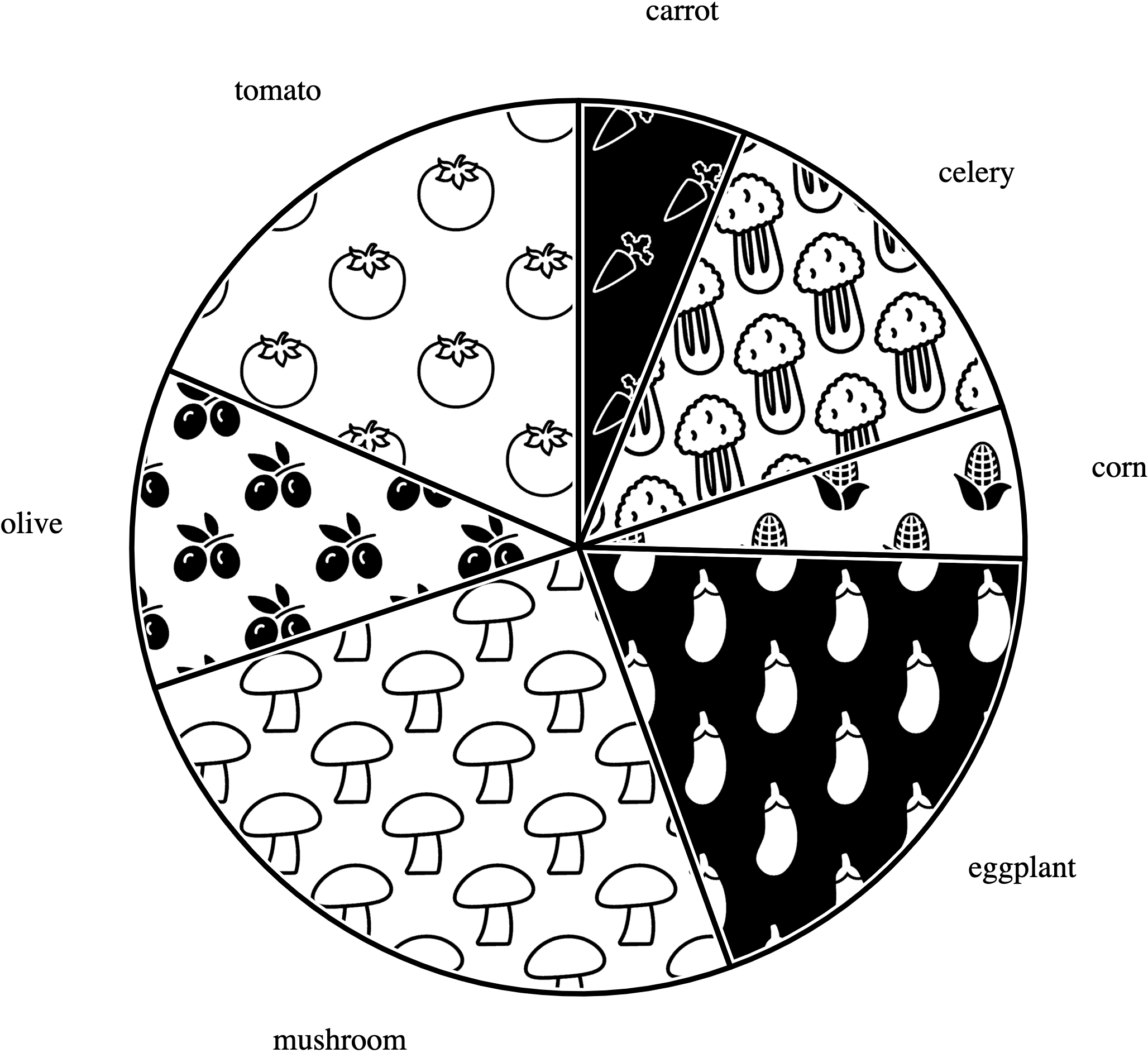}\\
	\caption{An iconic textured pie chart design (PI6) collected in our Experiment 1.}
  \label{fig:PI6}
\end{figure}

\begin{figure}[t] % htbp are optional placement specifiers (here, top, bottom, page)
	\centering % Centers the figure
	\includegraphics[width=\appendixfigurewidth]{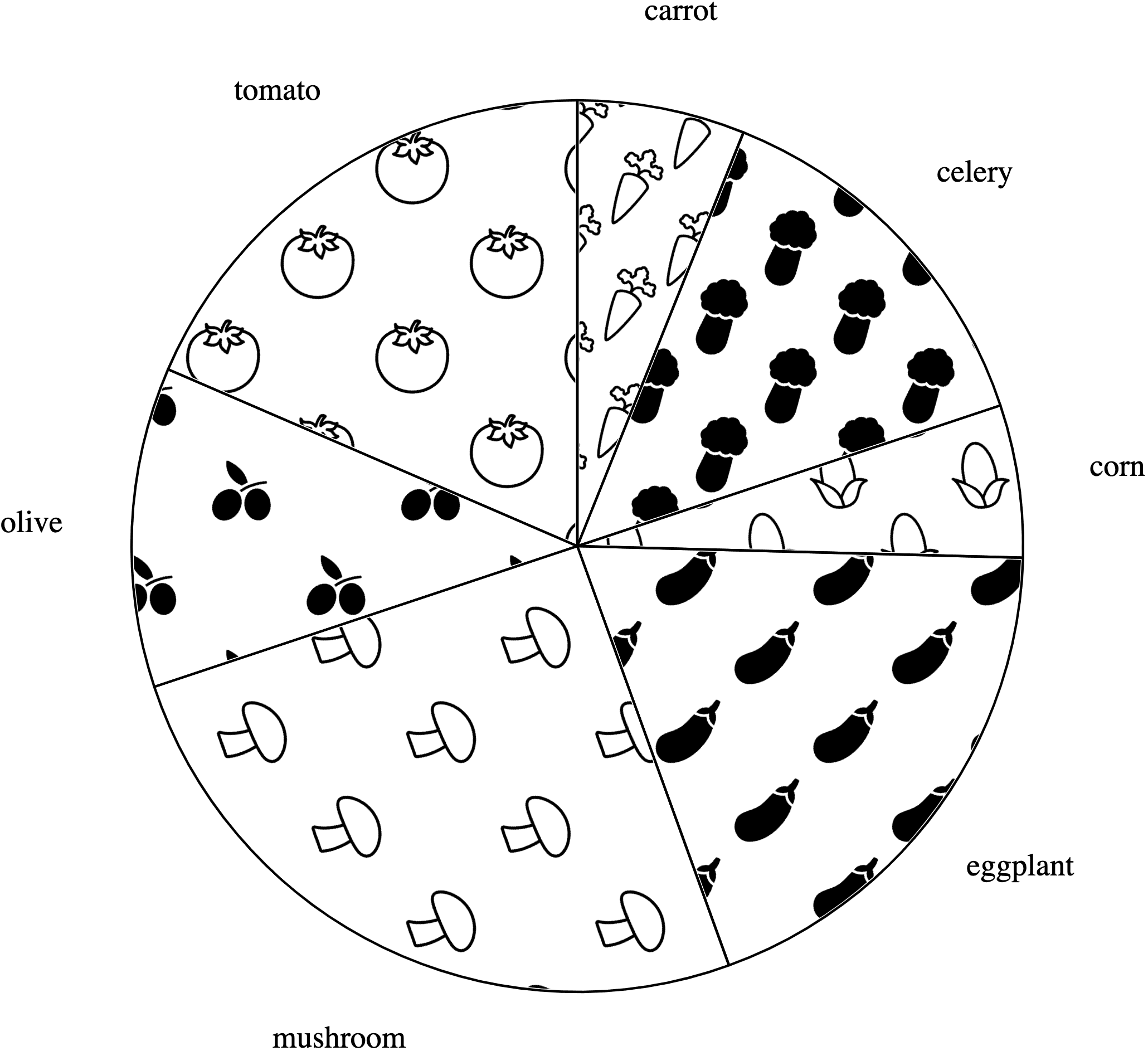}\\
	\caption{An iconic textured pie chart design (PI7) collected in our Experiment 1.}
  \label{fig:PI7}
\end{figure}

\clearpage

\begin{figure}[t] % htbp are optional placement specifiers (here, top, bottom, page)
	\centering % Centers the figure
	\includegraphics[width=\appendixfigurewidth]{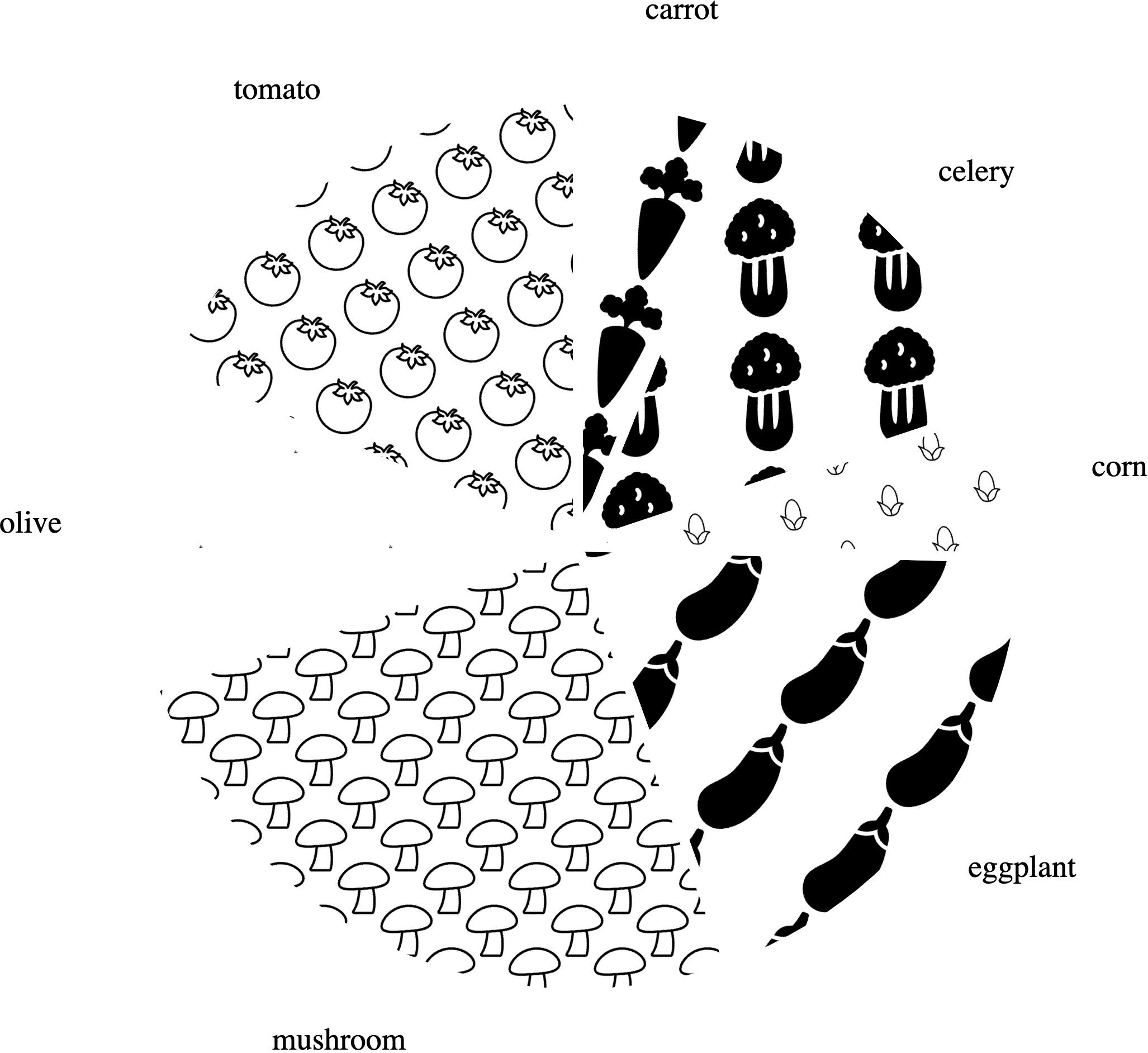}\\
	\caption{An iconic textured pie chart design (PI8) collected in our Experiment 1.}
  \label{fig:PI8}
\end{figure}

\begin{figure}[t] % htbp are optional placement specifiers (here, top, bottom, page)
	\centering % Centers the figure
	\includegraphics[width=\appendixfigurewidth]{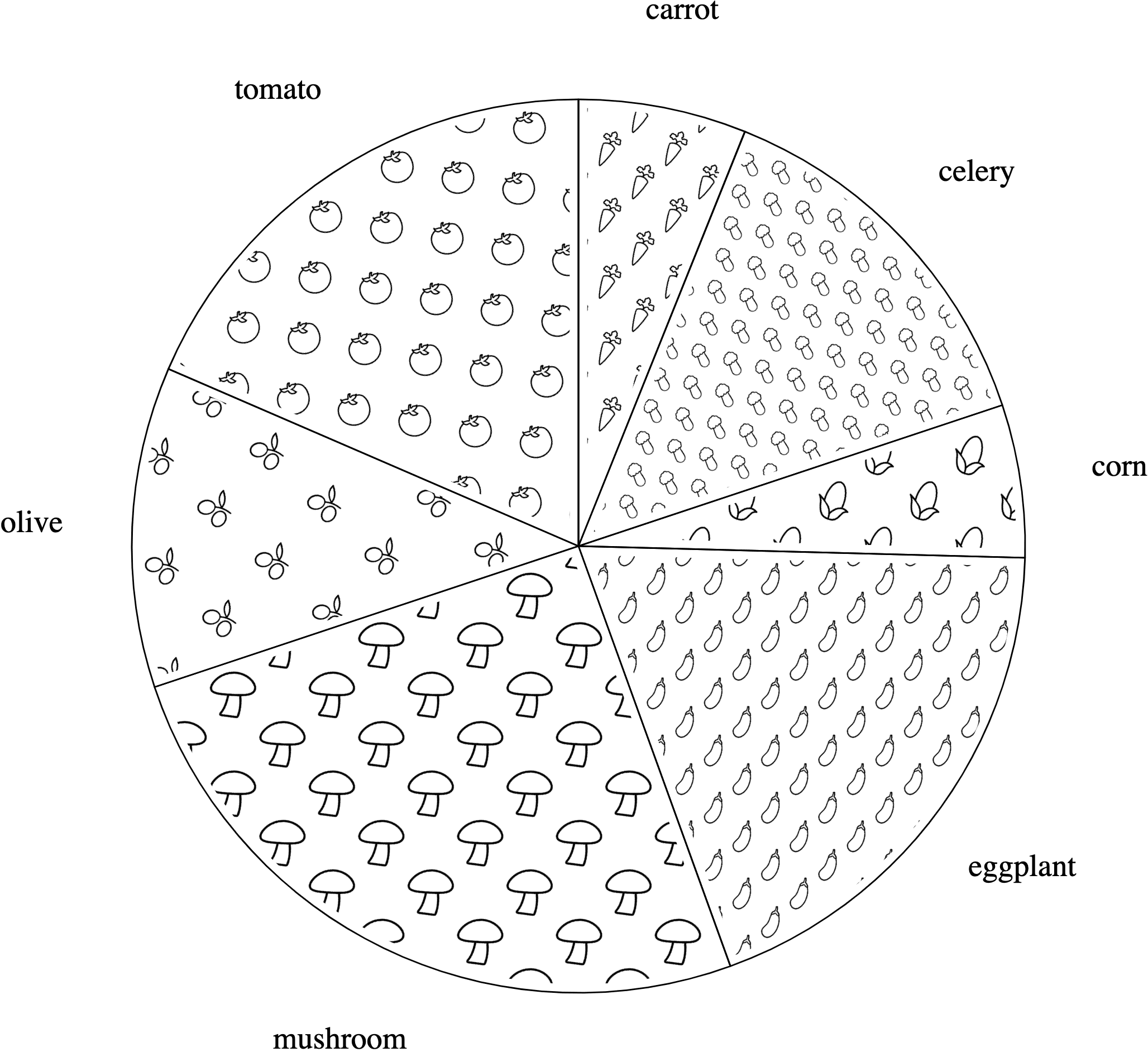}\\
	\caption{An iconic textured pie chart design (PI9) collected in our Experiment 1.}
  \label{fig:PI9}
\end{figure}

\begin{figure}[t] % htbp are optional placement specifiers (here, top, bottom, page)
	\centering % Centers the figure
	\includegraphics[width=\appendixfigurewidth]{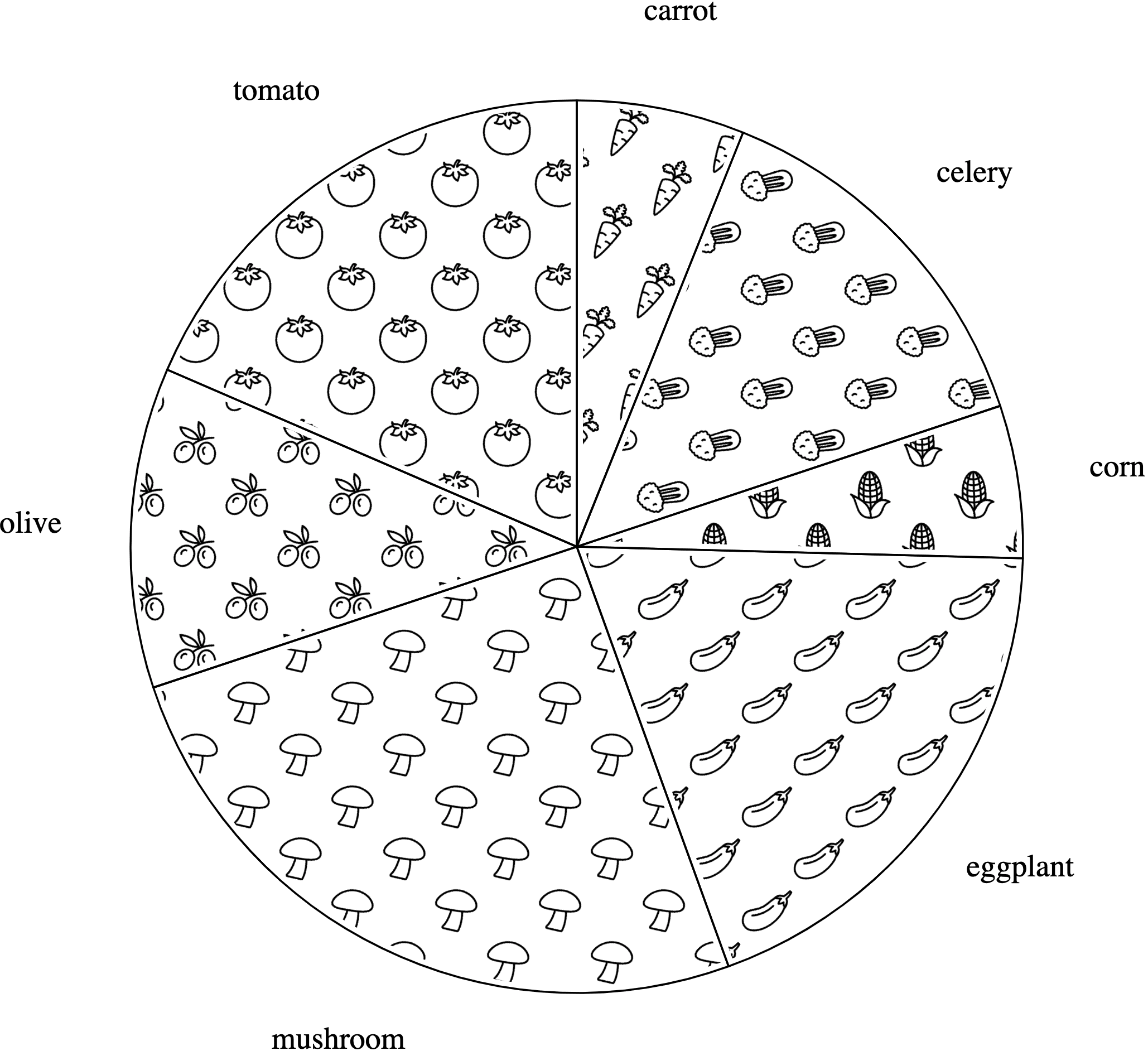}\\
	\caption{An iconic textured pie chart design (PI10) collected in our Experiment 1.}
  \label{fig:PI10}
\end{figure}

\begin{figure}[t] % htbp are optional placement specifiers (here, top, bottom, page)
	\centering % Centers the figure
	\includegraphics[width=\appendixfigurewidth]{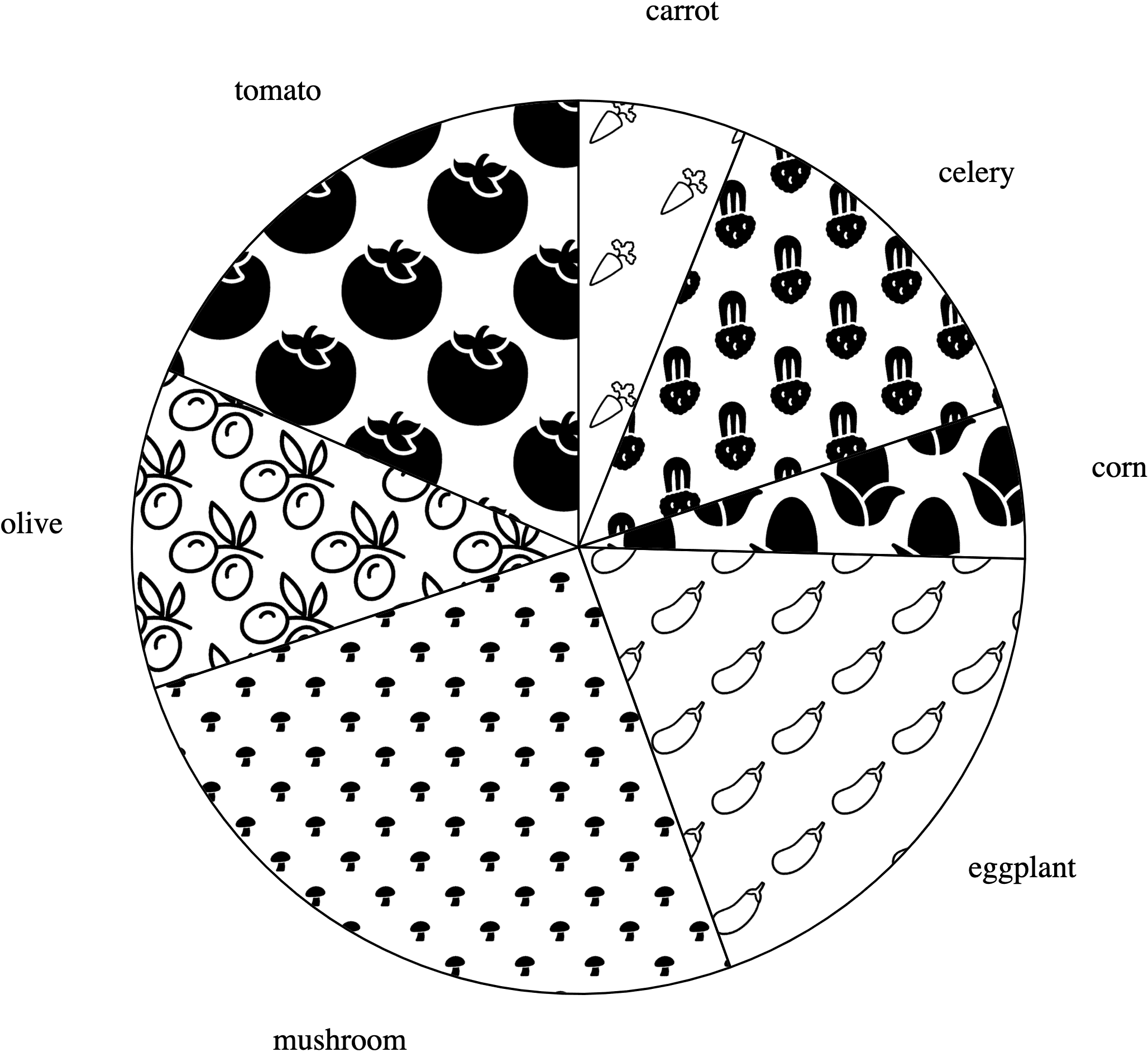}\\
	\caption{An iconic textured pie chart design (PI11) collected in our Experiment 1.}
  \label{fig:PI11}
\end{figure}

\begin{figure}[t] % htbp are optional placement specifiers (here, top, bottom, page)
	\centering % Centers the figure
	\includegraphics[width=\appendixfigurewidth]{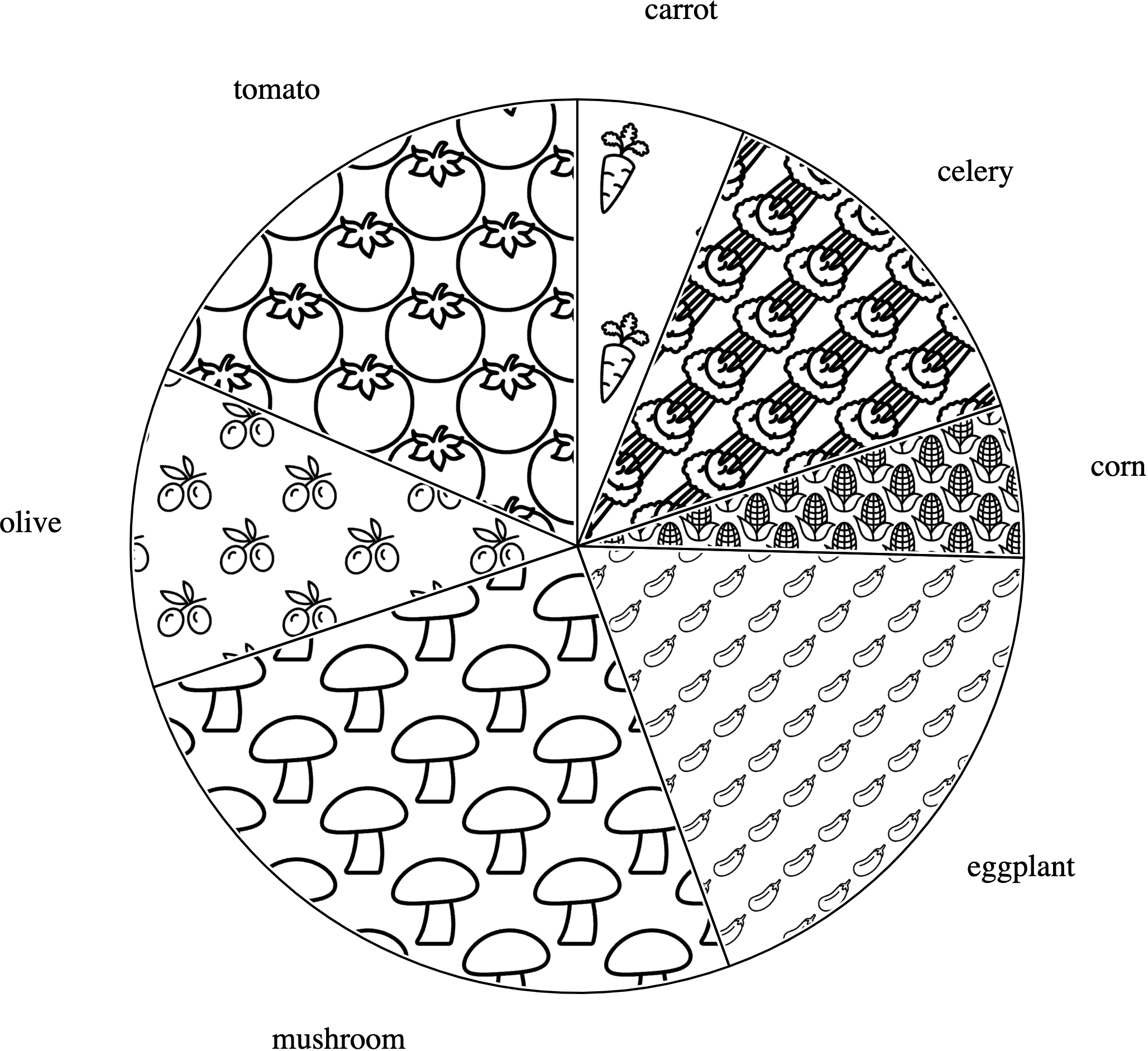}\\
	\caption{An iconic textured pie chart design (PI12) collected in our Experiment 1.}
  \label{fig:PI12}
\end{figure}

\begin{figure}[t] % htbp are optional placement specifiers (here, top, bottom, page)
	\centering % Centers the figure
	\includegraphics[width=\appendixfigurewidth]{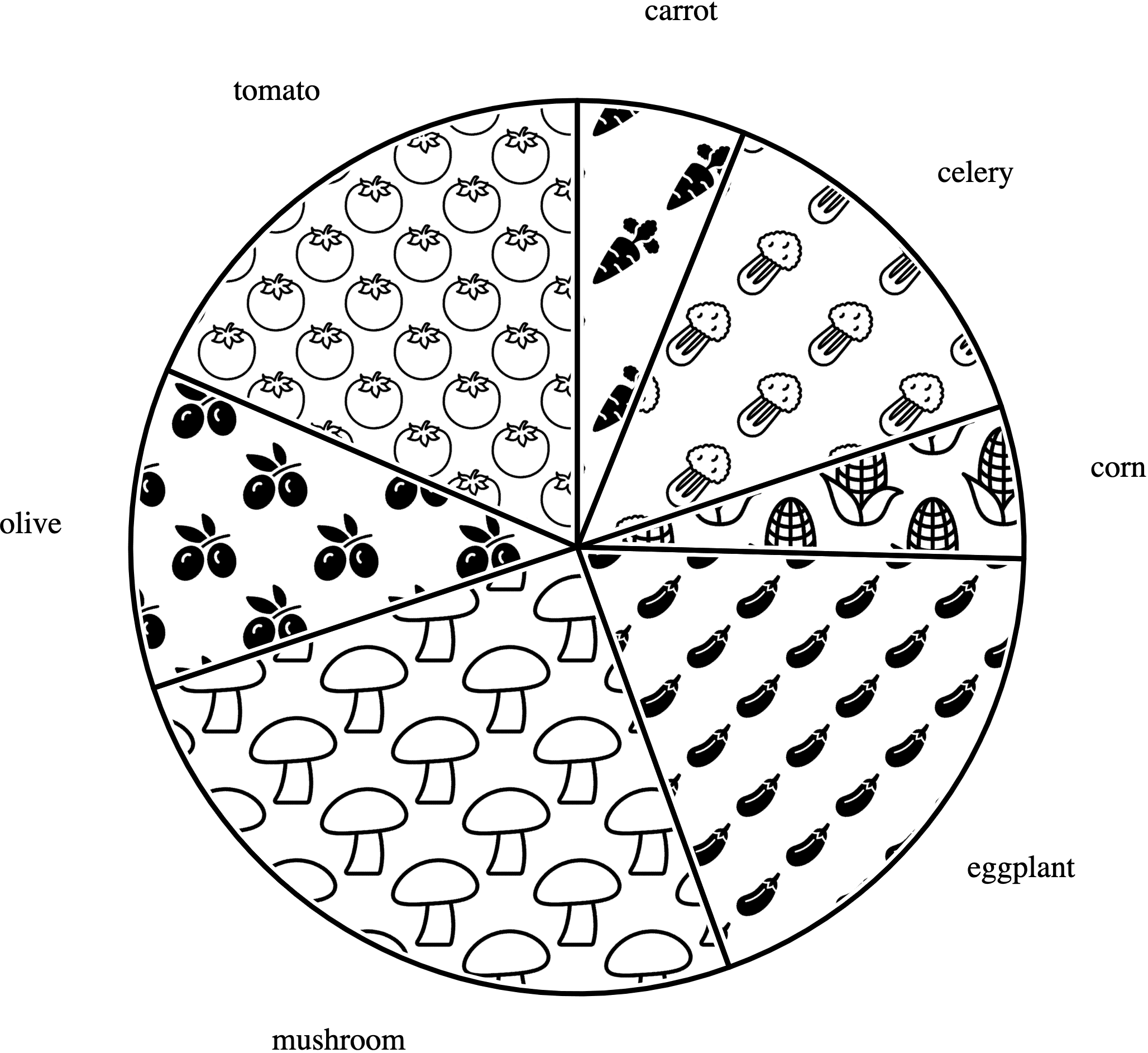}\\
	\caption{An iconic textured pie chart design (PI13) collected in our Experiment 1.}
  \label{fig:PI13}
\end{figure}

\clearpage

\begin{figure}[t] % htbp are optional placement specifiers (here, top, bottom, page)
	\centering % Centers the figure
	\includegraphics[width=\appendixfigurewidth]{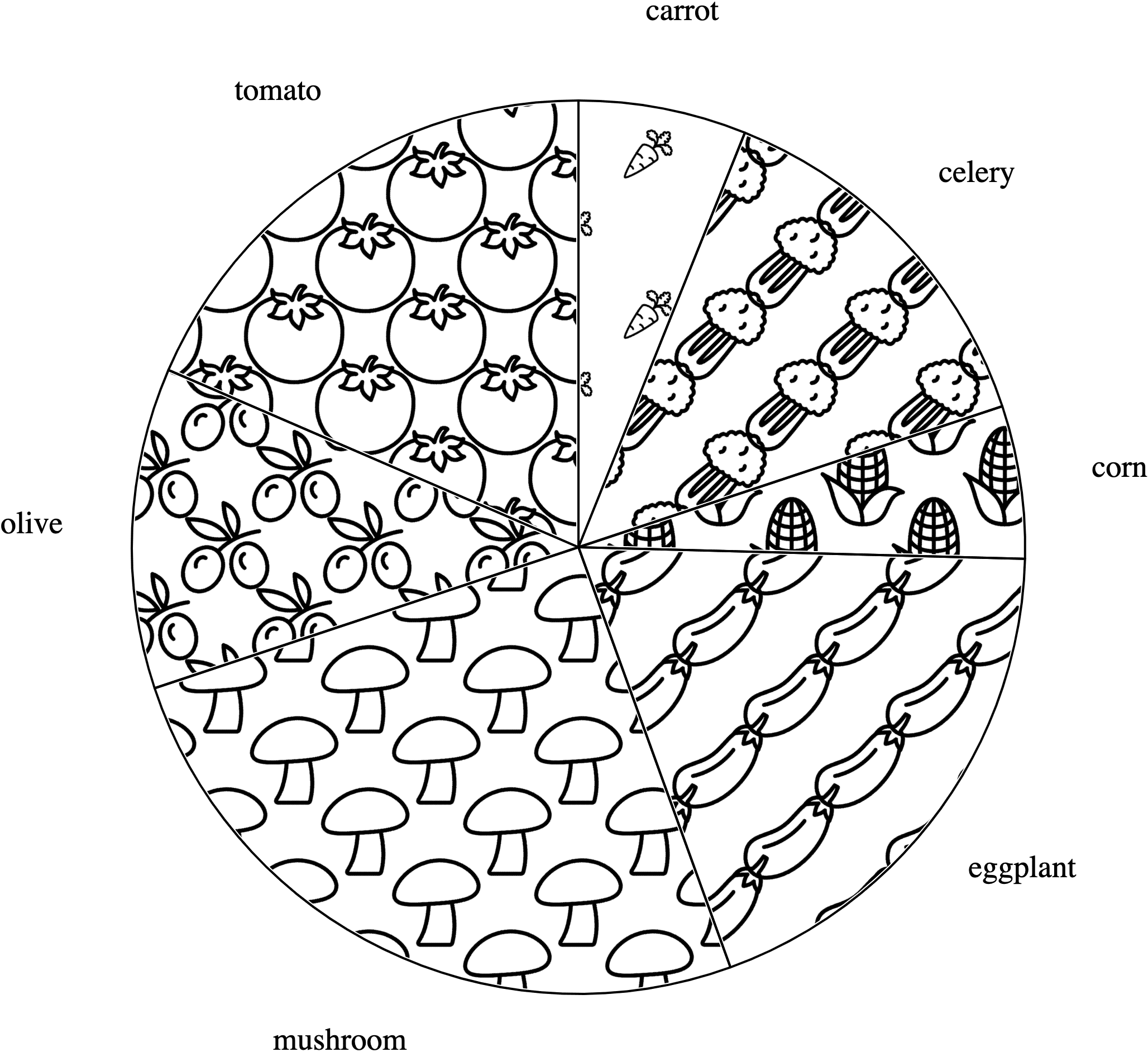}\\
	\caption{An iconic textured pie chart design (PI14) collected in our Experiment 1.}
  \label{fig:PI14}
\end{figure}

\begin{figure}[t] % htbp are optional placement specifiers (here, top, bottom, page)
	\centering % Centers the figure
	\includegraphics[width=\appendixfigurewidth]{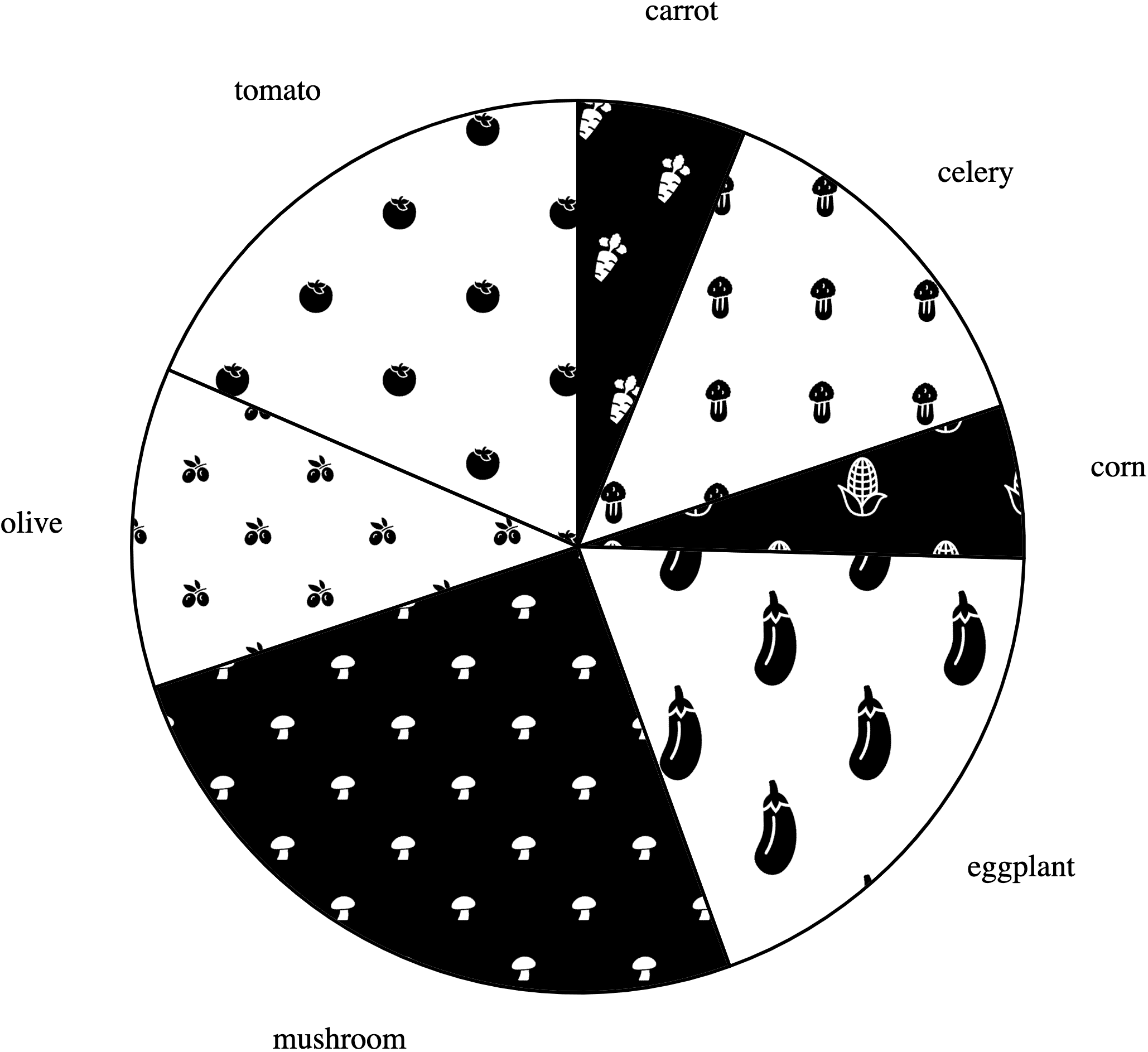}\\
	\caption{An iconic textured pie chart design (PI15) collected in our Experiment 1.}
  \label{fig:PI15}
\end{figure}

\begin{figure}[t] % htbp are optional placement specifiers (here, top, bottom, page)
	\centering % Centers the figure
	\includegraphics[width=\appendixfigurewidth]{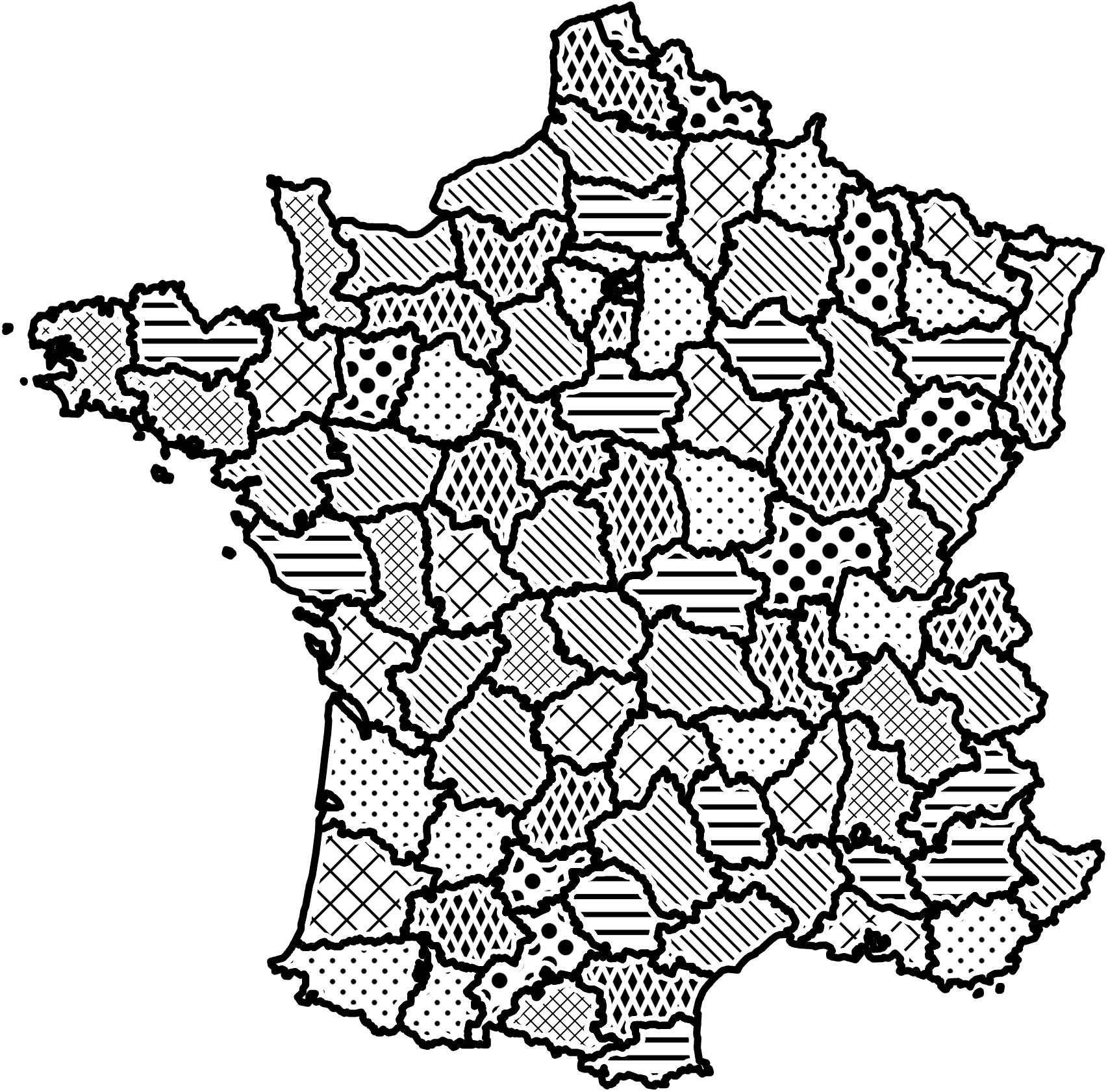}\\
	\caption{An geometric textured map design (MG1) collected in our Experiment 1. This is a larger version of the first image in \autoref{tab:exp2-map-geo}.}
  \label{fig:MG1}
\end{figure}

\begin{figure}[t] % htbp are optional placement specifiers (here, top, bottom, page)
	\centering % Centers the figure
	\includegraphics[width=\appendixfigurewidth]{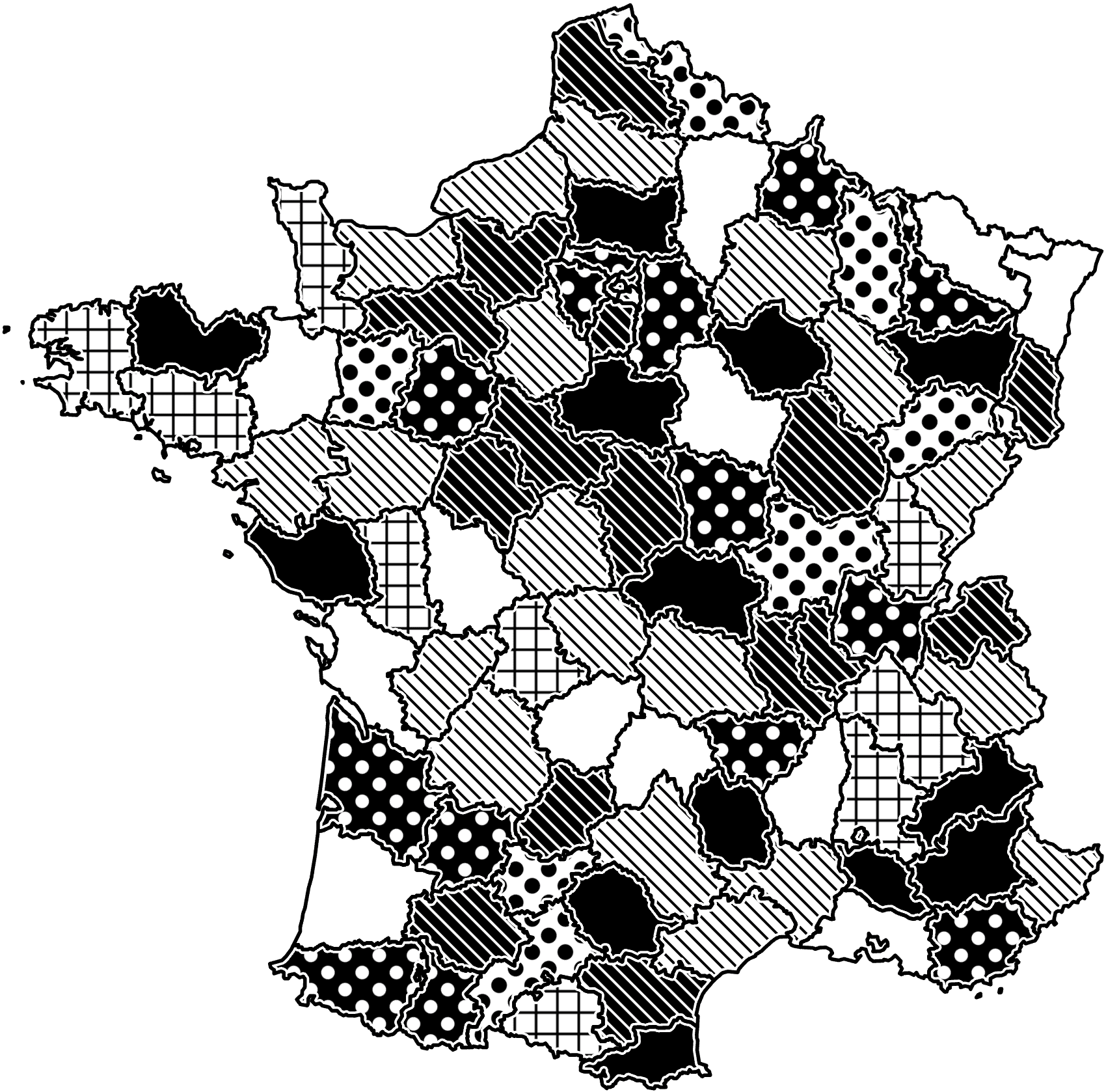}\\
	\caption{An geometric textured map design (MG2) collected in our Experiment 1. This is a larger version of the second image in \autoref{tab:exp2-map-geo}.}
  \label{fig:MG2}
\end{figure}

\begin{figure}[t] % htbp are optional placement specifiers (here, top, bottom, page)
	\centering % Centers the figure
	\includegraphics[width=\appendixfigurewidth]{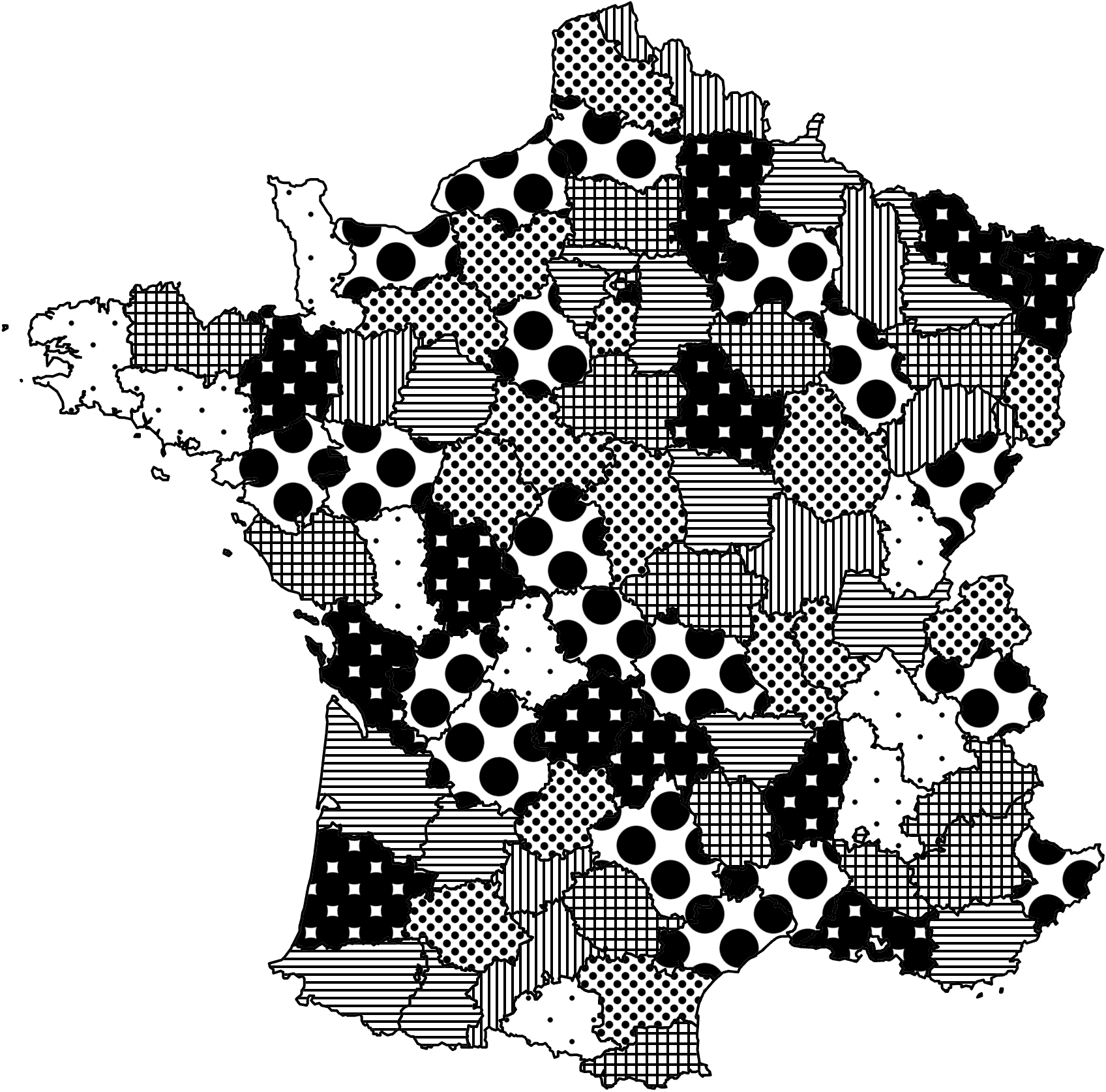}\\
	\caption{An geometric textured map design (MG3) collected in our Experiment 1. This is a larger version of the third image in \autoref{tab:exp2-map-geo}.}
  \label{fig:MG3}
\end{figure}

\begin{figure}[t] % htbp are optional placement specifiers (here, top, bottom, page)
	\centering % Centers the figure
	\includegraphics[width=\appendixfigurewidth]{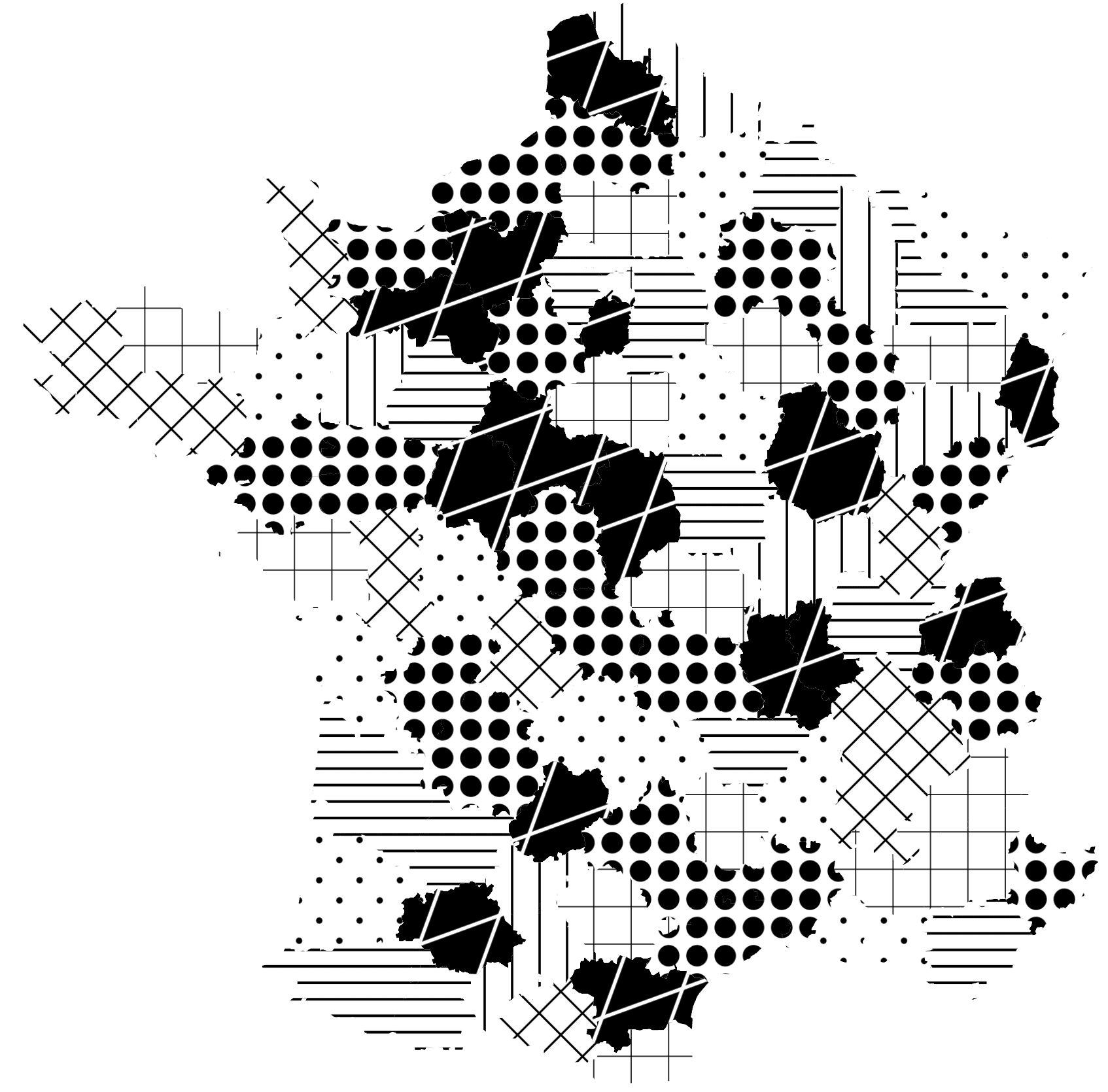}\\
	\caption{An geometric textured map design (MG4) collected in our Experiment 1. This is a larger version of the fourth image in \autoref{tab:exp2-map-geo}.}
  \label{fig:MG4}
\end{figure}

\clearpage

\begin{figure}[t] % htbp are optional placement specifiers (here, top, bottom, page)
	\centering % Centers the figure
	\includegraphics[width=\appendixfigurewidth]{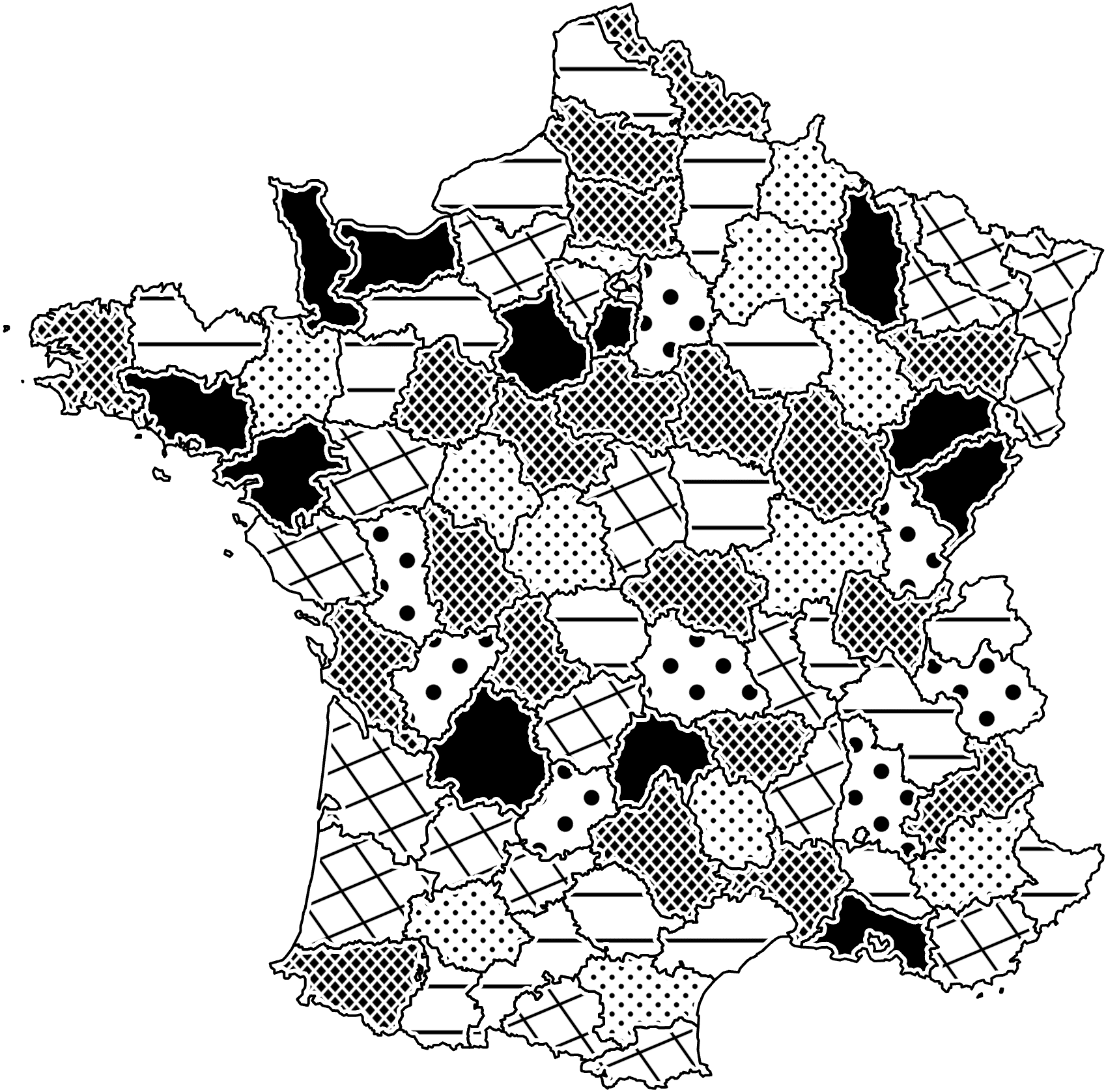}\\
	\caption{An geometric textured map design (MG5) collected in our Experiment 1.}
  \label{fig:MG5}
\end{figure}

\begin{figure}[t] % htbp are optional placement specifiers (here, top, bottom, page)
	\centering % Centers the figure
	\includegraphics[width=\appendixfigurewidth]{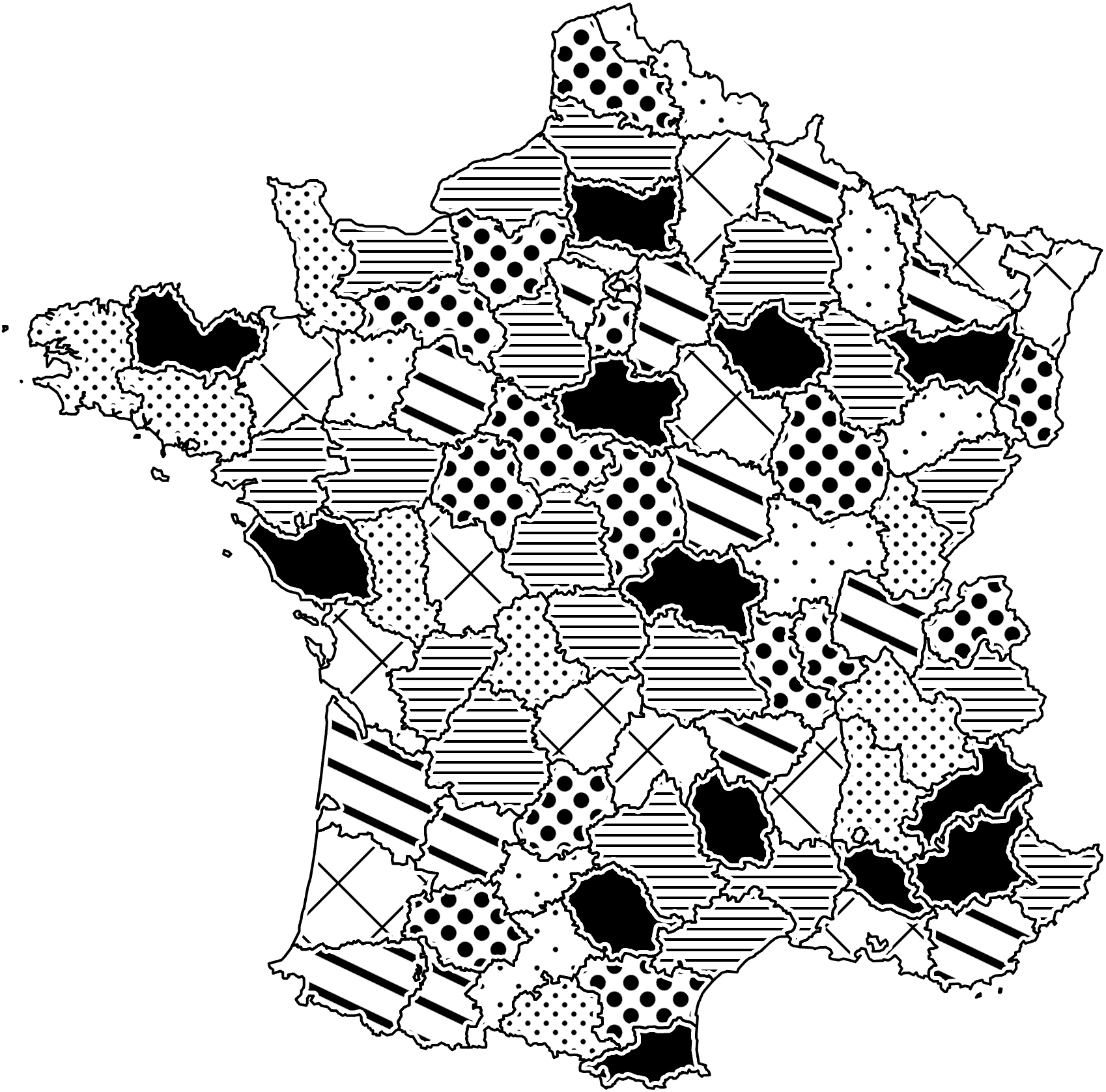}\\
	\caption{An geometric textured map design (MG6) collected in our Experiment 1.}
  \label{fig:MG6}
\end{figure}

\begin{figure}[t] % htbp are optional placement specifiers (here, top, bottom, page)
	\centering % Centers the figure
	\includegraphics[width=\appendixfigurewidth]{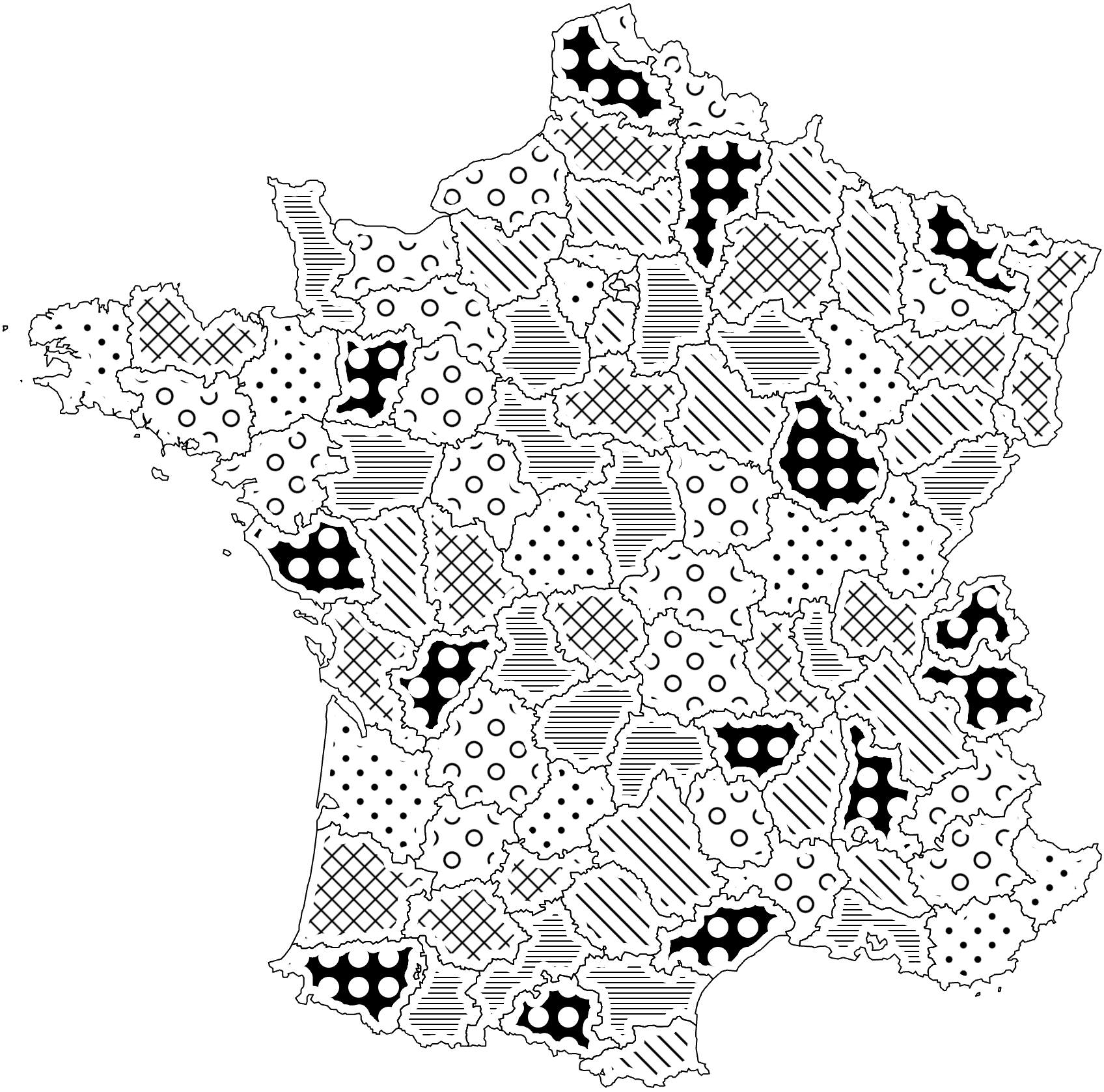}\\
	\caption{An geometric textured map design (MG7) collected in our Experiment 1.}
  \label{fig:MG7}
\end{figure}

\begin{figure}[t] % htbp are optional placement specifiers (here, top, bottom, page)
	\centering % Centers the figure
	\includegraphics[width=\appendixfigurewidth]{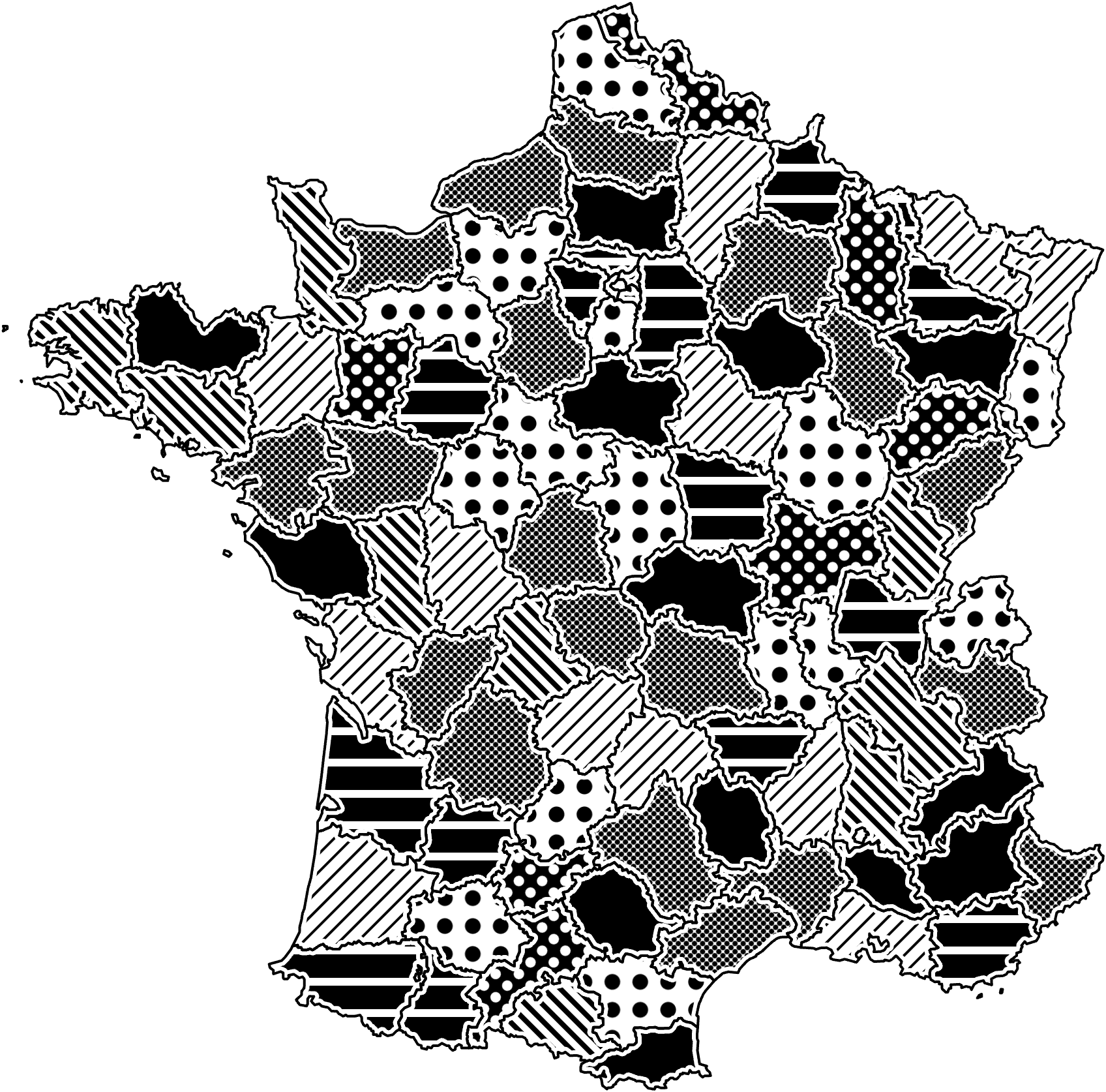}\\
	\caption{An geometric textured map design (MG8) collected in our Experiment 1.}
  \label{fig:MG8}
\end{figure}

\begin{figure}[t] % htbp are optional placement specifiers (here, top, bottom, page)
	\centering % Centers the figure
	\includegraphics[width=\appendixfigurewidth]{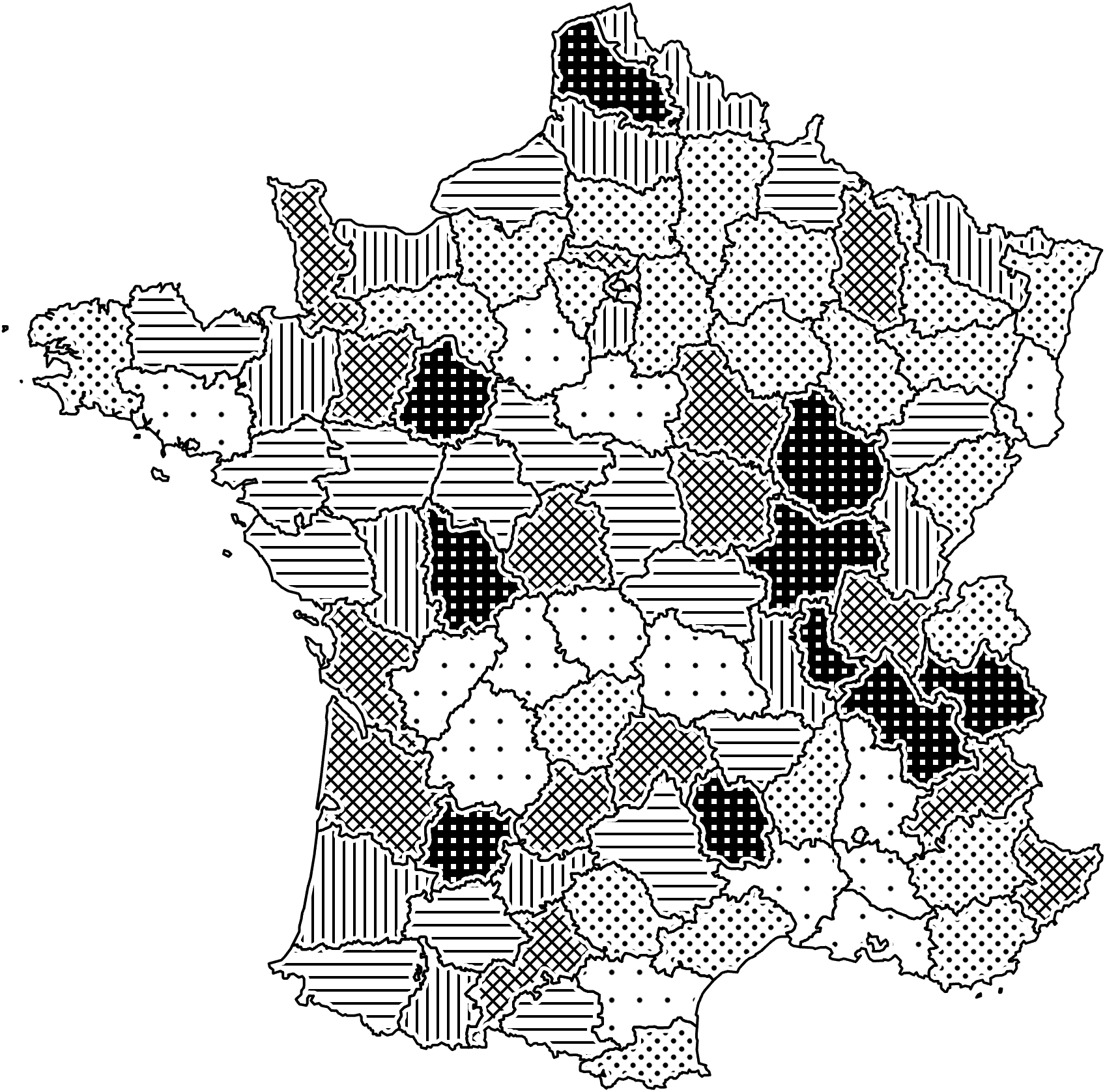}\\
	\caption{An geometric textured map design (MG9) collected in our Experiment 1.}
  \label{fig:MG9}
\end{figure}

\begin{figure}[t] % htbp are optional placement specifiers (here, top, bottom, page)
	\centering % Centers the figure
	\includegraphics[width=\appendixfigurewidth]{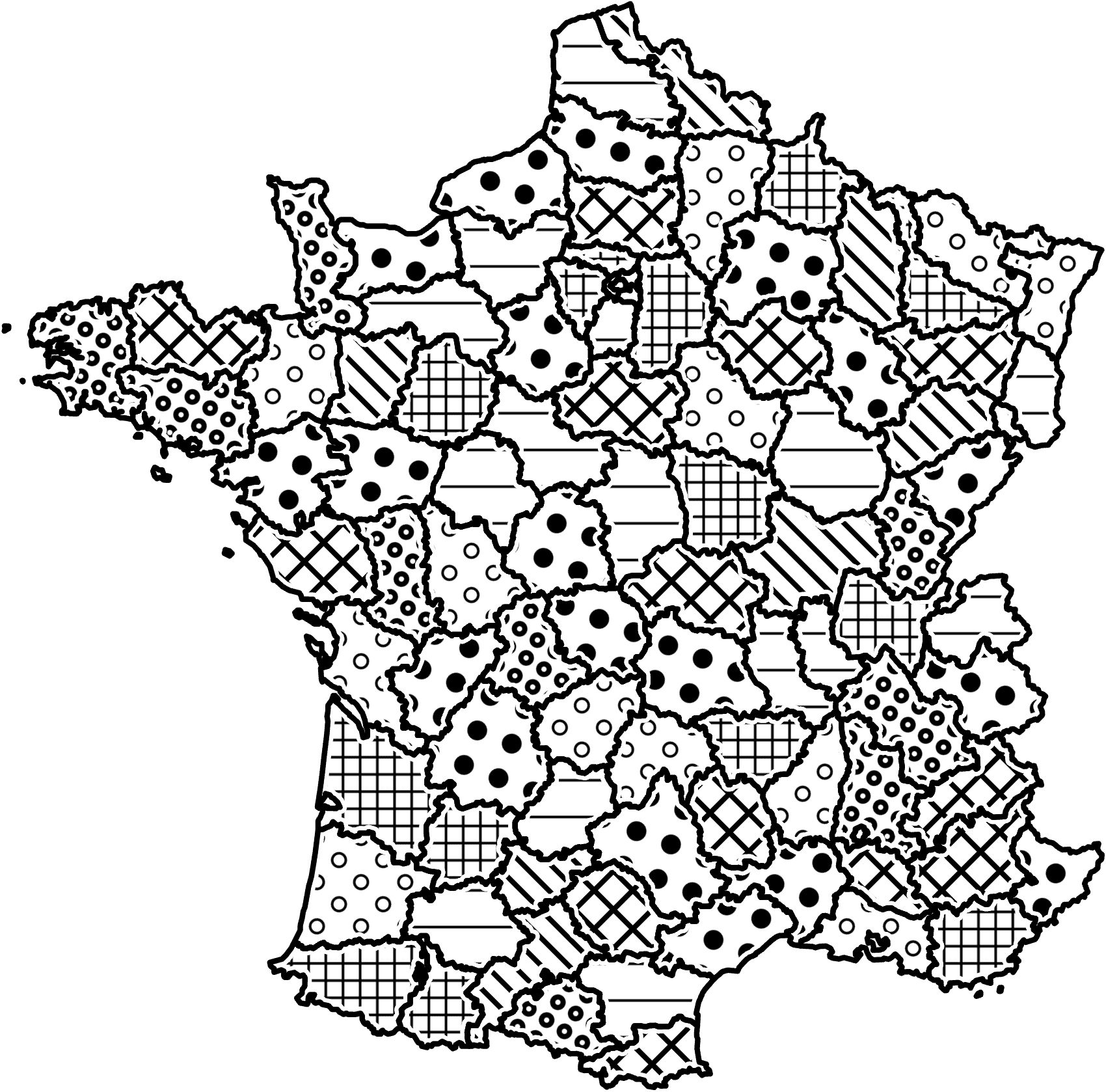}\\
	\caption{An geometric textured map design (MG10) collected in our Experiment 1.}
  \label{fig:MG10}
\end{figure}

\clearpage

\begin{figure}[t] % htbp are optional placement specifiers (here, top, bottom, page)
	\centering % Centers the figure
	\includegraphics[width=\appendixfigurewidth]{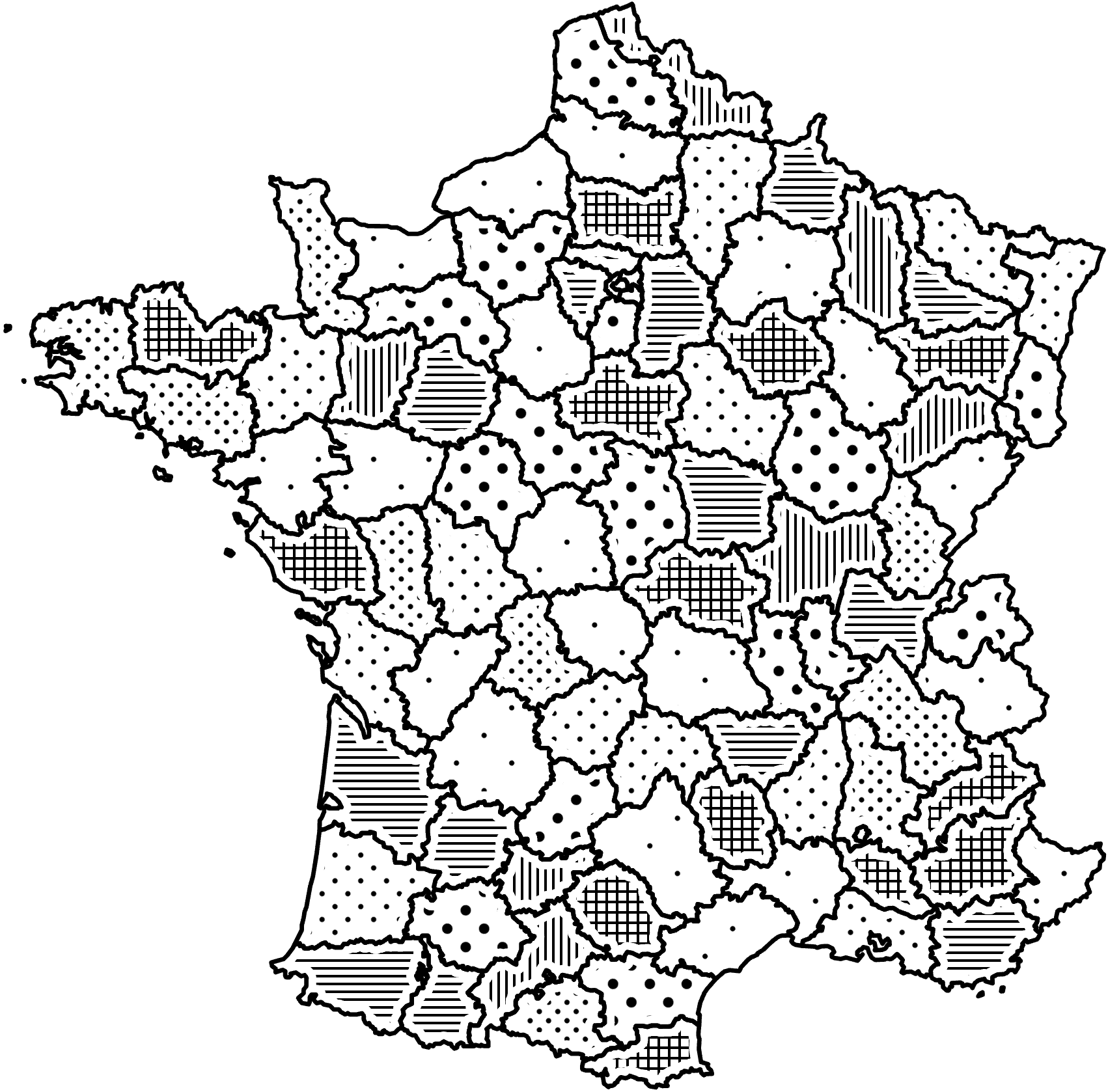}\\
	\caption{An geometric textured map design (MG11) collected in our Experiment 1.}
  \label{fig:MG11}
\end{figure}

\begin{figure}[t] % htbp are optional placement specifiers (here, top, bottom, page)
	\centering % Centers the figure
	\includegraphics[width=\appendixfigurewidth]{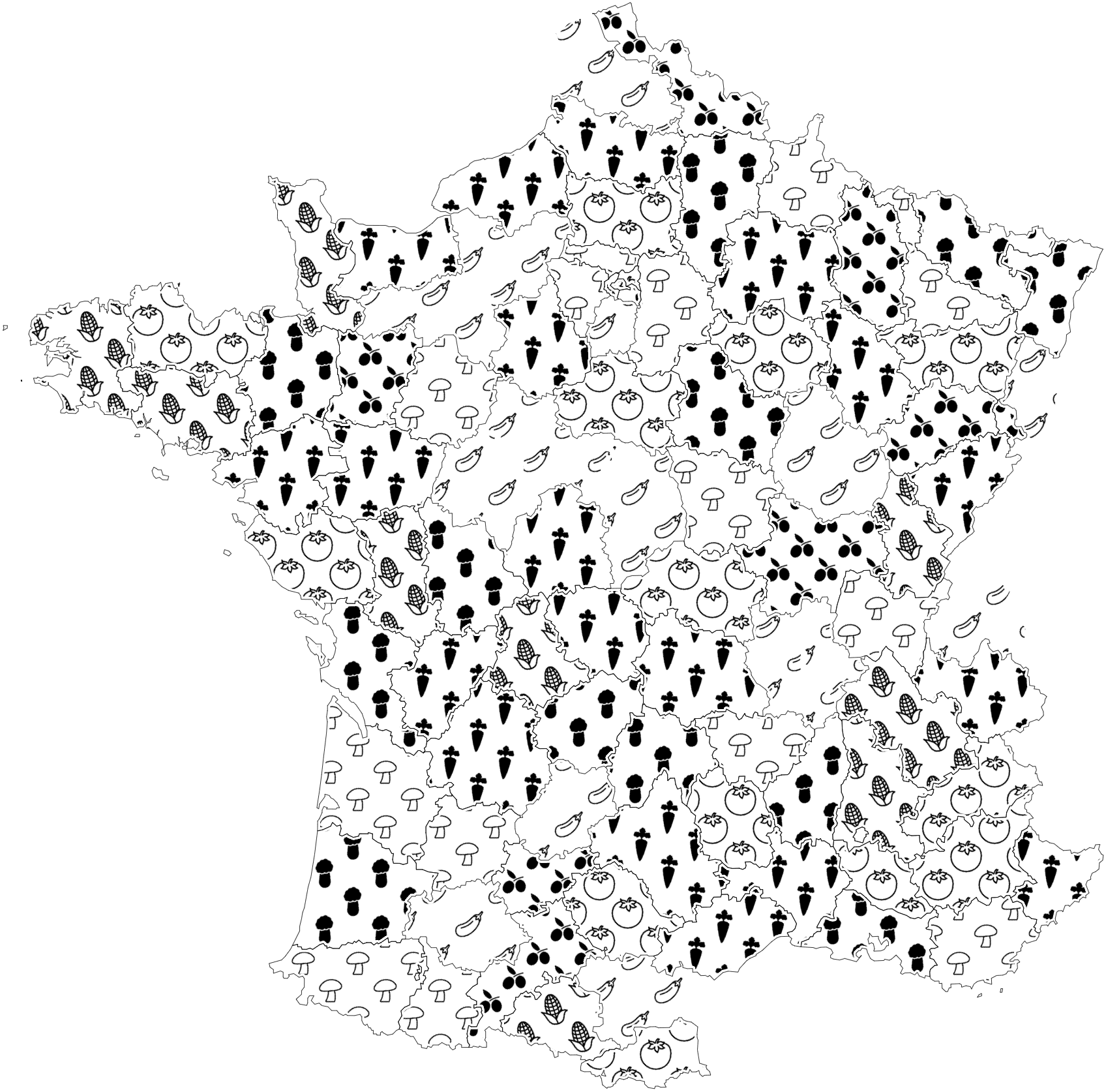}\\
	\caption{An iconic textured map design (MI1) collected in our Experiment 1. This is a larger version of the first image in \autoref{tab:exp2-map-icon}.}
  \label{fig:MI1}
\end{figure}

\begin{figure}[t] % htbp are optional placement specifiers (here, top, bottom, page)
	\centering % Centers the figure
	\includegraphics[width=\appendixfigurewidth]{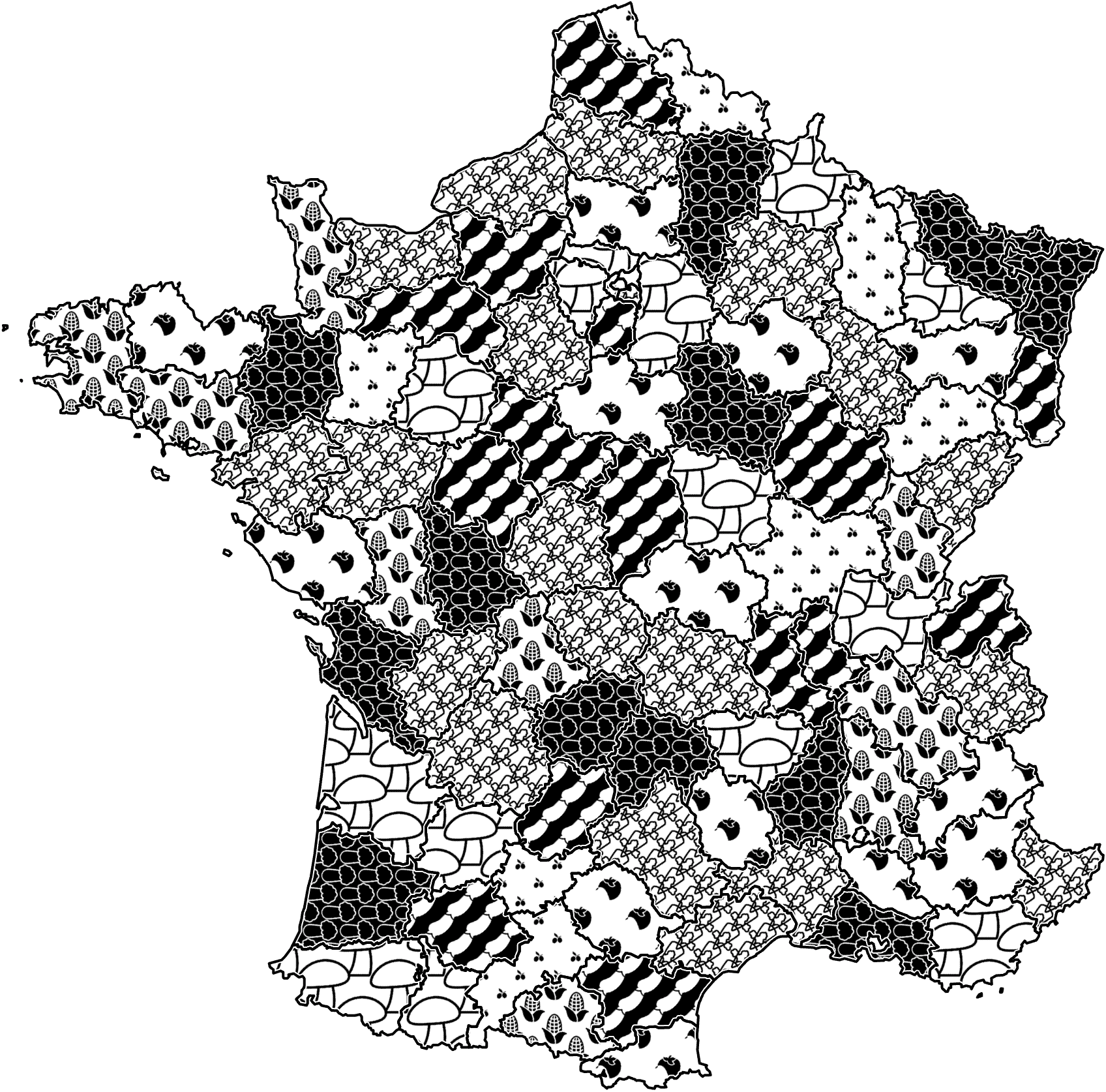}\\
	\caption{An iconic textured map design (MI2) collected in our Experiment 1. This is a larger version of the second image in \autoref{tab:exp2-map-icon}.}
  \label{fig:MI2}
\end{figure}

\begin{figure}[t] % htbp are optional placement specifiers (here, top, bottom, page)
	\centering % Centers the figure
	\includegraphics[width=\appendixfigurewidth]{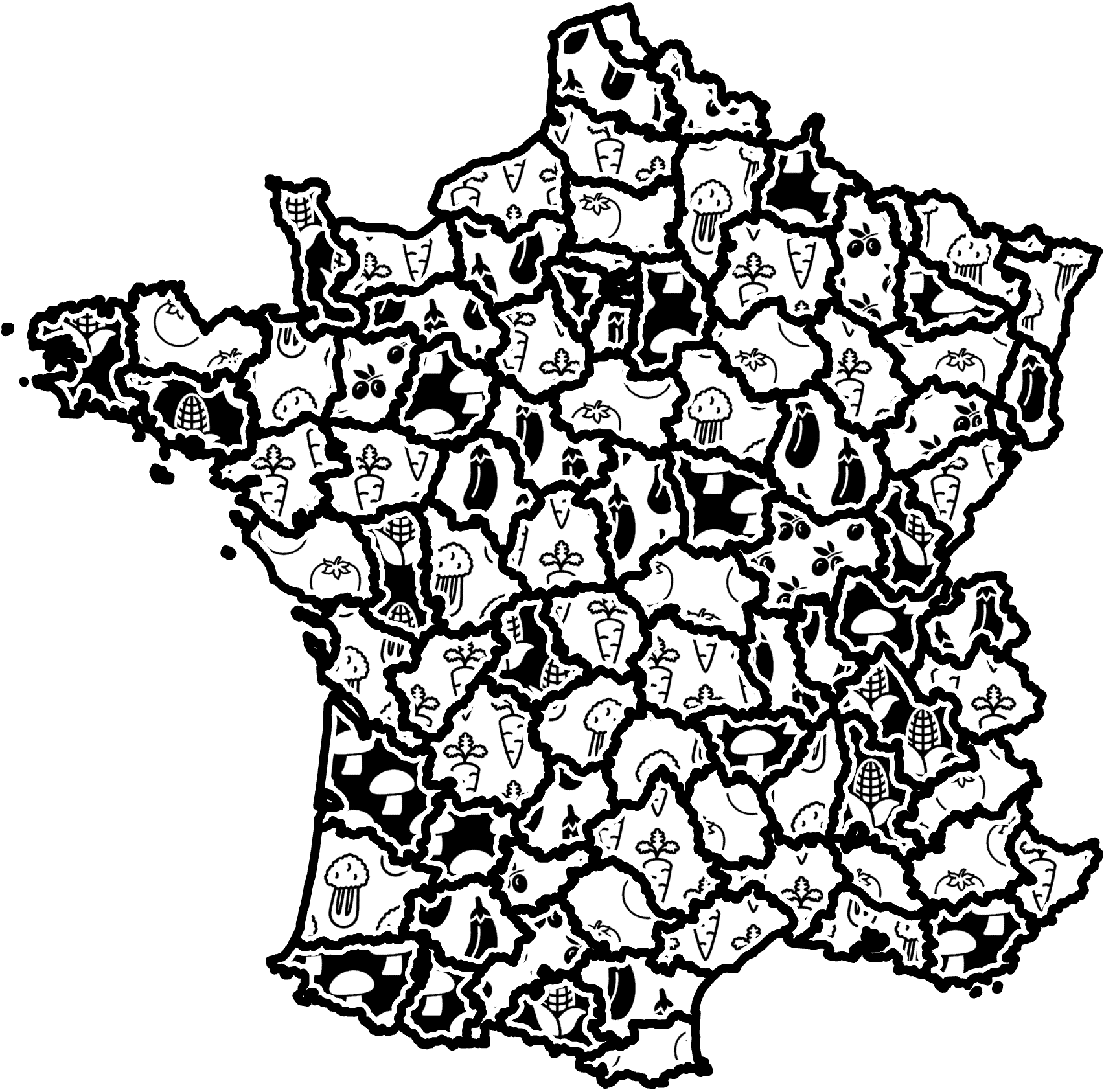}\\
	\caption{An iconic textured map design (MI3) collected in our Experiment 1. This is a larger version of the third image in \autoref{tab:exp2-map-icon}.}
  \label{fig:MI3}
\end{figure}

\begin{figure}[t] % htbp are optional placement specifiers (here, top, bottom, page)
	\centering % Centers the figure
	\includegraphics[width=\appendixfigurewidth]{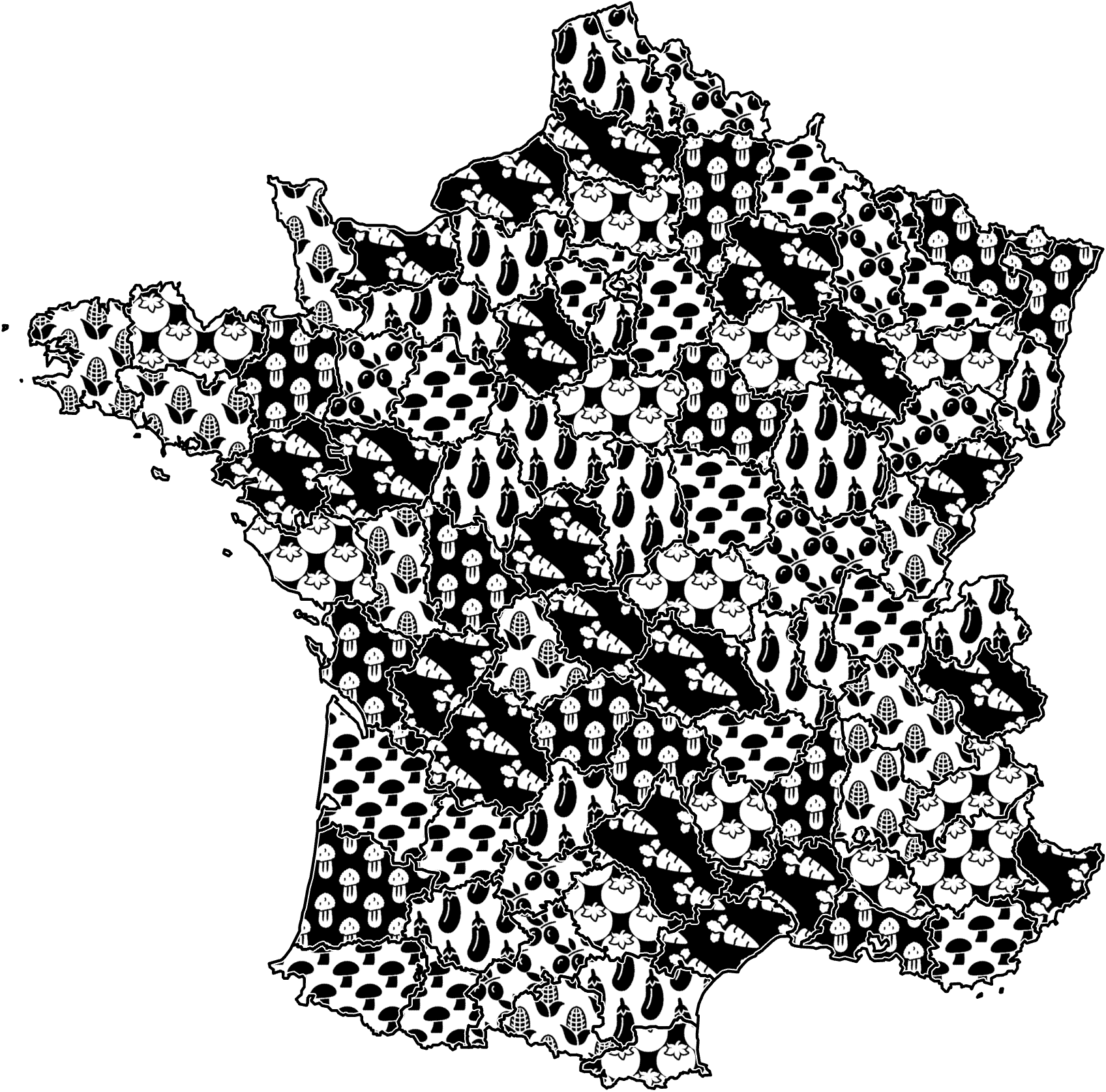}\\
	\caption{An iconic textured map design (MI4) collected in our Experiment 1. This is a larger version of the fourth image in \autoref{tab:exp2-map-icon}.}
  \label{fig:MI4}
\end{figure}

\begin{figure}[t] % htbp are optional placement specifiers (here, top, bottom, page)
	\centering % Centers the figure
	\includegraphics[width=\appendixfigurewidth]{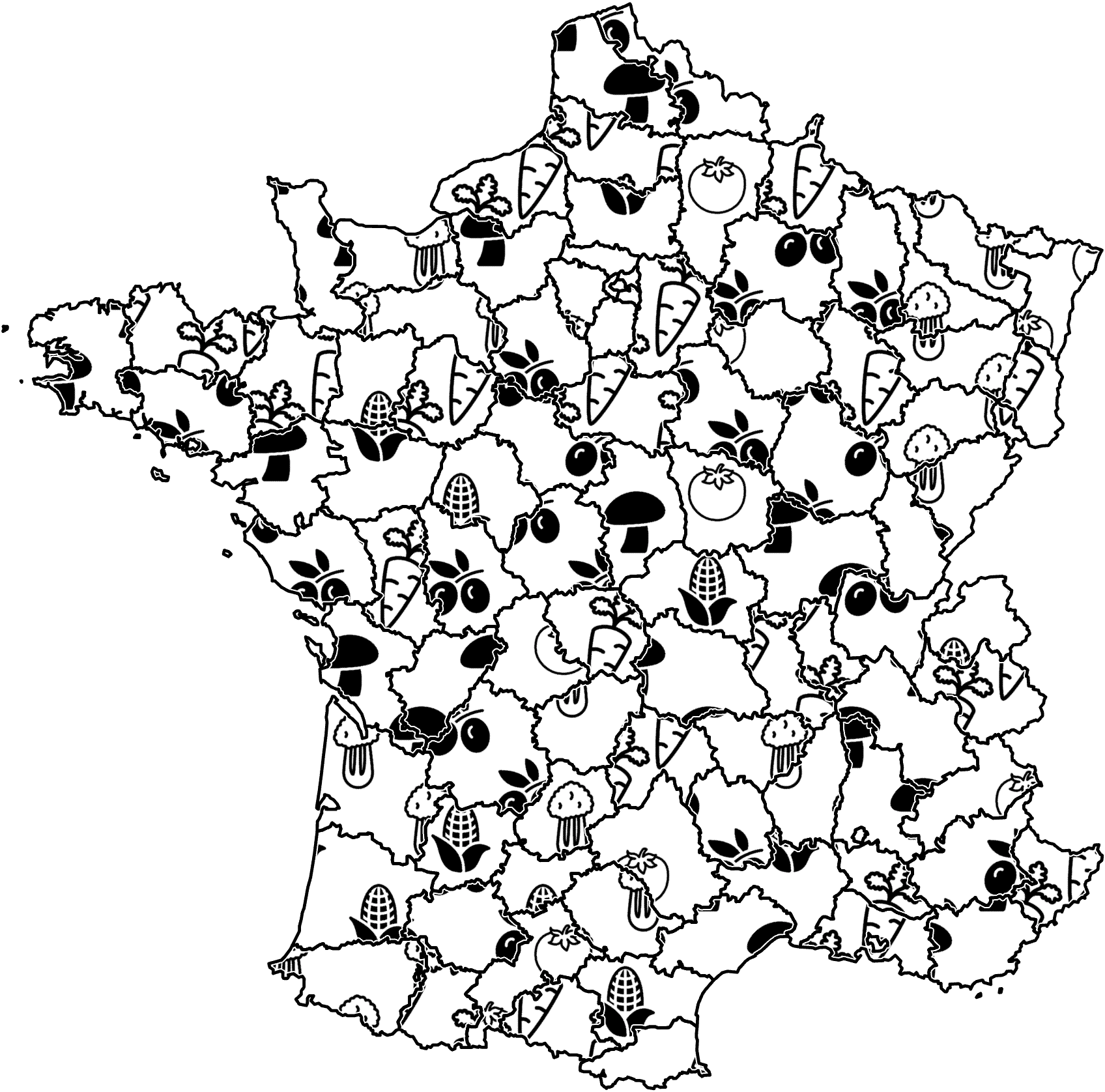}\\
	\caption{An iconic textured map design (MI5) collected in our Experiment 1.}
  \label{fig:MI5}
\end{figure}

\clearpage

\begin{figure}[t] % htbp are optional placement specifiers (here, top, bottom, page)
	\centering % Centers the figure
	\includegraphics[width=\appendixfigurewidth]{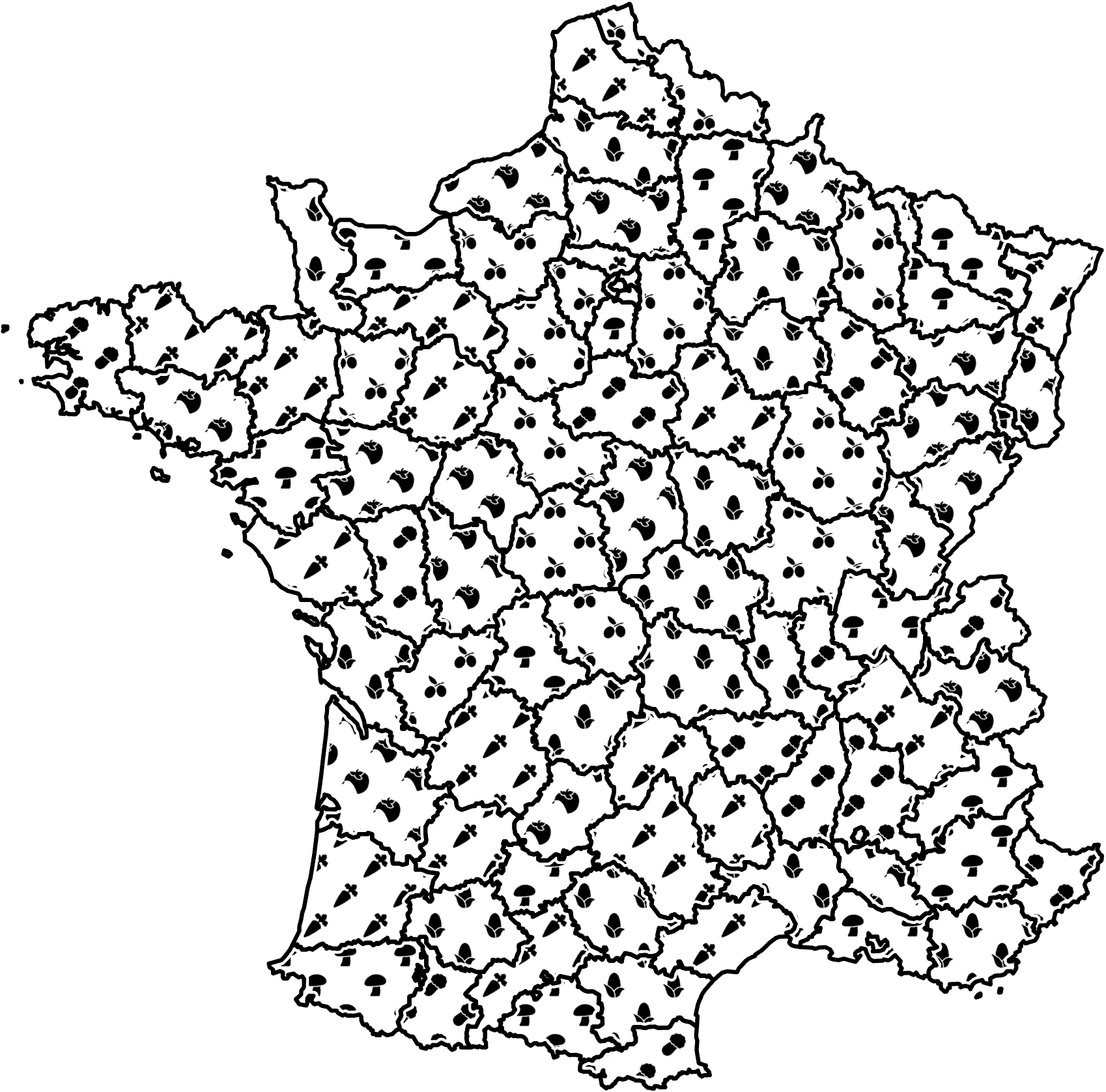}\\
	\caption{An iconic textured map design (MI6) collected in our Experiment 1.}
  \label{fig:MI6}
\end{figure}

\begin{figure}[t] % htbp are optional placement specifiers (here, top, bottom, page)
	\centering % Centers the figure
	\includegraphics[width=\appendixfigurewidth]{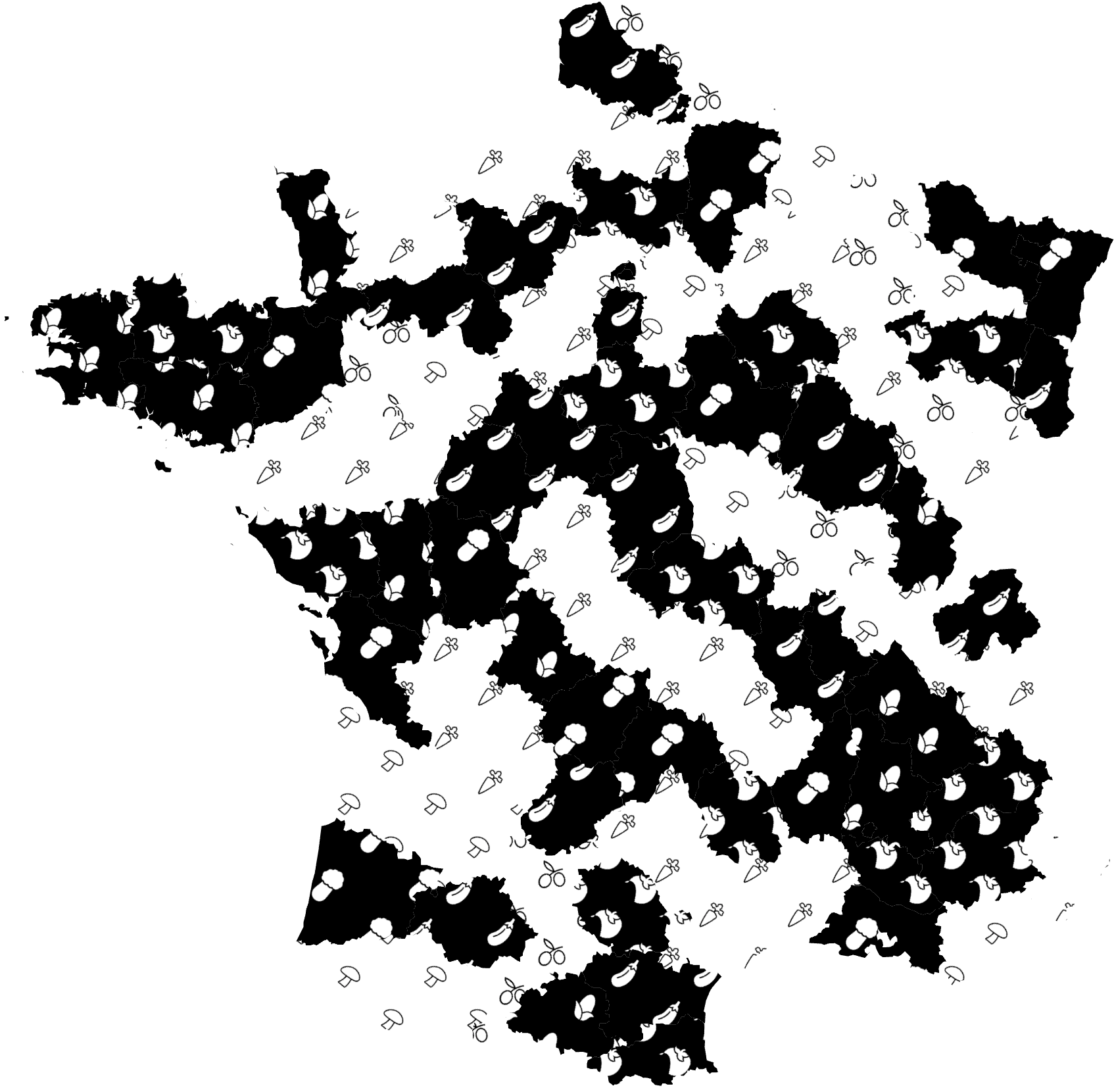}\\
	\caption{An iconic textured map design (MI7) collected in our Experiment 1.}
  \label{fig:MI7}
\end{figure}

\begin{figure}[t] % htbp are optional placement specifiers (here, top, bottom, page)
	\centering % Centers the figure
	\includegraphics[width=\appendixfigurewidth]{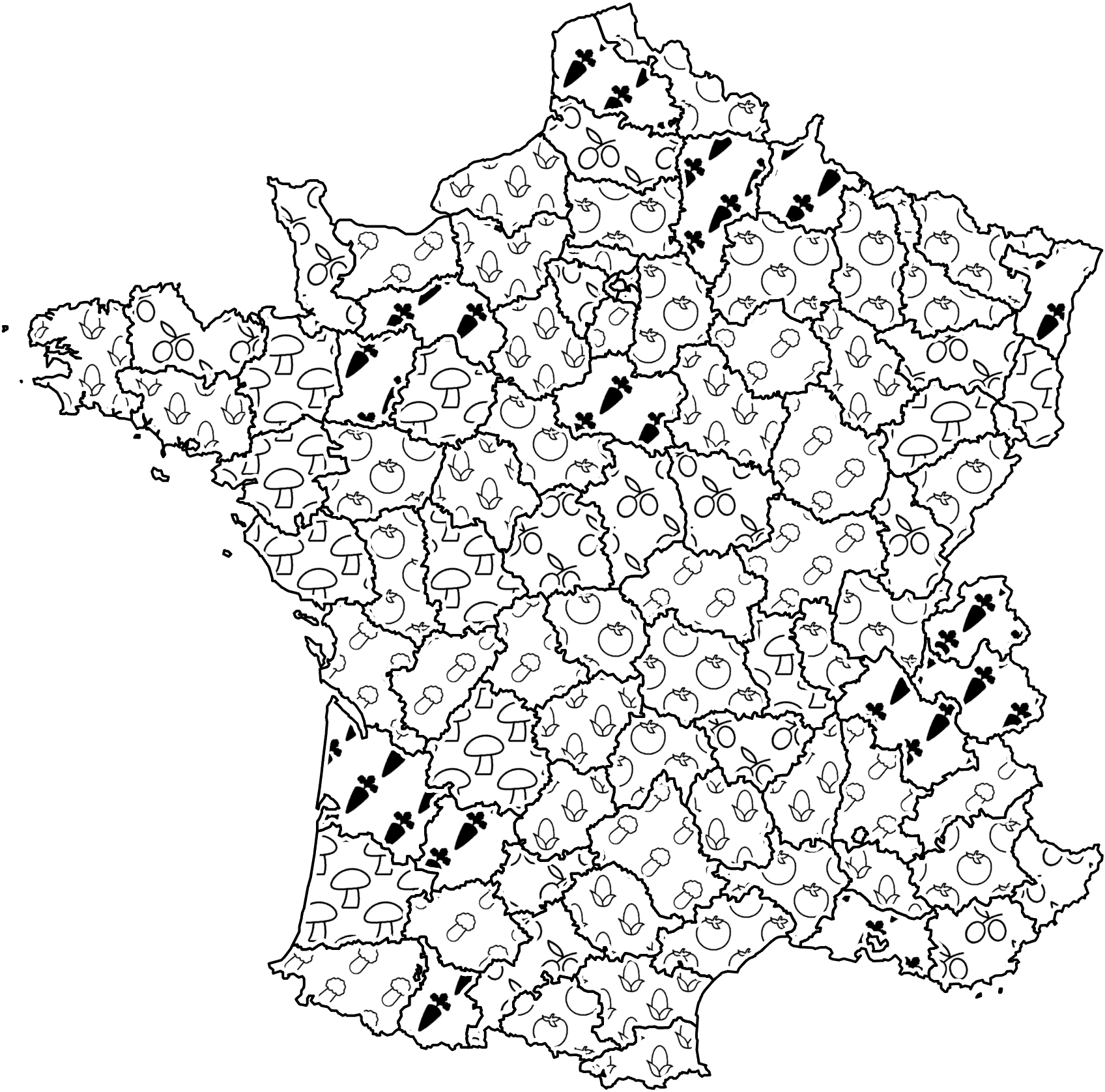}\\
	\caption{An iconic textured map design (MI8) collected in our Experiment 1.}
  \label{fig:MI8}
\end{figure}

\begin{figure}[t] % htbp are optional placement specifiers (here, top, bottom, page)
	\centering % Centers the figure
	\includegraphics[width=\appendixfigurewidth]{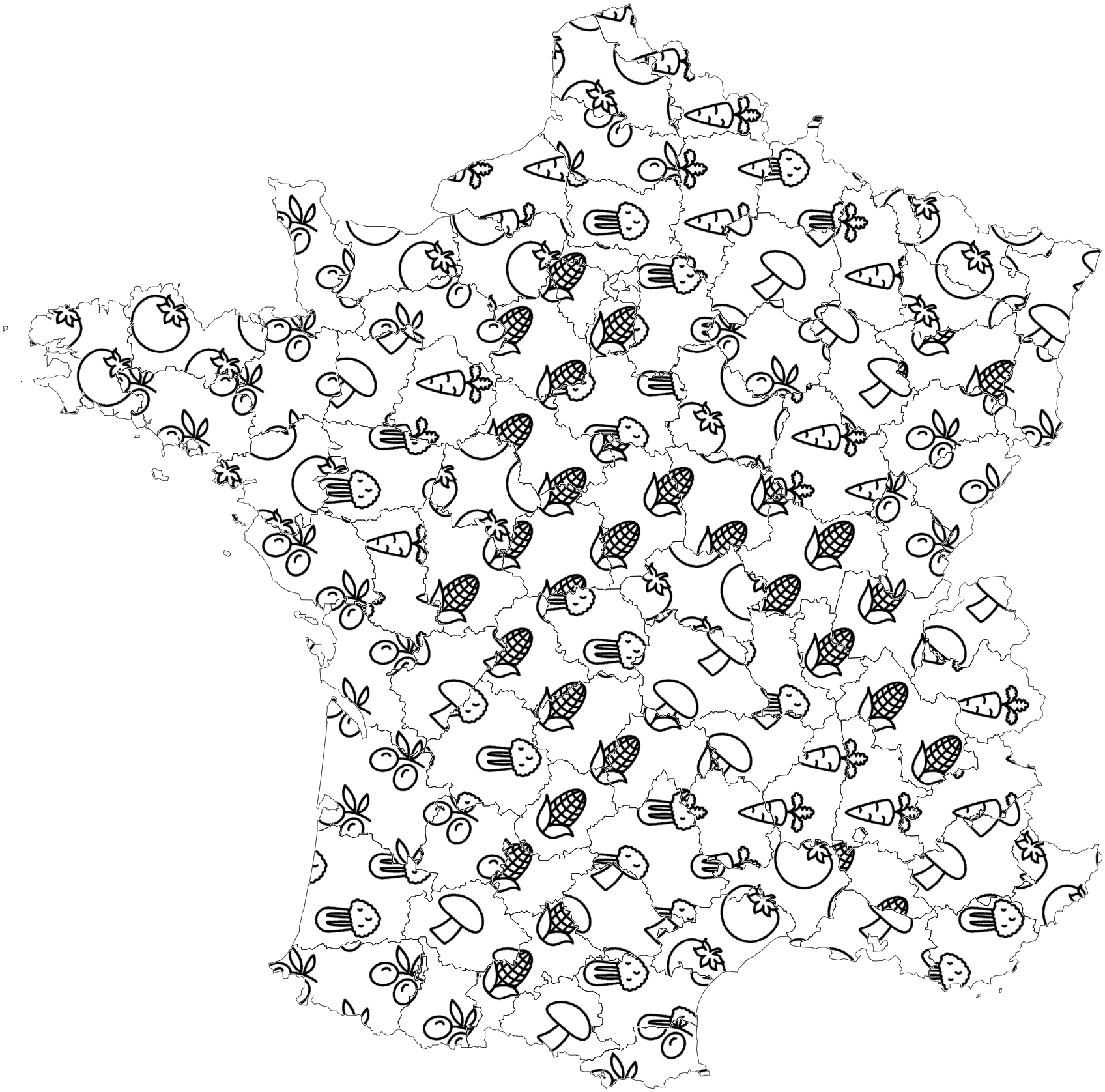}\\
	\caption{An iconic textured map design (MI9) collected in our Experiment 1.}
  \label{fig:MI9}
\end{figure}

\begin{figure}[t] % htbp are optional placement specifiers (here, top, bottom, page)
	\centering % Centers the figure
	\includegraphics[width=\appendixfigurewidth]{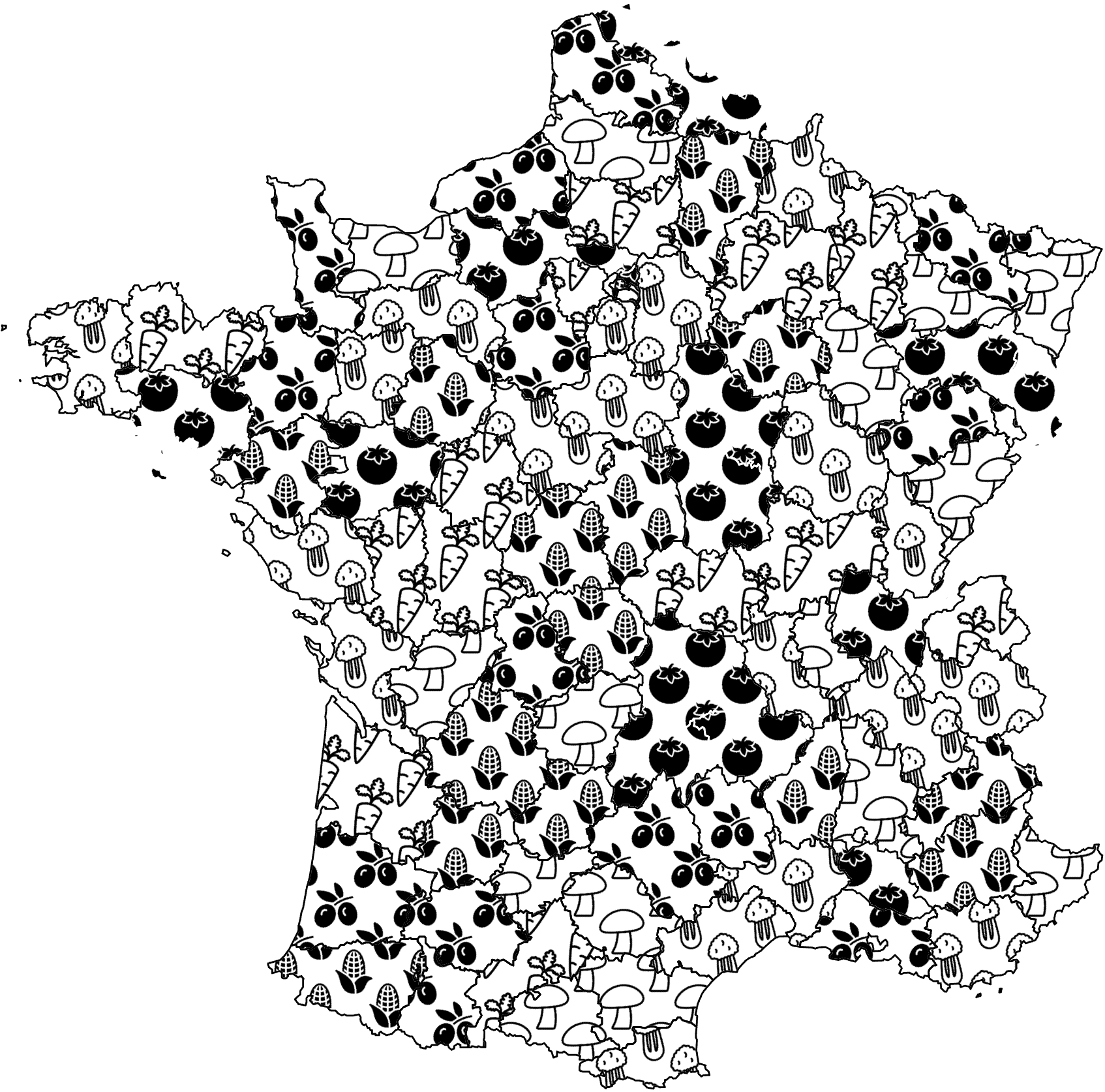}\\
	\caption{An iconic textured map design (MI10) collected in our Experiment 1.}
  \label{fig:MI10}
\end{figure}

\begin{figure}[t] % htbp are optional placement specifiers (here, top, bottom, page)
	\centering % Centers the figure
	\includegraphics[width=\appendixfigurewidth]{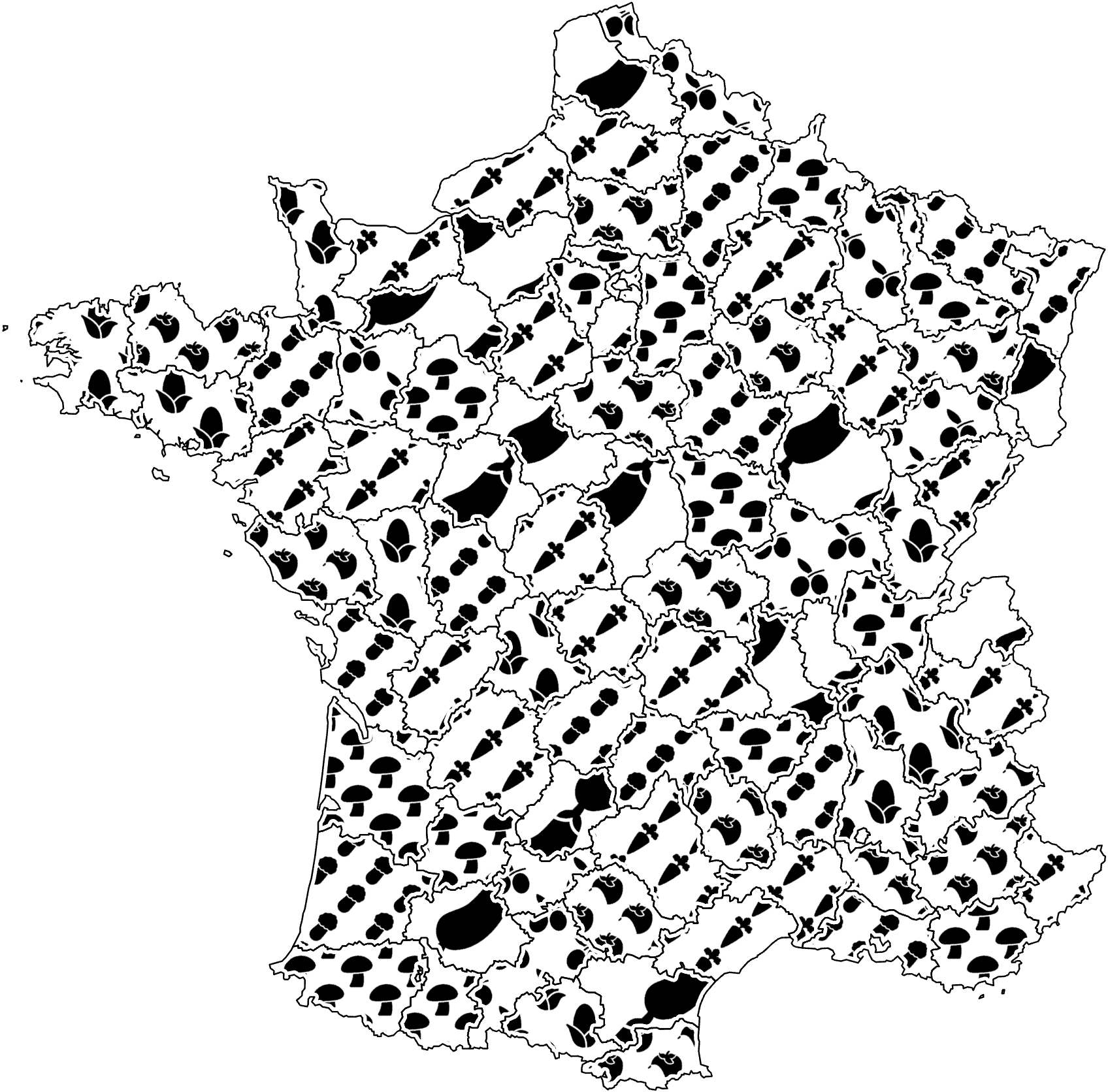}\\
	\caption{An iconic textured map design (MI11) collected in our Experiment 1.}
  \label{fig:MI11}
\end{figure}

%% file: template.bbl
\begin{thebibliography}{10}
\renewcommand*{\sfdefault}{PTSansNarrow-TLF}

\bibitem{datavissociety}
Data visualization society.
\newblock Global non-profit organization for data visualization practitioners
  and enthusiasts, url:
  \href{https://www.datavisualizationsociety.org/}{\texttt{www\discretionary{}{.}{.}datavisualizationsociety\discretionary{}{.}{.}org}}.
\newblock Last accessed: March 2023.

\bibitem{icons8}
Icons8.
\newblock Website and icons database, url:
  \href{https://icons8.com/}{\texttt{icons8.com}}.
\newblock Last accessed: March 2023.

\bibitem{amadasun:1989:textural}
M.~Amadasun and R.~King.
\newblock Textural features corresponding to textural properties.
\newblock {\em IEEE Trans Syst Man Cybern}, 19(5):1264--1274, 1989.
  \href{https://doi.org/10.1109/21.44046}
{doi: \textsf{%
10\hspace{.1pt}\discretionary{.}{%
}{.}\hspace{.4pt}1109\discretionary{/}{%
}{/}21\hspace{.1pt}\discretionary{.}{%
}{.}\hspace{.4pt}44046}}


\bibitem{Barla:2006:SPA}
P.~Barla, S.~Breslav, J.~Thollot, F.~Sillion, and L.~Markosian.
\newblock Stroke pattern analysis and synthesis.
\newblock {\em Comput Graph Forum}, 25(3):663--671, 2006.
  \href{https://doi.org/10.1111/j.1467-8659.2006.00986.x}
{doi: \textsf{%
10\hspace{.1pt}\discretionary{.}{%
}{.}\hspace{.4pt}1111\discretionary{/}{%
}{/}j\hspace{.1pt}\discretionary{.}{%
}{.}\hspace{.4pt}1467\discretionary{%
}{-}{-}8659\hspace{.1pt}\discretionary{.}{%
}{.}\hspace{.4pt}2006\hspace{.1pt}\discretionary{.}{%
}{.}\hspace{.4pt}00986\hspace{.1pt}\discretionary{.}{%
}{.}\hspace{.4pt}x}}


\bibitem{bateman:2010:useful}
S.~Bateman, R.~L. Mandryk, C.~Gutwin, A.~Genest, D.~McDine, and C.~Brooks.
\newblock Useful junk? {T}he effects of visual embellishment on comprehension
  and memorability of charts.
\newblock In {\em Proc.\ CHI}, pp. 2573--2582. ACM, New York, 2010.
  \href{https://doi.org/10.1145/1753326.1753716}
{doi: \textsf{%
10\hspace{.1pt}\discretionary{.}{%
}{.}\hspace{.4pt}1145\discretionary{/}{%
}{/}1753326\hspace{.1pt}\discretionary{.}{%
}{.}\hspace{.4pt}1753716}}


\bibitem{Bertin:1998:SG}
J.~Bertin.
\newblock {\em Sémiologie Graphique}.
\newblock Éd. de l'EHESS, Paris, 3\textsuperscript{rd} ed., 1998.
\newblock url:
  \href{http://editions.ehess.fr/ouvrages/ouvrage/semiologie-graphique/}{\texttt{editions\discretionary{}{.}{.}ehess\discretionary{}{.}{.}fr\discretionary{/}{}{/}ouvrages\discretionary{/}{}{/}ouvrage\discretionary{/}{}{/}semiologie\discretionary{}{-}{-}graphique/}}.

\bibitem{bertin:1983:semiology}
J.~Bertin.
\newblock {\em Semiology of Graphics: Diagrams, Networks, Maps}.
\newblock Esri Press, Redlands, 2011.
\newblock url:
  \href{https://www.esri.com/en-us/esri-press/browse/semiology-of-graphics-diagrams-networks-maps}{\texttt{esri\discretionary{}{.}{.}com/en\discretionary{}{-}{-}us\discretionary{/}{}{/}esri\discretionary{}{-}{-}press\discretionary{/}{}{/}browse\discretionary{/}{}{/}semiology-of\discretionary{}{-}{-}graphics\discretionary{}{-}{-}diagrams\discretionary{}{-}{-}networks\discretionary{}{-}{-}maps}}.

\bibitem{besanccon:2019:continued}
L.~Besan{\c{c}}on and P.~Dragicevic.
\newblock The continued prevalence of dichotomous inferences at {CHI}.
\newblock In {\em CHI Extended Abstracts}, pp. alt14:1--alt14:11. ACM, New
  York, 2019. \href{https://doi.org/10.1145/3290607.3310432}
{doi: \textsf{%
10\hspace{.1pt}\discretionary{.}{%
}{.}\hspace{.4pt}1145\discretionary{/}{%
}{/}3290607\hspace{.1pt}\discretionary{.}{%
}{.}\hspace{.4pt}3310432}}


\bibitem{blascheck:2023:studies}
T.~Blascheck, L.~Besan{\c{c}}on, A.~Bezerianos, B.~Lee, A.~Islam, T.~He, and
  P.~Isenberg.
\newblock Studies of part-to-whole glanceable visualizations on smartwatch
  faces.
\newblock In {\em Proc.\ PacificVis}, pp. 187--196. IEEE Comp.\ Soc., Los
  Alamitos, 2023. \href{https://doi.org/10.1109/PacificVis56936.2023.00028}
{doi: \textsf{%
10\hspace{.1pt}\discretionary{.}{%
}{.}\hspace{.4pt}1109\discretionary{/}{%
}{/}PacificVis56936\hspace{.1pt}\discretionary{.}{%
}{.}\hspace{.4pt}2023\hspace{.1pt}\discretionary{.}{%
}{.}\hspace{.4pt}00028}}


\bibitem{borgo:2013:glyph}
R.~Borgo, J.~Kehrer, D.~H. Chung, E.~Maguire, R.~S. Laramee, H.~Hauser,
  M.~Ward, and M.~Chen.
\newblock Glyph-based visualization: Foundations, design guidelines, techniques
  and applications.
\newblock In {\em Eurographics State of the Art Reports}, pp. 39--63. EG
  Assoc., Goslar, 2013.
  \href{https://doi.org/10.2312/conf/EG2013/stars/039-063}
{doi: \textsf{%
10\hspace{.1pt}\discretionary{.}{%
}{.}\hspace{.4pt}2312\discretionary{/}{%
}{/}conf\discretionary{/}{%
}{/}EG2013\discretionary{/}{%
}{/}stars\discretionary{/}{%
}{/}039\discretionary{%
}{-}{-}063}}


\bibitem{Borkin:2016:BMV}
M.~A. Borkin, Z.~Bylinskii, N.~W. Kim, C.~M. Bainbridge, C.~S. Yeh, D.~Borkin,
  H.~Pfister, and A.~Oliva.
\newblock Beyond memorability: Visualization recognition and recall.
\newblock {\em IEEE Trans Vis Comput Graph}, 22(1):519--528, 2016.
  \href{https://doi.org/10.1109/TVCG.2015.2467732}
{doi: \textsf{%
10\hspace{.1pt}\discretionary{.}{%
}{.}\hspace{.4pt}1109\discretionary{/}{%
}{/}TVCG\hspace{.1pt}\discretionary{.}{%
}{.}\hspace{.4pt}2015\hspace{.1pt}\discretionary{.}{%
}{.}\hspace{.4pt}2467732}}


\bibitem{Borkin:2013:WMV}
M.~A. Borkin, A.~A. Vo, Z.~Bylinskii, P.~Isola, S.~Sunkavalli, A.~Oliva, and
  H.~Pfister.
\newblock What makes a visualization memorable?
\newblock {\em IEEE Trans Vis Comput Graph}, 19(12):2306--2315, 2013.
  \href{https://doi.org/10.1109/TVCG.2013.234}
{doi: \textsf{%
10\hspace{.1pt}\discretionary{.}{%
}{.}\hspace{.4pt}1109\discretionary{/}{%
}{/}TVCG\hspace{.1pt}\discretionary{.}{%
}{.}\hspace{.4pt}2013\hspace{.1pt}\discretionary{.}{%
}{.}\hspace{.4pt}234}}


\bibitem{Borland:2007:RCM}
D.~Borland and R.~M. Taylor, II.
\newblock Rainbow color map (still) considered harmful.
\newblock {\em IEEE Comput Graph Appl}, 27(2):14--17, 2007.
  \href{https://doi.org/10.1109/MCG.2007.323435}
{doi: \textsf{%
10\hspace{.1pt}\discretionary{.}{%
}{.}\hspace{.4pt}1109\discretionary{/}{%
}{/}MCG\hspace{.1pt}\discretionary{.}{%
}{.}\hspace{.4pt}2007\hspace{.1pt}\discretionary{.}{%
}{.}\hspace{.4pt}323435}}


\bibitem{brinton:1919:graphic}
W.~C. Brinton.
\newblock {\em Graphic Methods for Presenting Facts}.
\newblock The Engineering Magazine Company, New York, 1914.
\newblock
  \href{https://archive.org/search.php?query=external-identifier:"urn:oclc:record:1045528209"}{urn:
  \textsf{urn:oclc:record:1045528209}}.

\bibitem{Brinton:1939:GP}
W.~C. Brinton.
\newblock {\em Graphic Presentation}.
\newblock Brinton Assoc., New York, 1939.
\newblock
  \href{https://archive.org/search.php?query=external-identifier:"urn:oclc:record:1045601113"}{urn:
  \textsf{urn:oclc:record:1045601113}}.

\bibitem{brodatz:1966:textures}
P.~Brodatz.
\newblock {\em Textures: A Photographic Album for Artists and Designers},
  vol.~2.
\newblock Dover Publications, New York, 1966.

\bibitem{burns:2021:designing}
A.~Burns, C.~Xiong, S.~Franconeri, A.~Cairo, and N.~Mahyar.
\newblock Designing with pictographs: Envision topics without sacrificing
  understanding.
\newblock {\em IEEE Trans Vis Comput Graph}, 28(12):4515--4530, 2022.
  \href{https://doi.org/10.1109/TVCG.2021.3092680}
{doi: \textsf{%
10\hspace{.1pt}\discretionary{.}{%
}{.}\hspace{.4pt}1109\discretionary{/}{%
}{/}TVCG\hspace{.1pt}\discretionary{.}{%
}{.}\hspace{.4pt}2021\hspace{.1pt}\discretionary{.}{%
}{.}\hspace{.4pt}3092680}}


\bibitem{chen:2013:analysis}
M.~Chen and L.~Floridi.
\newblock An analysis of information visualisation.
\newblock {\em Synthese}, 190(16):3421--3438, 2013.
  \href{https://doi.org/10.1007/s11229-012-0183-y}
{doi: \textsf{%
10\hspace{.1pt}\discretionary{.}{%
}{.}\hspace{.4pt}1007\discretionary{/}{%
}{/}s11229\discretionary{%
}{-}{-}012\discretionary{%
}{-}{-}0183\discretionary{%
}{-}{-}y}}


\bibitem{cho:2000:reliability}
R.~Y. Cho, V.~Yang, and P.~E. Hallett.
\newblock Reliability and dimensionality of judgments of visually textured
  materials.
\newblock {\em Percept Psychophys}, 62(4):735--752, 2000.
  \href{https://doi.org/10.3758/BF03206920}
{doi: \textsf{%
10\hspace{.1pt}\discretionary{.}{%
}{.}\hspace{.4pt}3758\discretionary{/}{%
}{/}BF03206920}}


\bibitem{cleveland:1985:graphical}
W.~S. Cleveland and R.~McGill.
\newblock Graphical perception and graphical methods for analyzing scientific
  data.
\newblock {\em Science}, 229(4716):828--833, 1985.
  \href{https://doi.org/10.1126/science.229.4716.828}
{doi: \textsf{%
10\hspace{.1pt}\discretionary{.}{%
}{.}\hspace{.4pt}1126\discretionary{/}{%
}{/}science\hspace{.1pt}\discretionary{.}{%
}{.}\hspace{.4pt}229\hspace{.1pt}\discretionary{.}{%
}{.}\hspace{.4pt}4716\hspace{.1pt}\discretionary{.}{%
}{.}\hspace{.4pt}828}}


\bibitem{cockburn:2020:threats}
A.~Cockburn, P.~Dragicevic, L.~Besan\c{c}on, and C.~Gutwin.
\newblock Threats of a replication crisis in empirical computer science.
\newblock {\em Commun. ACM}, 63(8):70–79, jul 2020.
  \href{https://doi.org/10.1145/3360311}
{doi: \textsf{%
10\hspace{.1pt}\discretionary{.}{%
}{.}\hspace{.4pt}1145\discretionary{/}{%
}{/}3360311}}


\bibitem{cumming:2013:understanding}
G.~Cumming.
\newblock {\em Understanding the New Statistics: Effect Sizes, Confidence
  Intervals, and Meta-Analysis}.
\newblock Routledge, New York, 2012.
  \href{https://doi.org/10.4324/9780203807002}
{doi: \textsf{%
10\hspace{.1pt}\discretionary{.}{%
}{.}\hspace{.4pt}4324\discretionary{/}{%
}{/}9780203807002}}


\bibitem{dragicevic:2016:fair}
P.~Dragicevic.
\newblock Fair statistical communication in {HCI}.
\newblock In J.~Robertson and M.~Kaptein, eds., {\em Modern Statistical Methods
  for HCI}, chap.~13, pp. 291--330. Springer, Cham, 2016.
  \href{https://doi.org/10.1007/978-3-319-26633-6_13}
{doi: \textsf{%
10\hspace{.1pt}\discretionary{.}{%
}{.}\hspace{.4pt}1007\discretionary{/}{%
}{/}978\discretionary{%
}{-}{-}3\discretionary{%
}{-}{-}319\discretionary{%
}{-}{-}26633\discretionary{%
}{-}{-}6\_13}}


\bibitem{Goffin:2014:ETP}
P.~Goffin, W.~Willett, J.-D. Fekete, and P.~Isenberg.
\newblock Exploring the placement and design of word-scale visualizations.
\newblock {\em IEEE Trans Vis Comput Graph}, 20(12):2291--2300, 2014.
  \href{https://doi.org/10.1109/TVCG.2014.2346435}
{doi: \textsf{%
10\hspace{.1pt}\discretionary{.}{%
}{.}\hspace{.4pt}1109\discretionary{/}{%
}{/}TVCG\hspace{.1pt}\discretionary{.}{%
}{.}\hspace{.4pt}2014\hspace{.1pt}\discretionary{.}{%
}{.}\hspace{.4pt}2346435}}


\bibitem{haralick:1979:statistical}
R.~M. Haralick.
\newblock Statistical and structural approaches to texture.
\newblock {\em Proc IEEE}, 67(5):786--804, 1979.
  \href{https://doi.org/10.1109/PROC.1979.11328}
{doi: \textsf{%
10\hspace{.1pt}\discretionary{.}{%
}{.}\hspace{.4pt}1109\discretionary{/}{%
}{/}PROC\hspace{.1pt}\discretionary{.}{%
}{.}\hspace{.4pt}1979\hspace{.1pt}\discretionary{.}{%
}{.}\hspace{.4pt}11328}}


\bibitem{haroz:2015:isotype}
S.~Haroz, R.~Kosara, and S.~L. Franconeri.
\newblock Isotype visualization: Working memory, performance, and engagement
  with pictographs.
\newblock In {\em Proc.\ CHI}, pp. 1191--1200. ACM, New York, 2015.
  \href{https://doi.org/10.1145/2702123.2702275}
{doi: \textsf{%
10\hspace{.1pt}\discretionary{.}{%
}{.}\hspace{.4pt}1145\discretionary{/}{%
}{/}2702123\hspace{.1pt}\discretionary{.}{%
}{.}\hspace{.4pt}2702275}}


\bibitem{hawkins:1970:textural}
J.~K. Hawkins.
\newblock Textural properties for pattern recognition.
\newblock In B.~C. Lipkin and A.~Rosenfeld, eds., {\em Picture Processing and
  Psychopictorics}, pp. 347--370. Academic Press, New York, 1970.
\newblock url:
  \href{https://books.google.com/books?id=vp-w_pC9JBAC&pg=PA347}{\texttt{books\discretionary{}{.}{.}google\discretionary{}{.}{.}com\discretionary{/}{}{/}books\discretionary{?}{}{?}id\discretionary{=}{}{=}vp-w\_pC9JBAC\discretionary{\&}{}{\&}pg\discretionary{=}{}{=}PA347}}.

\bibitem{He:2023:BVS}
T.~He, P.~Isenberg, R.~Dachselt, and T.~Isenberg.
\newblock {B}eau{V}is: A validated scale for measuring the aesthetic pleasure
  of visual representations.
\newblock {\em IEEE Trans Vis Comput Graph}, 29(1):363--373, 2023.
  \href{https://doi.org/10.1109/TVCG.2022.3209390}
{doi: \textsf{%
10\hspace{.1pt}\discretionary{.}{%
}{.}\hspace{.4pt}1109\discretionary{/}{%
}{/}TVCG\hspace{.1pt}\discretionary{.}{%
}{.}\hspace{.4pt}2022\hspace{.1pt}\discretionary{.}{%
}{.}\hspace{.4pt}3209390}}


\bibitem{healey:1998:building}
C.~G. Healey and J.~T. Enns.
\newblock Building perceptual textures to visualize multidimensional datasets.
\newblock In {\em Proc.\ Visualization}, pp. 111--118. IEEE Comp.\ Soc., Los
  Alamitos, 1998. \href{https://doi.org/10.1109/VISUAL.1998.745292}
{doi: \textsf{%
10\hspace{.1pt}\discretionary{.}{%
}{.}\hspace{.4pt}1109\discretionary{/}{%
}{/}VISUAL\hspace{.1pt}\discretionary{.}{%
}{.}\hspace{.4pt}1998\hspace{.1pt}\discretionary{.}{%
}{.}\hspace{.4pt}745292}}


\bibitem{higgins:2004:introduction}
J.~J. Higgins.
\newblock {\em An Introduction to Modern Nonparametric Statistics}.
\newblock Brooks/Cole, Pacific Grove, 2004.

\bibitem{Hutchinson:2003:TechProbe}
H.~Hutchinson, W.~Mackay, B.~Westerlund, B.~B. Bederson, A.~Druin, C.~Plaisant,
  M.~Beaudouin-Lafon, S.~Conversy, H.~Evans, H.~Hansen, N.~Roussel, and
  B.~Eiderb\"{a}ck.
\newblock Technology probes: Inspiring design for and with families.
\newblock In {\em Proc.\ CHI}, pp. 17--24. ACM, New York, 2003.
  \href{https://doi.org/10.1145/642611.642616}
{doi: \textsf{%
10\hspace{.1pt}\discretionary{.}{%
}{.}\hspace{.4pt}1145\discretionary{/}{%
}{/}642611\hspace{.1pt}\discretionary{.}{%
}{.}\hspace{.4pt}642616}}


\bibitem{julesz:1962:visual}
B.~Julesz.
\newblock Visual pattern discrimination.
\newblock {\em IRE Trans Inf Theory}, 8(2):84--92, 1962.
  \href{https://doi.org/10.1109/TIT.1962.1057698}
{doi: \textsf{%
10\hspace{.1pt}\discretionary{.}{%
}{.}\hspace{.4pt}1109\discretionary{/}{%
}{/}TIT\hspace{.1pt}\discretionary{.}{%
}{.}\hspace{.4pt}1962\hspace{.1pt}\discretionary{.}{%
}{.}\hspace{.4pt}1057698}}


\bibitem{julesz:1975:experiments}
B.~Julesz.
\newblock Experiments in the visual perception of texture.
\newblock {\em Sci Am}, 232(4):34--43, 1975.
  \href{https://doi.org/10.1038/scientificamerican0475-34}
{doi: \textsf{%
10\hspace{.1pt}\discretionary{.}{%
}{.}\hspace{.4pt}1038\discretionary{/}{%
}{/}scientificamerican0475\discretionary{%
}{-}{-}34}}


\bibitem{julesz:1981:textons}
B.~Julesz.
\newblock Textons, the elements of texture perception, and their interactions.
\newblock {\em Nature}, 290(5802):91--97, 1981.
  \href{https://doi.org/10.1038/290091a0}
{doi: \textsf{%
10\hspace{.1pt}\discretionary{.}{%
}{.}\hspace{.4pt}1038\discretionary{/}{%
}{/}290091a0}}


\bibitem{julesz:1981:theory}
B.~Julesz.
\newblock A theory of preattentive texture discrimination based on first-order
  statistics of textons.
\newblock {\em Biol Cybern}, 41(2):131--138, 1981.
  \href{https://doi.org/10.1007/BF00335367}
{doi: \textsf{%
10\hspace{.1pt}\discretionary{.}{%
}{.}\hspace{.4pt}1007\discretionary{/}{%
}{/}BF00335367}}


\bibitem{julesz:1983:human}
B.~Julesz and J.~R. Bergen.
\newblock Human factors and behavioral science: Textons, the fundamental
  elements in preattentive vision and perception of textures.
\newblock {\em Bell Syst Tech J}, 62(6):1619--1645, 1983.
  \href{https://doi.org/10.1002/j.1538-7305.1983.tb03502.x}
{doi: \textsf{%
10\hspace{.1pt}\discretionary{.}{%
}{.}\hspace{.4pt}1002\discretionary{/}{%
}{/}j\hspace{.1pt}\discretionary{.}{%
}{.}\hspace{.4pt}1538\discretionary{%
}{-}{-}7305\hspace{.1pt}\discretionary{.}{%
}{.}\hspace{.4pt}1983\hspace{.1pt}\discretionary{.}{%
}{.}\hspace{.4pt}tb03502\hspace{.1pt}\discretionary{.}{%
}{.}\hspace{.4pt}x}}


\bibitem{julesz:1973:inability}
B.~Julesz, E.~N. Gilbert, L.~A. Shepp, and H.~L. Frisch.
\newblock Inability of hu\-mans to discriminate between visual textures that
  agree in se\-cond-or\-der statistics---{R}evisited.
\newblock {\em Percept}, 2(4):391--405, 1973.
  \href{https://doi.org/10.1068/p020391}
{doi: \textsf{%
10\hspace{.1pt}\discretionary{.}{%
}{.}\hspace{.4pt}1068\discretionary{/}{%
}{/}p020391}}


\bibitem{kosslyn:2006:graph}
S.~M. Kosslyn.
\newblock {\em Graph Design for the Eye and Mind}.
\newblock Oxford University Press, 2006.
  \href{https://doi.org/10.1093/acprof:oso/9780195311846.001.0001}
{doi: \textsf{%
10\hspace{.1pt}\discretionary{.}{%
}{.}\hspace{.4pt}1093\discretionary{/}{%
}{/}acprof\discretionary{:}{%
}{:}oso\discretionary{/}{%
}{/}9780195311846\hspace{.1pt}\discretionary{.}{%
}{.}\hspace{.4pt}001\hspace{.1pt}\discretionary{.}{%
}{.}\hspace{.4pt}0001}}


\bibitem{lin:2013:selecting}
S.~Lin, J.~Fortuna, C.~Kulkarni, M.~Stone, and J.~Heer.
\newblock Selecting se\-man\-ti\-cal\-ly-resonant colors for data
  visualization.
\newblock {\em Comput Graph Forum}, 32(3):401--410, 2013.
  \href{https://doi.org/10.1111/cgf.12127}
{doi: \textsf{%
10\hspace{.1pt}\discretionary{.}{%
}{.}\hspace{.4pt}1111\discretionary{/}{%
}{/}cgf\hspace{.1pt}\discretionary{.}{%
}{.}\hspace{.4pt}12127}}


\bibitem{liu:1996:periodicity}
F.~Liu and R.~W. Picard.
\newblock Periodicity, directionality, and randomness: Wold features for image
  modeling and retrieval.
\newblock {\em IEEE Trans Pattern Anal Mach Intell}, 18(7):722--733, 1996.
  \href{https://doi.org/10.1109/34.506794}
{doi: \textsf{%
10\hspace{.1pt}\discretionary{.}{%
}{.}\hspace{.4pt}1109\discretionary{/}{%
}{/}34\hspace{.1pt}\discretionary{.}{%
}{.}\hspace{.4pt}506794}}


\bibitem{mackinlay:1986:automating}
J.~Mackinlay.
\newblock Automating the design of graphical presentations of relational
  information.
\newblock {\em ACM Trans Graph}, 5(2):110--141, 1986.
  \href{https://doi.org/10.1145/22949.22950}
{doi: \textsf{%
10\hspace{.1pt}\discretionary{.}{%
}{.}\hspace{.4pt}1145\discretionary{/}{%
}{/}22949\hspace{.1pt}\discretionary{.}{%
}{.}\hspace{.4pt}22950}}


\bibitem{Martin:2017:SDS}
D.~Mart{\'i}n, G.~Arroyo, A.~Rodr{\'i}guez, and T.~Isenberg.
\newblock A survey of digital stippling.
\newblock {\em Comput Graph}, 67:24--44, Oct. 2017.
  \href{https://doi.org/10.1016/j.cag.2017.05.001}
{doi: \textsf{%
10\hspace{.1pt}\discretionary{.}{%
}{.}\hspace{.4pt}1016\discretionary{/}{%
}{/}j\hspace{.1pt}\discretionary{.}{%
}{.}\hspace{.4pt}cag\hspace{.1pt}\discretionary{.}{%
}{.}\hspace{.4pt}2017\hspace{.1pt}\discretionary{.}{%
}{.}\hspace{.4pt}05\hspace{.1pt}\discretionary{.}{%
}{.}\hspace{.4pt}001}}


\bibitem{morais:2020:showing}
L.~Morais, Y.~Jansen, N.~Andrade, and P.~Dragicevic.
\newblock Showing data about people: A design space of anthropographics.
\newblock {\em IEEE Trans Vis Comput Graph}, 28(3):1661--1679, 2020.
  \href{https://doi.org/10.1109/TVCG.2020.3023013}
{doi: \textsf{%
10\hspace{.1pt}\discretionary{.}{%
}{.}\hspace{.4pt}1109\discretionary{/}{%
}{/}TVCG\hspace{.1pt}\discretionary{.}{%
}{.}\hspace{.4pt}2020\hspace{.1pt}\discretionary{.}{%
}{.}\hspace{.4pt}3023013}}


\bibitem{neurath:2010:hieroglyphics}
O.~Neurath.
\newblock {\em From Hieroglyphics to Isotype: A Visual Autobiography}.
\newblock Hyphen Press, London, 2010.
\newblock Edited by M. Eve and C. Burke, url:
  \href{https://perpensapress.com/books/from-hieroglyphics-to-isotype/}{\texttt{perpensapress\discretionary{}{.}{.}com\discretionary{/}{}{/}books\discretionary{/}{}{/}from\discretionary{}{-}{-}hieroglyphics\discretionary{}{-}{-}to\discretionary{}{-}{-}isotype}}.

\bibitem{rao:1996:towards}
A.~R. Rao and G.~L. Lohse.
\newblock Towards a texture naming system: Identifying relevant dimensions of
  texture.
\newblock {\em Vision Res}, 36(11):1649--1669, 1996.
  \href{https://doi.org/10.1016/0042-6989(95)00202-2}
{doi: \textsf{%
10\hspace{.1pt}\discretionary{.}{%
}{.}\hspace{.4pt}1016\discretionary{/}{%
}{/}0042\discretionary{%
}{-}{-}6989\discretionary{%
}{(}{(}95\discretionary{)}{%
}{)}00202\discretionary{%
}{-}{-}2}}


\bibitem{salisbury:1994:interactive}
M.~P. Salisbury, S.~E. Anderson, R.~Barzel, and D.~H. Salesin.
\newblock Interactive pen-and-ink illustration.
\newblock In {\em Proc.\ SIGGRAPH}, pp. 101--108. ACM, New York, 1994.
  \href{https://doi.org/10.1145/192161.192185}
{doi: \textsf{%
10\hspace{.1pt}\discretionary{.}{%
}{.}\hspace{.4pt}1145\discretionary{/}{%
}{/}192161\hspace{.1pt}\discretionary{.}{%
}{.}\hspace{.4pt}192185}}


\bibitem{shi:2022:supporting}
Y.~Shi, P.~Liu, S.~Chen, M.~Sun, and N.~Cao.
\newblock Supporting expressive and faithful pictorial visualization design
  with visual style transfer.
\newblock {\em IEEE Trans Vis Comput Graph}, 29(1):236--246, 2023.
  \href{https://doi.org/10.1109/TVCG.2022.3209486}
{doi: \textsf{%
10\hspace{.1pt}\discretionary{.}{%
}{.}\hspace{.4pt}1109\discretionary{/}{%
}{/}TVCG\hspace{.1pt}\discretionary{.}{%
}{.}\hspace{.4pt}2022\hspace{.1pt}\discretionary{.}{%
}{.}\hspace{.4pt}3209486}}


\bibitem{tamura:1978:textural}
H.~Tamura, S.~Mori, and T.~Yamawaki.
\newblock Textural features corresponding to visual perception.
\newblock {\em IEEE Trans Syst Man Cybern}, 8(6):460--473, 1978.
  \href{https://doi.org/10.1109/TSMC.1978.4309999}
{doi: \textsf{%
10\hspace{.1pt}\discretionary{.}{%
}{.}\hspace{.4pt}1109\discretionary{/}{%
}{/}TSMC\hspace{.1pt}\discretionary{.}{%
}{.}\hspace{.4pt}1978\hspace{.1pt}\discretionary{.}{%
}{.}\hspace{.4pt}4309999}}


\bibitem{tufte:1985:visual}
E.~R. Tufte.
\newblock {\em The Visual Display of Quantitative Information}.
\newblock Graphics Press, Cheshire, 2\textsuperscript{nd} ed., 2001.
\newblock url:
  \href{https://www.edwardtufte.com/tufte/books_vdqi}{\texttt{edwardtufte\discretionary{}{.}{.}com\discretionary{/}{}{/}tufte\discretionary{/}{}{/}books\_vdqi}}.

\bibitem{ware:2019:information}
C.~Ware.
\newblock {\em Information Visualization: Perception for Design}.
\newblock Elsevier, 2019. \href{https://doi.org/10.1016/C2016-0-02395-1}
{doi: \textsf{%
10\hspace{.1pt}\discretionary{.}{%
}{.}\hspace{.4pt}1016\discretionary{/}{%
}{/}C2016\discretionary{%
}{-}{-}0\discretionary{%
}{-}{-}02395\discretionary{%
}{-}{-}1}}


\bibitem{ware:1992:orderable}
C.~Ware and W.~Knight.
\newblock Orderable dimensions of visual texture for data display: Orientation,
  size and contrast.
\newblock In {\em Proc.\ CHI}, pp. 203--209. ACM, New York, 1992.
  \href{https://doi.org/10.1145/142750.142791}
{doi: \textsf{%
10\hspace{.1pt}\discretionary{.}{%
}{.}\hspace{.4pt}1145\discretionary{/}{%
}{/}142750\hspace{.1pt}\discretionary{.}{%
}{.}\hspace{.4pt}142791}}


\bibitem{zhang:2020:dataquilt}
J.~E. Zhang, N.~Sultanum, A.~Bezerianos, and F.~Chevalier.
\newblock {DataQuilt}: Extracting visual elements from images to craft
  pictorial visualizations.
\newblock In {\em Proc.\ CHI}, pp. 45:1--45:13. ACM, New York, 2020.
  \href{https://doi.org/10.1145/3313831.3376172}
{doi: \textsf{%
10\hspace{.1pt}\discretionary{.}{%
}{.}\hspace{.4pt}1145\discretionary{/}{%
}{/}3313831\hspace{.1pt}\discretionary{.}{%
}{.}\hspace{.4pt}3376172}}


\bibitem{zhong:2020:BWT}
Y.~Zhong, T.~Isenberg, and P.~Isenberg.
\newblock Black-and-white textures for visualization on e-ink displays.
\newblock In {\em Posters at IEEE VIS}, 2020.
\newblock Extended abstract and poster, url:
  \href{https://hal.science/hal-02944212}{\texttt{hal.science/hal-02944212}}.

\end{thebibliography}
